\documentclass[twocolumn,tighten]{aastex631}
\usepackage{natbib}
\usepackage{color}
\usepackage{booktabs}
\usepackage{wrapfig}
\usepackage{appendix}
\definecolor{LightCyan}{rgb}{0.80,1,1}
\usepackage{color, colortbl}
\usepackage[first=0,last=9]{lcg}

\newcommand{\gaia}{{\em{Gaia}}}
\newcommand{\galex}{{GALEX}}
\newcommand{\porb}{{P_{\rm{orb}}}}
\newcommand{\kpc}{{\rm{kpc}}}
\newcommand{\days}{{\rm{day}}}
\newcommand{\Teff}{{T_{\rm{eff}}}}
\newcommand{\vosa}{{{\bf VOSA}}}

\newcommand{\mdr}{{M_{\rm{DR}}}}
\newcommand{\mlc}{{M_{\rm{LC}}}}
\newcommand{\mch}{{M_{\rm{Ch}}}}

\newcommand{\teff}{{T_{\rm{eff}}}}
\newcommand{\tefflc}{{T_{\rm{eff},LC}}}
\newcommand{\teffwd}{{T_{\rm{eff,WD}}}}
\newcommand{\MIST}{{{\tt MIST}}}

\newcommand{\Gyr}{{{\rm Gyr}}}
\newcommand{\acceleration}{{\rm{cm\ s^{-2}}}}
\newcommand{\msun}{{M_{\odot}}}
\newcommand{\lbol}{{L_{\rm{bol}}}}
\newcommand{\lbollc}{{L_{\rm{bol},LC}}}
\newcommand{\logglc}{{\log\ g_{\rm{LC}}}}
\newcommand{\lbolwd}{{L_{\rm{bol,WD}}}}

\graphicspath{{./}{figures/}}

\begin{document}

\title{White Dwarfs Revealed in Gaia’s Candidate Compact Object Binaries}

\author[0000-0002-0991-8438]{Anindya Ganguly}
\affiliation{Tata Institute of Fundamental Research, Department of Astronomy and Astrophysics, Homi Bhabha Road, Navy Nagar, Colaba, Mumbai, 400005, India}
\email{anindya.ganguly@tifr.res.in}
\author[0000-0002-4638-1035]{Prasanta K. Nayak}
\affiliation{Tata Institute of Fundamental Research, Department of Astronomy and Astrophysics, Homi Bhabha Road, Navy Nagar, Colaba, Mumbai, 400005, India}
\author[0000-0002-3680-2684]{Sourav Chatterjee}
\affiliation{Tata Institute of Fundamental Research, Department of Astronomy and Astrophysics, Homi Bhabha Road, Navy Nagar, Colaba, Mumbai, 400005, India}
\email{souravchatterjee.tifr@gmail.com}

\begin{abstract}
Discovery and characterisation of black holes (BHs), neutron stars (NSs), and white dwarfs (WDs) with detached luminous companions (LCs) in wide orbits are exciting because they are important test beds for dark remnant (DR) formation physics as well as binary stellar evolution models. Recently, 187 candidates have been identified from \gaia's non-single star catalog as wide orbit ($\porb/\days>45$), detached binaries hosting DRs. We identify UV counterparts for 49 of these sources in the archival \galex\ data. Modeling the observed spectral energy distribution (SED) spanning FUV-NUV to IR for these sources and stellar evolution models, we constrain the LC properties including mass, bolometric luminosity, and effective temperature for these 49 sources. Using the LC masses, and the astrometric mass function constrained by \gaia, we constrain the DR masses for these sources. We find that 9 have masses clearly in the NS or BH mass range. Fifteen sources exhibit significant NUV excess and 4 show excess both in FUV and NUV. The simplest explanation for these excess UV fluxes is that the DRs in these sources are WDs. Using SED modeling we constrain the effective temperature and bolometric luminosity for these 15 sources. Our estimated DR masses for all of these 15 sources are lower than the Chandrasekhar mass limit for WDs. Interestingly, five of these sources had been wrongly identified as neutron stars in literature.
\end{abstract}


\section{Introduction} \label{sec:intro}
Understanding the formation details of dark stellar remnants (DRs) and binary stellar evolution, especially for interacting binaries, are among the most interesting questions in today's stellar astrophysics. Identifying and characterizing the properties of DRs with luminous companions (LCs) in wide detached orbits in large numbers can be instrumental in improving our understanding in this regard \citep[e.g.,][]{Chawla2022}. If created from a primordial binary in the field, the age and metallicity of the DR's progenitor and the LC must be the same. In a small fraction of cases, the DR--LC binary may have been created inside star clusters via dynamical processes and ejected from them into the field \citep[e.g.,][]{Chatterjee2017,Khurana_2023}. For dynamically created DR--LC binaries, although the DR's progenitor and LC may not have been together by birth, all stars in a star cluster are roughly coeval with usually small spreads in metallicity and age \citep[e.g.,][]{Milone2020}. As a result, constraining the age and metallicity of the DR's progenitor from the constraints of the LC is possible even in dynamically produced DR--LC binaries \citep{Chawla2022}. Outside star clusters, the possibility of dynamical exchange or capture is of course vanishingly low. 

Several studies have pointed out that \gaia\, is expected to astrometrically identify hundreds to thousands of black holes (BHs) and neutron stars (NSs) in detached binaries simply from the motion of the LCs \citep{Mashian2017, Breivik2017, Yamaguchi2018, Chawla2022}. If so, the mass of the LC can also be constrained from parallax, magnitude ($G$) and colors ($BP$, $RP$), and hence the mass of the DR can be constrained as well via \gaia's astrometry \citep[e.g.,][]{Gould2002, Andrews2019, Chawla2022}, photometric variation \citep[e.g.,][]{Shakura_1987, Masuda2019}, or radial velocity (RV) followup \citep[e.g.,][]{Zeldovich_1966, Trimble_1969, Chawla2022}. Thus these sources can directly establish a much coveted map between the DR mass and the progenitor properties. 

DR--LC binaries in detached orbits are also expected to be instrumental to put constraints on the details of the supernova process and binary interaction physics. For example, it is expected that common envelope evolution is an important channel for the production of astrometrically detectable DR--LC binaries and hence the details of the common envelope physics can be constrained based on the orbital properties of the DR--LC binaries  \citep[e.g.,][]{Yamaguchi_2018, Shikauchi_2022}. In addition, the distribution of formation kicks that BHs and NSs receive may leave its imprints on the distribution of DR--LC orbital properties, as well as the total number of DR--LC binaries in astrometrically resolvable orbits \citep[e.g.,][]{Breivik2017, Chawla2022}. Furthermore, it is expected that the majority of the BH and NS binaries in nature are in wide detached orbits, a population complimentary to those detected through X-ray, radio, and gravitational wave (GW) observations.  

In this context, the sources identified by \citet[][ ATF22 hereafter]{Andrews2022} and \citet[][ SHA23 hereafter]{Shahaf2023} as candidate BH, NS, or WD binaries from the non-single star (NSS) catalog \citep{gaia_nss} of \gaia's third data release (DR3) provide a really exciting group for further investigation and characterisation. Interestingly, these are not the only candidates for DR--LC binaries identified from \gaia\ DRs. For example, \citet{jay_2023} provided a catalog of 80 DR--LC candidates with high mass function ($f_{M}$) from the spectroscopic binary catalog in \gaia's, early DR3. Although, \citet{el_badry_2022} showed that at least some of the \cite{jay_2023} candidates are likely Algol-type binaries near the end of the mass transfer process and not BH or NS binaries with LCs. Furthermore, \citet{Gomel2022} identified 6,306 short-period binaries as candidates to host massive unseen companions from \gaia's ellipsoidal variables catalog. Clearly, the demography of possible DR--LC binaries is expanding fast and it is interesting to constrain the component properties of these candidates. Even when the existence of a DR is inferred and mass can be constrained with some accuracy, it remains challenging to identify the nature of the DR. For example, the Chandrasekhar mass limit for WDs ($\mch$) cannot clearly demarcate the boundary between NSs and WDs because of mass loss during the SN explosion \citep[e.g.,][]{fryer_2012}. As a result, demarcation is attempted using expectations from population synthesis and gaussian-mixture assumptions (e.g., SHA23) which can be highly uncertain and model dependent. Hence, in addition to mass, constraining stellar properties such as the bolometric luminosity ($\lbol$), effective temperature ($\teff$), and radius can be crucial in characterisation as well as confirmation.   

\begin{figure}
    \epsscale{1.2}
    \plotone{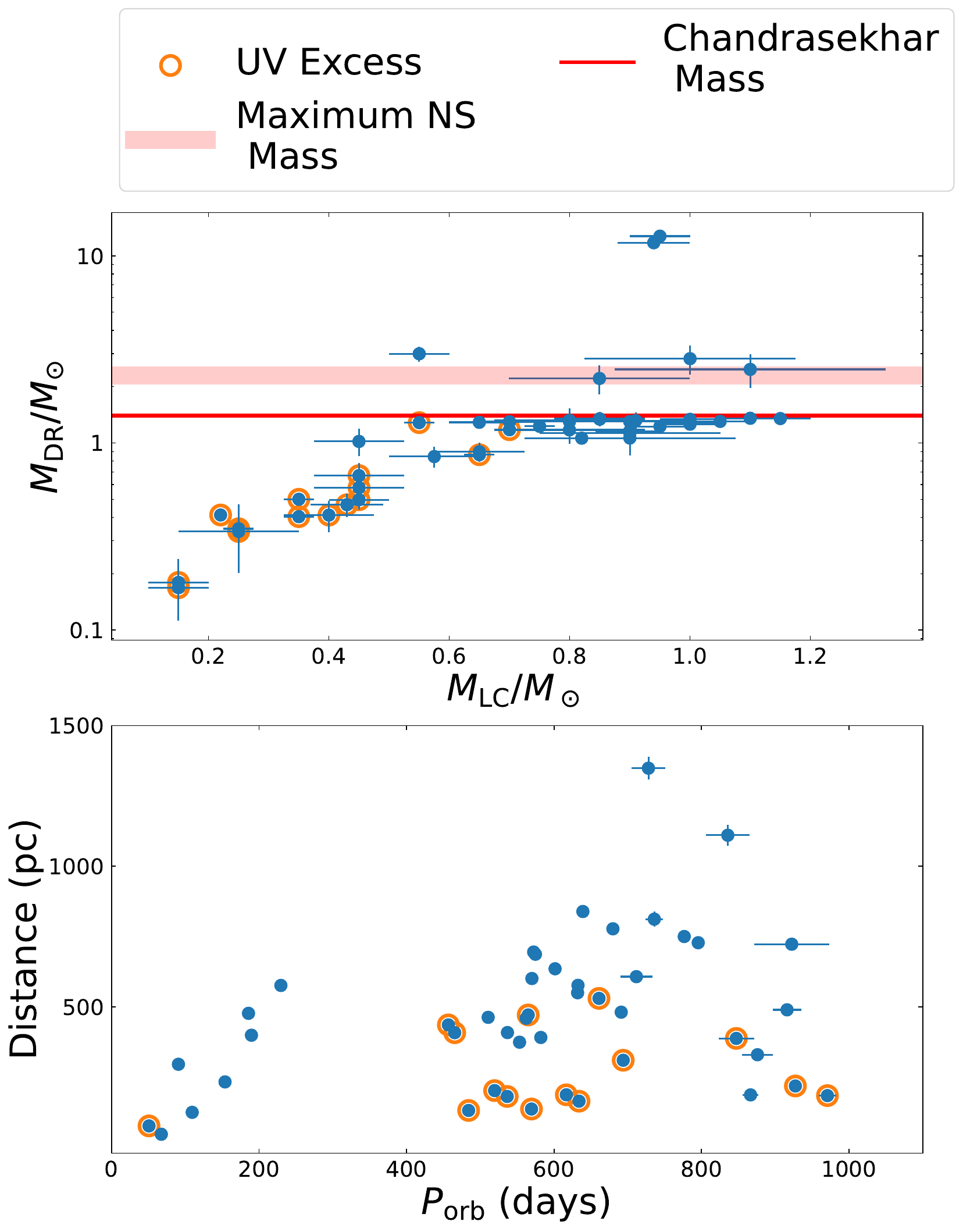}
    \caption{{\em Top}: $\mlc$ vs $\mdr$. Each blue dot is a candidate from the ATF22 and SHA23 catalogs. Orange open circles show the candidates for which we found UV excess. The red line shows the Chandrasekhar mass limit and the red-shaded region is the expected limiting mass for NSs. 
    {\em Bottom}: $\porb$ vs Distance for the same sources. All of them are in detached wide orbits and relatively close from us. Blue dots and orange circles have the same meaning. }
     
    \label{fig:combined_andrews}
\end{figure}
In this study, we investigate the spectral energy distribution (SED) of selected candidate DR--LC binaries from the ATF22 and SHA23 catalogues in wavelengths spanning ultraviolet (UV), optical, and infrared (IR), and aim to constrain the stellar properties of the components. All sources in these catalogues are expected to have a MS star as the LC in wide orbits ($\porb>45\,\days$) to a DR. Such wide orbits make it very unlikely for active ongoing mass transfer at present. Interestingly, all of these sources are within $1.5\,\kpc$ making them easy targets for follow-up studies (\autoref{fig:combined_andrews}). 

A SED is the end result of a compilation of images across the electromagnetic spectrum. When such multi-wavelength coverage is available, SEDs have been commonly used to reveal the individual components of an unresolved source. SED analysis can be particularly powerful if the components have significantly different $\teff$ \citep{WDS5_Ren2020, Rebassa2021}. If the DRs in the candidate binaries are non-accreting NSs or BHs, then the SED is not expected to show any statistically significant excess in the UV flux relative to what can be explained by the SED of the LC. On the other hand, if instead, the DR is a WD, then depending on the age, the WD can contribute significantly to the UV flux. As a result, the source's SED would exhibit significant UV excess compared to what is expected of the SED of the LC alone. In either case, SED analysis can constrain the LC's stellar properties such as mass ($\mlc$), effective temperature ($\tefflc$), metallicity, and bolometric luminosity ($\lbollc$). In combination with \gaia's astrometric constraints, e.g., for $f_M$ or the astrometric mass ratio function (AMRF) the estimated $\mlc$ can also put constraints on the mass of the DR ($\mdr$). In case, a significant UV excess is found which can be fitted by a WD, then in addition, we can also estimate the $\teff$ and $\lbol$ of the candidate WD.\footnote{Chromospheric activity may also create an excess in the UV flux. However, significant excess is unlikely for FGK stars as old ($>3\,\Gyr$) as the LCs in our list of candidates \citep{lorenzo_2016, gomes_2021}.}

We discuss how we select candidates and observational data for our analysis, model the SEDs and estimate masses for each source in \autoref{sec:sed_mass}. 
We show our key results in \autoref{sec:results}. Finally, we discuss the implications of our results and future avenues to improve the characterisation of similar sources in \autoref{sec:discussion}. 

\begin{deluxetable*}{c|c|cc|cccc|c|cccc|c}
\tabletypesize{\tiny}
\tablecaption{Candidates from ATF22 and SHA23 catalogs with \galex\ UV counterparts. } 
\tablehead{
 \colhead{Object}\vline & \colhead{\gaia\, DR3 ID}\vline & \colhead{RA} & \colhead{DEC}\vline &
 \multicolumn{4}{c}{Parameters from \gaia\, DR3}\vline&
 \colhead{$A_v$}\vline&
 \multicolumn{5}{c}{Our estimates} \\
 \cline{10-14}
 \colhead{}\vline & \colhead{}\vline & \colhead{(deg)} &\colhead{(deg)}\vline &
 \colhead{}&\colhead{}  & \colhead{}&\colhead{}\vline & \colhead{}\vline &
 \multicolumn{4}{c}{LC}\vline &
 \multicolumn{1}{c}{DR} \\
 \cline{5-8}
 \cline{10-14}
 \colhead{}\vline & \colhead{}\vline & \colhead{} &\colhead{}\vline   & \colhead{$\porb$} &
 \colhead{e}&
 \colhead{AMRF} &\colhead{[Fe/H]}\vline &\colhead{}\vline &
 \colhead{$\Teff$}&\colhead{$L_{\rm{bol}}$}  & \colhead{[Fe/H]} &
 \colhead{$\mlc$}\vline& 
  \colhead{$\mdr$} \\
 \colhead{}\vline & \colhead{}\vline & \colhead{} &\colhead{}\vline  & \colhead{($10^2\,\days$)} &
 \colhead{}&
 \colhead{}  &\colhead{(dex)}\vline& \colhead{}\vline &
 \colhead{($10^3\,\rm{K}$)} &\colhead{$L_{\odot}$}& \colhead{(dex)}  &\colhead{($M_\odot$)}\vline & 
   \colhead{($M_\odot$)}
 }
\startdata
    \noalign{\vskip 1mm} 
    \multicolumn{13}{c}{\footnotesize sources with significant observed UV excess} \\
    \noalign{\vskip 1mm}
    0812$+$7046 & 1098489704734299648 & 123.04 &  70.78 &    9.71 $\pm$          0.12 &          0.10 $\pm$                0.01 &  0.68 $\pm$       0.03 &   -0.43 & 0.08 &    3.50 $\pm$   0.12 & 0.02 $\pm$   0.00 &  -0.5 $\pm$   0.25 &     0.15 $\pm$      0.05 &     0.17 $\pm$      0.06 \\
    1220$+$5841 & 1581117310088807552 & 185.05 &  58.69 &    9.27 $\pm$          0.06 &          0.52 $\pm$                0.01 &  1.05 $\pm$       0.03 &  \nodata & 0.04 &    4.25$\pm$    0.12&  0.08$\pm$    0.00 &   0.2$\pm$    0.12 &     0.55$\pm$       0.03 &     1.28$\pm$       0.06 \\
    2106$-$5218 & 6476764747694402560 & 316.53 & -52.31 &    8.47 $\pm$          0.24 &          0.08 $\pm$                0.04 &  0.76 $\pm$       0.04 &   -1.82 & 0.11 &    3.75 $\pm$   0.12 & 0.05 $\pm$   0.00 &  -2.0 $\pm$   0.25 &     0.25 $\pm$      0.10 &     0.34 $\pm$      0.13 \\
   0824$+$2300 &  678085695778294784 & 126.06 &  23.00 &    6.94 $\pm$          0.06 &          0.13 $\pm$                0.07 &  0.67 $\pm$       0.04 &   -0.58 & 0.10 &    3.75 $\pm$   0.12 & 0.04 $\pm$   0.00 &  -0.5 $\pm$   0.25 &     0.43 $\pm$      0.06 &     0.47 $\pm$      0.07 \\
   0709$+$7052 & 1109902566711892352 & 107.26 &  70.87 &    6.61 $\pm$          0.06 &          0.36 $\pm$                0.03 &  0.81 $\pm$       0.04 &    0.02 & 0.16 &    4.25 $\pm$   0.12 & 0.09 $\pm$   0.01 &   0.0 $\pm$   0.18 &     0.45 $\pm$      0.08 &     0.67 $\pm$      0.11 \\
   2338$-$7152 & 6380360186645909760 & 354.60 & -71.88 &    6.34 $\pm$          0.02 &          0.01 $\pm$                0.02 &  0.69 $\pm$       0.03 &  \nodata & 0.08 &    3.50$\pm$    0.12&  0.02$\pm$    0.00 &  -0.5$\pm$    0.25 &     0.35$\pm$       0.03 &     0.40$\pm$       0.03 \\
   0124$+$0758 & 2566461354152574976 &  21.05 &   7.98 &    6.17 $\pm$          0.04 &          0.05 $\pm$                0.04 &  0.71 $\pm$       0.04 &   -0.30 & 0.15 &    3.50 $\pm$   0.12 & 0.02 $\pm$   0.00 &  -0.5 $\pm$   0.25 &     0.15 $\pm$      0.05 &     0.18 $\pm$      0.06 \\
   0358$-$8154 & 4616146191642331008 &  59.50 & -81.90 &    5.69 $\pm$          0.02 &          0.09 $\pm$                0.02 &  0.78 $\pm$       0.05 &  \nodata & 0.19 &    3.75$\pm$    0.12&  0.01$\pm$    0.00 &  -1.0$\pm$    0.25 &     0.25$\pm$       0.03 &     0.35$\pm$       0.03 \\
   0327$-$4342 & 4847718871053268480 &  51.83 & -43.71 &    5.65 $\pm$          0.05 &          0.33 $\pm$                0.04 &  0.76 $\pm$       0.03 &  \nodata & 0.04 &    4.50$\pm$    0.12&  0.17$\pm$    0.01 &   0.2$\pm$    0.12 &     0.65$\pm$       0.03 &     0.87$\pm$       0.03 \\
   1143$-$2807 & 3484138291549376128 & 175.97 & -28.12 &    5.37 $\pm$          0.03 &          0.09 $\pm$                0.04 &  0.93 $\pm$       0.07 &   -0.74 & 0.30 &    3.50 $\pm$   0.12 & 0.01 $\pm$   0.00 &  -0.5 $\pm$   0.25 &     0.22 $\pm$      0.00 &     0.41 $\pm$      0.00 \\
   2001$-$4438 & 6685604337007194368 & 300.31 & -44.63 &    5.19 $\pm$          0.02 &          0.06 $\pm$                0.03 &  0.64 $\pm$       0.02 &  \nodata & 0.15 &    3.75$\pm$    0.12&  0.04$\pm$    0.00 &  -0.5$\pm$    0.25 &     0.40$\pm$       0.08 &     0.41$\pm$       0.08 \\
   0338$+$3913 &  224549450109569536 &  54.63 &  39.23 &    4.84 $\pm$          0.01 &          0.03 $\pm$                0.01 &  0.67 $\pm$       0.03 &    0.07 & 0.47 &    3.75 $\pm$   0.12 & 0.03 $\pm$   0.00 &   0.2 $\pm$   0.12 &     0.45 $\pm$      0.05 &     0.50 $\pm$      0.06 \\
   1330$+$2827 & 1449731030688880512 & 202.65 &  28.46 &    4.65 $\pm$          0.02 &          0.29 $\pm$                0.03 &  0.74 $\pm$       0.02 &   -0.15 & 0.05 &    4.50 $\pm$   0.12 & 0.25 $\pm$   0.01 &  -0.5 $\pm$   0.25 &     0.45 $\pm$      0.07 &     0.58 $\pm$      0.10 \\
   0640$-$2621 & 2919995917769953408 & 100.03 & -26.36 &    4.57 $\pm$          0.02 &          0.75 $\pm$                0.05 &  0.87 $\pm$       0.04 &    0.10 & 0.40 &    5.25 $\pm$   0.12 & 0.21 $\pm$   0.01 &   0.2 $\pm$   0.12 &     0.70 $\pm$      0.03 &     1.18 $\pm$      0.04 \\
   1606$+$6120 & 1626845895609073536 & 241.56 &  61.35 &    0.51 $\pm$          0.00 &          0.02 $\pm$                0.02 &  0.79 $\pm$       0.04 &  \nodata & 0.04 &    3.50$\pm$    0.12&  0.02$\pm$    0.00 &  -1.5$\pm$    0.25 &     0.35$\pm$       0.02 &     0.50$\pm$       0.04 \\
    \bottomrule
   \noalign{\vskip 1mm}
   \multicolumn{13}{c}{\footnotesize other sources} \\
   \noalign{\vskip 1mm}
   1007$+$4453 &  809741149368202752 & 151.79 &  44.90 &      9.22 $\pm$            0.51 &          0.35 $\pm$                0.03 &  0.96 $\pm$        0.04 &   -0.09 & 0.03 &    5.25 $\pm$   0.12 &  0.45 $\pm$   0.03 &   0.0 $\pm$   0.18 &     0.65 $\pm$      0.05 &     1.29 $\pm$    0.10 \\
    2244$-$2236 & 2397135910639986304 & 341.20 & -22.60 &      9.16 $\pm$            0.19 &          0.56 $\pm$                0.03 &  0.84 $\pm$        0.03 &   -0.17 & 0.07 &    6.25 $\pm$   0.12 &  0.85 $\pm$   0.03 &   0.0 $\pm$   0.18 &     0.85 $\pm$      0.05 &     1.34 $\pm$    0.08 \\
    0336$+$1419 &   41408333753757056 &  54.13 &  14.32 &      8.76 $\pm$            0.21 &          0.53 $\pm$                0.03 &  0.84 $\pm$        0.03 &    0.00 & 1.54 &    6.75 $\pm$   0.12 &  0.31 $\pm$   0.01 &   0.0 $\pm$   0.18 &   \nodata$\pm$     \nodata &   \nodata$\pm$     \nodata \\
    1012$-$3537 & 5446310318525312768 & 153.10 & -35.62 &      8.67 $\pm$            0.11 &          0.25 $\pm$                0.00 &  0.70 $\pm$        0.02 &  \nodata & 0.38 &    7.00$\pm$    0.12 &  2.41$\pm$    0.05 &   0.2$\pm$    0.12 &     1.15$\pm$       0.05 &     1.35$\pm$     0.06 \\
    1048$+$6547 & 1058875159778407808 & 162.25 &  65.80 &      8.36 $\pm$            0.29 &          0.42 $\pm$                0.04 &  0.85 $\pm$        0.04 &   -0.19 & 0.04 &    6.00 $\pm$   0.12 &  1.45 $\pm$   0.10 &   0.0 $\pm$   0.18 &     0.80 $\pm$      0.10 &     1.30 $\pm$    0.16 \\
    1205$+$6914 & 1683575679079854848 & 181.35 &  69.24 &      7.96 $\pm$            0.08 &          0.62 $\pm$                0.03 &  0.72 $\pm$        0.02 &    0.42 & 0.05 &    5.75 $\pm$   0.12 &  0.76 $\pm$   0.02 &   0.5 $\pm$   0.15 &     1.10 $\pm$      0.05 &     1.36 $\pm$    0.06 \\
   1622$+$1647 & 4466767229088016256 & 245.63 &  16.80 &      7.77 $\pm$            0.05 &          0.15 $\pm$                0.03 &  0.70 $\pm$        0.03 &   -0.26 & 0.14 &    5.75 $\pm$   0.12 &  1.44 $\pm$   0.08 &  -0.5 $\pm$   0.25 &     0.90 $\pm$      0.18 &     1.06 $\pm$    0.21 \\
   1432$-$1021 & 6328149636482597888 & 218.09 & -10.37 &      7.36 $\pm$            0.12 &          0.14 $\pm$                0.04 &  1.03 $\pm$        0.04 &   -1.50 & 0.28 &    6.00 $\pm$   0.12 &  3.02 $\pm$   0.22 &  -1.5 $\pm$   0.25 &     1.10 $\pm$      0.23 &     2.47 $\pm$    0.51 \\
   1812$+$2409 & 4578398926673187328 & 273.06 &  24.15 &      7.28 $\pm$            0.23 &          0.56 $\pm$                0.05 &  0.80 $\pm$        0.04 &   -0.14 & 0.39 &    6.00 $\pm$   0.12 &  1.80 $\pm$   0.11 &   0.0 $\pm$   0.18 &     0.80 $\pm$      0.13 &     1.18 $\pm$    0.18 \\
   0036$-$0932 & 2426116249713980416 &   9.05 &  -9.54 &      7.12 $\pm$            0.21 &          0.40 $\pm$                0.03 &  0.80 $\pm$        0.03 &   -0.56 & 0.11 &    5.75 $\pm$   0.12 &  1.85 $\pm$   0.13 &  -0.5 $\pm$   0.25 &     0.91 $\pm$      0.09 &     1.32 $\pm$    0.14 \\
   1949$+$0129 & 4240540718818313984 & 297.43 &   1.49 &      6.91 $\pm$            0.02 &          0.62 $\pm$                0.02 &  0.86 $\pm$        0.02 &   -0.07 & 0.85 &    5.75 $\pm$   0.12 &  0.48 $\pm$   0.02 &   0.0 $\pm$   0.18 &     0.75 $\pm$      0.03 &     1.23 $\pm$    0.04 \\
   2228$-$3943 & 6593763230249162112 & 337.21 & -39.72 &      6.80 $\pm$            0.03 &          0.61 $\pm$                0.07 &  0.86 $\pm$        0.04 &   -0.06 & 0.05 &    6.00 $\pm$   0.12 &  1.76 $\pm$   0.09 &   0.0 $\pm$   0.18 &     0.80 $\pm$      0.13 &     1.32 $\pm$    0.21 \\
   0217$-$7541 & 4637171465304969216 &  34.46 & -75.70 &      6.39 $\pm$            0.04 &          0.37 $\pm$                0.03 &  0.80 $\pm$        0.02 &    0.05 & 0.15 &    5.75 $\pm$   0.12 &  1.45 $\pm$   0.04 &   0.0 $\pm$   0.18 &     0.90 $\pm$      0.05 &     1.30 $\pm$    0.07 \\
   1150$-$2203 & 3494029910469026432 & 177.72 & -22.06 &      6.33 $\pm$            0.02 &          0.55 $\pm$                0.02 &  0.74 $\pm$        0.02 &    0.21 & 0.13 &    6.00 $\pm$   0.12 &  2.30 $\pm$   0.06 &   0.2 $\pm$   0.12 &     1.00 $\pm$      0.05 &     1.29 $\pm$    0.06 \\
   1449$+$6919 & 1694708646628402048 & 222.33 &  69.32 &      6.32 $\pm$            0.03 &          0.26 $\pm$                0.02 &  0.73 $\pm$        0.02 &   -0.71 & 0.09 &    6.00 $\pm$   0.12 &  1.28 $\pm$   0.04 &  -0.5 $\pm$   0.25 &     1.00 $\pm$      0.05 &     1.26 $\pm$    0.06 \\
   1310$+$6016 & 1579254496872812032 & 197.71 &  60.28 &      6.01 $\pm$            0.03 &          0.31 $\pm$                0.03 &  0.74 $\pm$        0.02 &   -0.69 & 0.04 &    5.25 $\pm$   0.12 &  0.53 $\pm$   0.01 &  -0.5 $\pm$   0.25 &     0.82 $\pm$      0.06 &     1.06 $\pm$    0.08 \\
   0109$-$1034 & 2469926638416055168 &  17.37 & -10.58 &      5.82 $\pm$            0.04 &          0.74 $\pm$                0.06 &  1.03 $\pm$        0.07 &   -0.17 & 0.08 &    4.75 $\pm$   0.12 &  0.09 $\pm$   0.00 &  -0.5 $\pm$   0.25 &     0.45 $\pm$      0.08 &     1.02 $\pm$    0.17 \\
   2100$-$2535$^*$ & 6802561484797464832 & 315.11 & -25.59 &      5.75 $\pm$            0.06 &          0.83 $\pm$                0.07 &  1.11 $\pm$        0.16 &   -0.22 & 0.25 &    6.25 $\pm$   0.12 &  2.98 $\pm$   0.13 &   0.0 $\pm$   0.18 &     0.85 $\pm$      0.15 &     2.21 $\pm$    0.39 \\
   1733$+$5808 & 1434445448240677376 & 263.40 &  58.15 &      5.72 $\pm$            0.02 &          0.30 $\pm$                0.02 &  0.73 $\pm$        0.02 &   -0.21 & 0.13 &    5.75 $\pm$   0.12 &  1.37 $\pm$   0.03 &  -0.5 $\pm$   0.25 &     0.90 $\pm$      0.15 &     1.13 $\pm$    0.19 \\
   1046$+$1002 & 3869650535947137920 & 161.52 &  10.05 &      5.70 $\pm$            0.05 &          0.18 $\pm$                0.04 &  0.84 $\pm$        0.04 &   -0.09 & 0.08 &    6.25 $\pm$   0.12 &  1.87 $\pm$   0.08 &   0.0 $\pm$   0.18 &     0.85 $\pm$      0.08 &     1.35 $\pm$    0.12 \\
   0003$-$5604 & 4922744974687373440 &   0.86 & -56.08 &      5.62 $\pm$            0.02 &          0.81 $\pm$                0.03 &  0.92 $\pm$        0.07 &  \nodata & 0.04 &    5.00$\pm$    0.12 &  0.29$\pm$    0.01 &   0.0$\pm$    0.18 &     0.70$\pm$       0.05 &     1.30$\pm$     0.09 \\
   0119$-$2526 & 5039979680444075392 &  19.79 & -25.44 &      5.53 $\pm$            0.01 &          0.18 $\pm$                0.01 &  0.73 $\pm$        0.02 &    0.03 & 0.05 &    5.75 $\pm$   0.12 &  0.87 $\pm$   0.02 &   0.2 $\pm$   0.12 &     1.05 $\pm$      0.05 &     1.30 $\pm$    0.06 \\
   0152$-$2049 & 5136025521527939072 &  28.21 & -20.82 &      5.37 $\pm$            0.01 &          0.65 $\pm$                0.01 &  0.74 $\pm$        0.02 &   -1.39 & 0.05 &    6.50 $\pm$   0.12 &  2.07 $\pm$   0.04 &  -1.0 $\pm$   0.25 &     0.90 $\pm$      0.06 &     1.16 $\pm$    0.07 \\
   0334$+$0009 & 3263804373319076480 &  53.73 &   0.15 &      5.11 $\pm$            0.05 &          0.28 $\pm$                0.02 &  1.15 $\pm$        0.11 &   -1.26 & 0.35 &    5.75 $\pm$   0.12 &  1.89 $\pm$   0.08 &  -1.0 $\pm$   0.25 &     1.00 $\pm$      0.18 &     2.82 $\pm$    0.49 \\
   2057$-$4742 & 6481502062263141504 & 314.49 & -47.70 &      2.30 $\pm$            0.01 &          0.30 $\pm$                0.04 &  0.74 $\pm$        0.02 &  \nodata & 0.14 &    6.25$\pm$    0.12 &  0.98$\pm$    0.03 &   0.2$\pm$    0.12 &     0.95$\pm$       0.05 &     1.23$\pm$     0.06 \\
   0553$-$1349 & 2995961897685517312 &  88.47 & -13.83 &      1.90 $\pm$            0.00 &          0.37 $\pm$                0.07 &  0.76 $\pm$        0.03 &   -0.46 & 1.07 &    7.25 $\pm$   0.12 &  1.78 $\pm$   0.03 &  -0.5 $\pm$   0.25 &     1.00 $\pm$      0.05 &     1.34 $\pm$    0.07 \\
   1728$-$0034 & 4373465352415301632 & 262.17 &  -0.58 &      1.86 $\pm$            0.00 &          0.49 $\pm$                0.07 &  2.26 $\pm$        0.17 &   -1.07 & 1.23 &    6.25 $\pm$   0.12 &  1.43 $\pm$   0.03 &  -1.0 $\pm$   0.25 &     0.95 $\pm$      0.05 &    12.74 $\pm$    0.67 \\
   1452$-$1922 & 6281177228434199296 & 223.21 & -19.37 &      1.54 $\pm$            0.00 &          0.18 $\pm$                0.04 &  2.21 $\pm$        0.11 &   -0.54 & 0.29 &    5.75 $\pm$   0.12 &  1.65 $\pm$   0.03 &  -0.5 $\pm$   0.25 &     0.94 $\pm$      0.06 &    11.75 $\pm$    0.75 \\
   1301$-$1852 & 3509370326763016704 & 195.32 & -18.87 &      1.09 $\pm$            0.00 &          0.24 $\pm$                0.02 &  1.57 $\pm$        0.04 &   -0.19 & 0.32 &    4.75 $\pm$   0.12 &  0.18 $\pm$   0.00 &   0.0 $\pm$   0.18 &     0.55 $\pm$      0.05 &     3.00 $\pm$    0.27 \\
   0632$-$6614 & 5283631903842076032 &  98.17 & -66.24 &      0.91 $\pm$            0.00 &          0.31 $\pm$                0.04 &  0.77 $\pm$        0.03 &   -0.22 & 0.17 &    5.00 $\pm$   0.12 &  0.39 $\pm$   0.00 &  -0.5 $\pm$   0.25 &     0.65 $\pm$      0.08 &     0.90 $\pm$    0.10 \\
   0156$+$1228 & 2574867704662509568 &  29.15 &  12.47 &      0.68 $\pm$            0.00 &          0.56 $\pm$                0.03 &  0.81 $\pm$        0.03 &  \nodata & 0.25 &    4.25$\pm$    0.12 &  0.20$\pm$    0.00 &  -1.5$\pm$    0.25 &     0.57$\pm$       0.08 &     0.85$\pm$     0.11 \\
   \bottomrule
    \colhead{} & \colhead{} & \colhead{} &\colhead{}   & \colhead{} &
    \colhead{}&
    \colhead{$f_{M}$($\msun$)} &\colhead{}
    &\colhead{} &
     \colhead{}&\colhead{}  & \colhead{} &
    \colhead{}& 
    \colhead{} \\
   \bottomrule
   \noalign{\vskip 1mm} 
   1007$+$3408 &  747174436620510976 &  151.84 &  34.14  & 9.99$\pm$0.26&0.71$\pm$0.02&  $0.56 ^{+0.05}_{-0.04}$  &-0.02 &  0.04 &   5.5$\pm$0.1&$0.30\pm0.02$ &  0.0$\pm$0.2  & $0.6\pm 0.1$ & $1.2^{+0.2}_{-0.1}$\\
   1433$-$0114 & 3649963989549165440 & 218.38 & -1.25 & 8.93$\pm$0.60&0.36$\pm$0.14   & 0.8 $^{+0.5}_{-0.2}$ &  0.01   & 0.13 & 24$\pm$1&$8.1\pm0.1$  & 0.0$\pm$0.2 & \nodata& \nodata\\
   2033$+$0758 &  1749013354127453696 &  308.31 &   7.98  &    9.32$\pm$0.78& 0.51$\pm$ 0.08&  0.8 $^{+0.6}_{-0.4}$   &-0.27&  0.30 &   6.0$\pm$0.1&$4.0\pm0.1$  &  0.0$\pm$0.2  &$1.0\pm0.5$ & $1.9^{+1.2}_{-0.9}$\\
\enddata
\tablecomments{
\scriptsize
Properties of 49 sources from the ATF22 and SHA23 catalogs with \galex\ counterparts. The sources are sorted by their $\porb$.
Parameters we estimate and those found in \gaia's DR3 and its non-single star catalog are noted separately. Errors in AMRF (for SHA23), $f_M$ ( for ATF22) and $\porb$ denote $95\%$ confidence interval. Our estimated for $\Teff$ and [Fe/H] for the LC come from SED modelling with uncertainty denoting half of the grid spacing for the SED models. $\lbol$\ measurement and their corresponding errors come from the observed flux and \gaia-estimated distances. We did not model 1433$-$0114, a hot sub-dwarf \citep{sub_dwarf_2017_2,sub_dwarf_2017}. * denotes low (6.82) significance for the astrometric solution. }

\label{tab:all_data}
\end{deluxetable*}
\section{Data Selection and Mass estimation} 
\label{sec:sed_mass}
\subsection{Data Selection and SED fitting}\label{sec:sed_fit}
\begin{figure}
\plotone{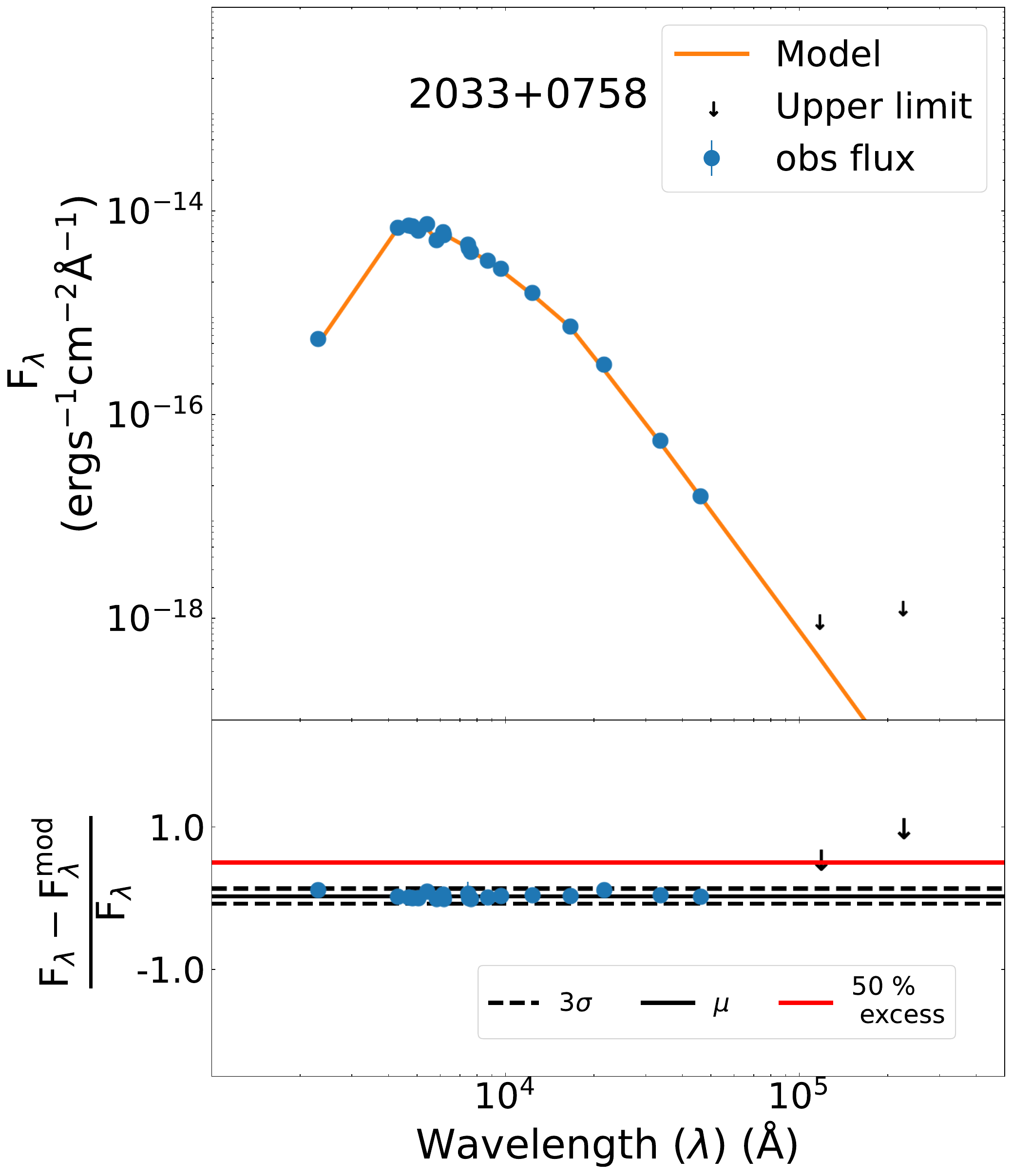}
\caption{Flux density (per unit wavelength, $F_\lambda$) vs wavelength ($\lambda$) for source 2033$+$0758 (\autoref{tab:all_data}), as an example of a source exhibiting no significant (c.f., \autoref{sec:sed_fit}) UV excess. Blue dots and orange line show the observed and best-fit synthetic $F_\lambda$ assuming a single-component MS star SED. Black arrows for IR flux denote upper limits. {\em Bottom:} Fractional residuals (blue dots) vs $\lambda$. We show the mean of the residual flux ratios (blue solid), $3\sigma$ limits centering the mean (black dashed), and $50\%$ excess from the mean (red solid line). Similar figures with single-component SEDs for the other sources are available online. }\label{fig:sed_example_no_excess}
\end{figure}

\figsetgrpstart
\figsetgrpnum{2.1}
\figsetgrptitle{Image for figure 2_1}
\figsetplot{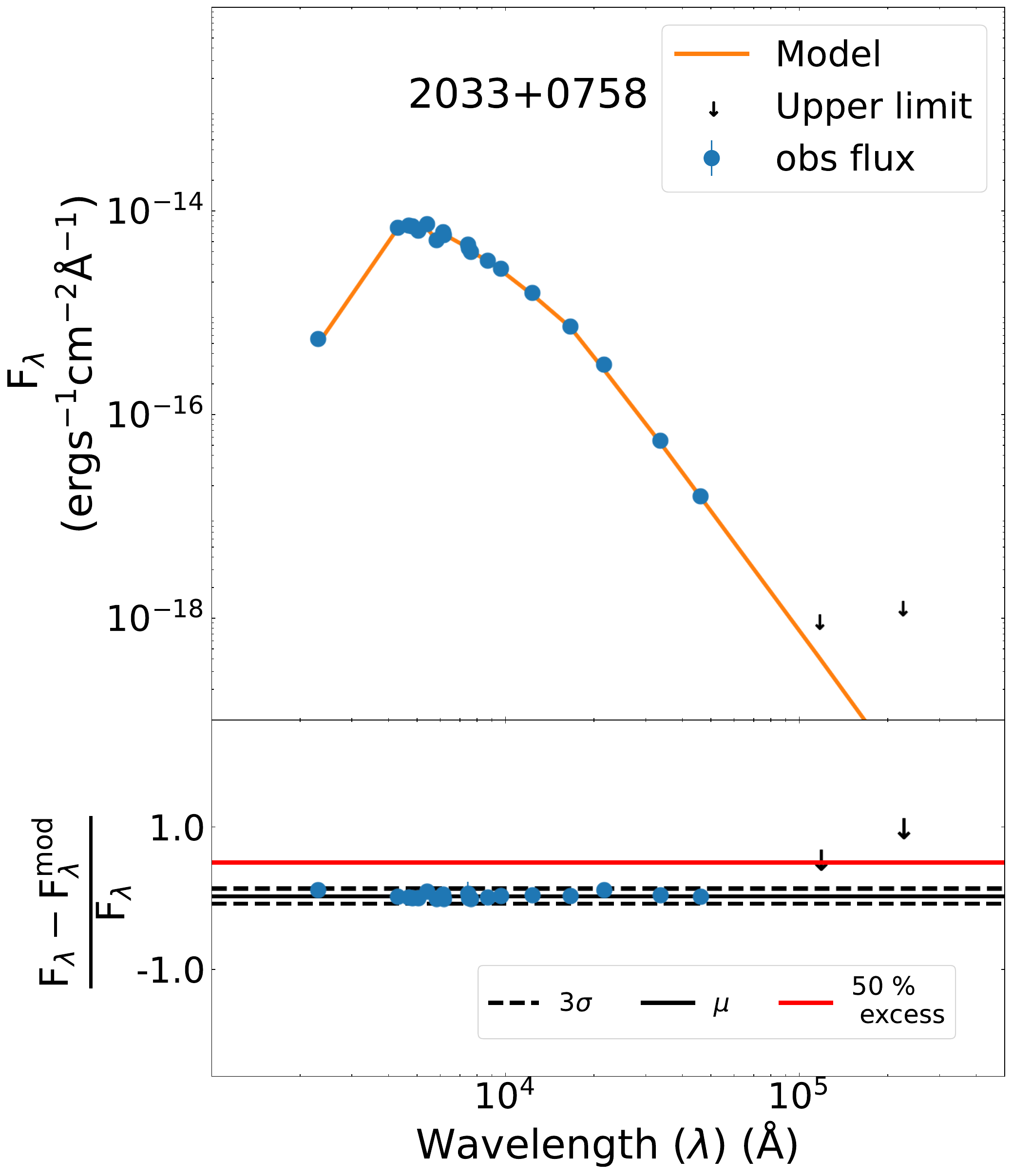}
\figsetgrpnote{Same as \autoref{fig:sed_example_no_excess},but for source 2033$+$0758}
\label{fig:sed_example_no_excess_1}

\figsetgrpend

\figsetgrpstart
\figsetgrpnum{2.2}
\figsetgrptitle{Image for figure 2_2}
\figsetplot{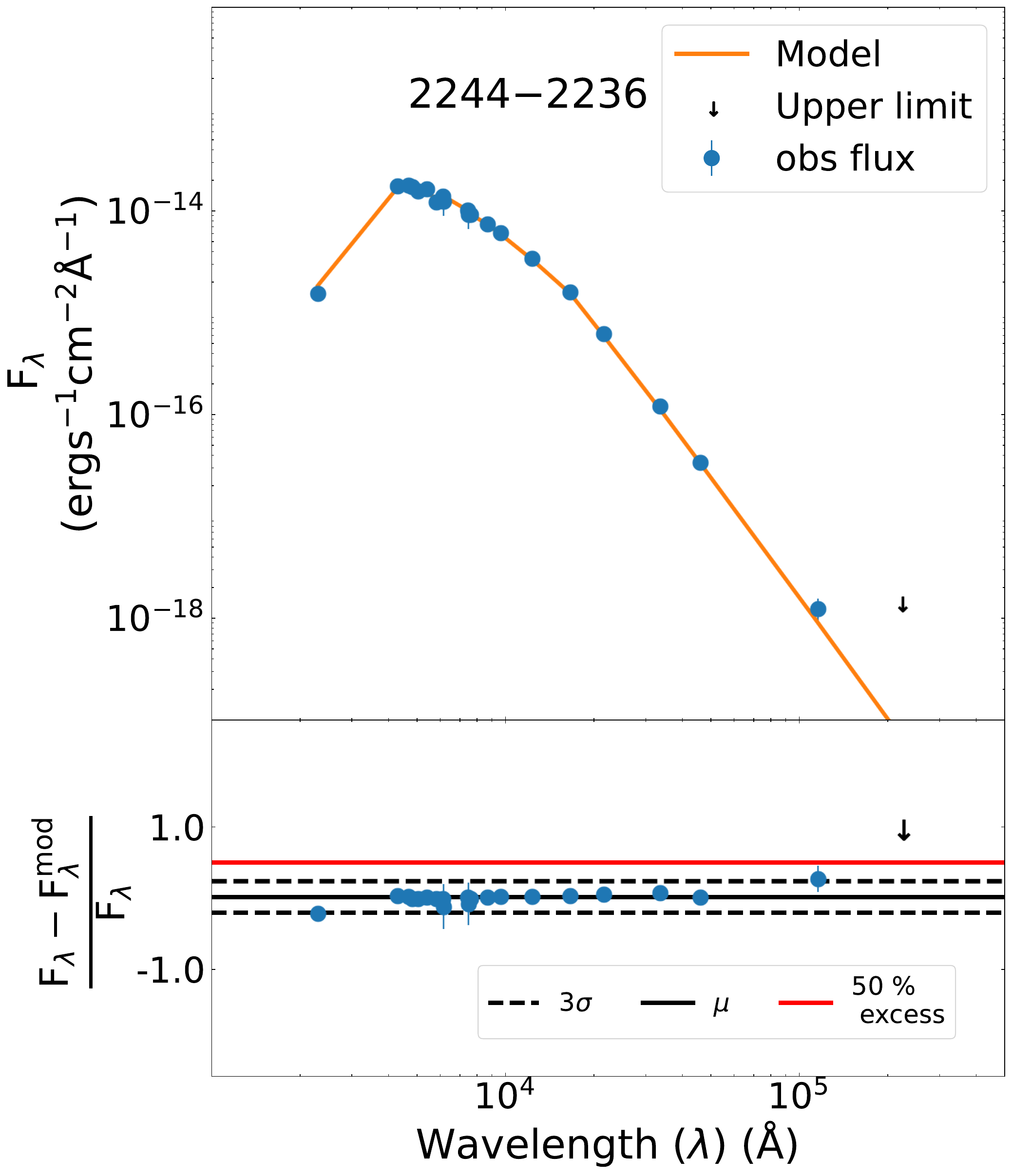}
\figsetgrpnote{Same as \autoref{fig:sed_example_no_excess},but for source 2244$-$2236}
\label{fig:sed_example_no_excess_2}

\figsetgrpend

\figsetgrpstart
\figsetgrpnum{2.3}
\figsetgrptitle{Image for figure 2_3}
\figsetplot{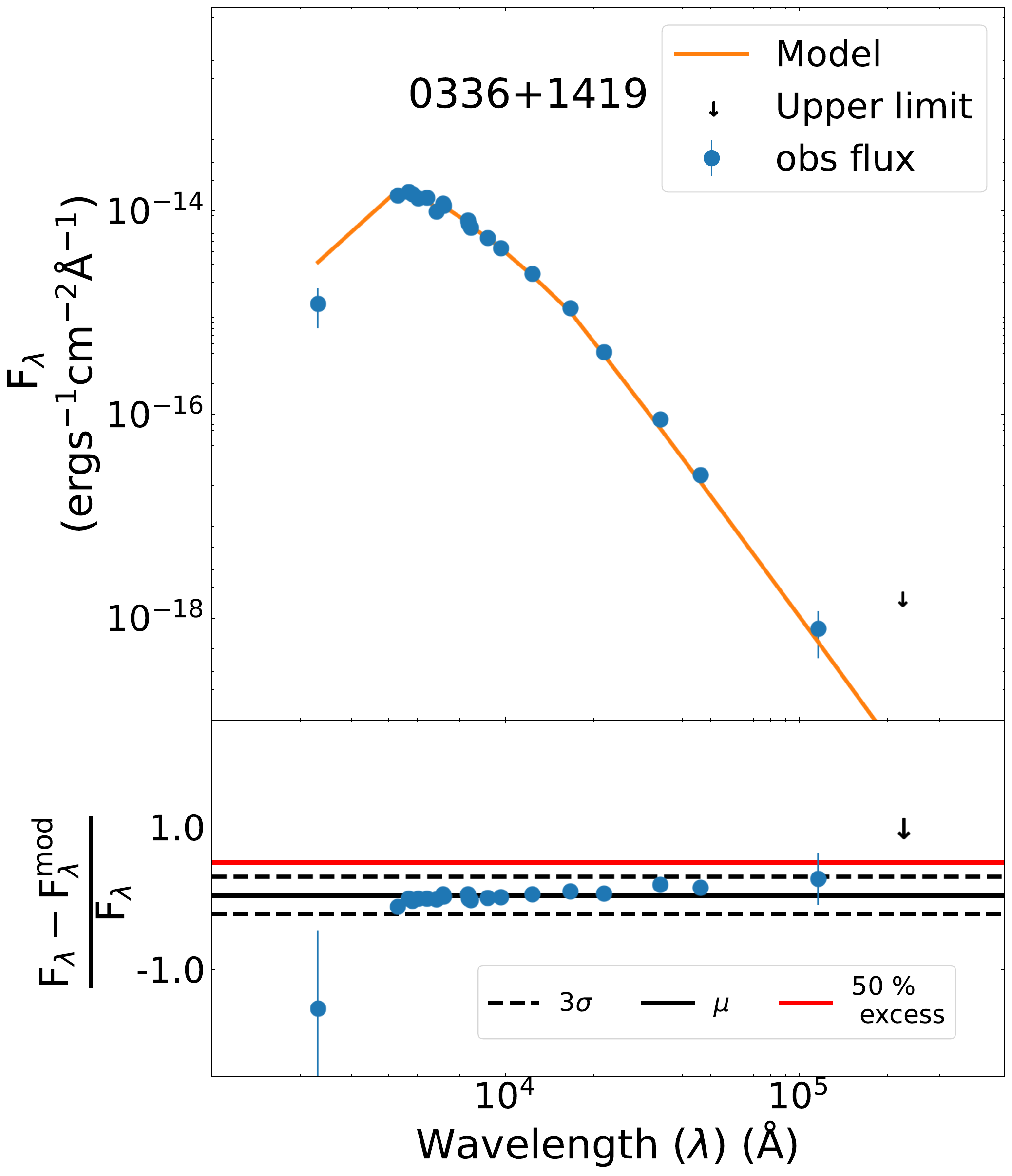}
\figsetgrpnote{Same as \autoref{fig:sed_example_no_excess},but for source 0336$+$1419}
\label{fig:sed_example_no_excess_3}

\figsetgrpend

\figsetgrpstart
\figsetgrpnum{2.4}
\figsetgrptitle{Image for figure 2_4}
\figsetplot{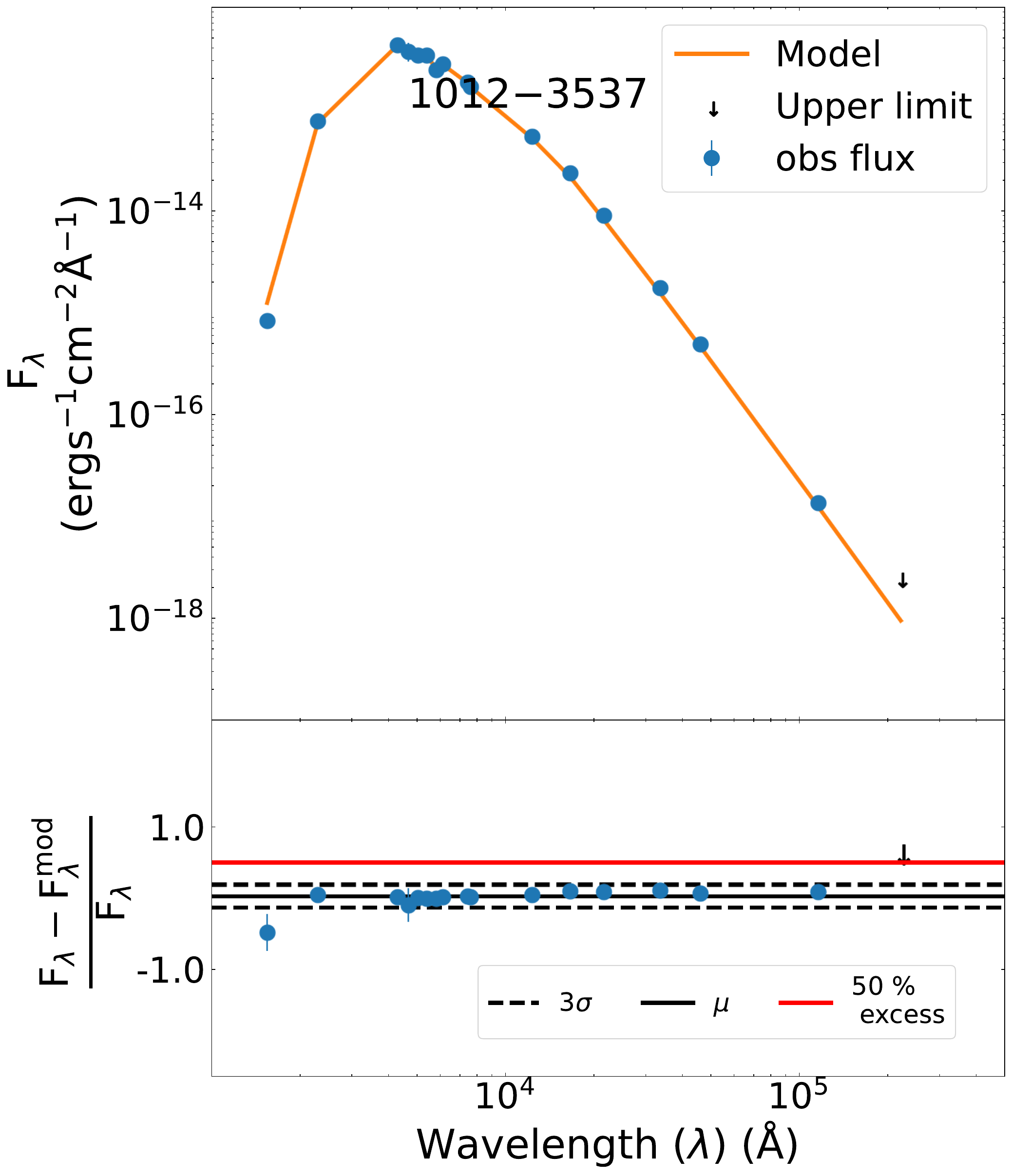}
\figsetgrpnote{Same as \autoref{fig:sed_example_no_excess},but for source 1012$-$3537}
\label{fig:sed_example_no_excess_4}

\figsetgrpend

\figsetgrpstart
\figsetgrpnum{2.5}
\figsetgrptitle{Image for figure 2_5}
\figsetplot{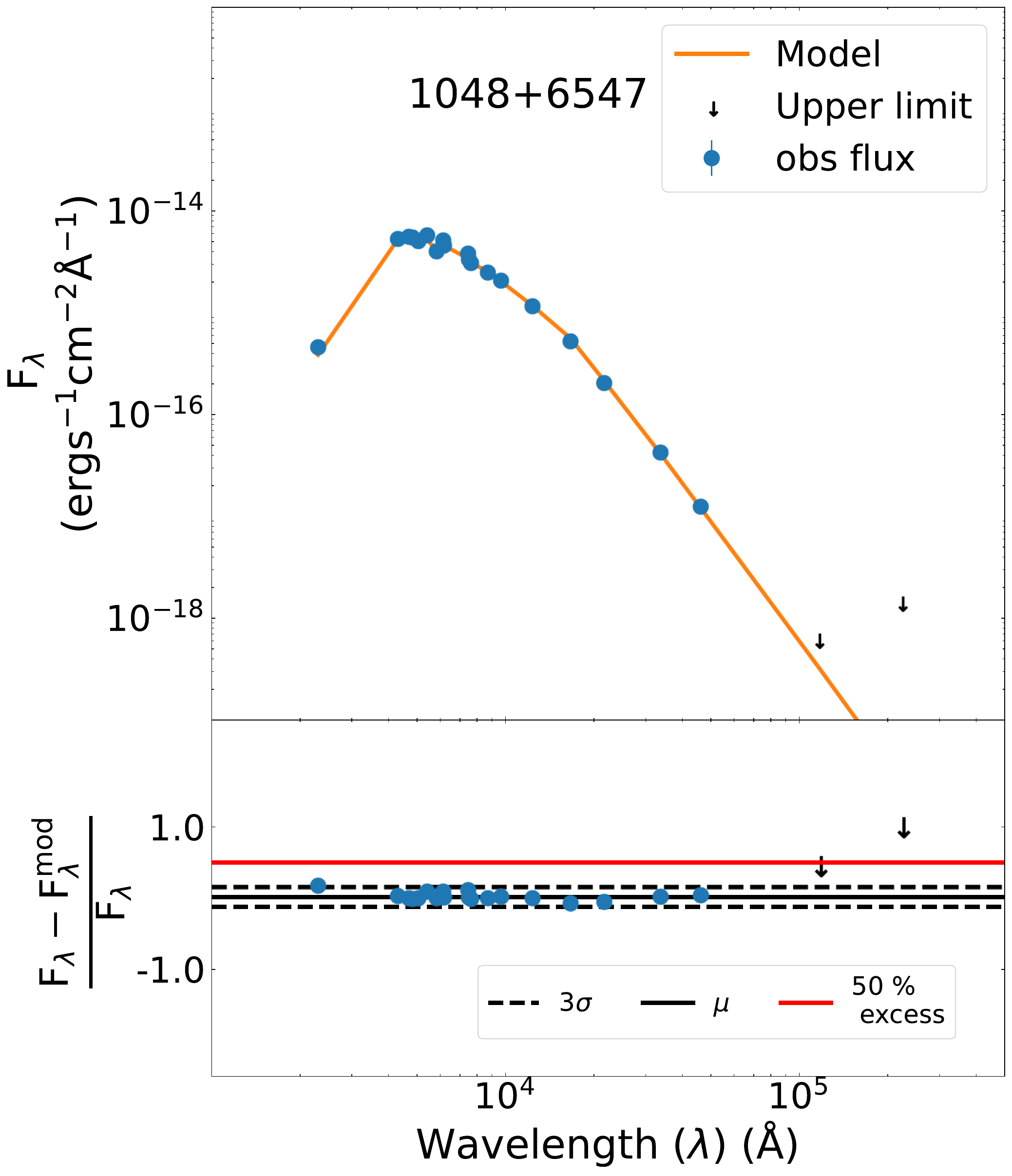}
\figsetgrpnote{Same as \autoref{fig:sed_example_no_excess},but for source 1048$+$6547}
\label{fig:sed_example_no_excess_5}

\figsetgrpend

\figsetgrpstart
\figsetgrpnum{2.6}
\figsetgrptitle{Image for figure 2_6}
\figsetplot{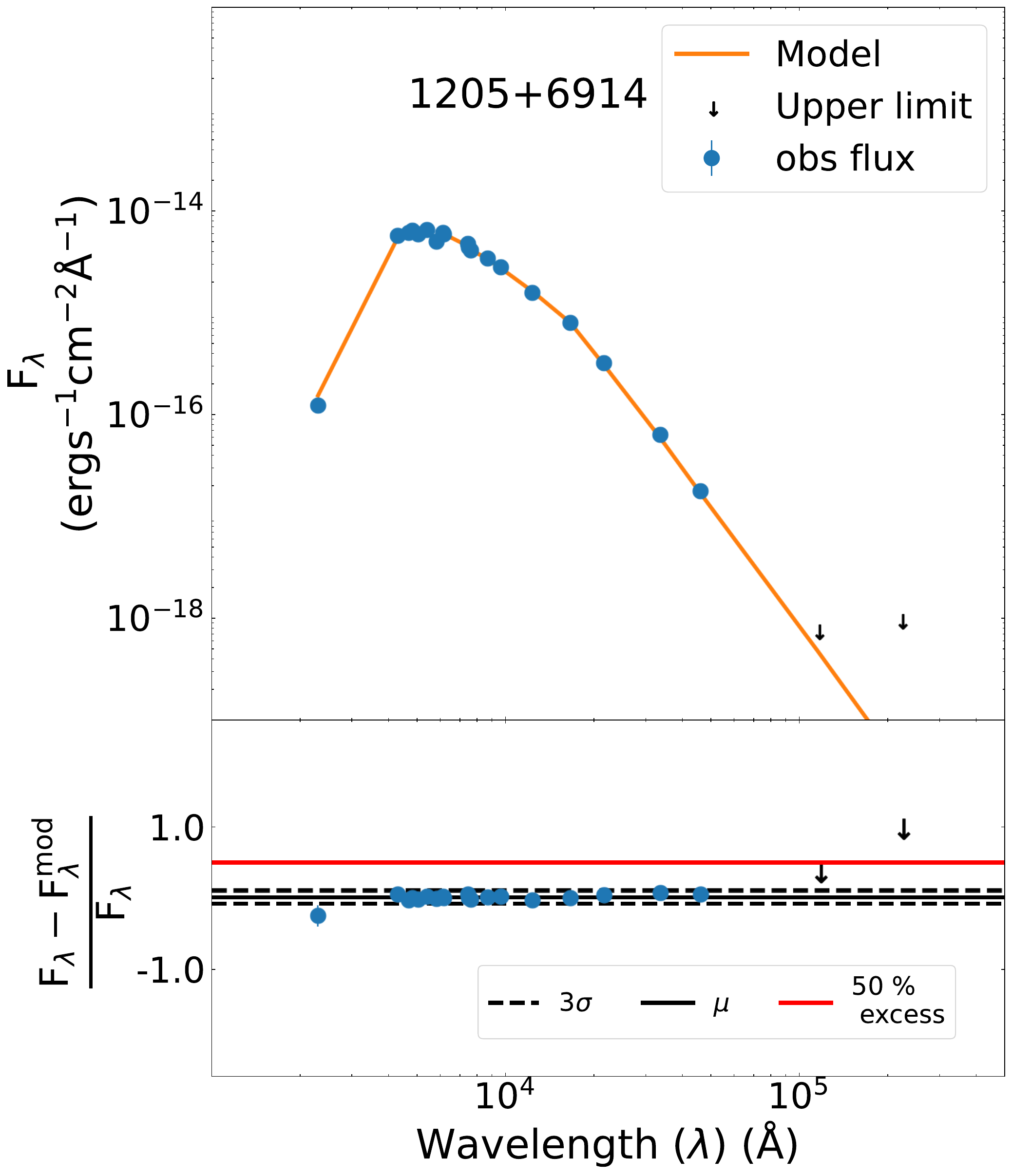}
\figsetgrpnote{Same as \autoref{fig:sed_example_no_excess},but for source 1205$+$6914}
\label{fig:sed_example_no_excess_6}

\figsetgrpend

\figsetgrpstart
\figsetgrpnum{2.7}
\figsetgrptitle{Image for figure 2_7}
\figsetplot{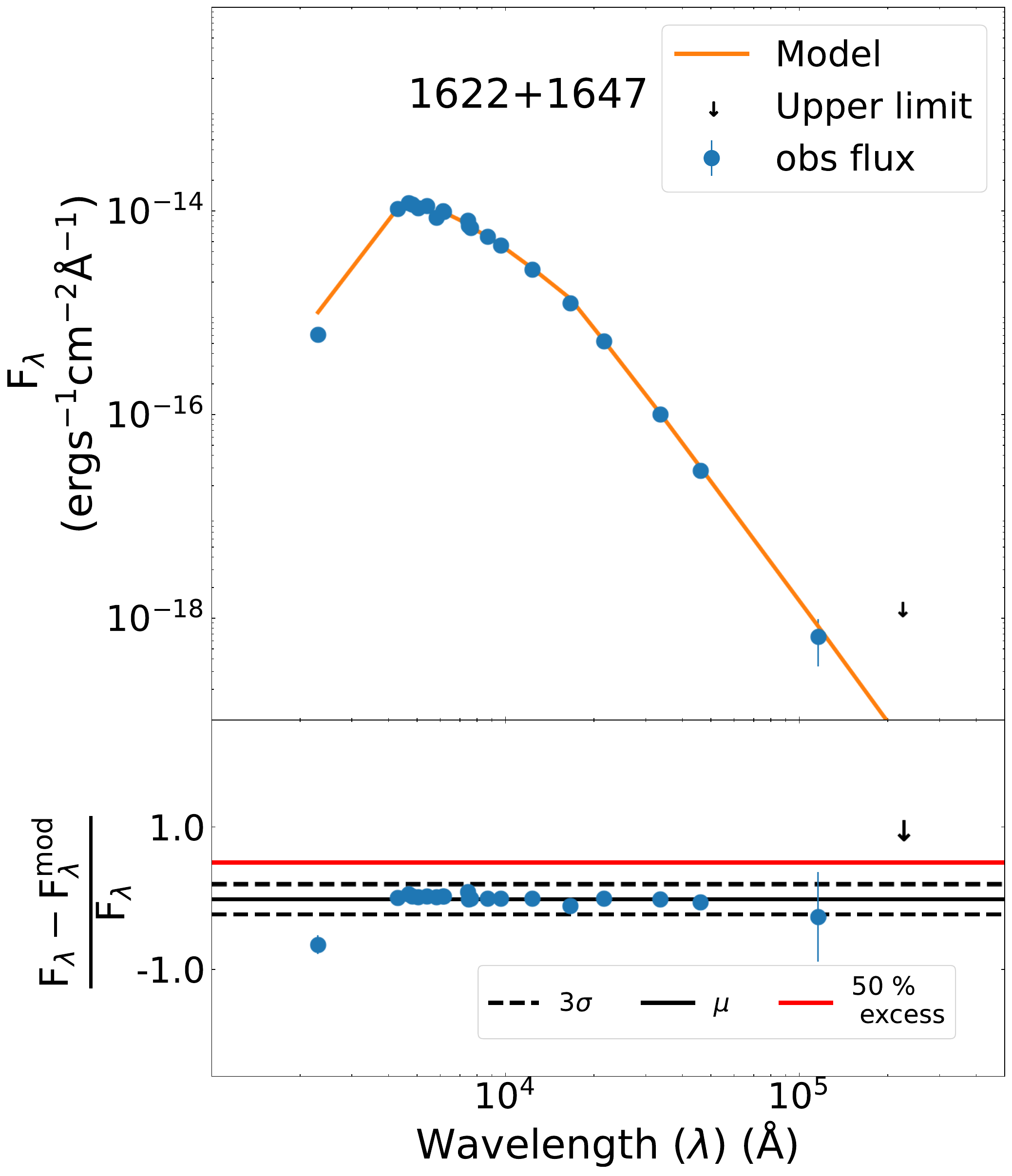}
\figsetgrpnote{Same as \autoref{fig:sed_example_no_excess},but for source 1622$+$1647}
\label{fig:sed_example_no_excess_7}

\figsetgrpend

\figsetgrpstart
\figsetgrpnum{2.8}
\figsetgrptitle{Image for figure 2_8}
\figsetplot{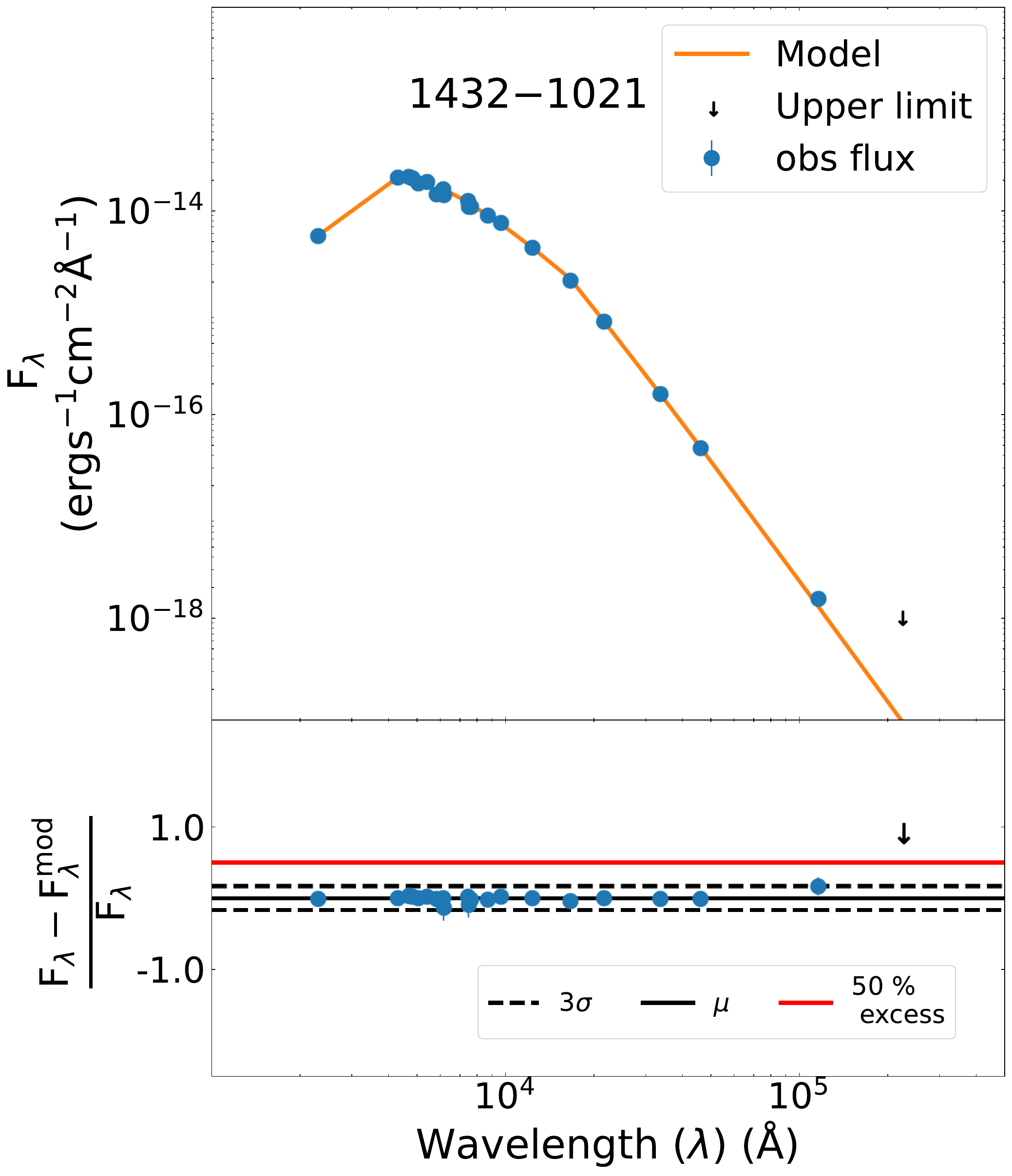}
\figsetgrpnote{Same as \autoref{fig:sed_example_no_excess},but for source 1432$-$1021}
\label{fig:sed_example_no_excess_8}

\figsetgrpend

\figsetgrpstart
\figsetgrpnum{2.9}
\figsetgrptitle{Image for figure 2_9}
\figsetplot{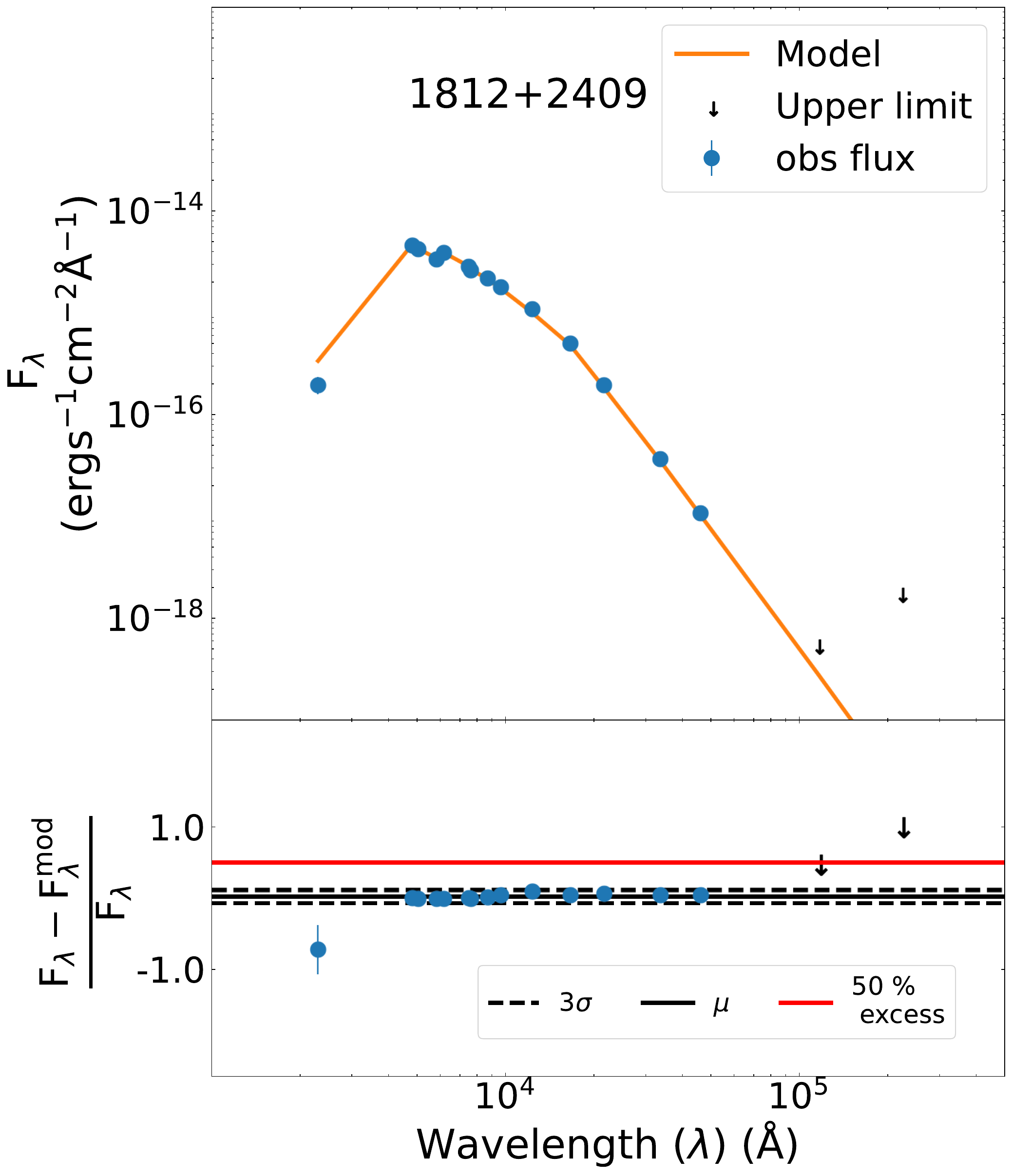}
\figsetgrpnote{Same as \autoref{fig:sed_example_no_excess},but for source 1812$+$2409}
\label{fig:sed_example_no_excess_9}

\figsetgrpend

\figsetgrpstart
\figsetgrpnum{2.10}
\figsetgrptitle{Image for figure 2_10}
\figsetplot{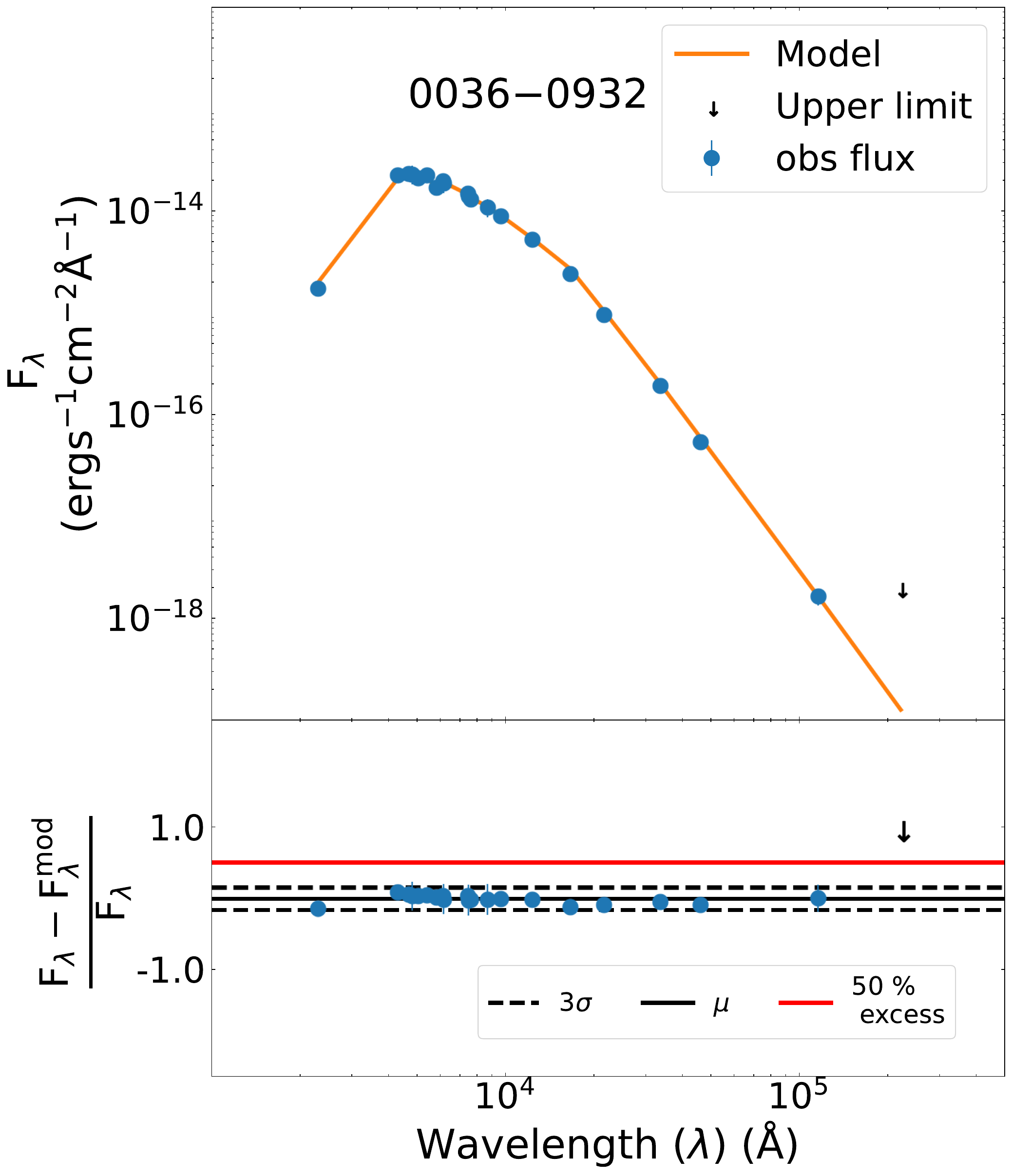}
\figsetgrpnote{Same as \autoref{fig:sed_example_no_excess},but for source 0036$-$0932}
\label{fig:sed_example_no_excess_10}

\figsetgrpend

\figsetgrpstart
\figsetgrpnum{2.11}
\figsetgrptitle{Image for figure 2_11}
\figsetplot{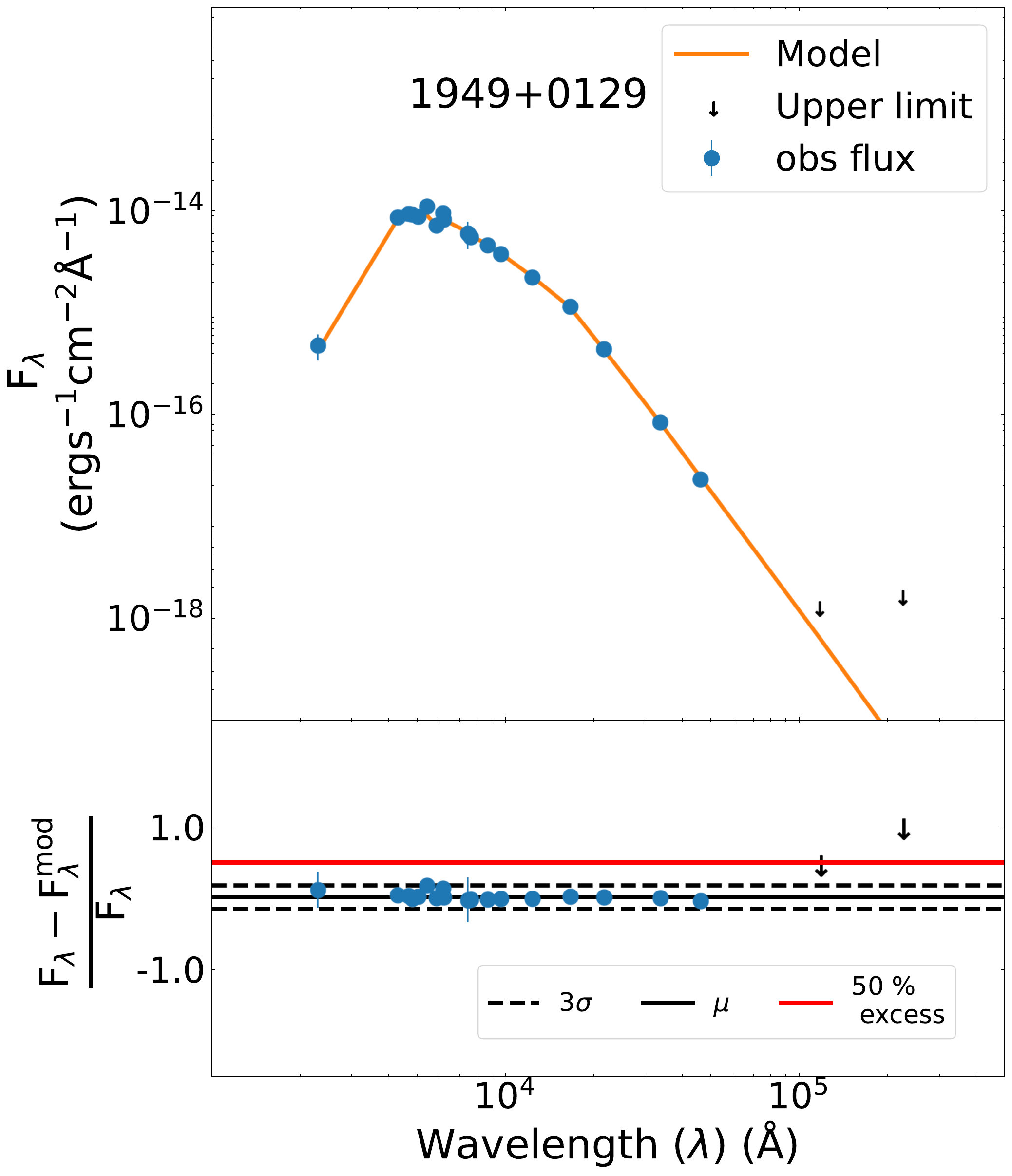}
\figsetgrpnote{Same as \autoref{fig:sed_example_no_excess},but for source 1949$+$0129}
\label{fig:sed_example_no_excess_11}

\figsetgrpend

\figsetgrpstart
\figsetgrpnum{2.12}
\figsetgrptitle{Image for figure 2_12}
\figsetplot{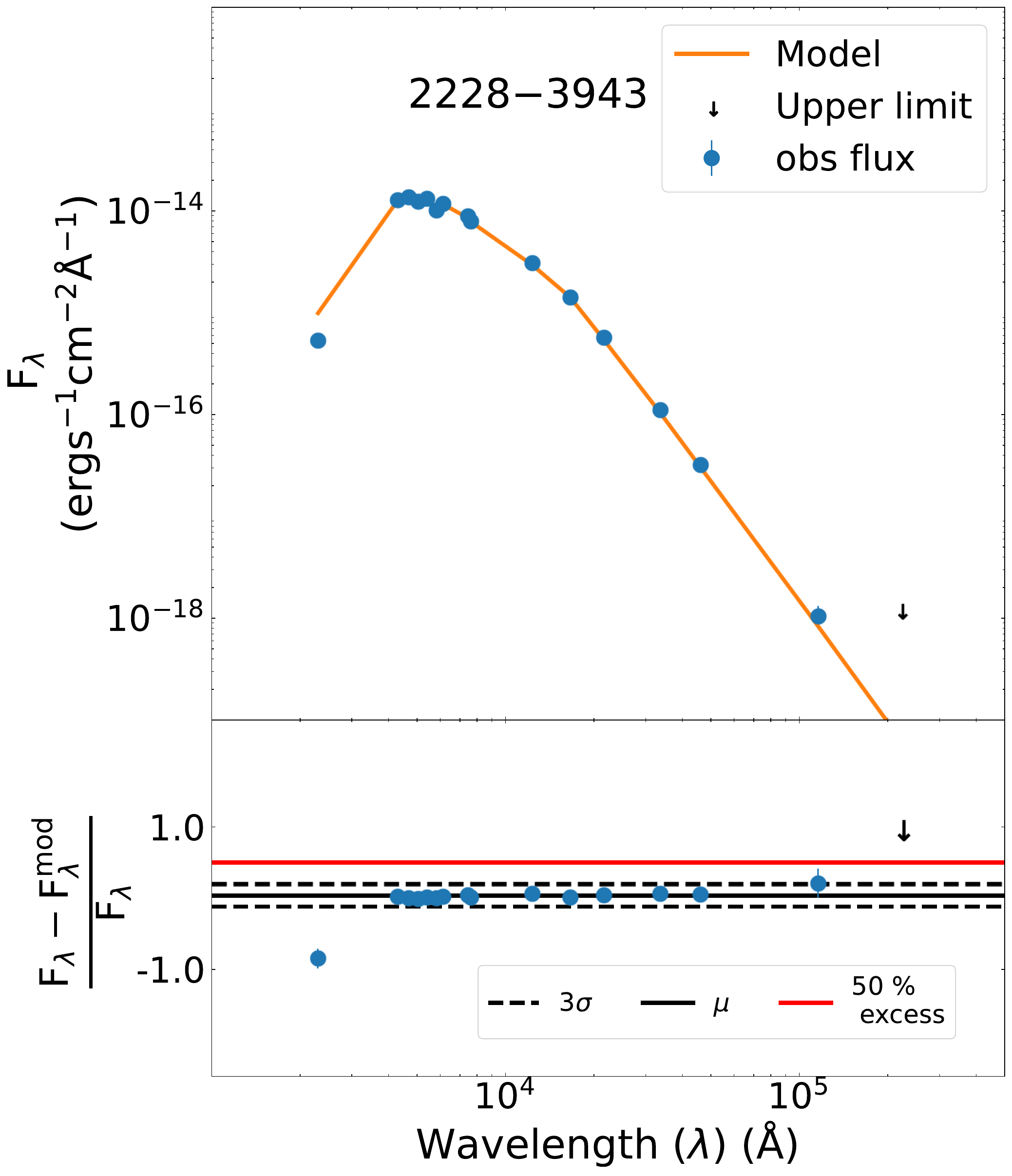}
\figsetgrpnote{Same as \autoref{fig:sed_example_no_excess},but for source 2228$-$3943}
\label{fig:sed_example_no_excess_12}

\figsetgrpend

\figsetgrpstart
\figsetgrpnum{2.13}
\figsetgrptitle{Image for figure 2_13}
\figsetplot{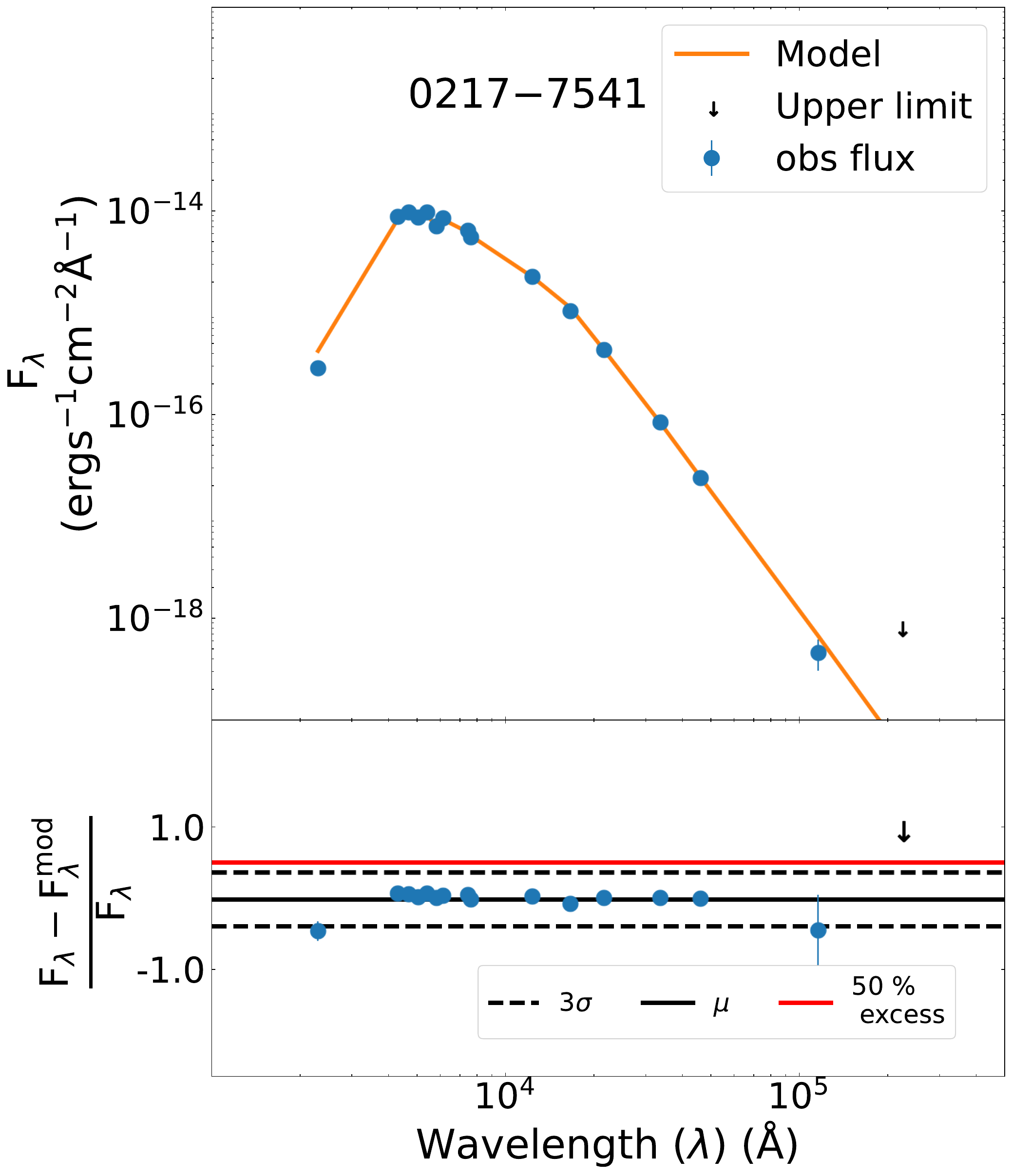}
\figsetgrpnote{Same as \autoref{fig:sed_example_no_excess},but for source 0217$-$7541}
\label{fig:sed_example_no_excess_13}

\figsetgrpend

\figsetgrpstart
\figsetgrpnum{2.14}
\figsetgrptitle{Image for figure 2_14}
\figsetplot{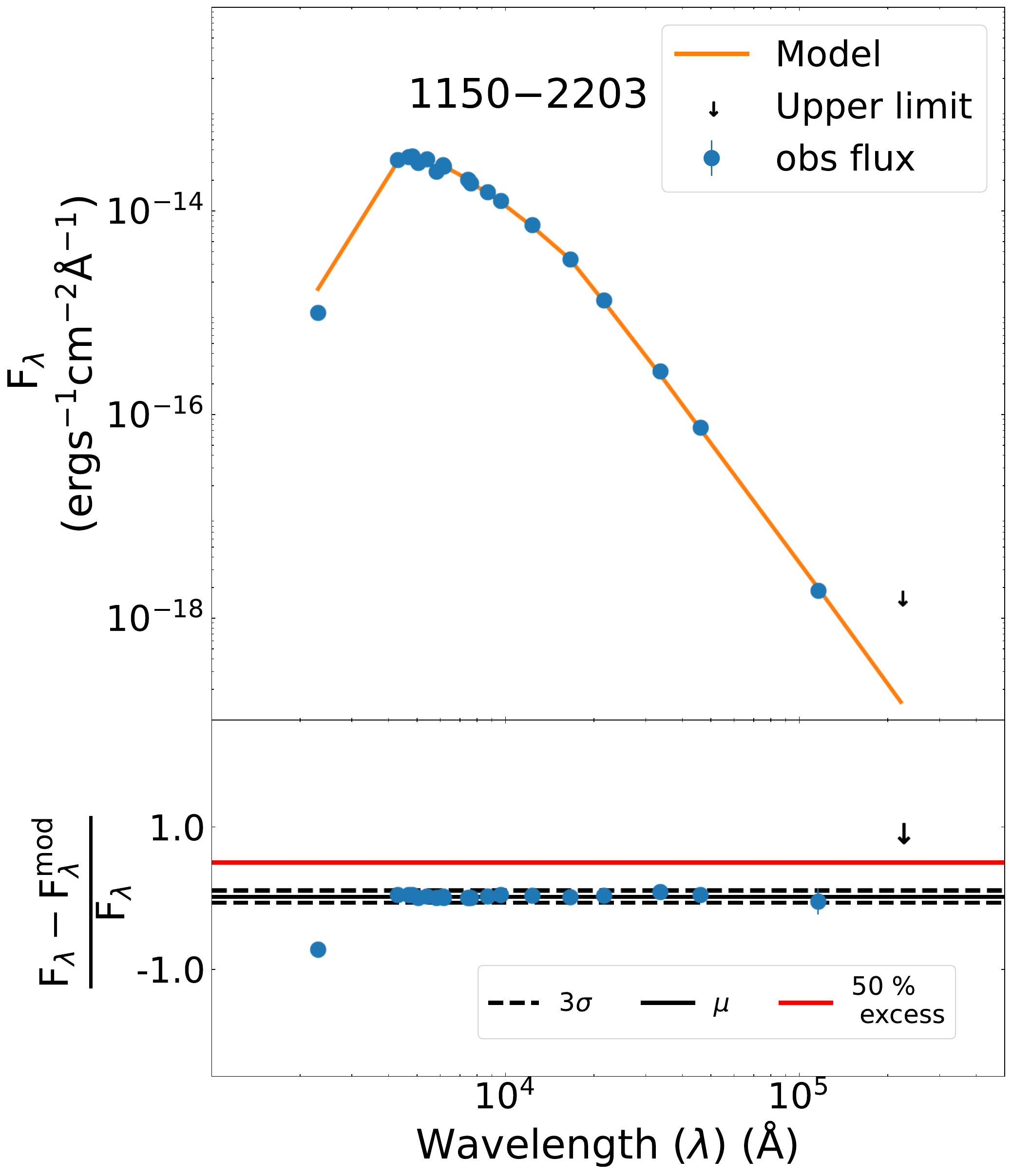}
\figsetgrpnote{Same as \autoref{fig:sed_example_no_excess},but for source 1150$-$2203}
\label{fig:sed_example_no_excess_14}

\figsetgrpend

\figsetgrpstart
\figsetgrpnum{2.15}
\figsetgrptitle{Image for figure 2_15}
\figsetplot{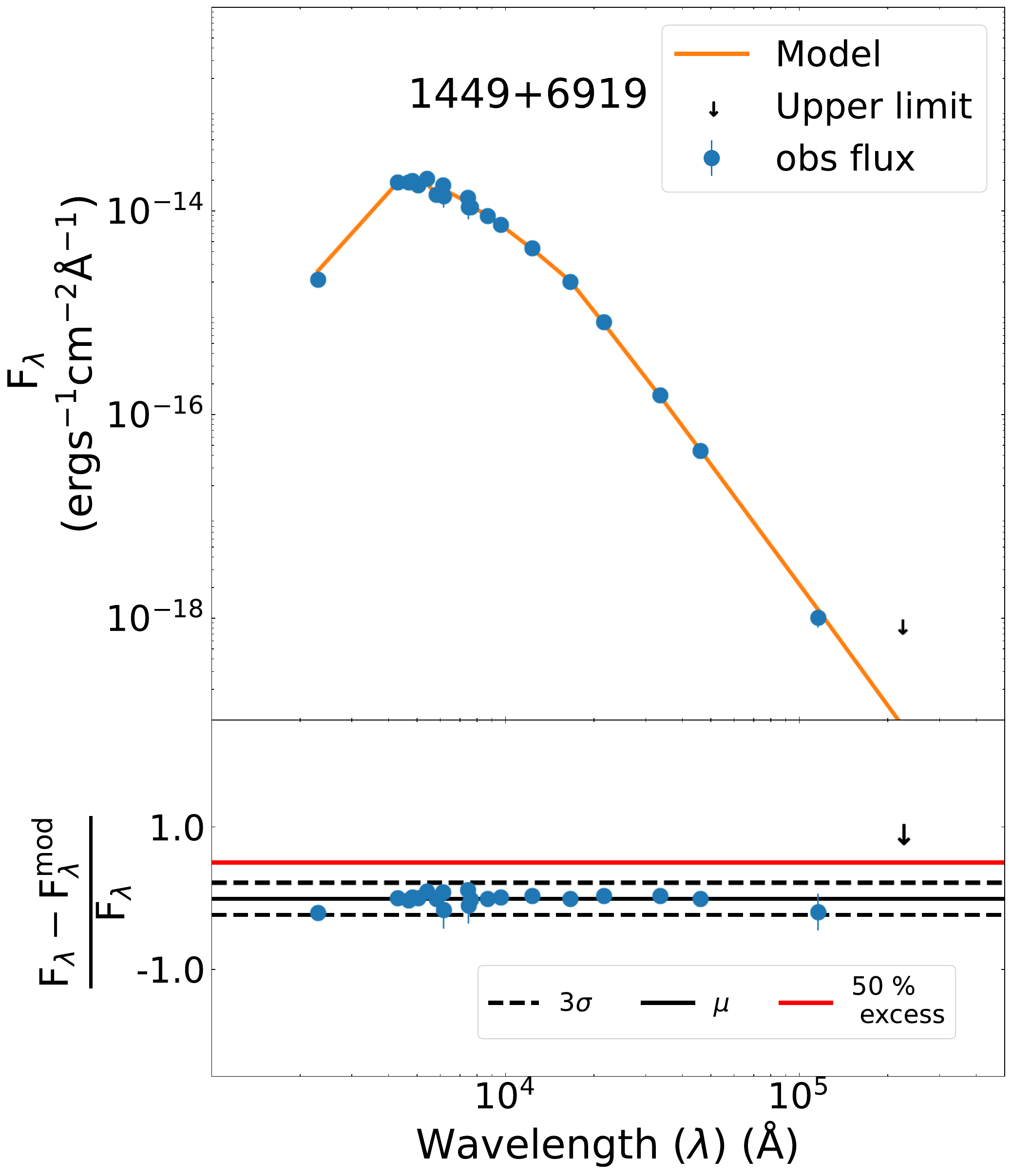}
\figsetgrpnote{Same as \autoref{fig:sed_example_no_excess},but for source 1449$+$6919}
\label{fig:sed_example_no_excess_15}

\figsetgrpend

\figsetgrpstart
\figsetgrpnum{2.16}
\figsetgrptitle{Image for figure 2_16}
\figsetplot{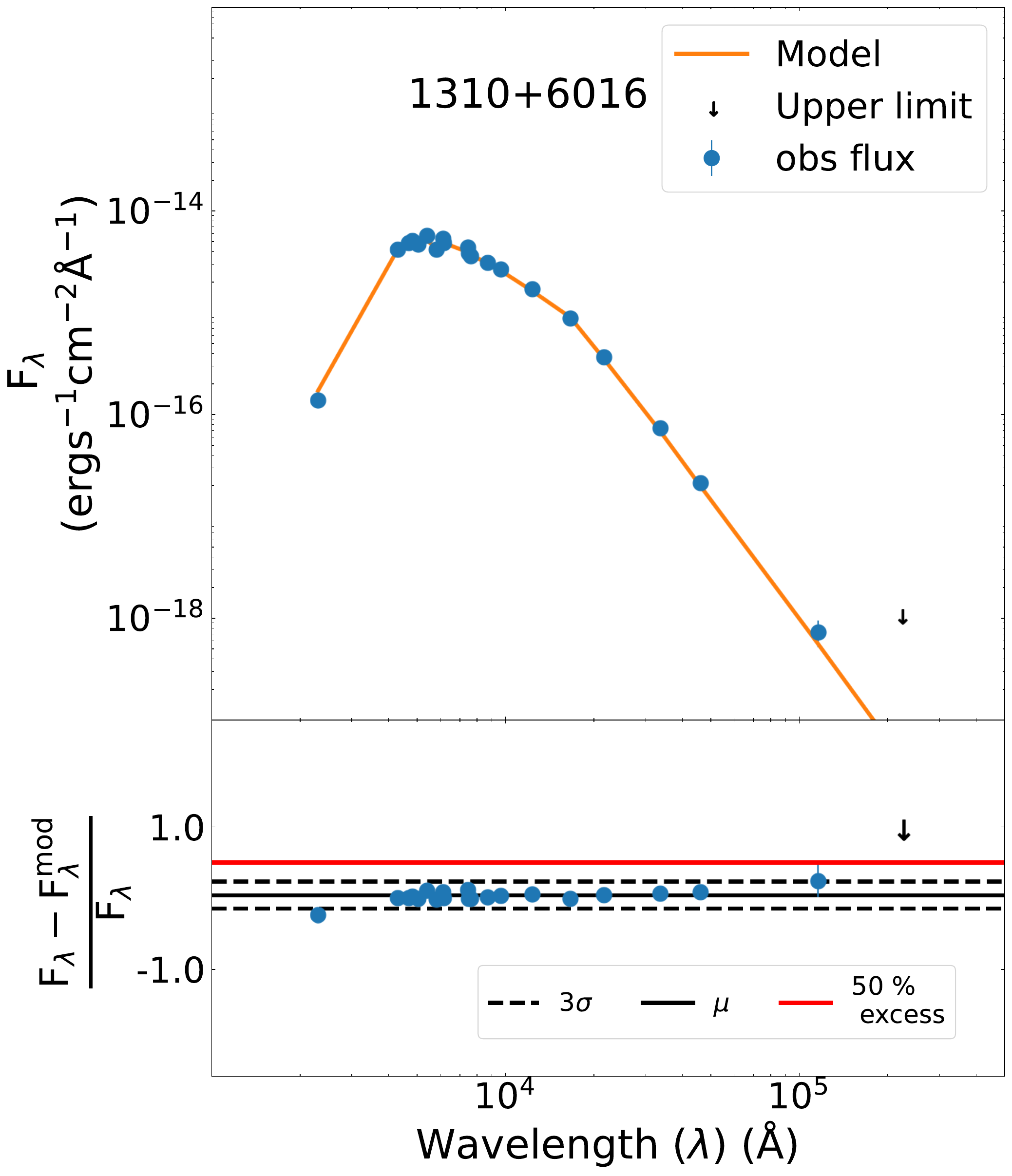}
\figsetgrpnote{Same as \autoref{fig:sed_example_no_excess},but for source 1310$+$6016}
\label{fig:sed_example_no_excess_16}

\figsetgrpend

\figsetgrpstart
\figsetgrpnum{2.17}
\figsetgrptitle{Image for figure 2_17}
\figsetplot{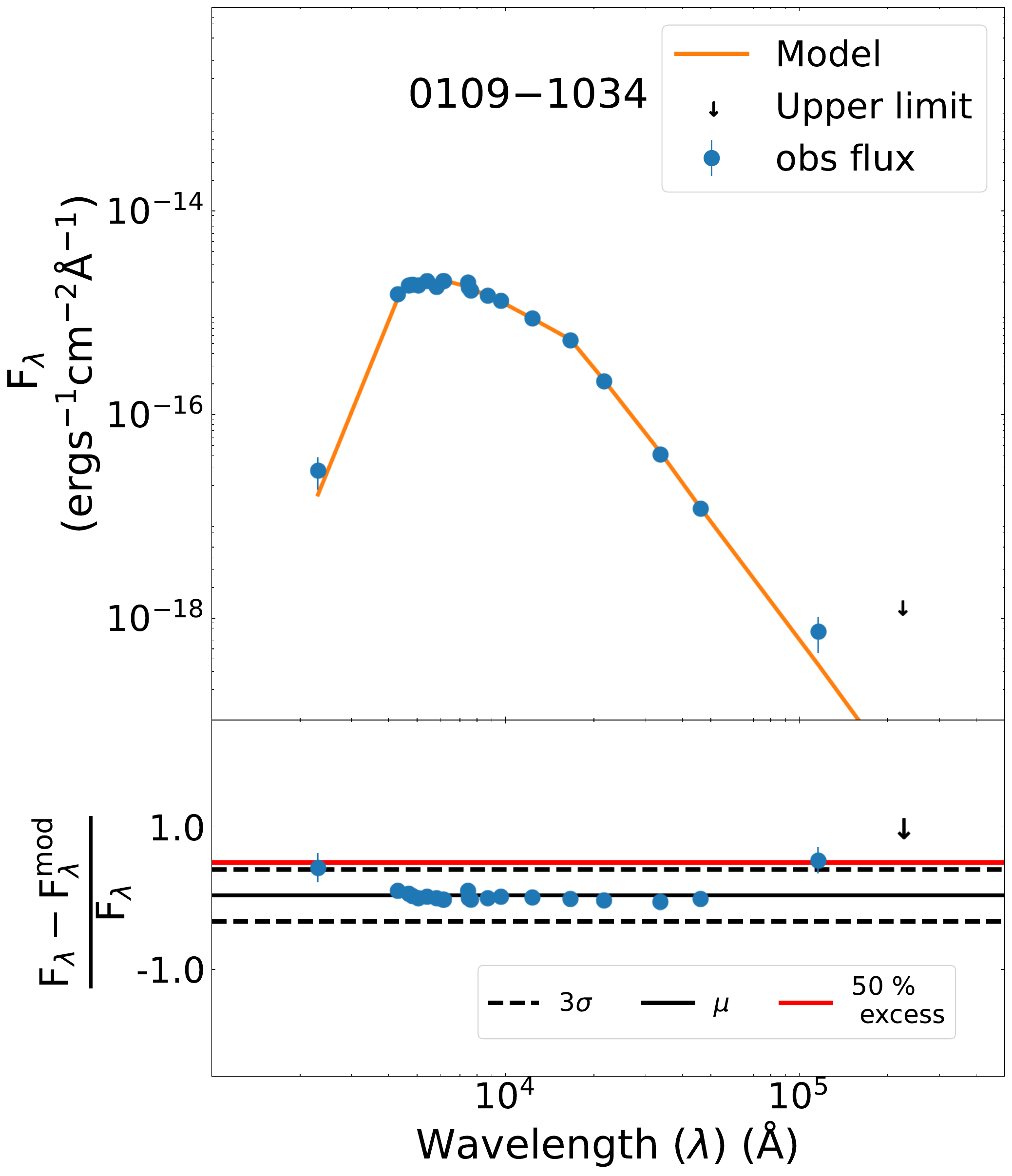}
\figsetgrpnote{Same as \autoref{fig:sed_example_no_excess},but for source 0109$-$1034}
\label{fig:sed_example_no_excess_17}

\figsetgrpend

\figsetgrpstart
\figsetgrpnum{2.18}
\figsetgrptitle{Image for figure 2_18}
\figsetplot{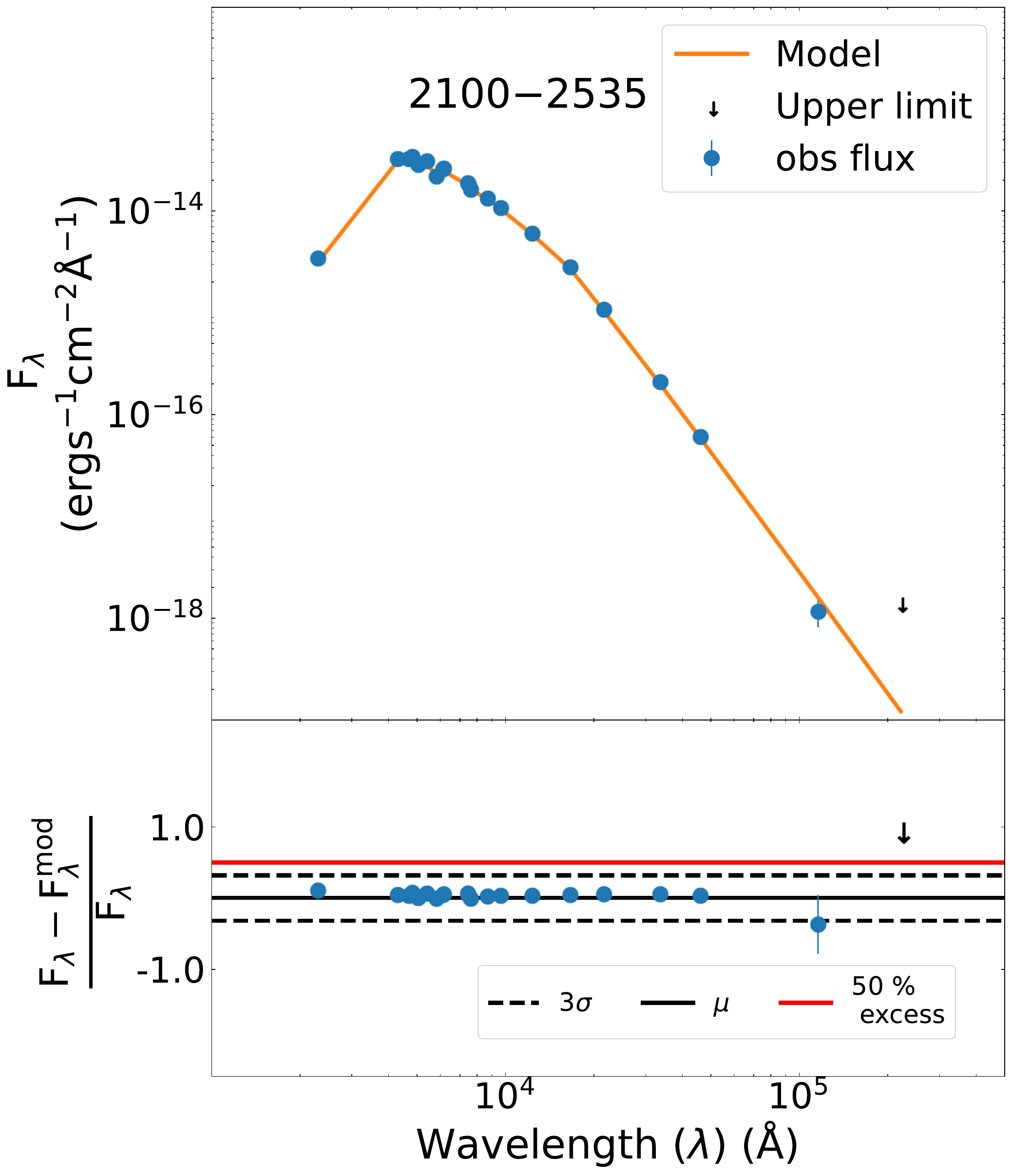}
\figsetgrpnote{Same as \autoref{fig:sed_example_no_excess},but for source 2100$-$2535}
\label{fig:sed_example_no_excess_18}

\figsetgrpend

\figsetgrpstart
\figsetgrpnum{2.19}
\figsetgrptitle{Image for figure 2_19}
\figsetplot{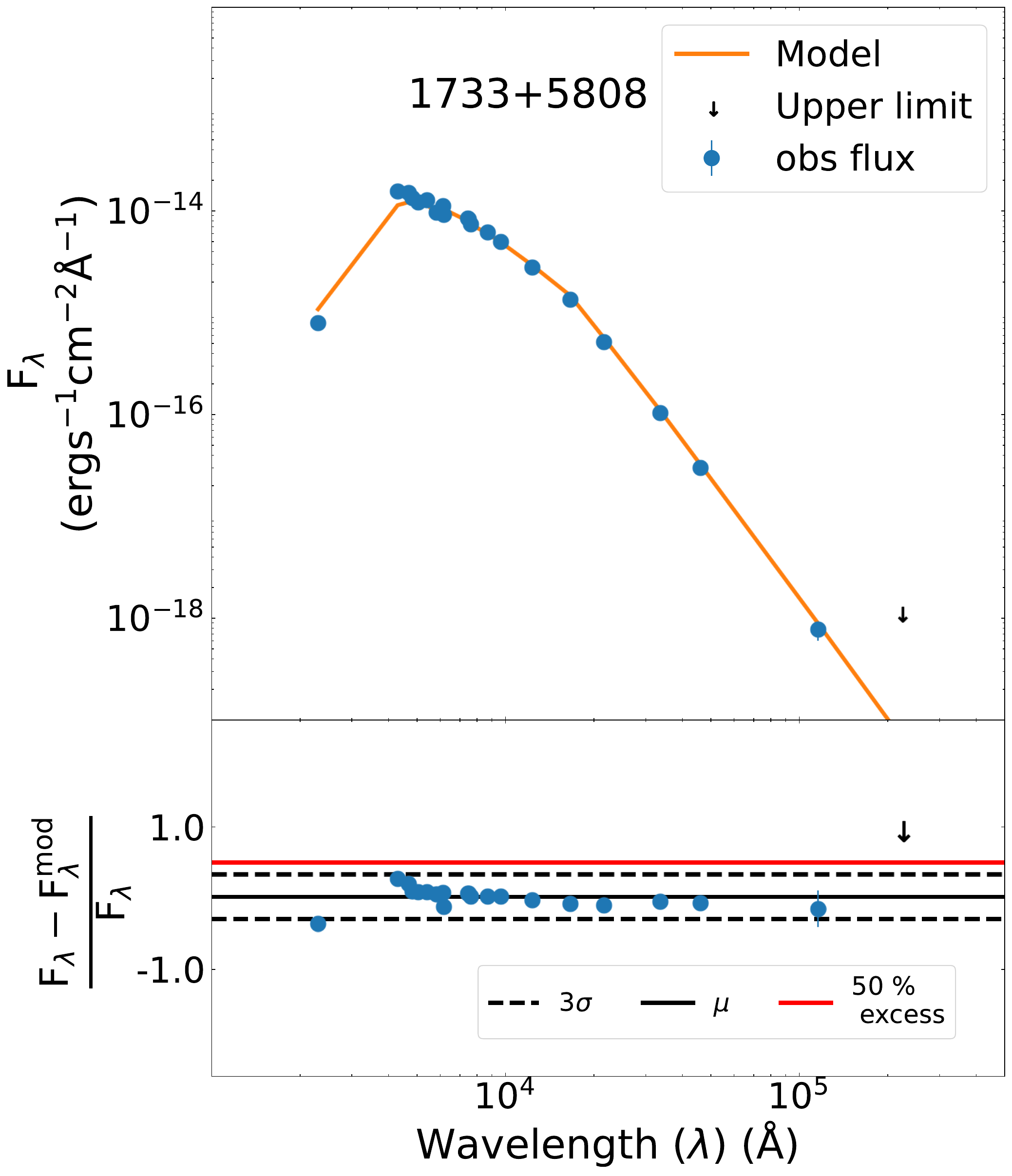}
\figsetgrpnote{Same as \autoref{fig:sed_example_no_excess},but for source 1733$+$5808}
\label{fig:sed_example_no_excess_19}

\figsetgrpend

\figsetgrpstart
\figsetgrpnum{2.20}
\figsetgrptitle{Image for figure 2_20}
\figsetplot{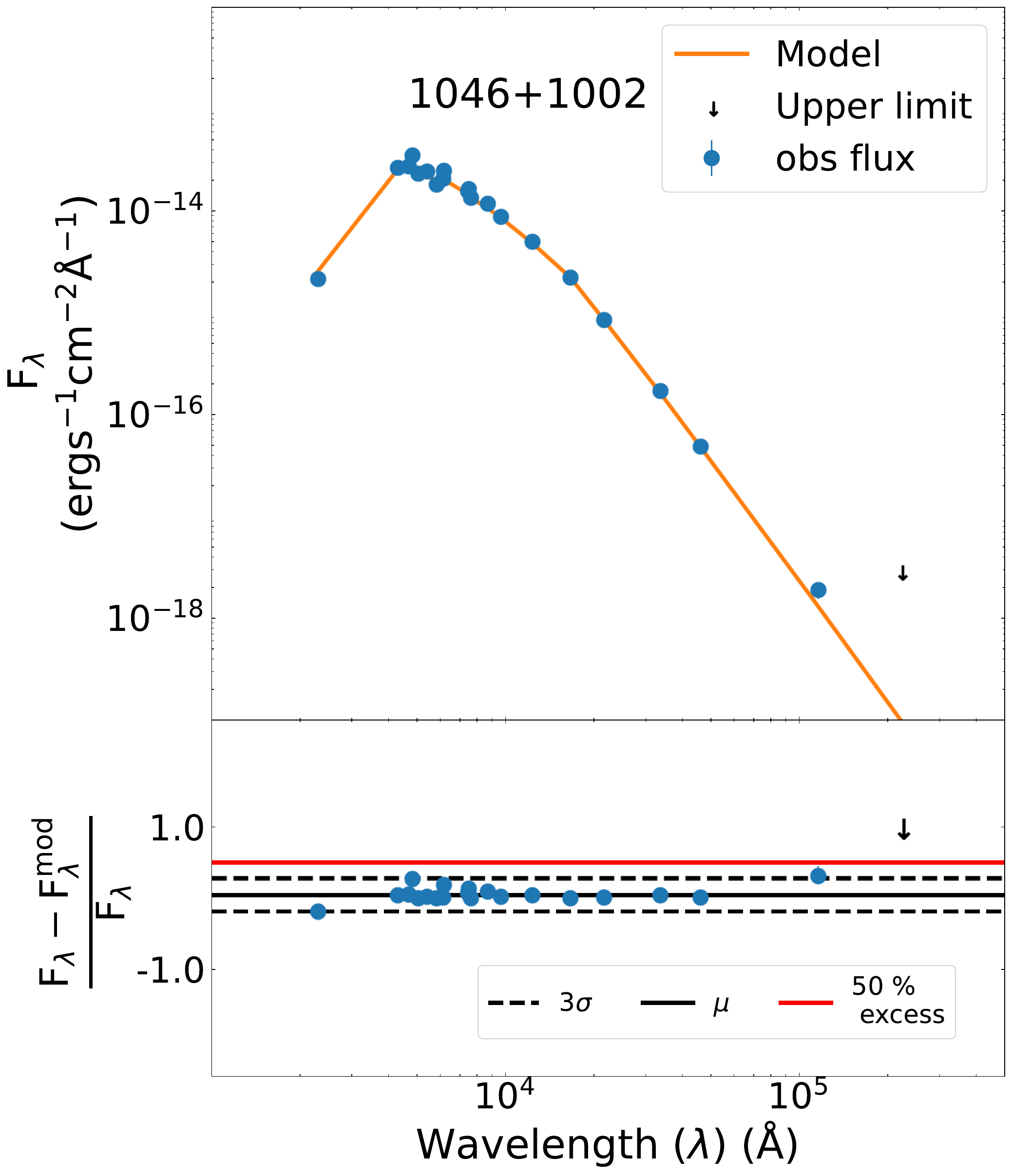}
\figsetgrpnote{Same as \autoref{fig:sed_example_no_excess},but for source 1046$+$1002}
\label{fig:sed_example_no_excess_20}

\figsetgrpend

\figsetgrpstart
\figsetgrpnum{2.21}
\figsetgrptitle{Image for figure 2_21}
\figsetplot{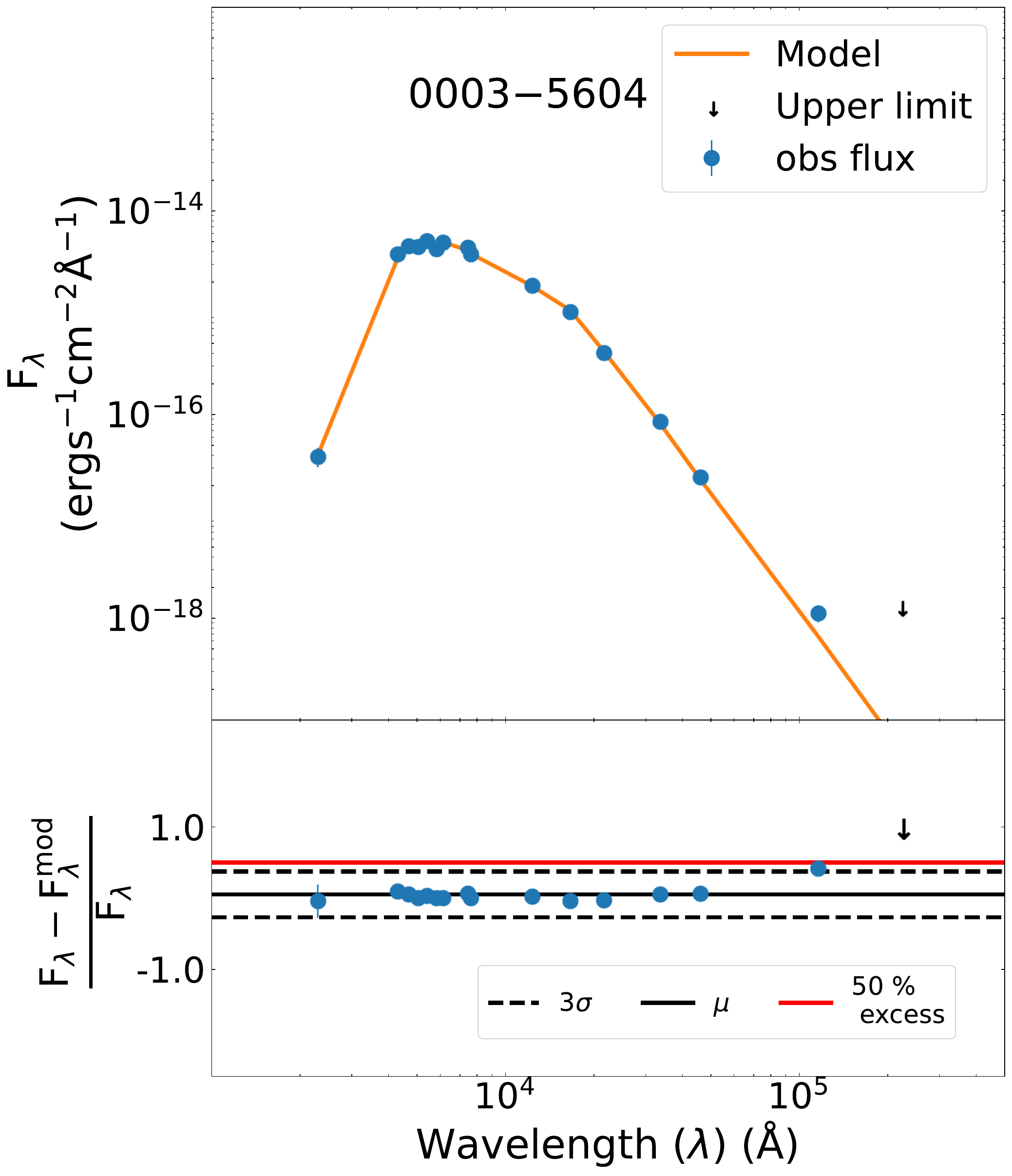}
\figsetgrpnote{Same as \autoref{fig:sed_example_no_excess},but for source 0003$-$5604}
\label{fig:sed_example_no_excess_21}

\figsetgrpend

\figsetgrpstart
\figsetgrpnum{2.22}
\figsetgrptitle{Image for figure 2_22}
\figsetplot{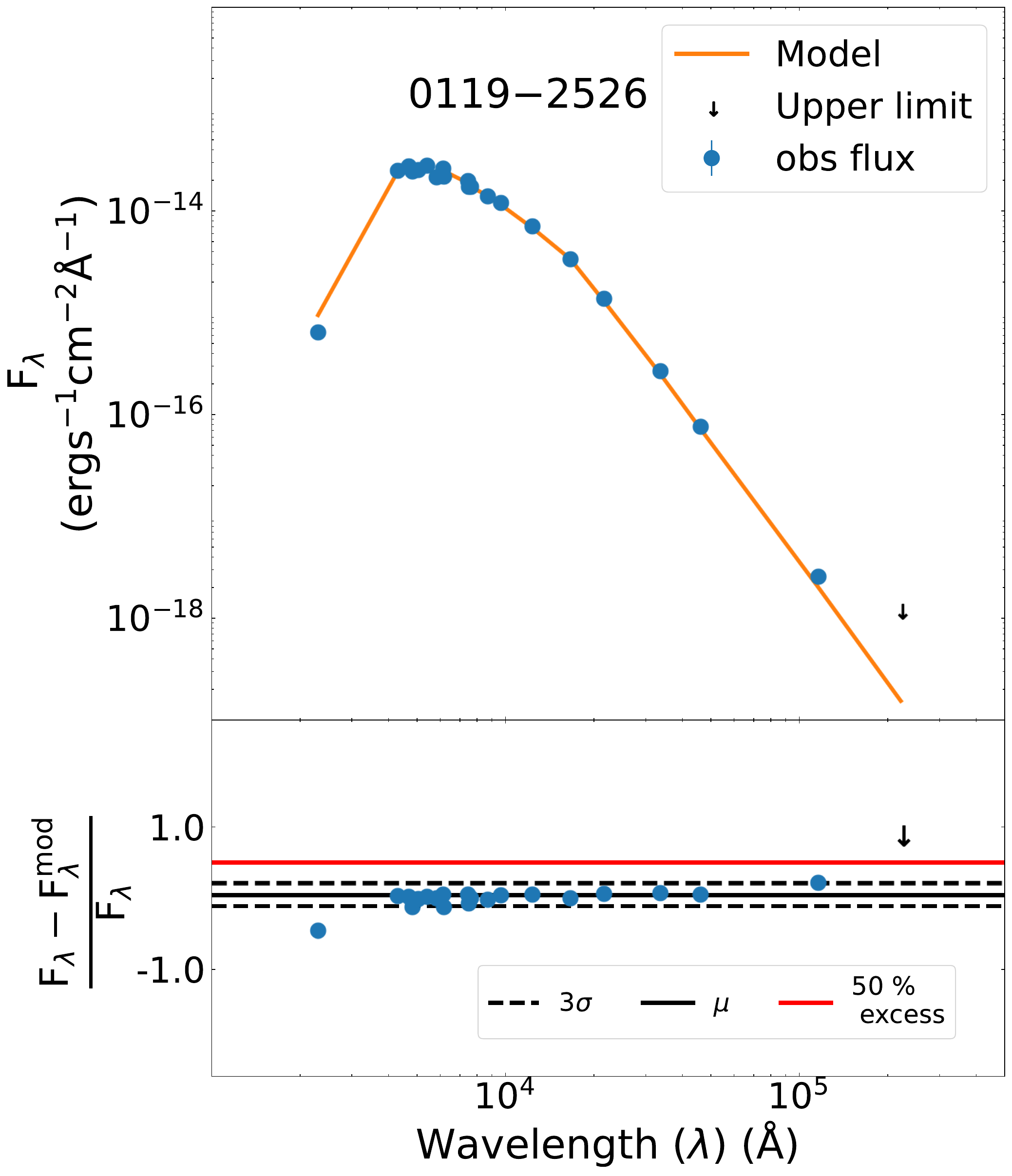}
\figsetgrpnote{Same as \autoref{fig:sed_example_no_excess},but for source 0119$-$2526}
\label{fig:sed_example_no_excess_22}

\figsetgrpend

\figsetgrpstart
\figsetgrpnum{2.23}
\figsetgrptitle{Image for figure 2_23}
\figsetplot{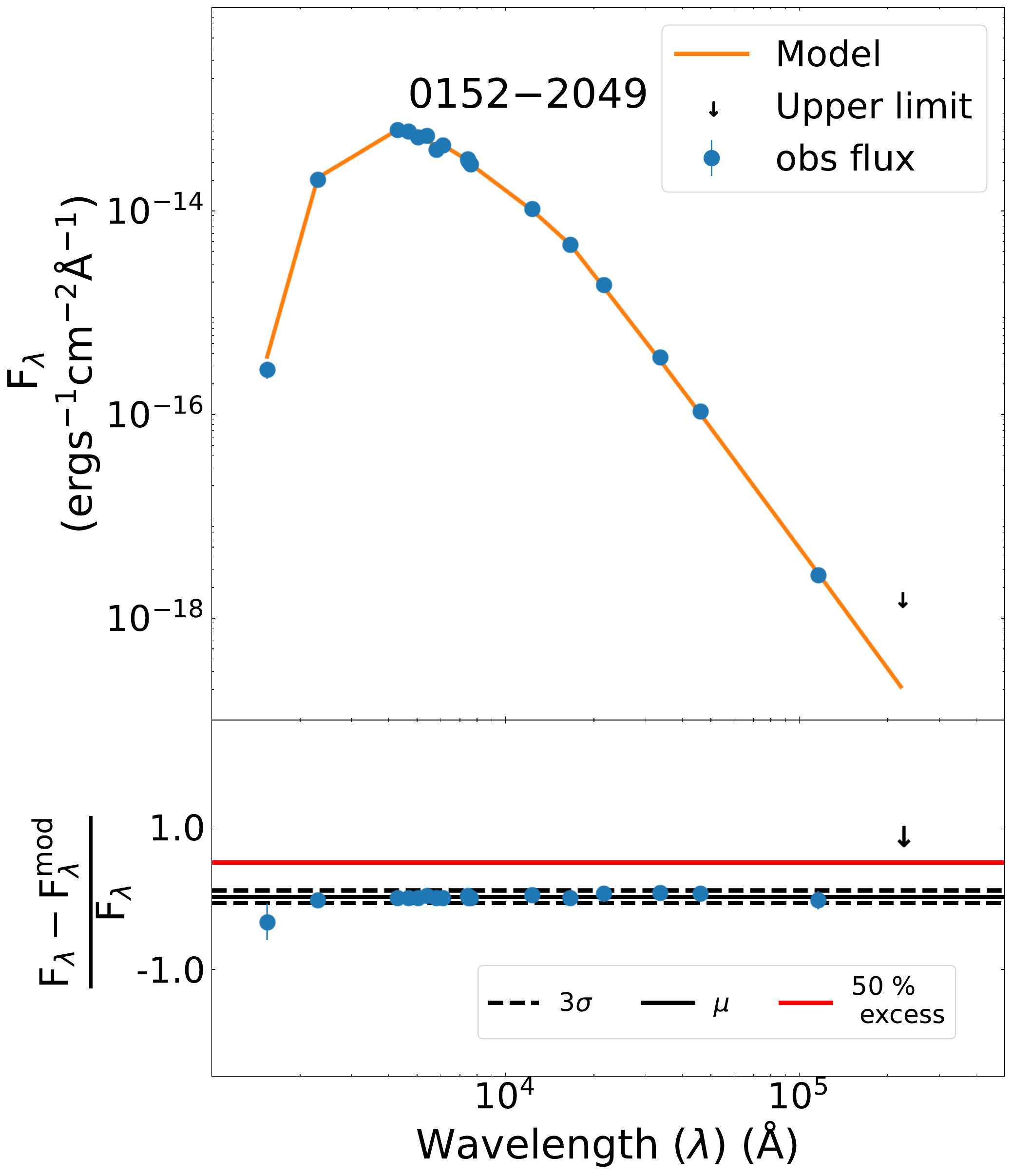}
\figsetgrpnote{Same as \autoref{fig:sed_example_no_excess},but for source 0152$-$2049}
\label{fig:sed_example_no_excess_23}

\figsetgrpend

\figsetgrpstart
\figsetgrpnum{2.24}
\figsetgrptitle{Image for figure 2_24}
\figsetplot{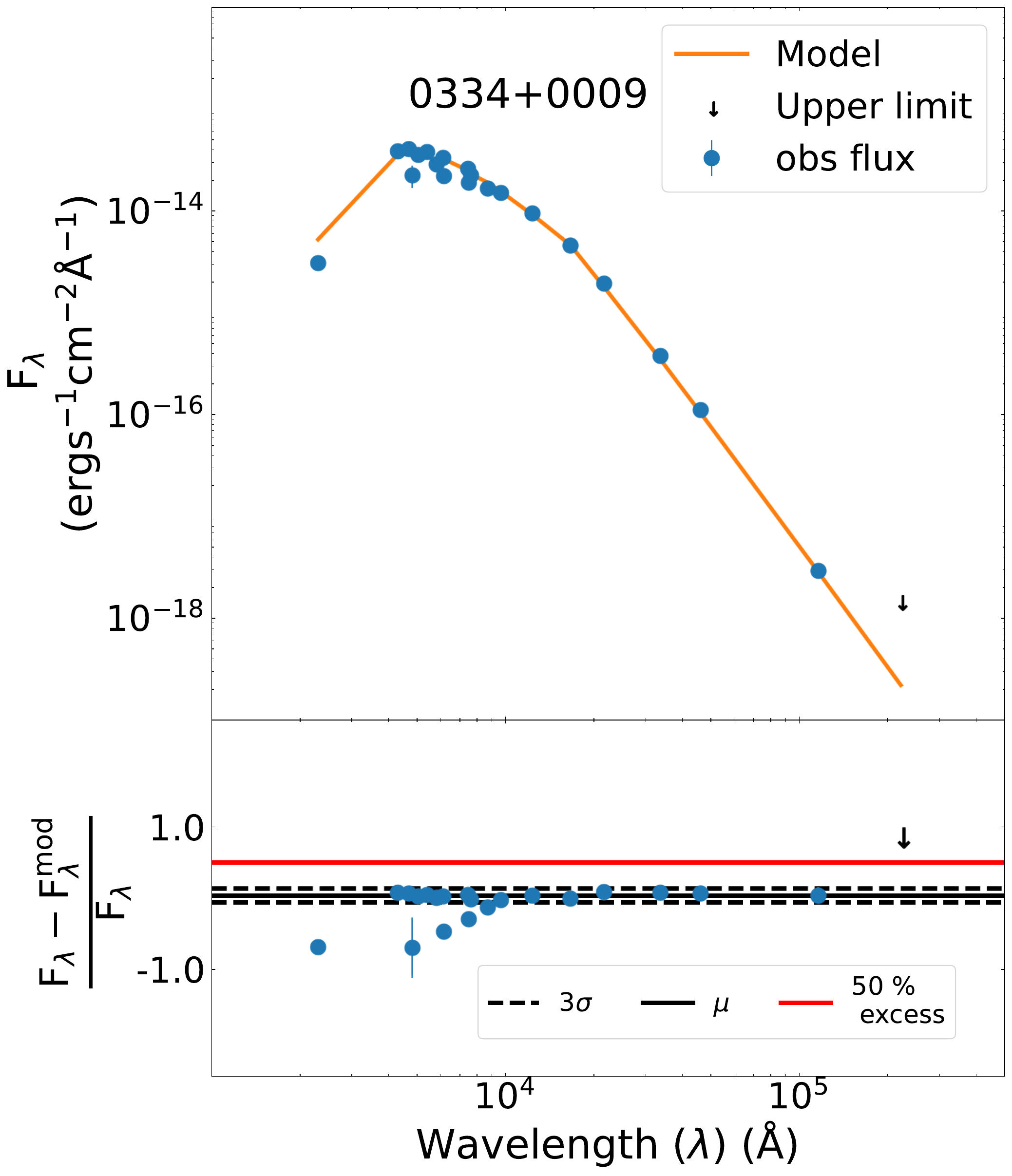}
\figsetgrpnote{Same as \autoref{fig:sed_example_no_excess},but for source 0334$+$0009}
\label{fig:sed_example_no_excess_24}

\figsetgrpend

\figsetgrpstart
\figsetgrpnum{2.25}
\figsetgrptitle{Image for figure 2_25}
\figsetplot{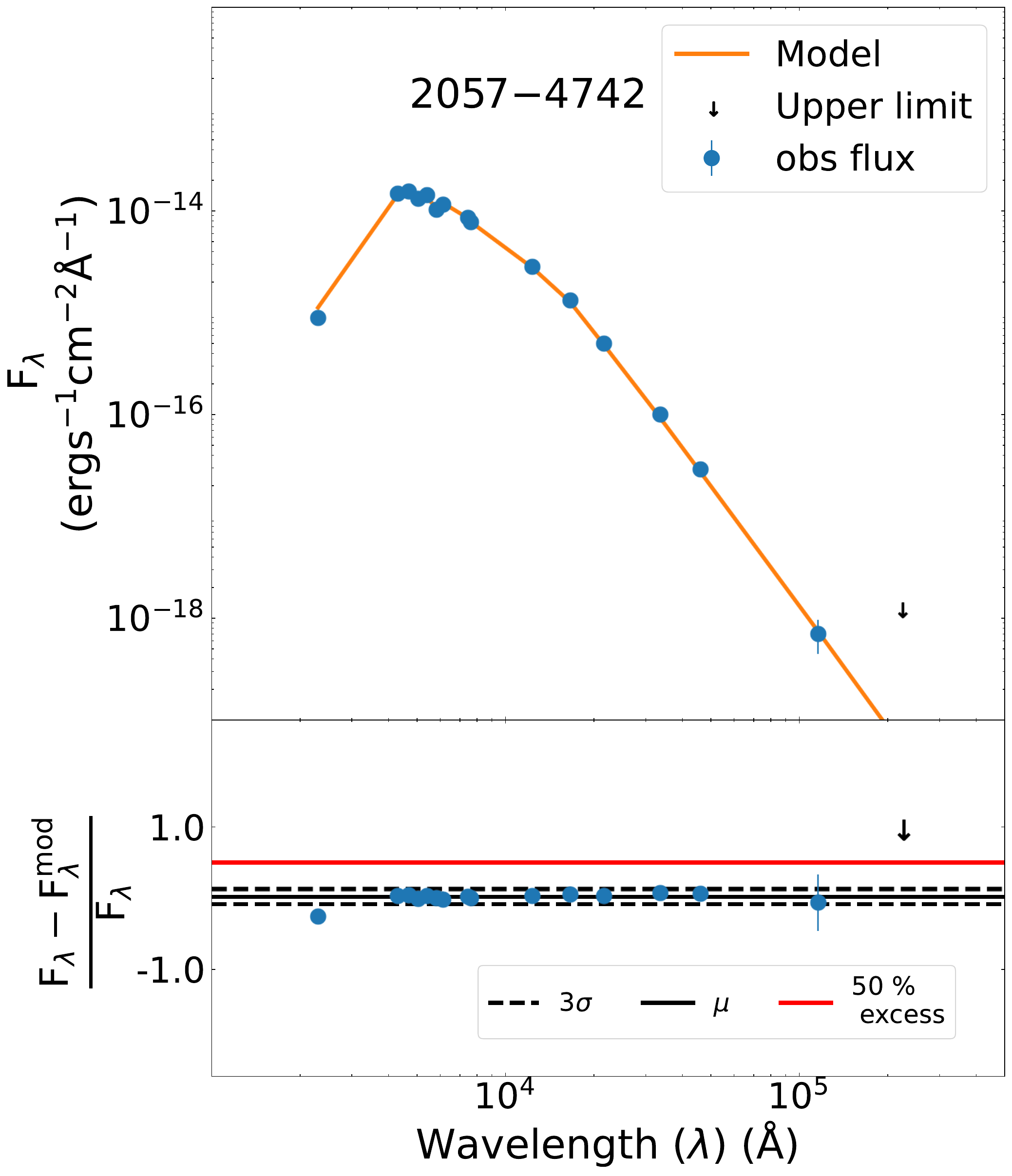}
\figsetgrpnote{Same as \autoref{fig:sed_example_no_excess},but for source 2057$-$4742}
\label{fig:sed_example_no_excess_25}

\figsetgrpend

\figsetgrpstart
\figsetgrpnum{2.26}
\figsetgrptitle{Image for figure 2_26}
\figsetplot{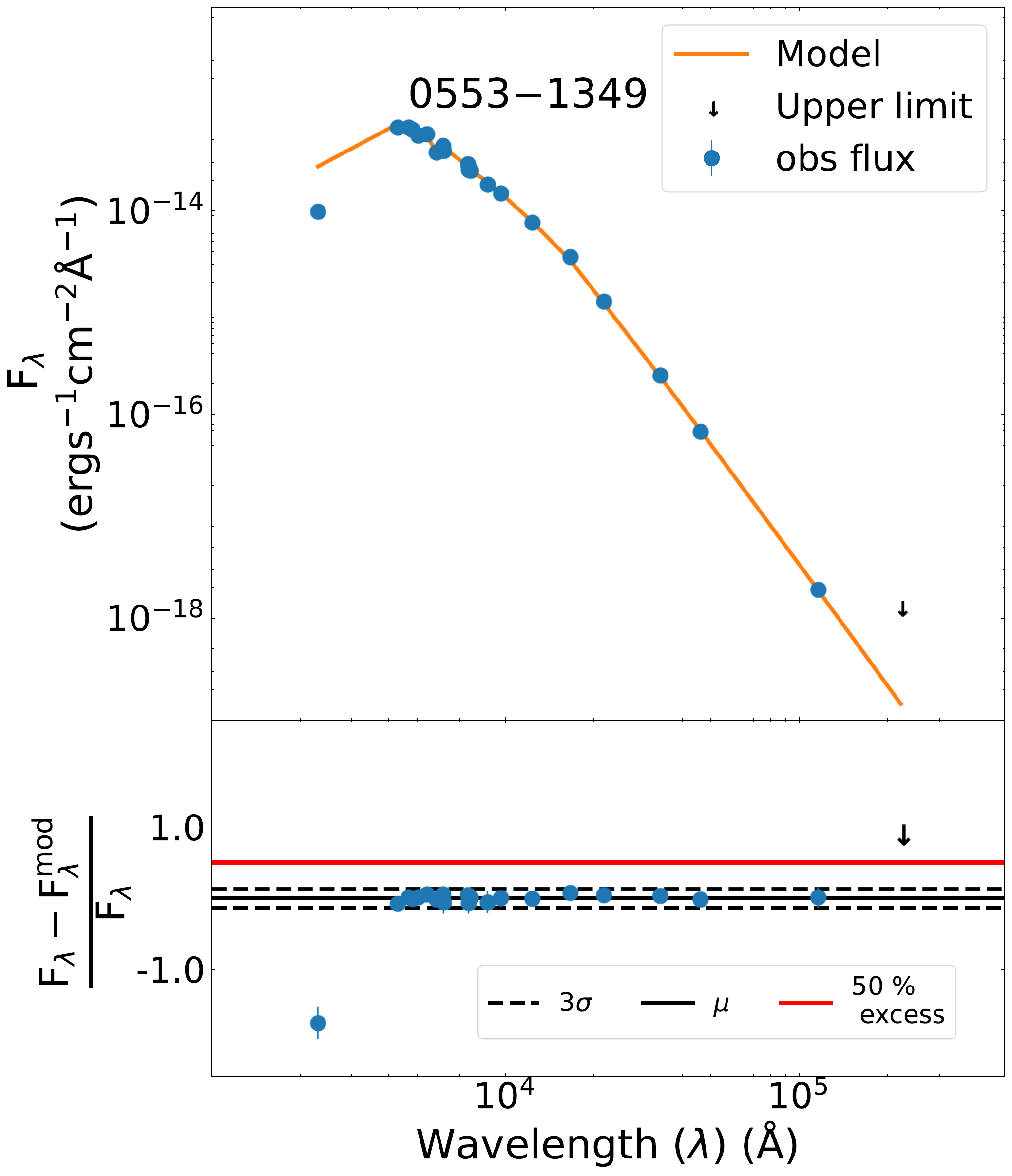}
\figsetgrpnote{Same as \autoref{fig:sed_example_no_excess},but for source 0553$-$1349}
\label{fig:sed_example_no_excess_26}

\figsetgrpend

\figsetgrpstart
\figsetgrpnum{2.27}
\figsetgrptitle{Image for figure 2_27}
\figsetplot{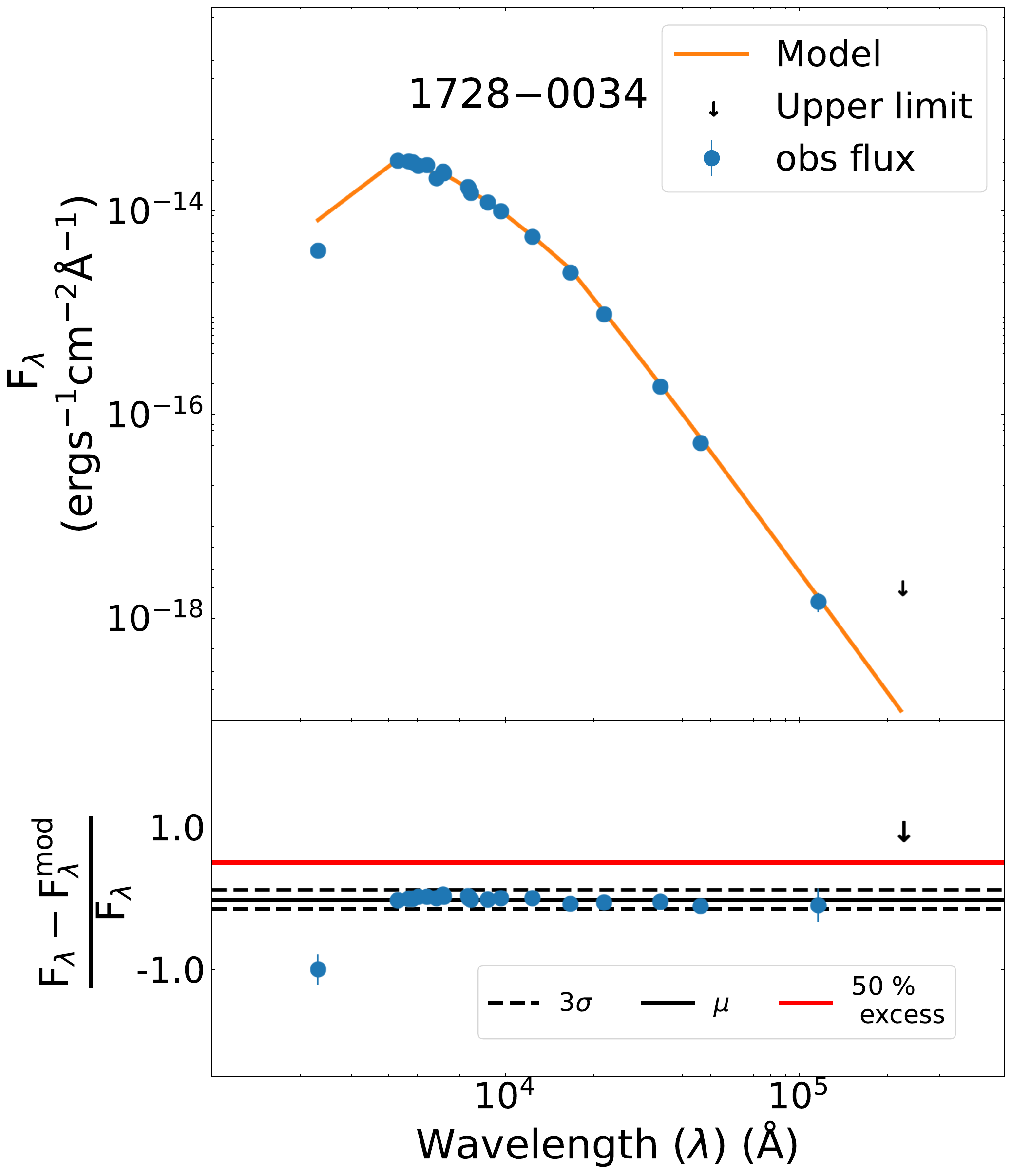}
\figsetgrpnote{Same as \autoref{fig:sed_example_no_excess},but for source 1728$-$0034}
\label{fig:sed_example_no_excess_27}

\figsetgrpend

\figsetgrpstart
\figsetgrpnum{2.28}
\figsetgrptitle{Image for figure 2_28}
\figsetplot{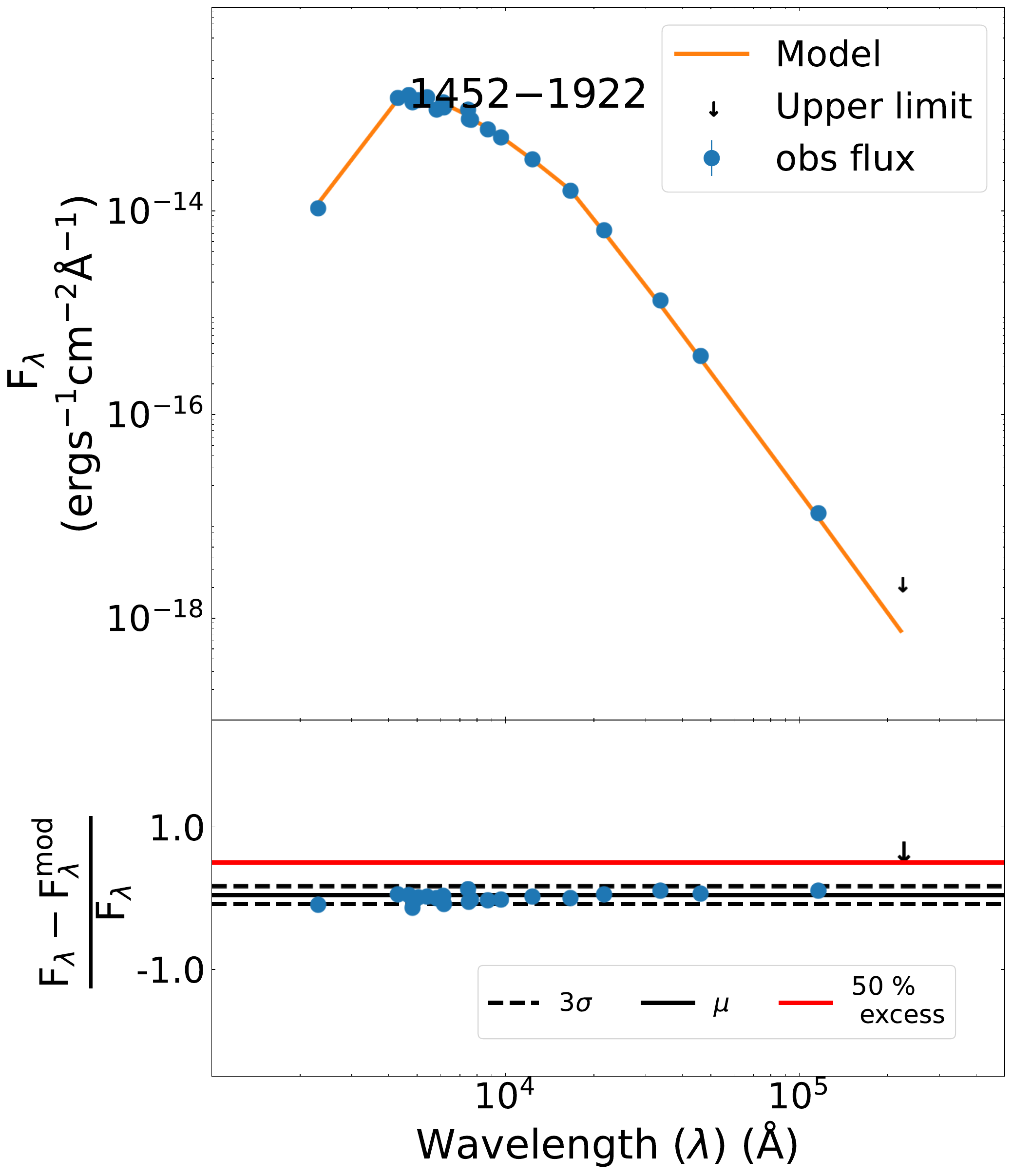}
\figsetgrpnote{Same as \autoref{fig:sed_example_no_excess},but for source 1452$-$1922}
\label{fig:sed_example_no_excess_28}

\figsetgrpend

\figsetgrpstart
\figsetgrpnum{2.29}
\figsetgrptitle{Image for figure 2_29}
\figsetplot{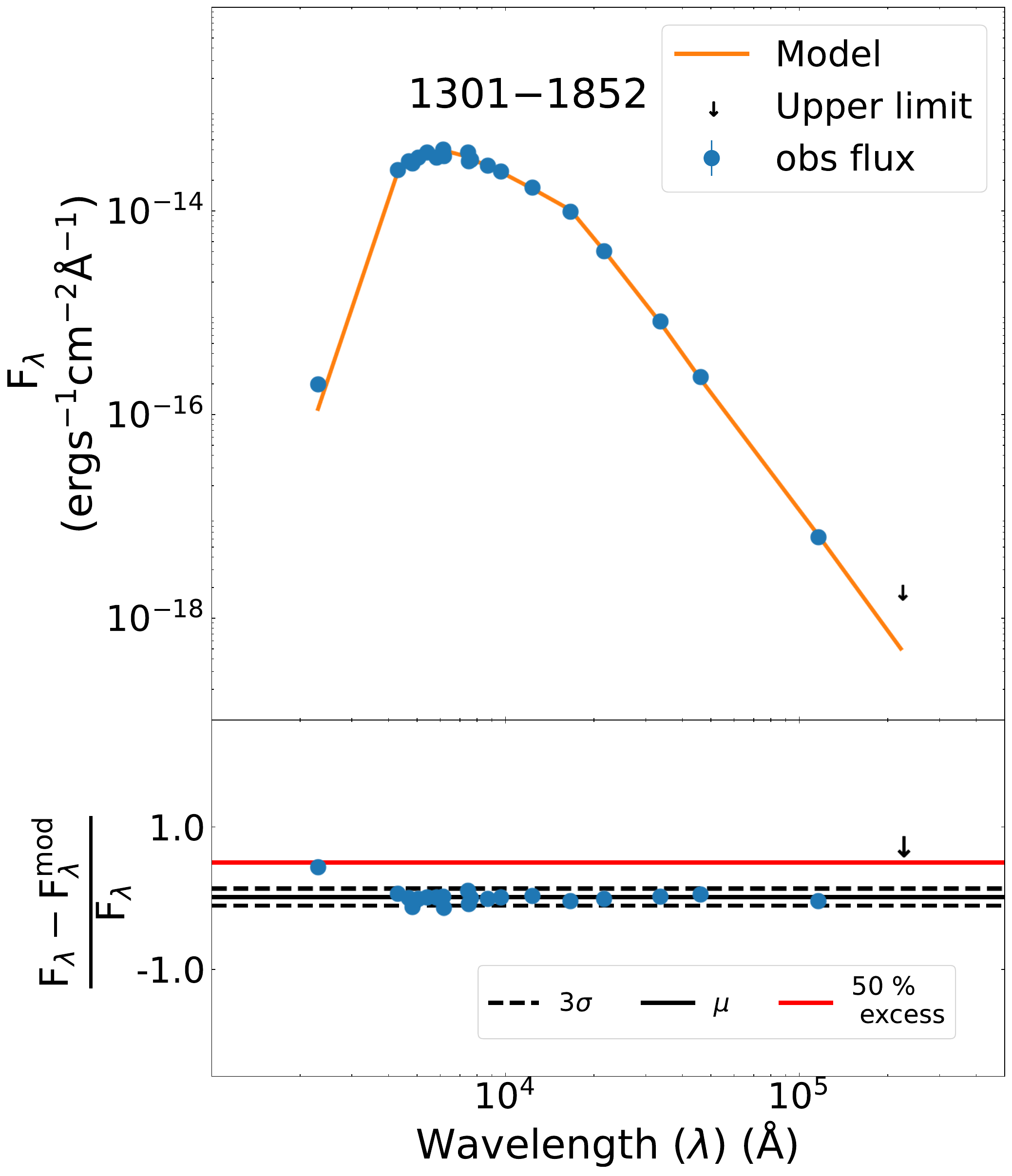}
\figsetgrpnote{Same as \autoref{fig:sed_example_no_excess},but for source 1301$-$1852}
\label{fig:sed_example_no_excess_29}

\figsetgrpend

\figsetgrpstart
\figsetgrpnum{2.30}
\figsetgrptitle{Image for figure 2_30}
\figsetplot{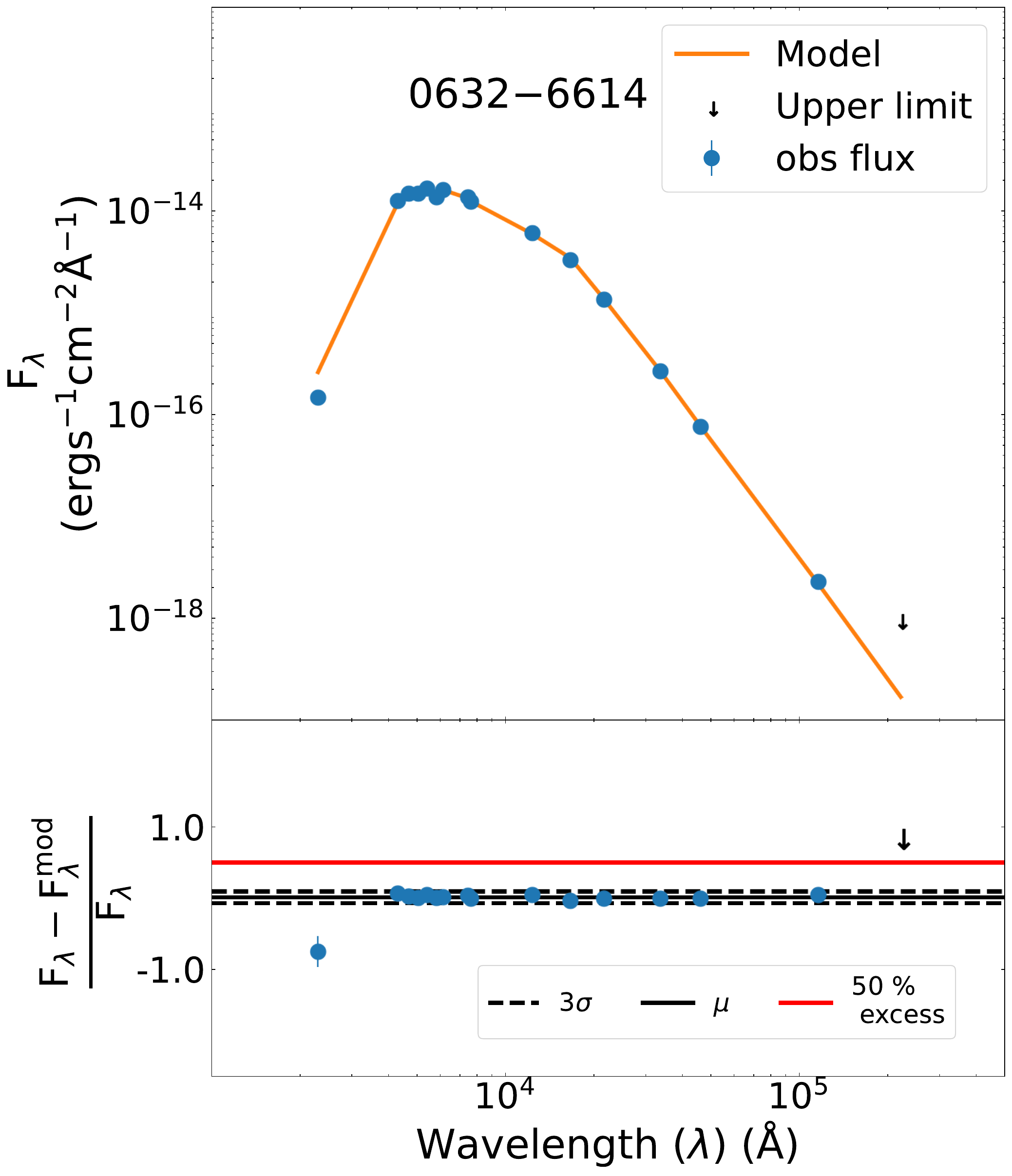}
\figsetgrpnote{Same as \autoref{fig:sed_example_no_excess},but for source 0632$-$6614}
\label{fig:sed_example_no_excess_30}

\figsetgrpend

\figsetgrpstart
\figsetgrpnum{2.31}
\figsetgrptitle{Image for figure 2_31}
\figsetplot{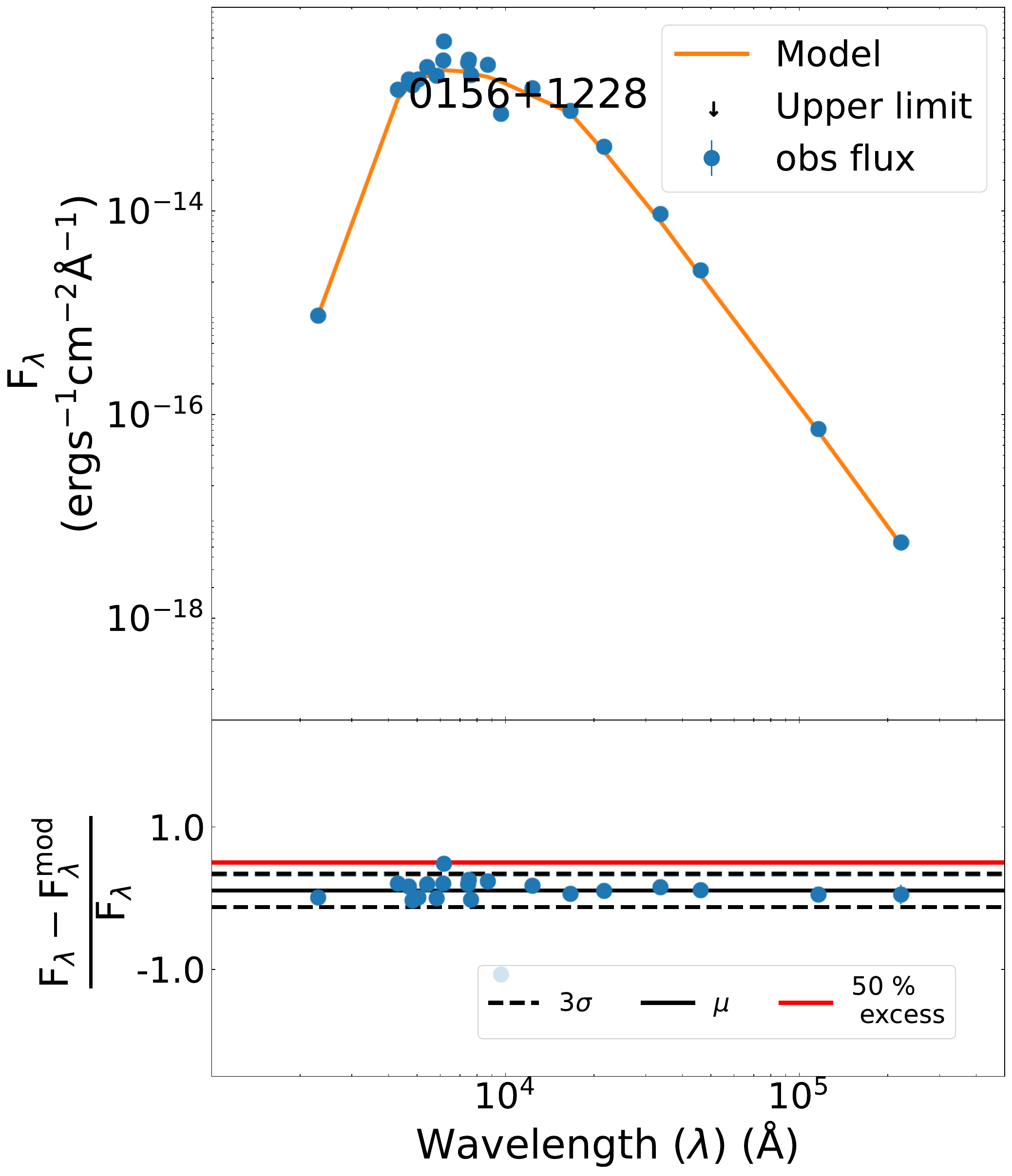}
\figsetgrpnote{Same as \autoref{fig:sed_example_no_excess},but for source 0156$+$1228}
\label{fig:sed_example_no_excess_31}

\figsetgrpend

\figsetgrpstart
\figsetgrpnum{2.32}
\figsetgrptitle{Image for figure 2_32}
\figsetplot{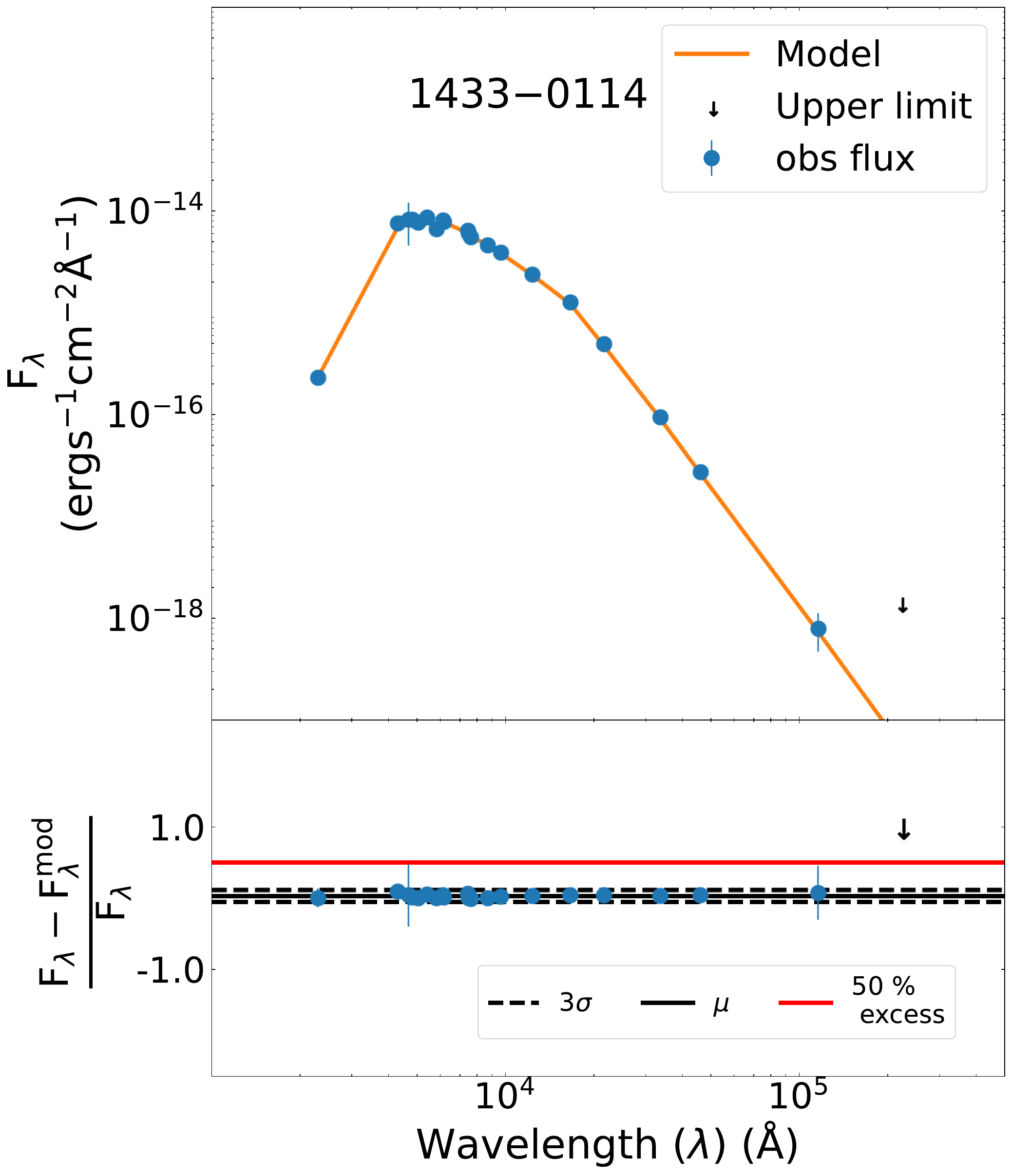}
\figsetgrpnote{Same as \autoref{fig:sed_example_no_excess},but for source 1433$-$0114}
\label{fig:sed_example_no_excess_32}

\figsetgrpend

\figsetgrpstart
\figsetgrpnum{2.33}
\figsetgrptitle{Image for figure 2_33}
\figsetplot{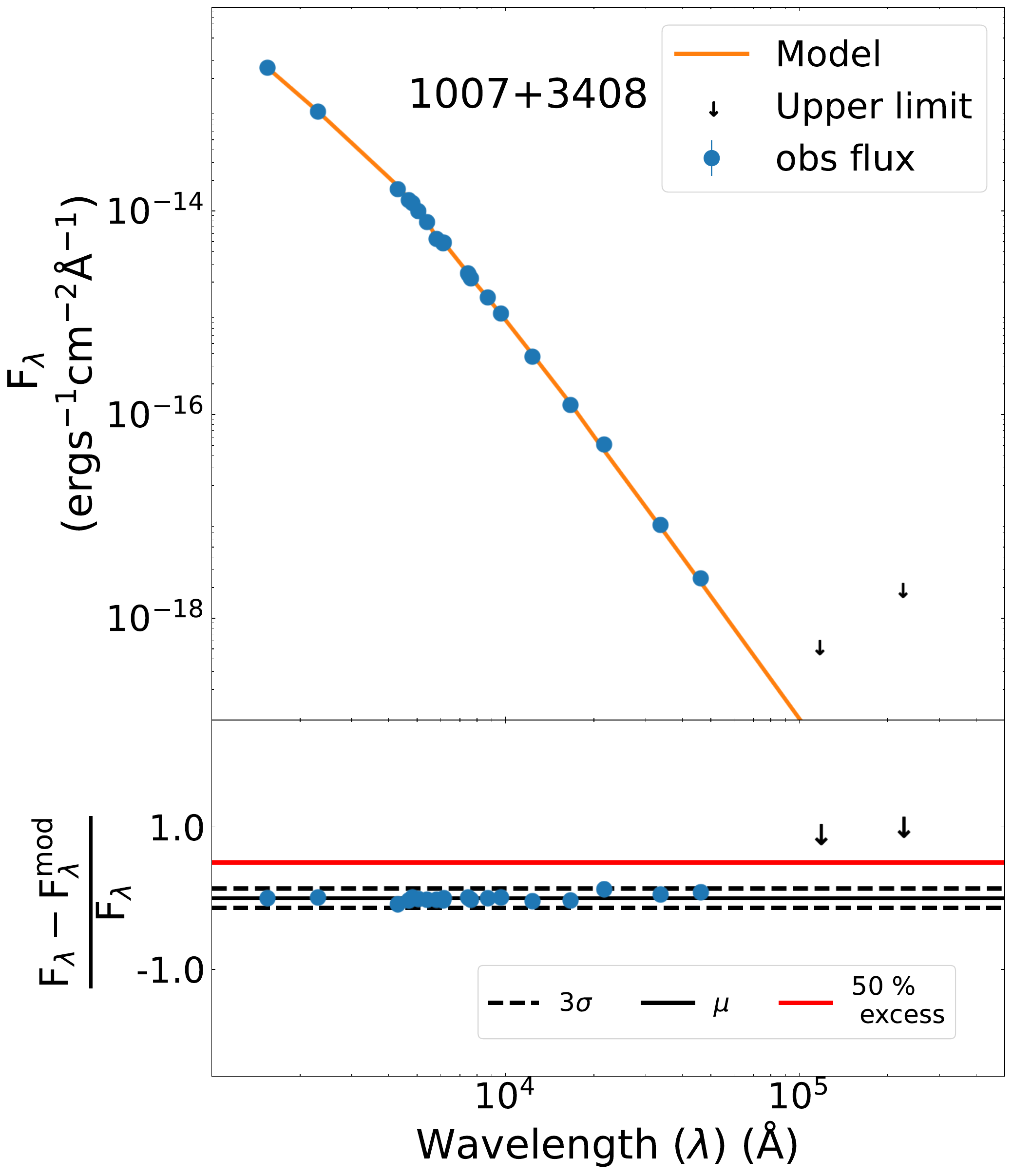}
\figsetgrpnote{Same as \autoref{fig:sed_example_no_excess},but for source 1007$+$3408}
\label{fig:sed_example_no_excess_33}

\figsetgrpend

\figsetend

\begin{figure}
\plotone{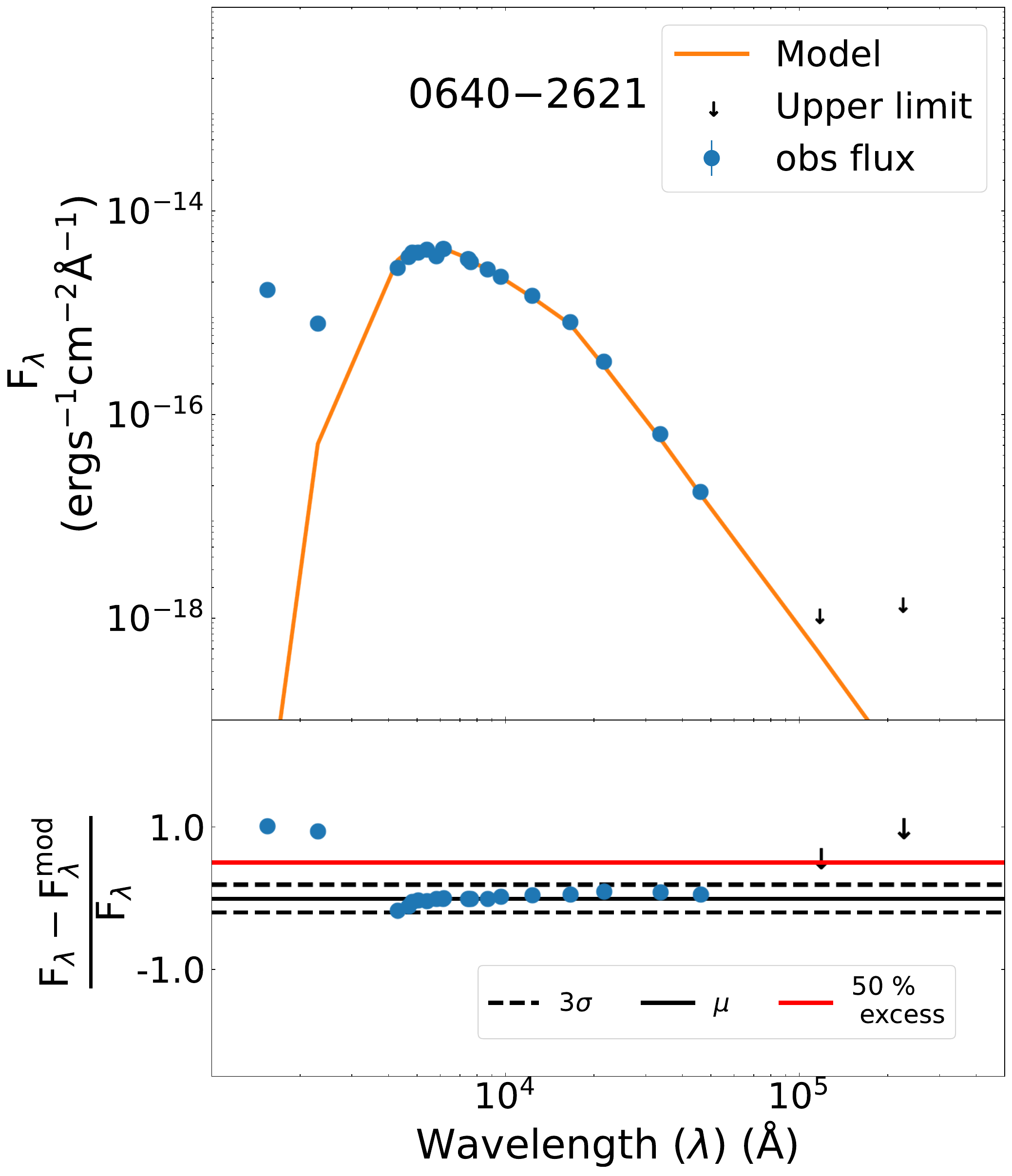}
\caption{Same as \autoref{fig:sed_example_no_excess}, but for source 0640$-$2621 which exhibits significant UV excess relative to the best-fit MS star SED. }
\label{fig:sed_example_excess}
\end{figure}

We cross match 187 DR--LC candidate sources listed in the ATF22 and SHA23 catalogs with the \galex\ archival data \citep{bianchi2017} using 3$''$ radius and find unique UV counterparts for 49 sources. The properties and call names for these sources are listed in \autoref{tab:all_data}.    
We cross-match these 49 sources with APASS DR9 \citep{Henden2015} and PanSTARRS-DR2 \citep{magnier2020} in the optical, and 2MASS \citep{Skrutskie2006}, ALLWISE \citep{wright2010} in the IR to construct the SED from UV to Near-IR (NIR). Often only the limiting IR flux is available, in these cases, we use the IR flux as an upper limit but do not use it for the SED fitting. 
We use the Kurucz model spectra \citep{castelli1997} in a large range of $\log(g/\acceleration)=4$--$5$ and effective temperature $\Teff/{\rm K}=3,500$--$50,000$ to fit the observed SEDs by synthetic MS star SEDs spanning NUV to NIR using the widely used and publicly available \vosa\ utility which estimates the best-fit stellar parameters based on astronomical input and available observed fluxes using $\chi^2$ optimization \citep{Bayo2008,vosa_doc}. 
We provide astronomical input to the \vosa\ utility while fitting the SED of each source. For example, we adopt the interstellar extinction values listed in the \galex\, catalog \citep[$A_v$; ][]{Schlegel1998}\footnote{We find that the $A_v$ values listed in the \galex\ catalog match reasonably well with those estimated using a three-dimensional dust map using the publicly available package mwdust.Green19 \citep{bovy_2016,green_2019}.}, the \gaia-estimated parallax for each source \citep{gaia_nss}, and use bounds around the \gaia-estimated photometric metallicity \citep{gaia_dr3}. 
\autoref{fig:sed_example_no_excess} shows source 2033$+$0758 as an example where the observed SED can be fitted with a single-component MS star. The best-fit SED provides us with a variety of stellar parameters for the LC including metallicity, $\tefflc$, $\lbollc$, and $\logglc$. 
\figsetgrpstart
\figsetgrpnum{3.1}
\figsetgrptitle{Image for figure 3_1}
\figsetplot{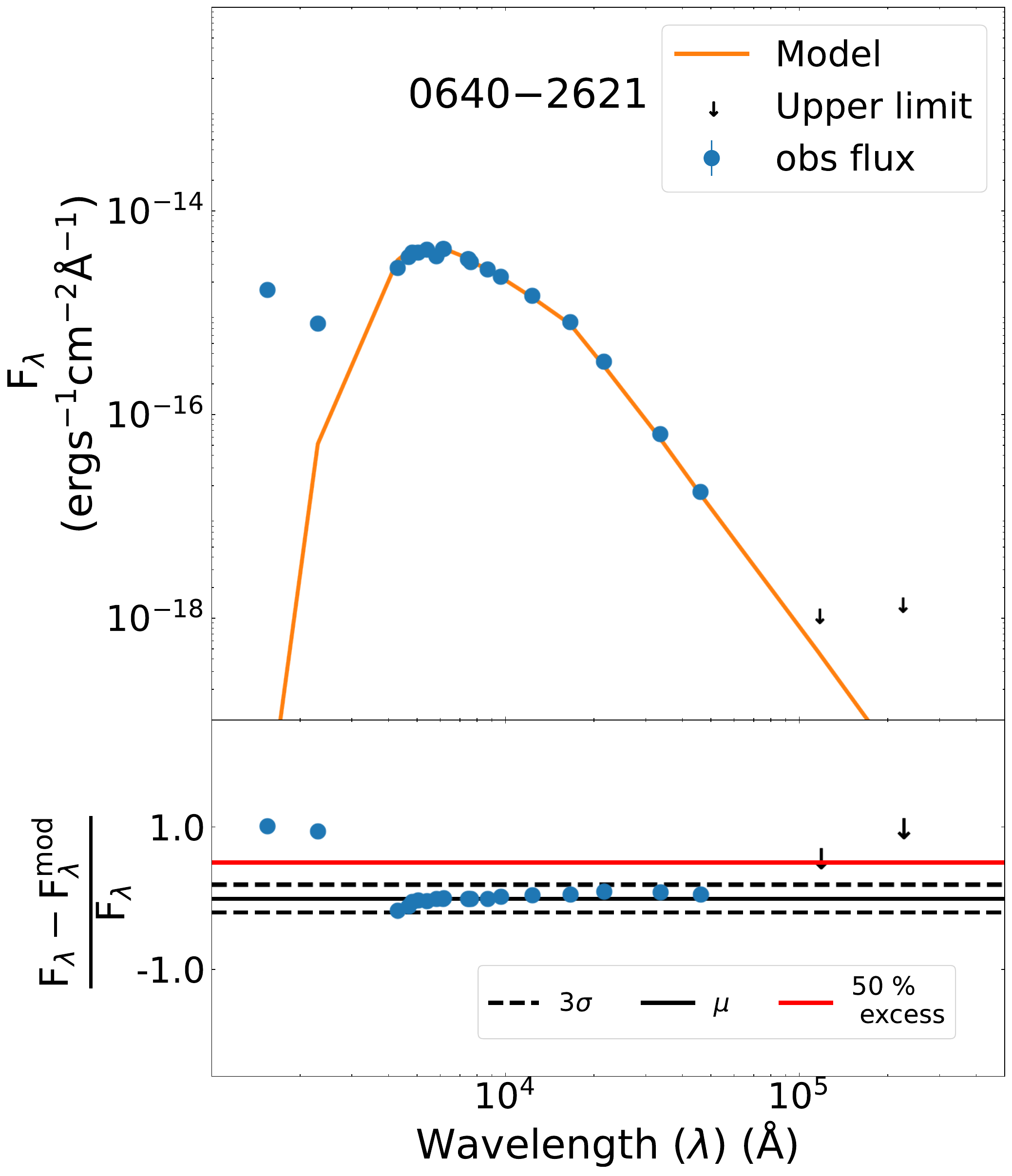}
\figsetgrpnote{Same as \autoref{fig:sed_example_excess},but for source 0640$-$2621}
\label{fig:sed_example_excess_1}

\figsetgrpend

\figsetgrpstart
\figsetgrpnum{3.2}
\figsetgrptitle{Image for figure 3_2}
\figsetplot{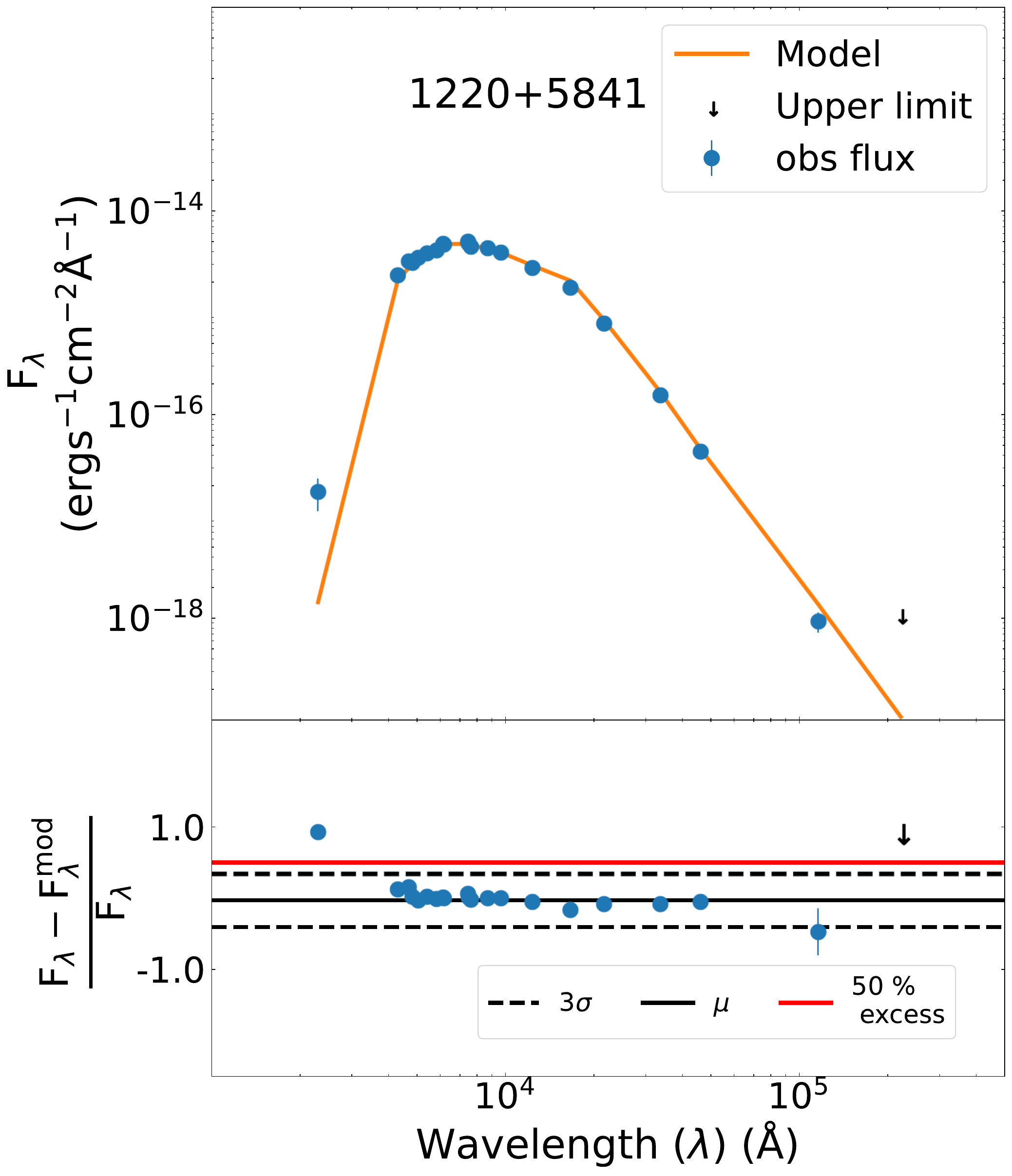}
\figsetgrpnote{Same as \autoref{fig:sed_example_excess},but for source 1220$+$5841}
\label{fig:sed_example_excess_2}

\figsetgrpend

\figsetgrpstart
\figsetgrpnum{3.3}
\figsetgrptitle{Image for figure 3_3}
\figsetplot{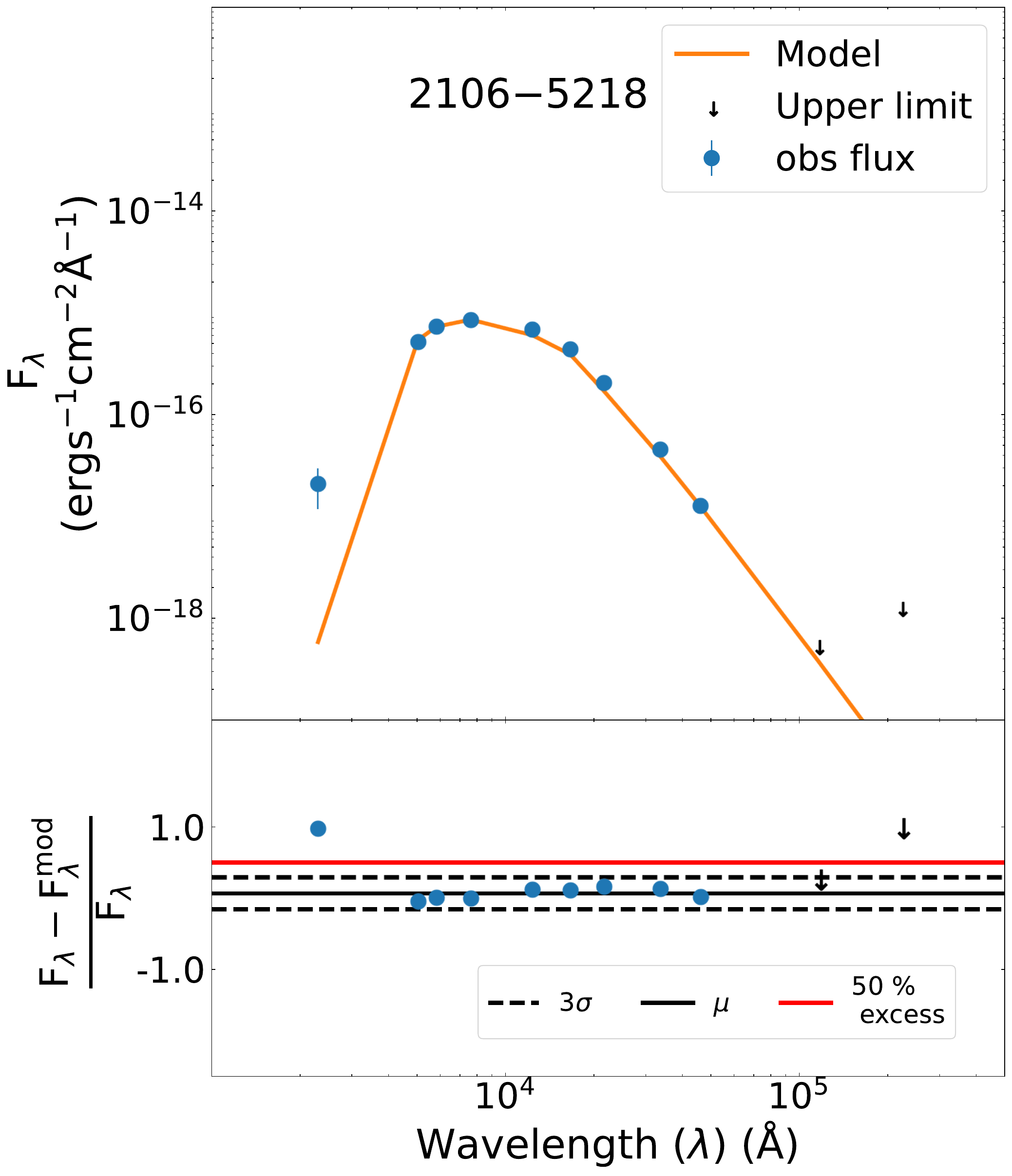}
\figsetgrpnote{Same as \autoref{fig:sed_example_excess},but for source 2106$-$5218}
\label{fig:sed_example_excess_3}

\figsetgrpend

\figsetgrpstart
\figsetgrpnum{3.4}
\figsetgrptitle{Image for figure 3_4}
\figsetplot{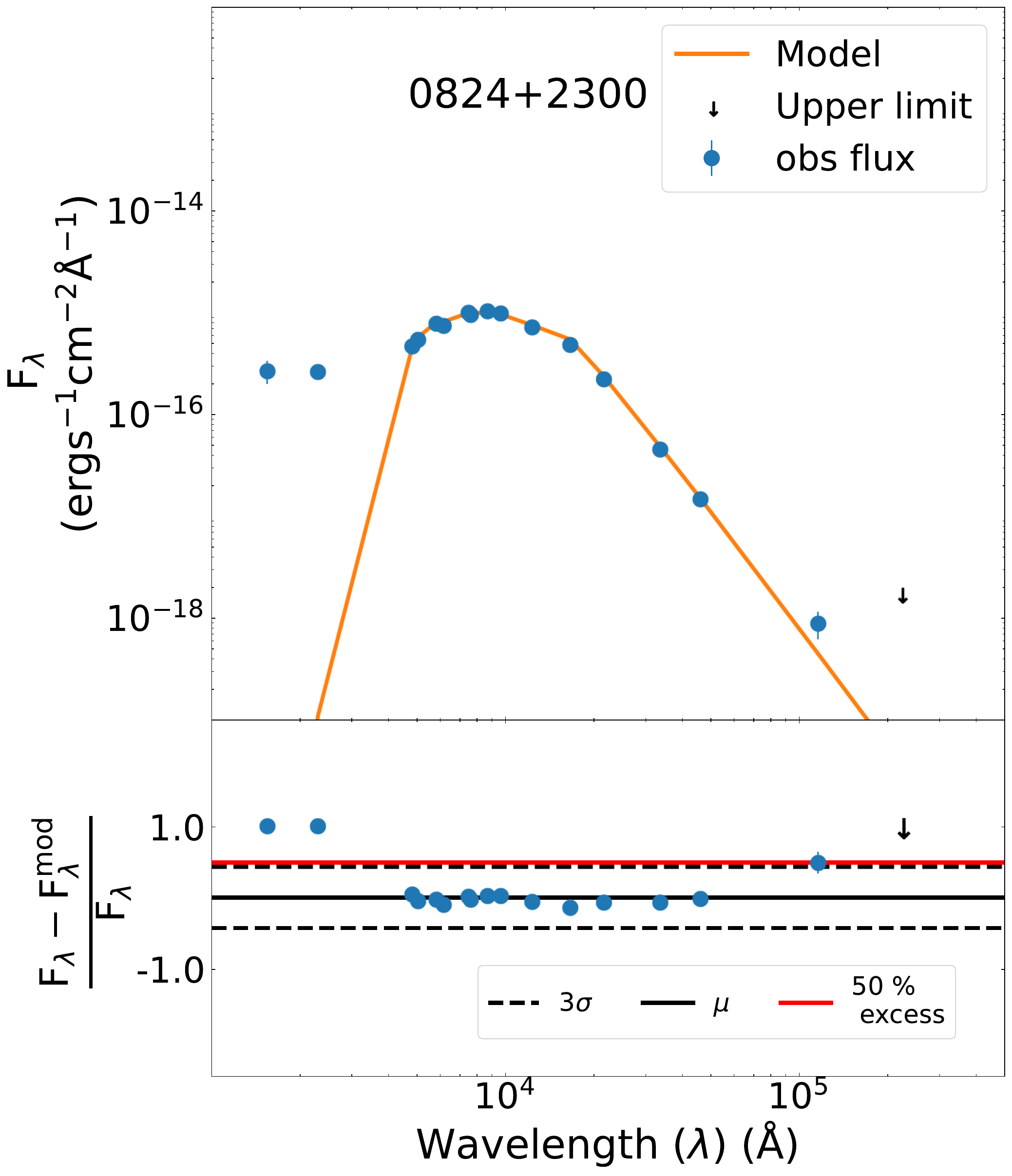}
\figsetgrpnote{Same as \autoref{fig:sed_example_excess},but for source 0824$+$2300}
\label{fig:sed_example_excess_4}

\figsetgrpend

\figsetgrpstart
\figsetgrpnum{3.5}
\figsetgrptitle{Image for figure 3_5}
\figsetplot{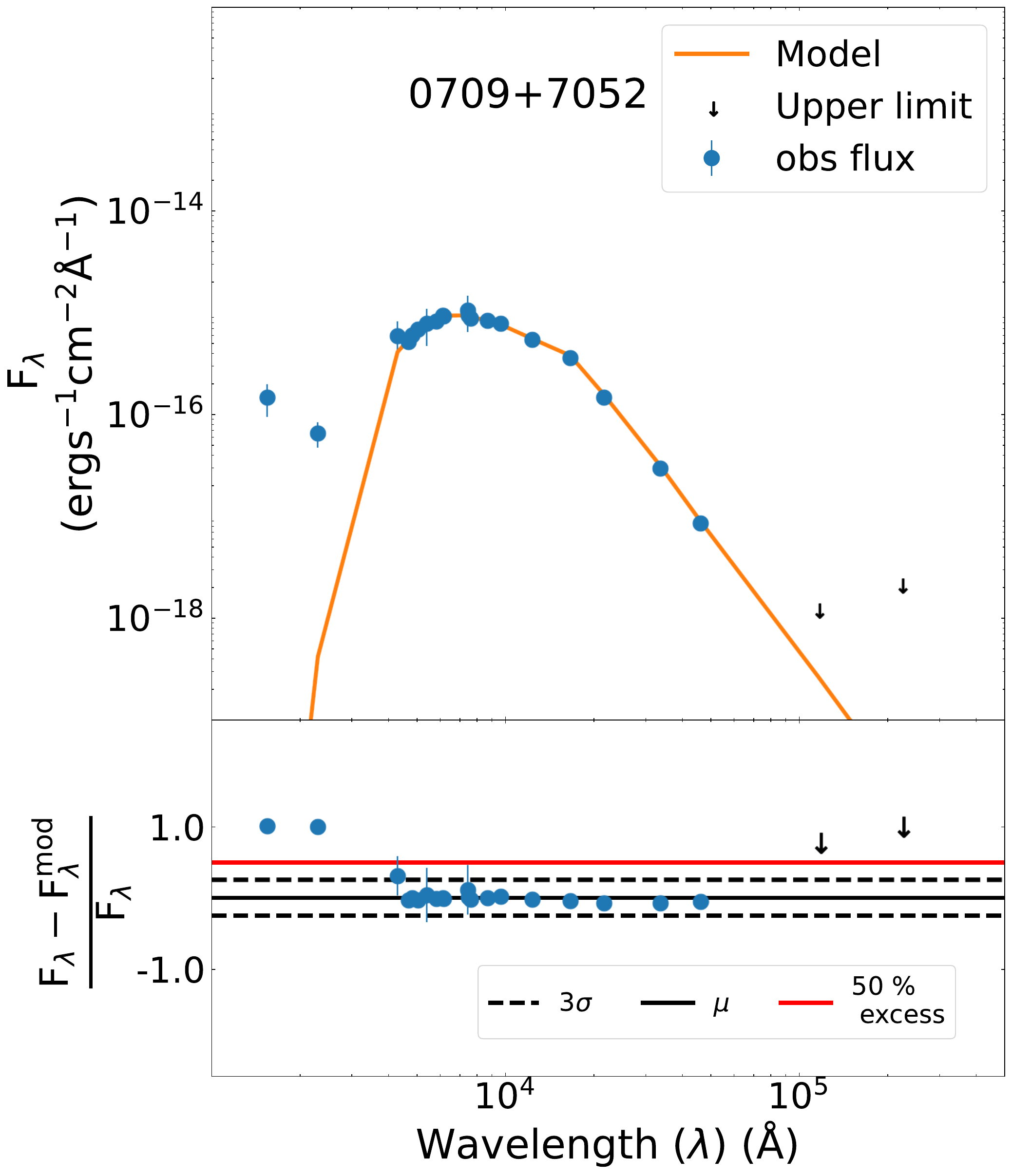}
\figsetgrpnote{Same as \autoref{fig:sed_example_excess},but for source 0709$+$7052}
\label{fig:sed_example_excess_5}

\figsetgrpend

\figsetgrpstart
\figsetgrpnum{3.6}
\figsetgrptitle{Image for figure 3_6}
\figsetplot{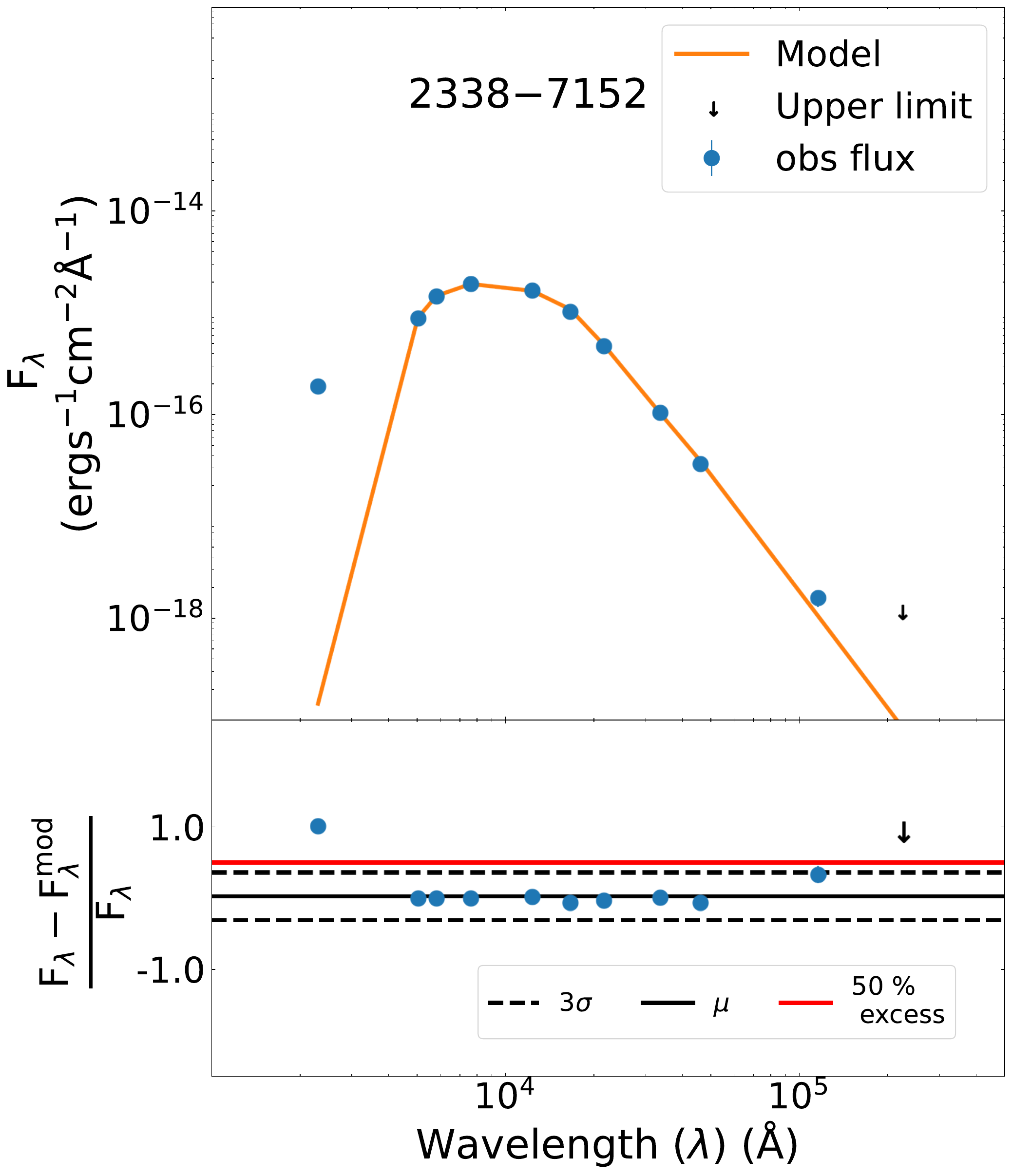}
\figsetgrpnote{Same as \autoref{fig:sed_example_excess},but for source 2338$-$7152}
\label{fig:sed_example_excess_6}

\figsetgrpend

\figsetgrpstart
\figsetgrpnum{3.7}
\figsetgrptitle{Image for figure 3_7}
\figsetplot{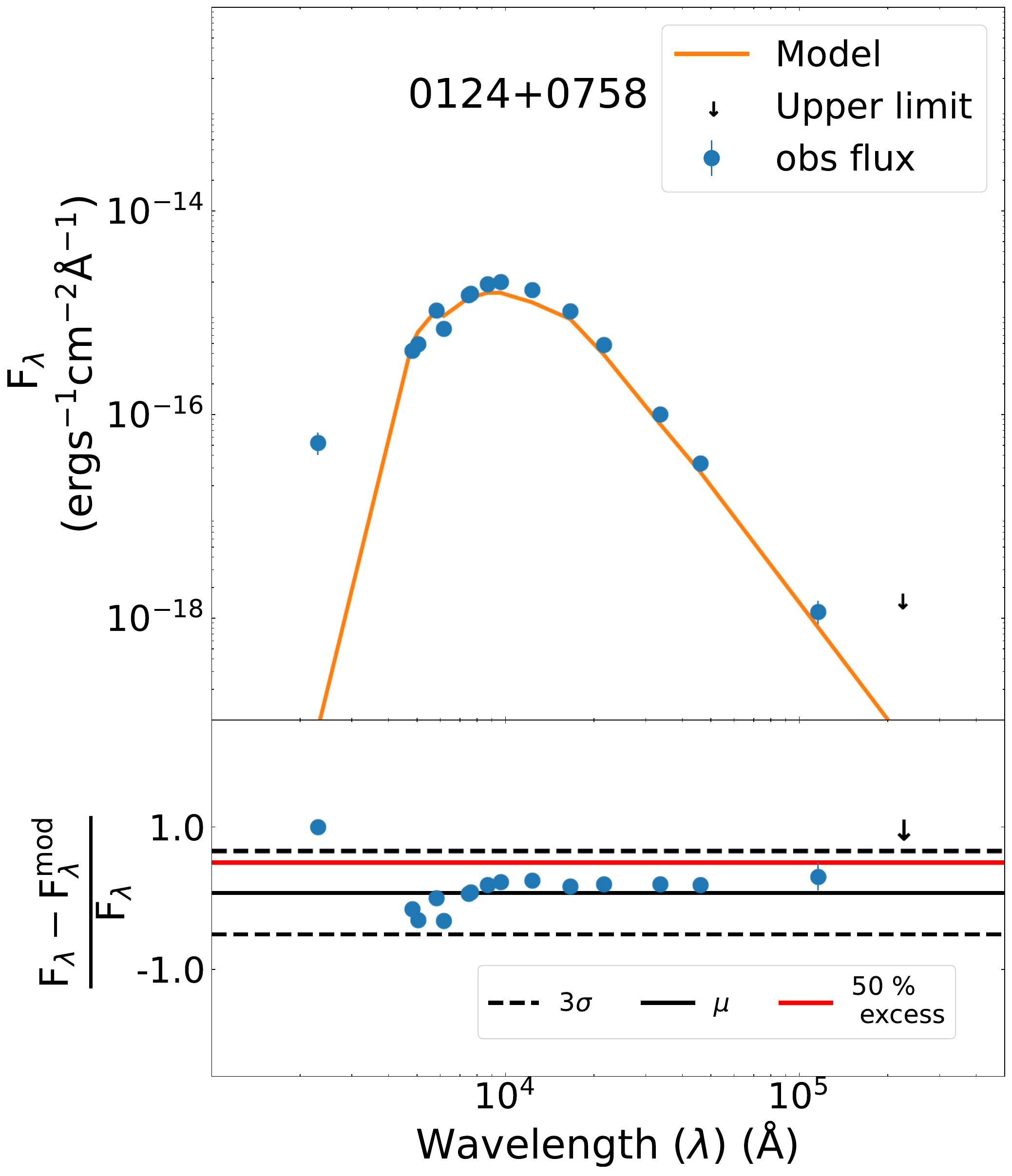}
\figsetgrpnote{Same as \autoref{fig:sed_example_excess},but for source 0124$+$0758}
\label{fig:sed_example_excess_7}

\figsetgrpend

\figsetgrpstart
\figsetgrpnum{3.8}
\figsetgrptitle{Image for figure 3_8}
\figsetplot{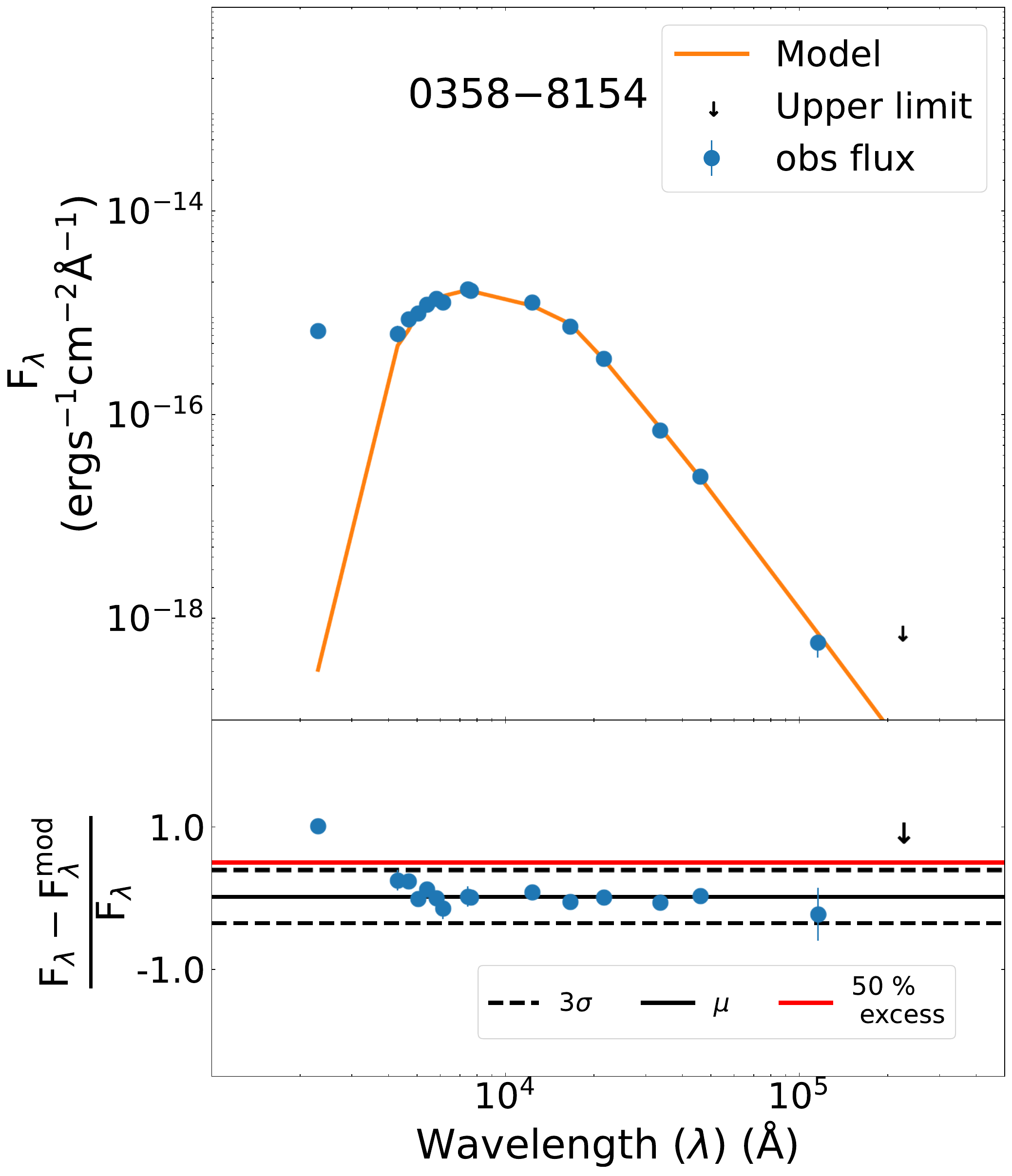}
\figsetgrpnote{Same as \autoref{fig:sed_example_excess},but for source 0358$-$8154}
\label{fig:sed_example_excess_8}

\figsetgrpend

\figsetgrpstart
\figsetgrpnum{3.9}
\figsetgrptitle{Image for figure 3_9}
\figsetplot{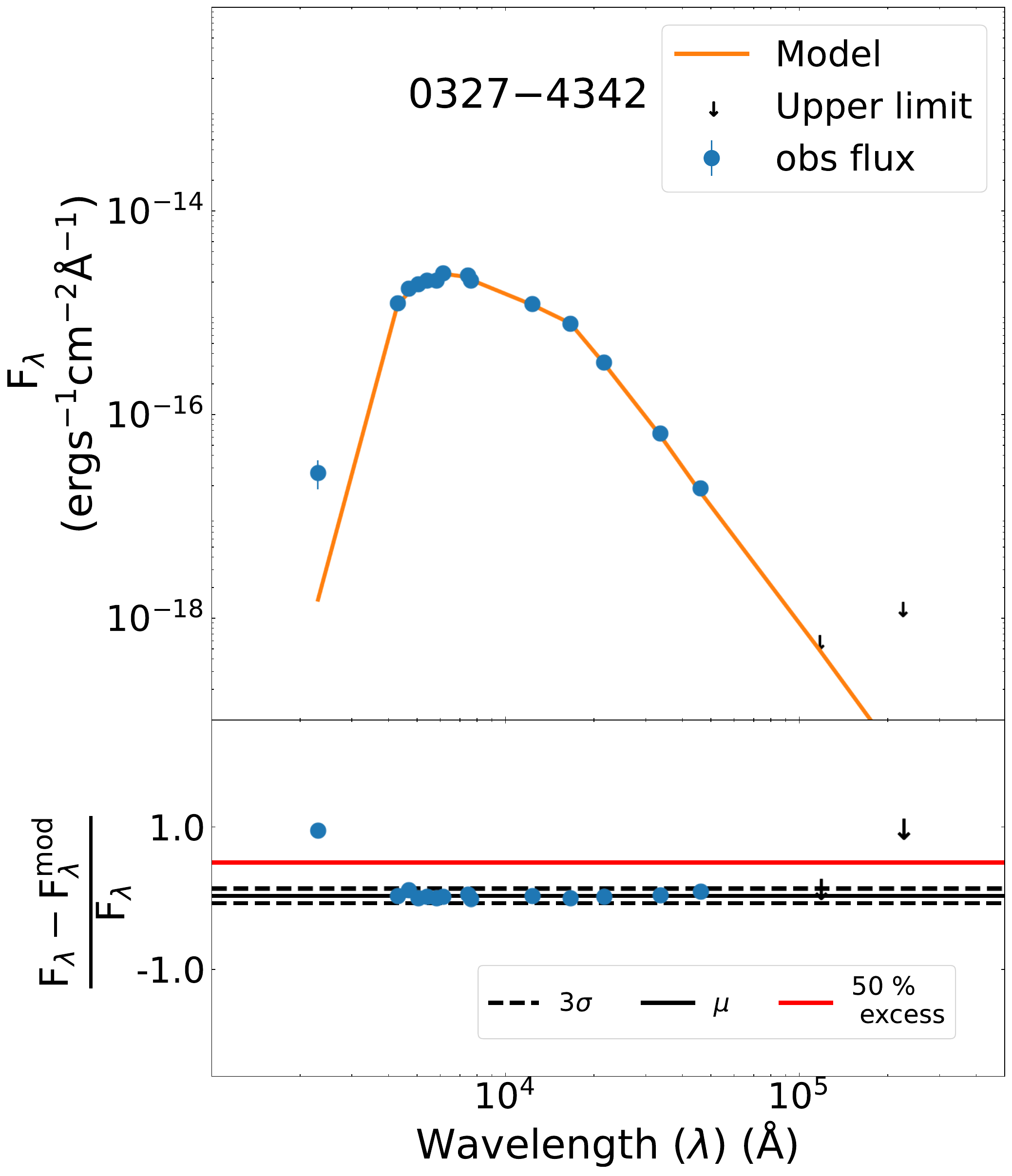}
\figsetgrpnote{Same as \autoref{fig:sed_example_excess},but for source 0327$-$4342}
\label{fig:sed_example_excess_9}

\figsetgrpend

\figsetgrpstart
\figsetgrpnum{3.10}
\figsetgrptitle{Image for figure 3_10}
\figsetplot{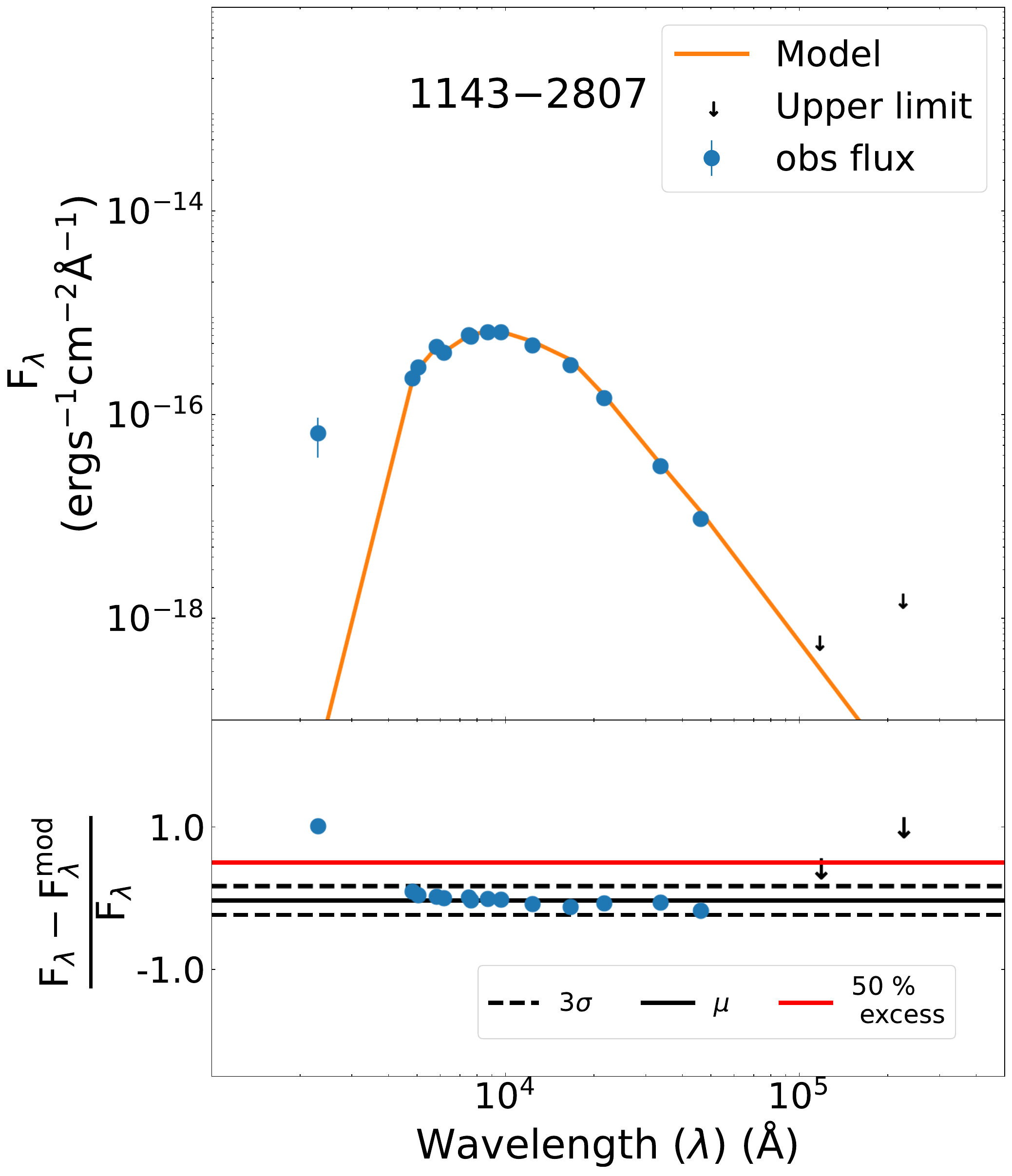}
\figsetgrpnote{Same as \autoref{fig:sed_example_excess},but for source 1143$-$2807}
\label{fig:sed_example_excess_10}

\figsetgrpend

\figsetgrpstart
\figsetgrpnum{3.11}
\figsetgrptitle{Image for figure 3_11}
\figsetplot{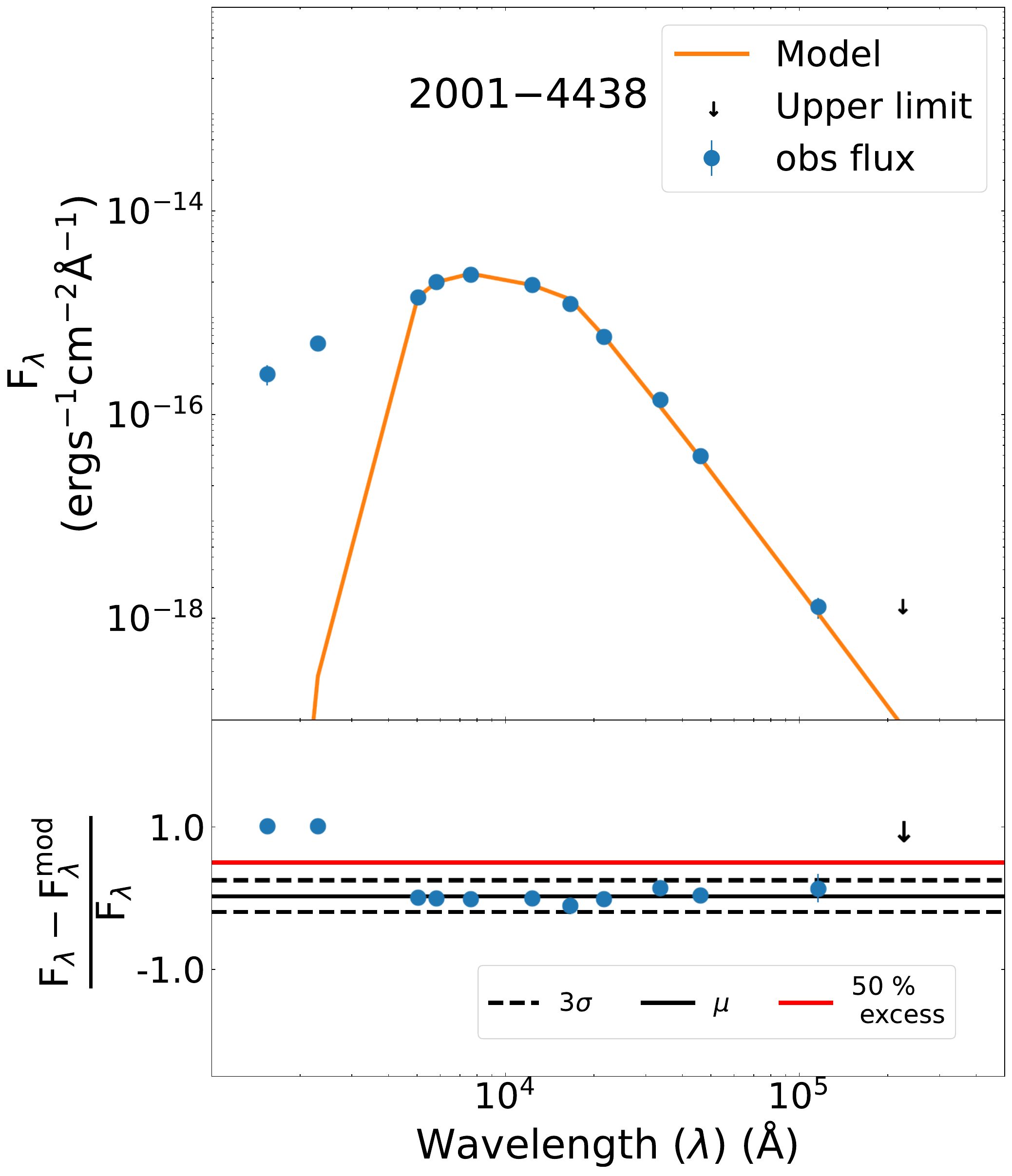}
\figsetgrpnote{Same as \autoref{fig:sed_example_excess},but for source 2001$-$4438}
\label{fig:sed_example_excess_11}

\figsetgrpend

\figsetgrpstart
\figsetgrpnum{3.12}
\figsetgrptitle{Image for figure 3_12}
\figsetplot{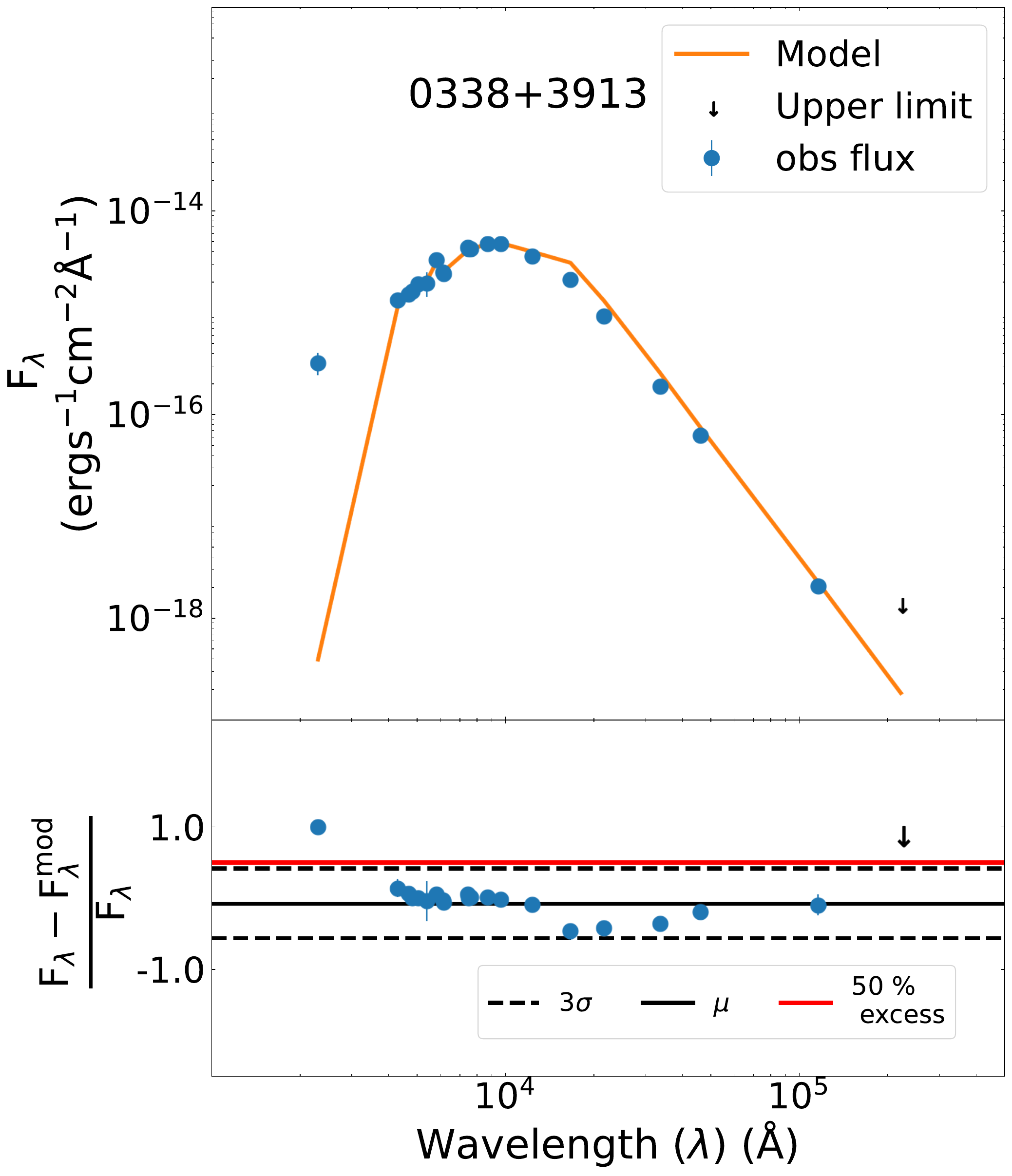}
\figsetgrpnote{Same as \autoref{fig:sed_example_excess},but for source 0338$+$3913}
\label{fig:sed_example_excess_12}

\figsetgrpend

\figsetgrpstart
\figsetgrpnum{3.13}
\figsetgrptitle{Image for figure 3_13}
\figsetplot{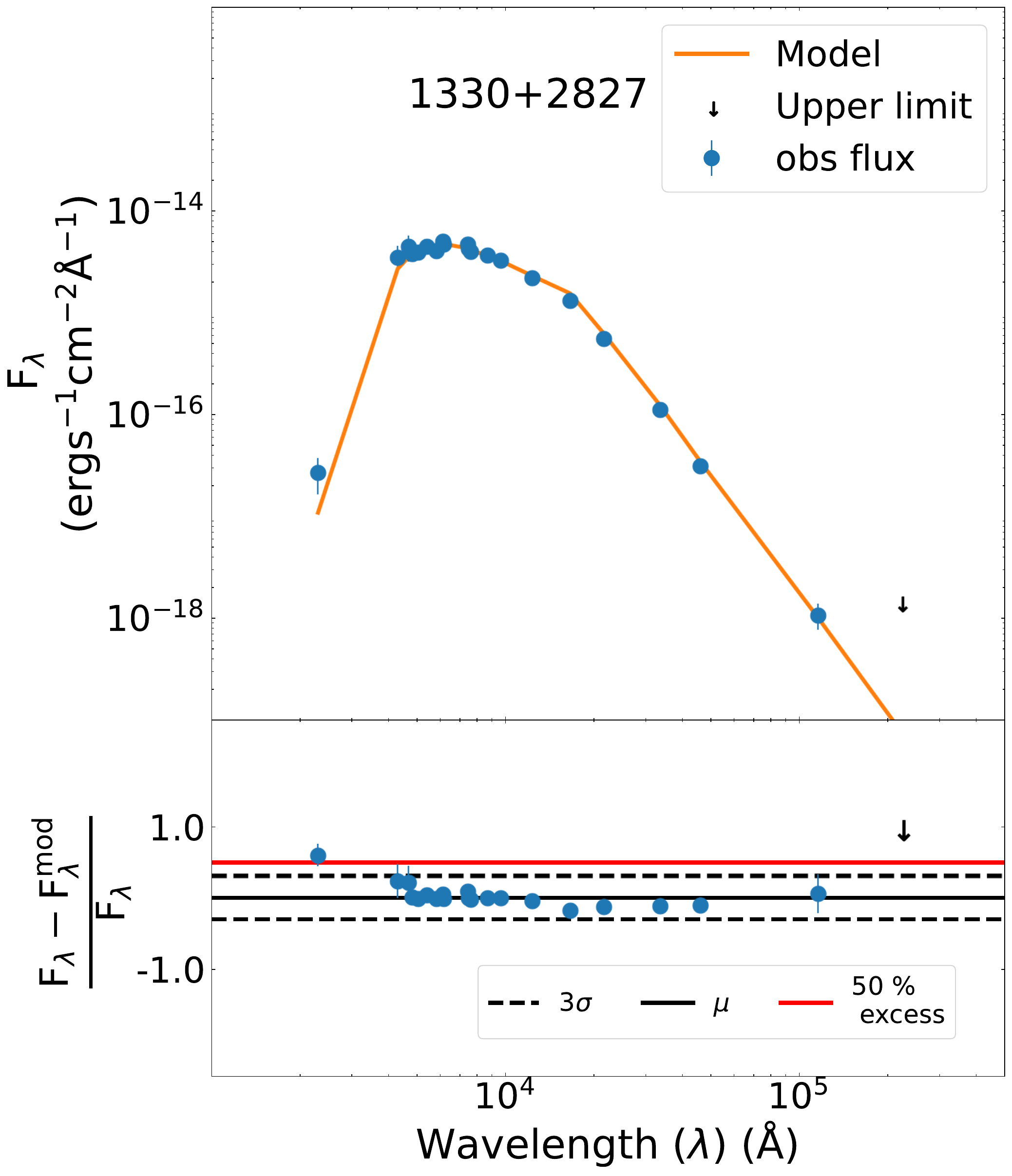}
\figsetgrpnote{Same as \autoref{fig:sed_example_excess},but for source 1330$+$2827}
\label{fig:sed_example_excess_13}

\figsetgrpend

\figsetgrpstart
\figsetgrpnum{3.14}
\figsetgrptitle{Image for figure 3_14}
\figsetplot{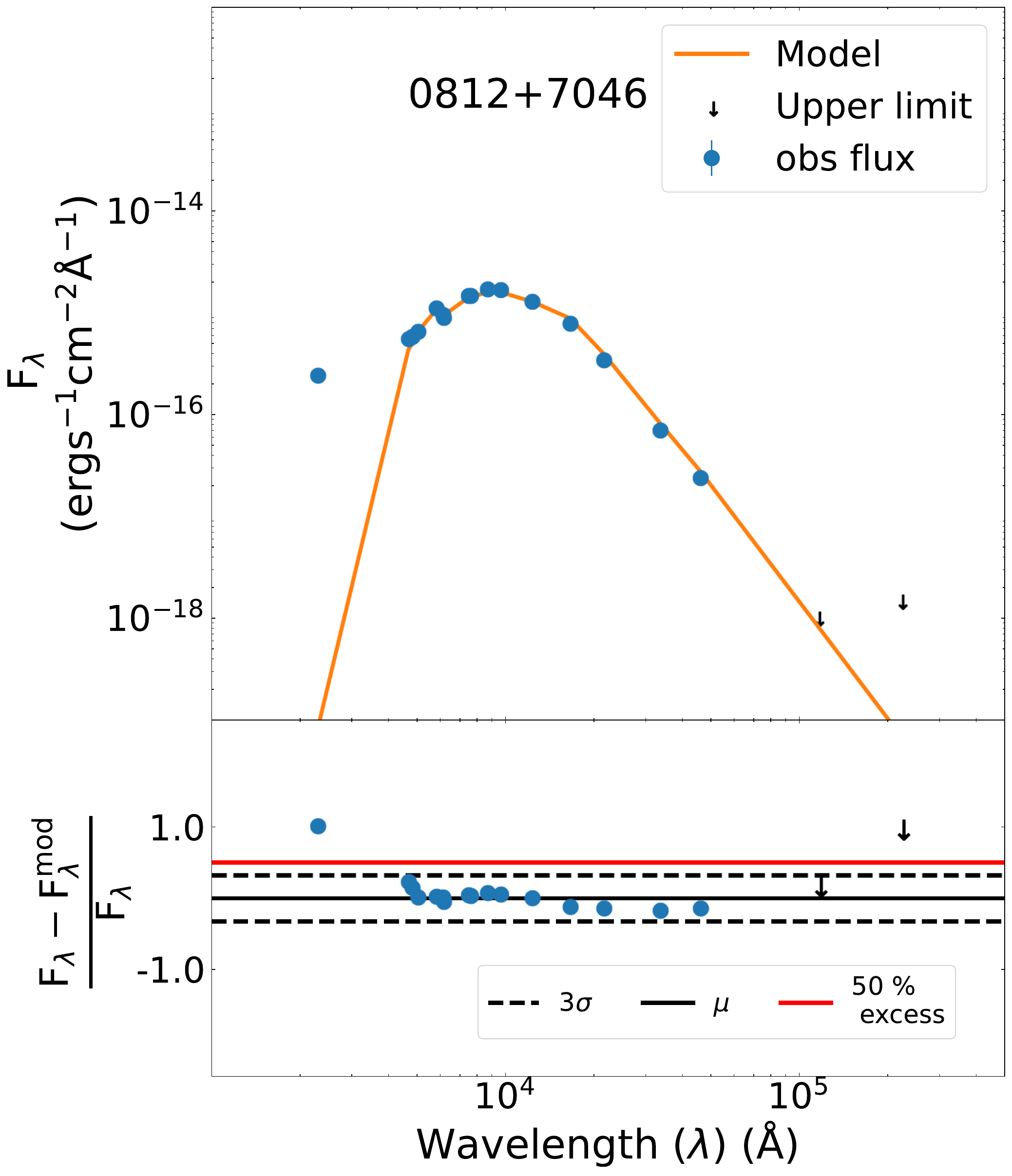}
\figsetgrpnote{Same as \autoref{fig:sed_example_excess},but for source 0812$+$7046}
\label{fig:sed_example_excess_14}

\figsetgrpend

\figsetgrpstart
\figsetgrpnum{3.15}
\figsetgrptitle{Image for figure 3_15}
\figsetplot{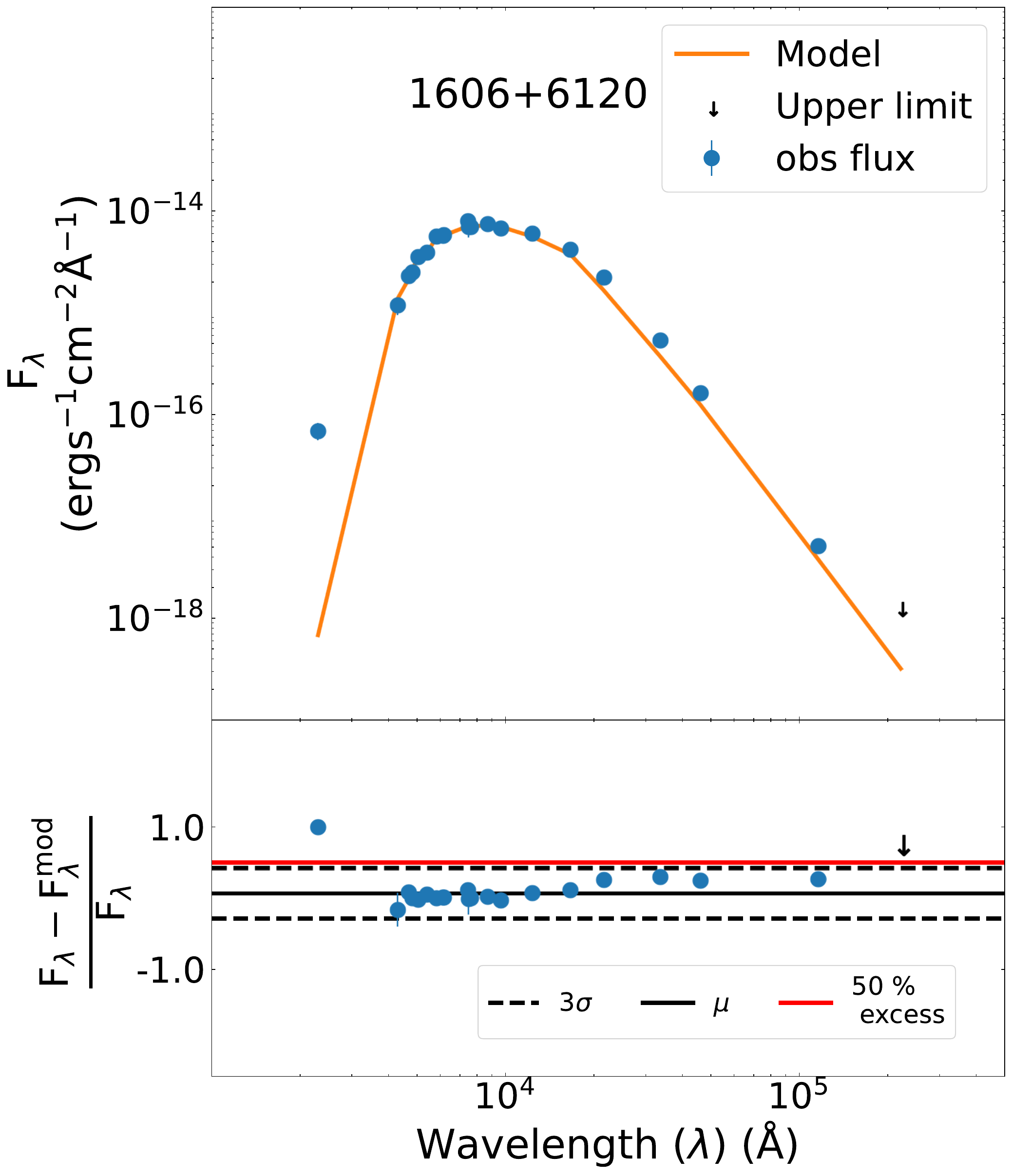}
\figsetgrpnote{Same as \autoref{fig:sed_example_excess},but for source 1606$+$6120}
\label{fig:sed_example_excess_15}

\figsetgrpend
\figsetend

For each source, we analyze the residuals in the observed SED with respect to the best-fit model SED of a MS star obtained from \vosa. 
If the UV flux is higher than three times the standard deviation found in the residuals across optical and IR, and shows a fractional excess of above $50\%$ in the UV, we consider that the source exhibits a significant UV excess. 
Note that these are stringent requirements; the latter ensures significant excess in UV flux and the former ensures that the excess is significantly higher compared to residuals in other bands. \autoref{fig:sed_example_excess} shows source 0640$-$2621 as an example with significant excess in FUV and NUV. 
If significant UV excess is found, we fit a composite MS--WD model to the observed SED to extract the WD properties including $\teffwd$ and $\lbolwd$, and update the LC properties. 
To model the contribution of the WD to the combined SED, we adopt the widely used Koester DA-type WD model spectra \citep{koester2010} in the range $\Teff/{\rm K}=5,000$--$80,000$ and $\log(g/\acceleration)=6.5$--$9.5$. 
We use \vosa-recommended Visual goodness of fit (Vgf$_{\mathrm{b}}$), similar to reduced $\chi^2$, but adopted for unknown or wrongly reported small flux errors, to justify composite SED fits \cite{vosa_doc}. When excess is found, the composite fit significantly improves $\rm{Vgf_b}$ (\autoref{tab:wd_data}). 
\subsection{Mass Estimates}
\label{S:methods_mass}
Single component (MS star only if there is no UV excess) or 
multi-component (MS--WD if there is significant UV excess) SED modelling provides us with excellent constraints on $\lbol$, $\Teff$, and the radius of the source (and components). However, we notice that the constraint on mass from the SED modelling is rather weak. This is because SEDs are not sufficiently sensitive to $\log\ g$ and this results in large errors in the estimated mass \citep{vosa_doc}. Hence, we constrain $\mlc$ using stellar evolution models created via \MIST\ \citep{MIST0, MIST1, paxton1, paxton2, paxton3, paxton4} and the constraints on the metallicity, $\lbollc$, and $\tefflc$ found from SED modeling. In particular, using \MIST, we evolve stars within a reasonable range ($0.10$--$1.45\,\msun$ with a grid resolution of $0.015\,\msun$) in zero-age MS mass 
from pre-MS to the MS turn-off adopting metallicity from the best-fit parameters of the SED. We create small 2D boxes around the SED-estimated $\lbollc$ and $\tefflc$ values for each source. The width and height of these small boxes are adopted to be $\delta\lbol/\lbol=\delta\Teff/\Teff=\pm5\%$. These adopted fractional errors are typical of SED modelling errors obtained from \vosa. We collect all stellar tracks passing through this box and assign $\mlc$ to the median value of the stellar masses for these tracks. The error in $\mlc$ is assigned to be the difference between the $25$ and $75$th percentiles for these masses. 

Astrometric solutions are thought to be acceptable if goodness of fit $<25$ and significance $>12$ \citep{halbwachs_2022}. We find that the maximum goodness of fit for our sources is 11.3. On the other hand, except for source 2100$-$2535 with a significance 6.82 all other sources have a significance $>12$, median, 5th, and 95th percentiles are 47.75, 14.03, 119.43. We use our estimated $\mlc$ and 
\begin{equation}
    \mathrm{AMRF}=\frac{\mdr/\mlc}{(1+\mdr/\mlc)^{2/3}}
    \label{eq:amrf}
\end{equation}
or 
\begin{equation}
    f_M = \frac{\mdr^{3}}{(\mdr + \mlc)^{2}}
    \label{eq:fm}
\end{equation}
calculated from \gaia's astrometric solution to estimate $\mdr$ for sources in the SHA23 or ATF22 catalogs, respectively. Hence, the $\mdr$ estimates depend on the accuracy of the astrometric solutions.

\section{Results}
\label{sec:results}
%
\subsection{Sources showing significant UV excess}
\label{sec:uv_excess}
\begin{figure}
\plotone{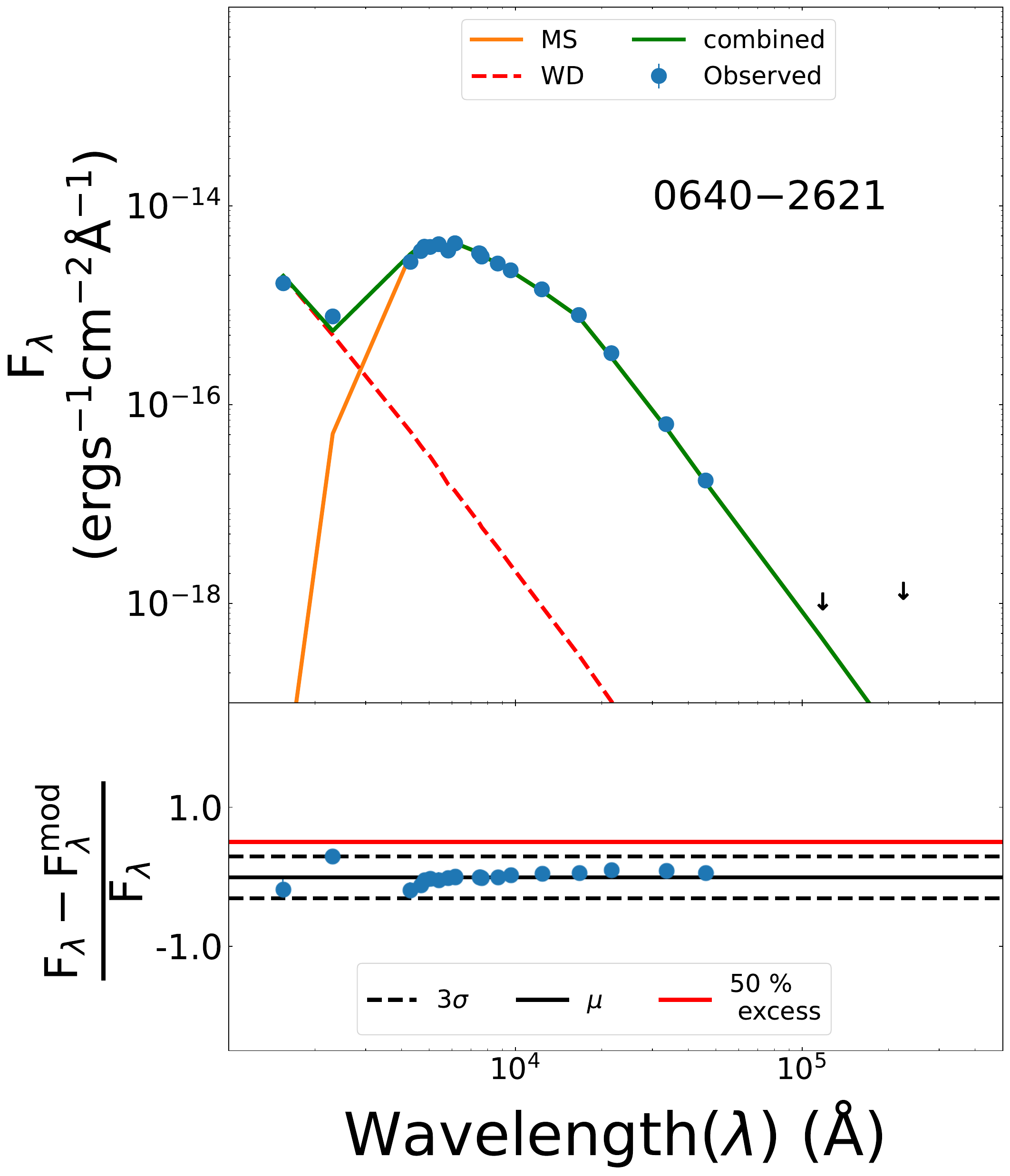}
\caption{SEDs and residuals for 0640$-$2621 showing significant UV excess. In the SED panels, the orange solid (red dashed) line shows the best-fit synthetic SED for the MS (WD) star. The solid green line shows the combined MS--WD flux. Blue dots denote observed flux (\autoref{sec:sed_fit}). Lines and symbols in the residual panels have the same meaning as in \autoref{fig:sed_example_excess}. The complete figure set (15 images) is available in the online journal.}
\label{fig:sed_wd_all}
\end{figure}

%
\figsetgrpstart
\figsetgrpnum{4.1}
\figsetgrptitle{Image for figure 4_1}
\figsetplot{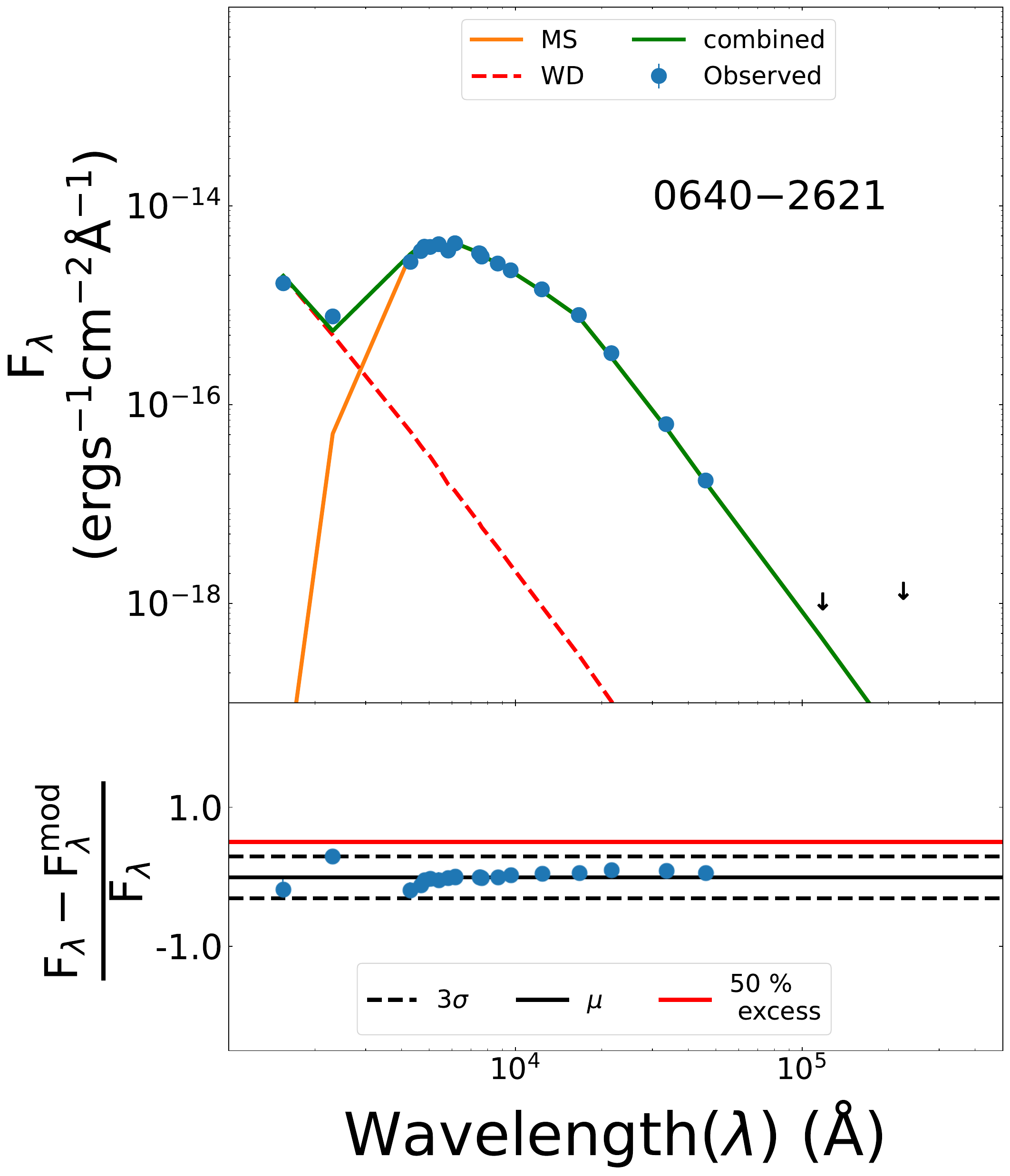}
\figsetgrpnote{Same as \autoref{fig:sed_wd_all},but for source 0640$-$2621}
\label{fig:sed_wd_all_1}

\figsetgrpend

\figsetgrpstart
\figsetgrpnum{4.2}
\figsetgrptitle{Image for figure 4_2}
\figsetplot{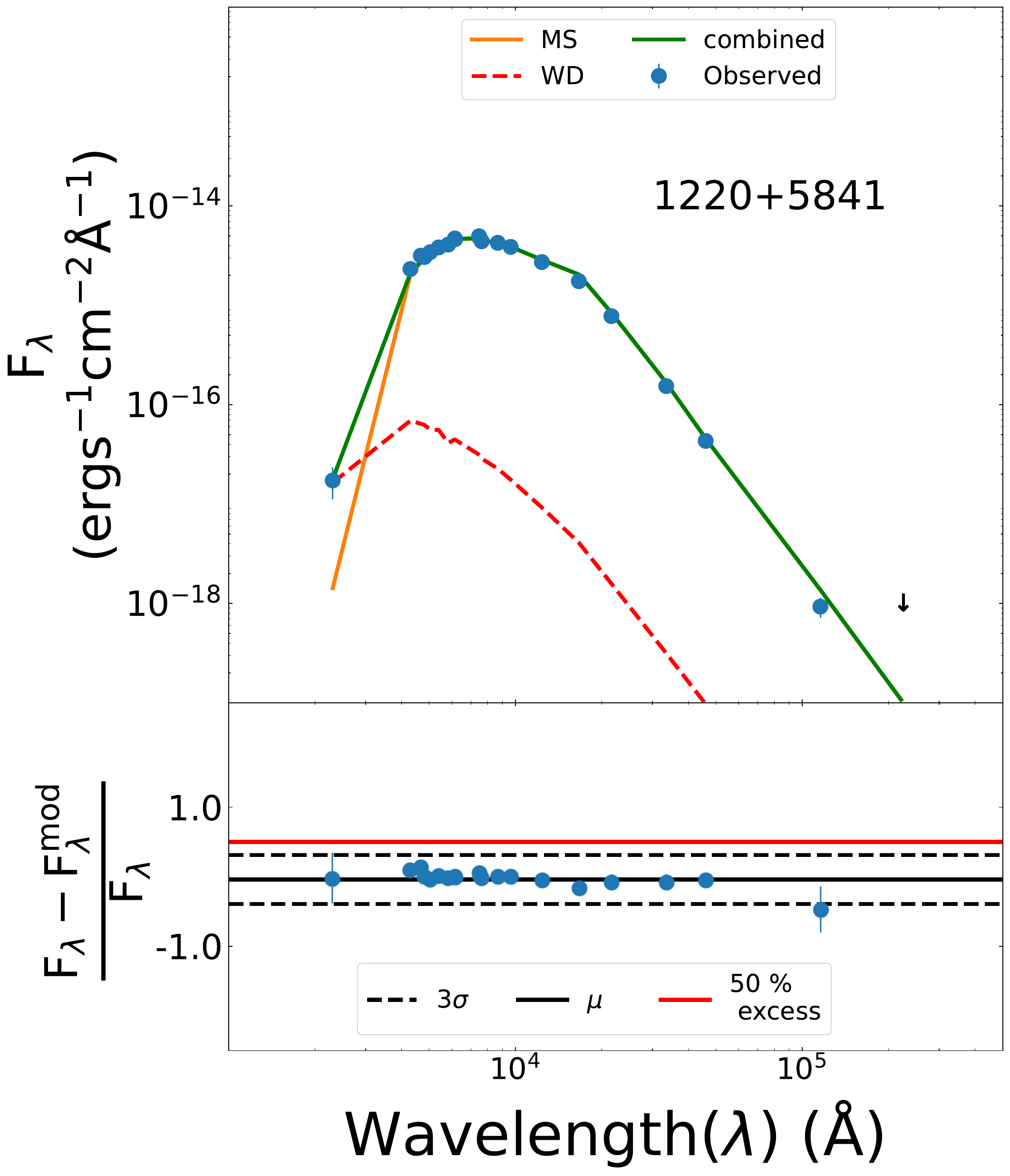}
\figsetgrpnote{Same as \autoref{fig:sed_wd_all},but for source 1220$+$5841}
\label{fig:sed_wd_all_2}

\figsetgrpend

\figsetgrpstart
\figsetgrpnum{4.3}
\figsetgrptitle{Image for figure 4_3}
\figsetplot{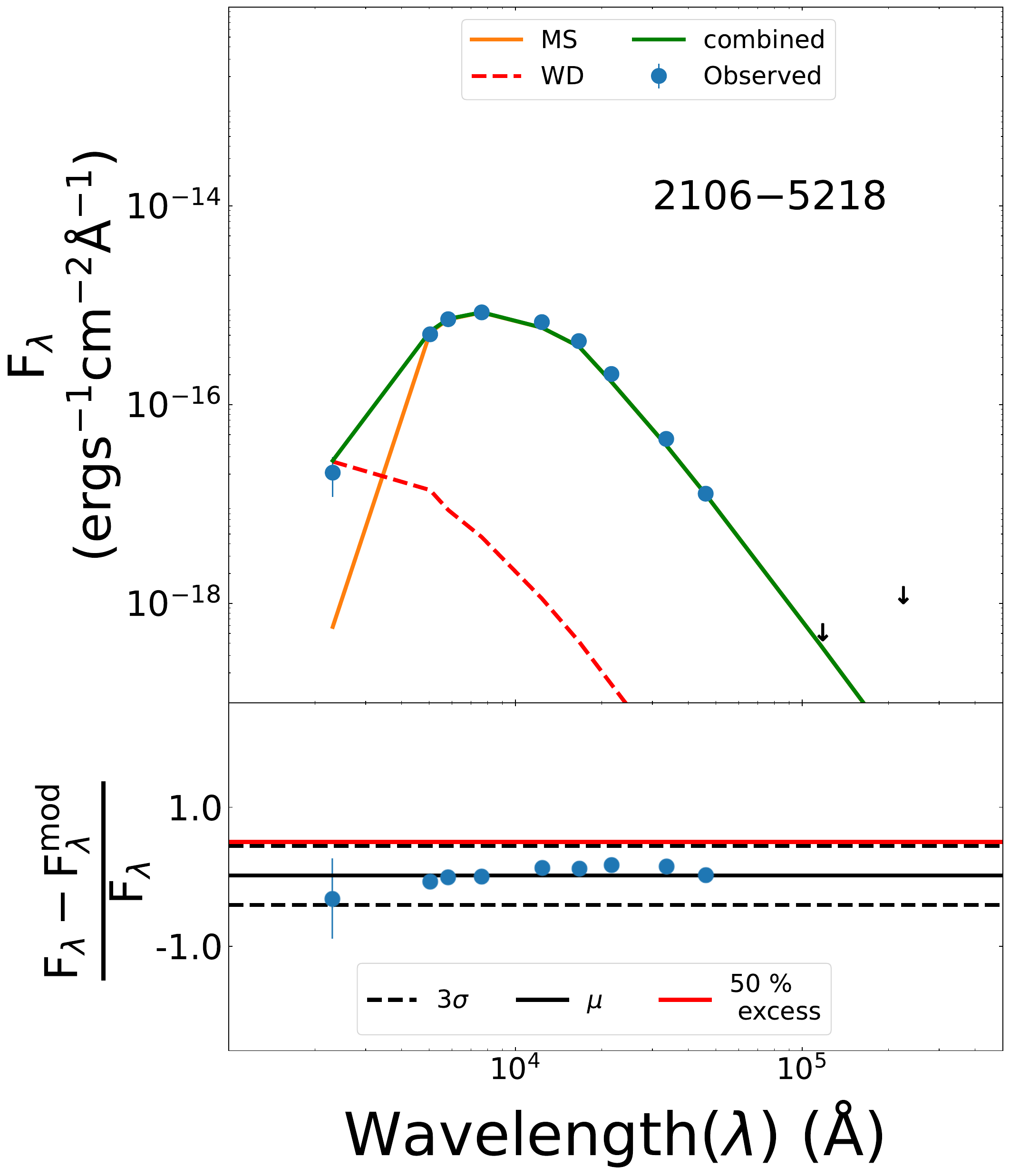}
\figsetgrpnote{Same as \autoref{fig:sed_wd_all},but for source 2106$-$5218}
\label{fig:sed_wd_all_3}

\figsetgrpend

\figsetgrpstart
\figsetgrpnum{4.4}
\figsetgrptitle{Image for figure 4_4}
\figsetplot{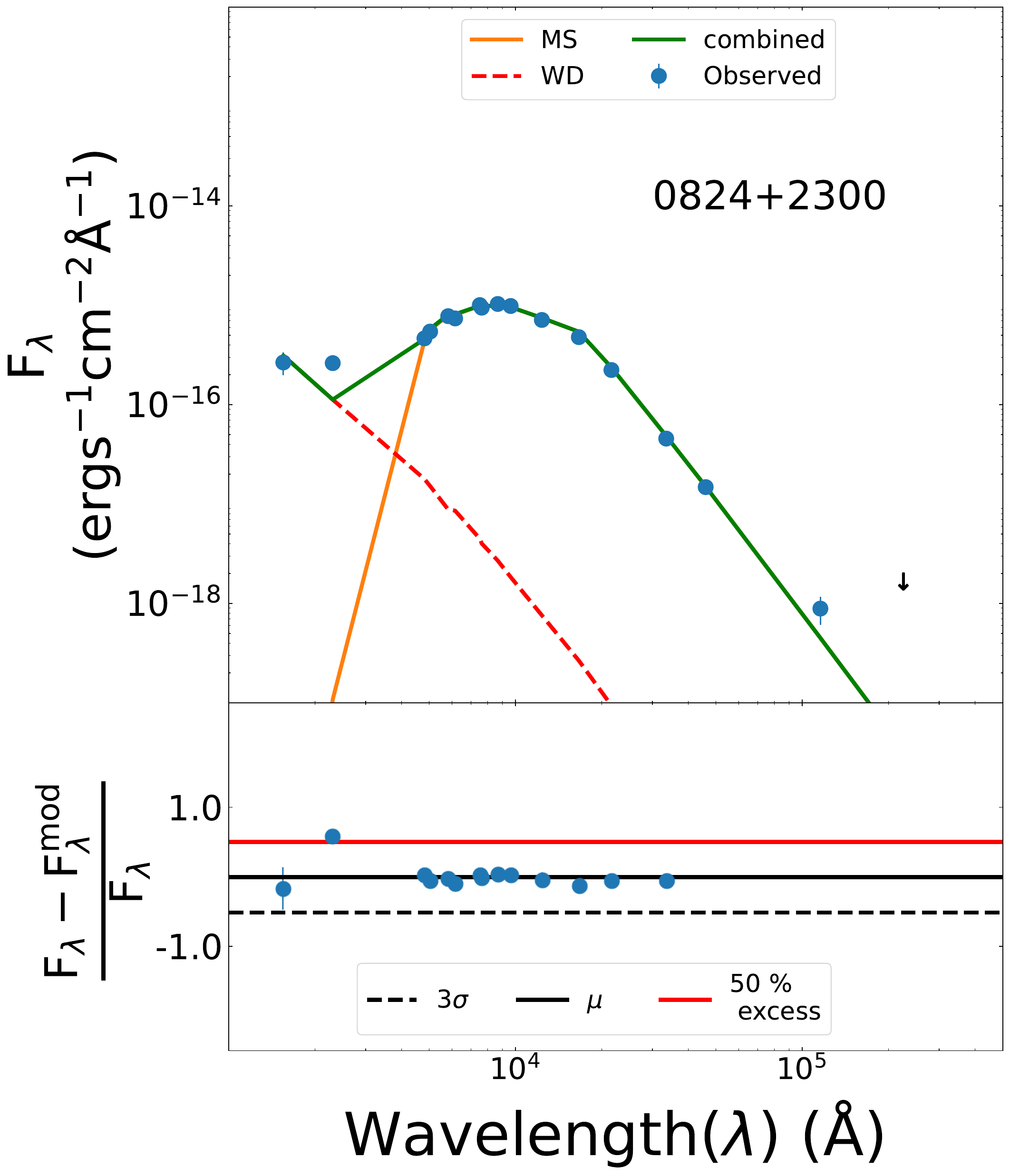}
\figsetgrpnote{Same as \autoref{fig:sed_wd_all},but for source 0824$+$2300}
\label{fig:sed_wd_all_4}

\figsetgrpend

\figsetgrpstart
\figsetgrpnum{4.5}
\figsetgrptitle{Image for figure 4_5}
\figsetplot{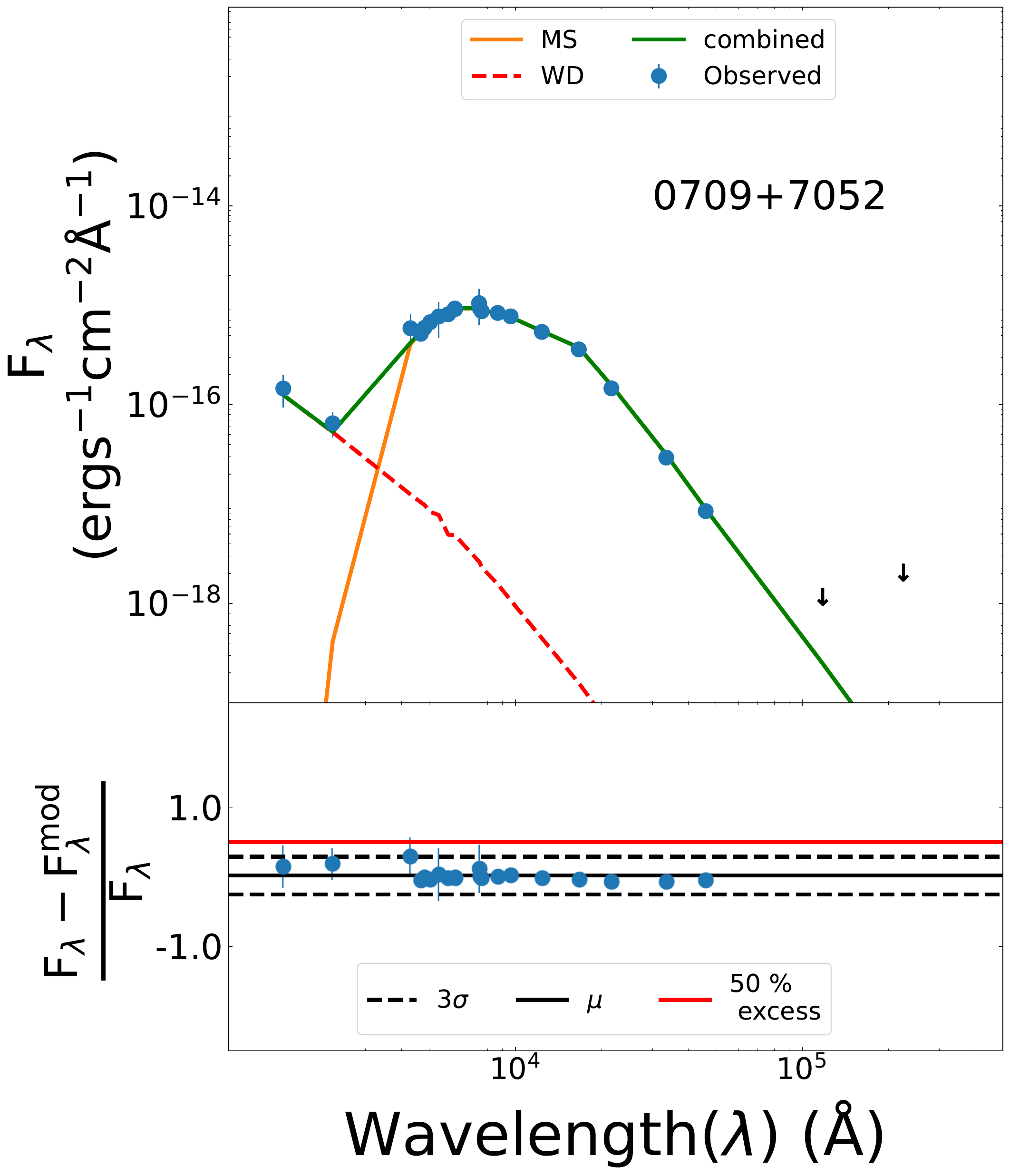}
\figsetgrpnote{Same as \autoref{fig:sed_wd_all},but for source 0709$+$7052}
\label{fig:sed_wd_all_5}

\figsetgrpend

\figsetgrpstart
\figsetgrpnum{4.6}
\figsetgrptitle{Image for figure 4_6}
\figsetplot{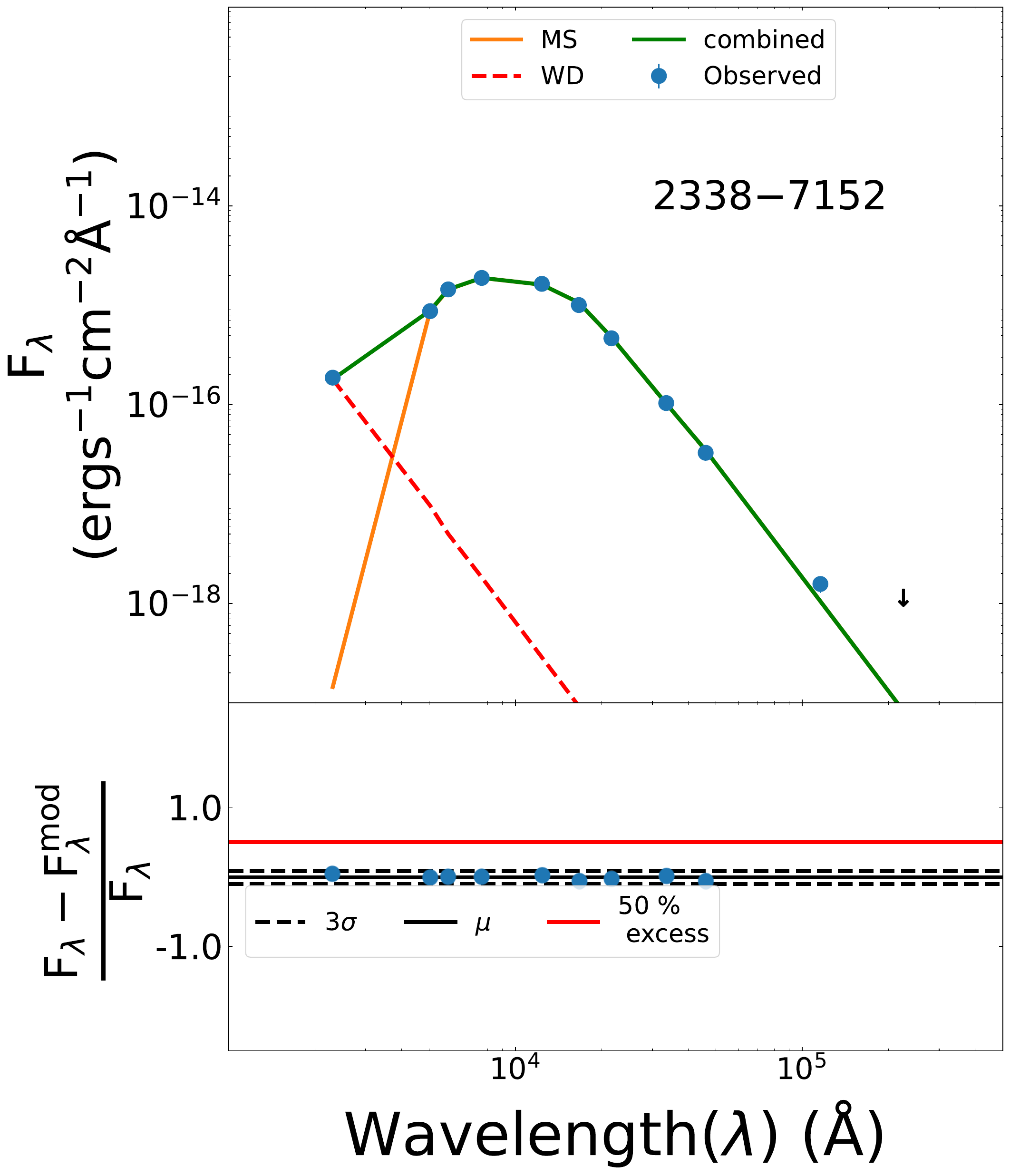}
\figsetgrpnote{Same as \autoref{fig:sed_wd_all},but for source 2338$-$7152}
\label{fig:sed_wd_all_6}

\figsetgrpend

\figsetgrpstart
\figsetgrpnum{4.7}
\figsetgrptitle{Image for figure 4_7}
\figsetplot{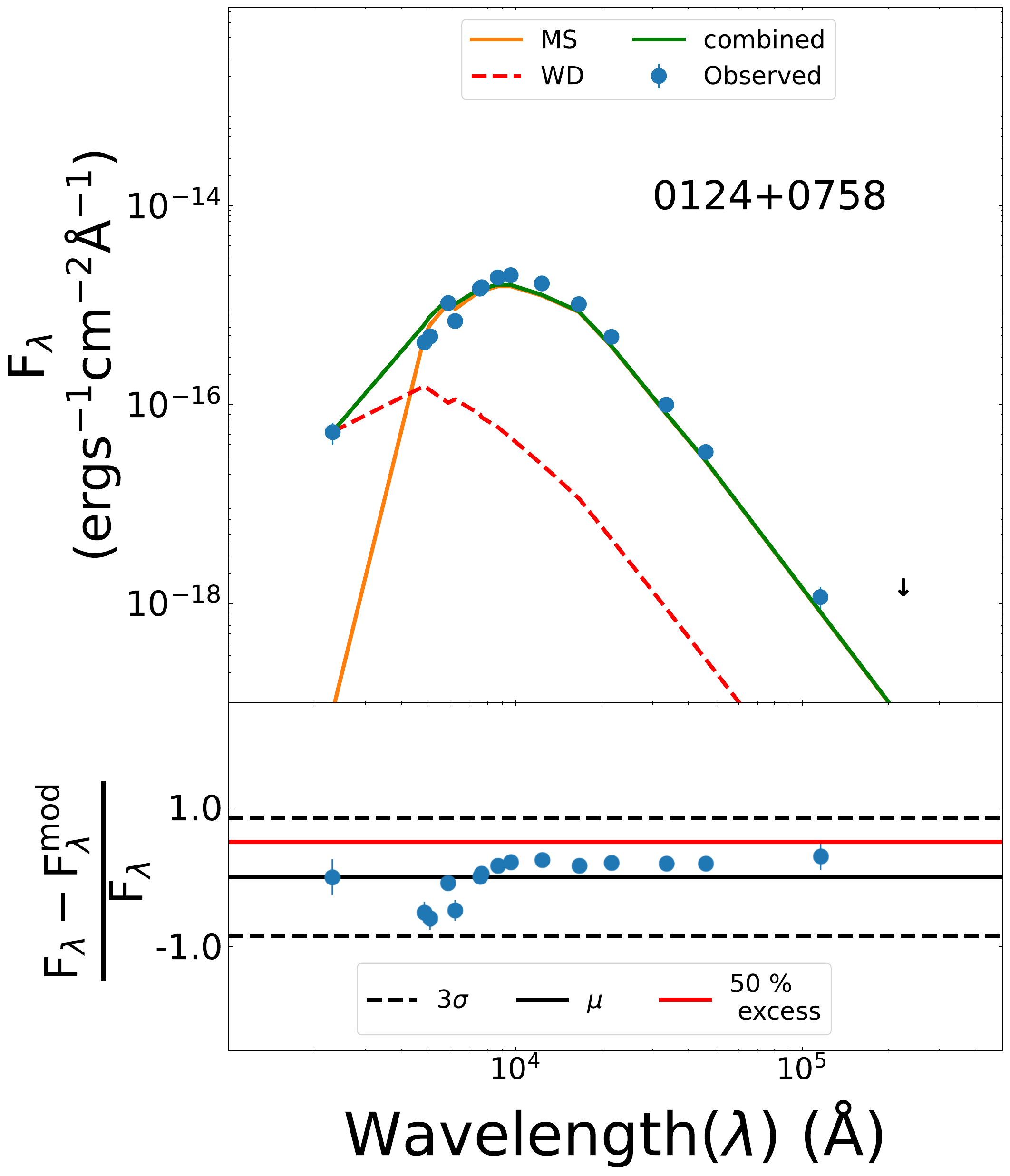}
\figsetgrpnote{Same as \autoref{fig:sed_wd_all},but for source 0124$+$0758}
\label{fig:sed_wd_all_7}

\figsetgrpend

\figsetgrpstart
\figsetgrpnum{4.8}
\figsetgrptitle{Image for figure 4_8}
\figsetplot{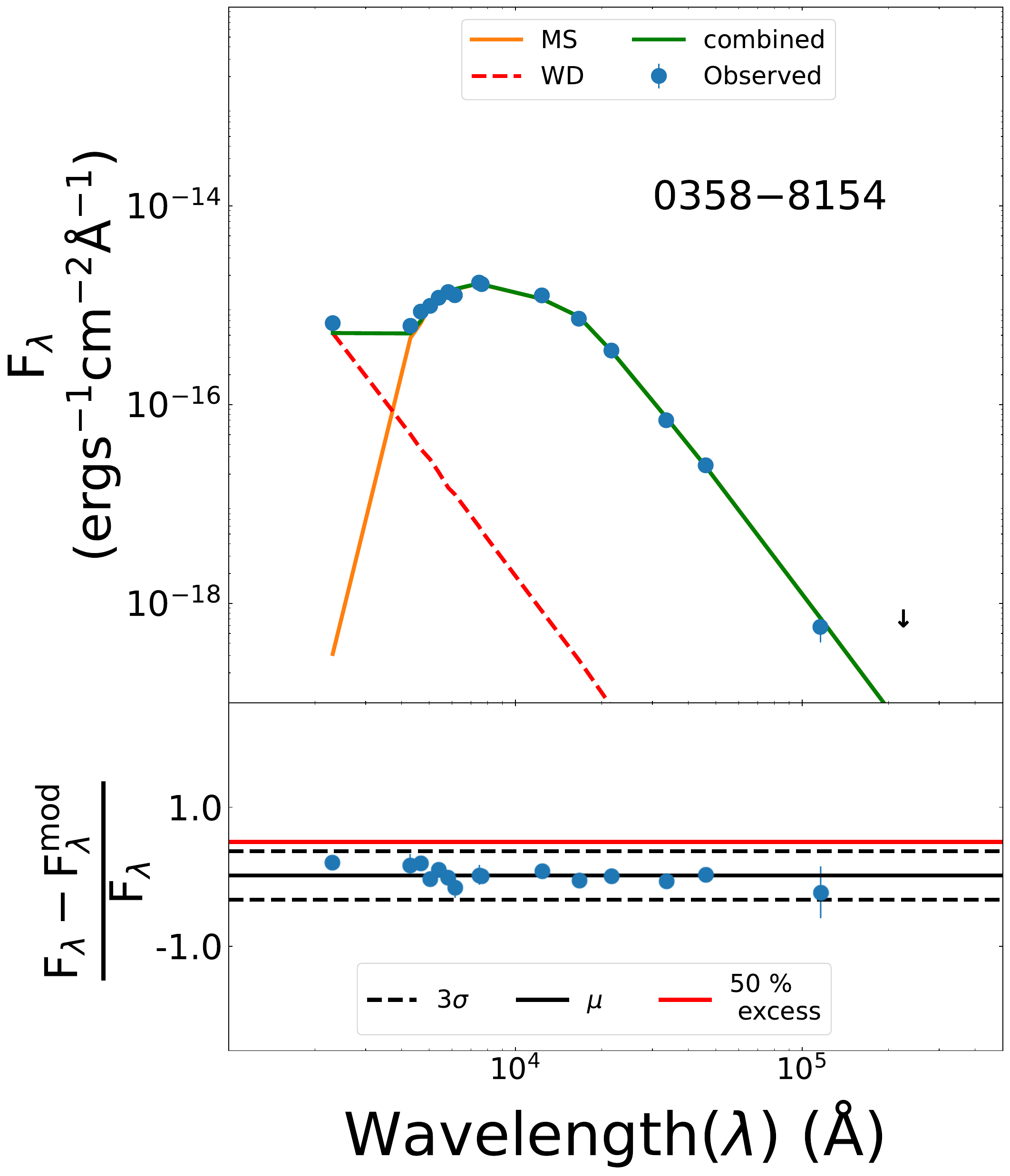}
\figsetgrpnote{Same as \autoref{fig:sed_wd_all},but for source 0358$-$8154}
\label{fig:sed_wd_all_8}

\figsetgrpend

\figsetgrpstart
\figsetgrpnum{4.9}
\figsetgrptitle{Image for figure 4_9}
\figsetplot{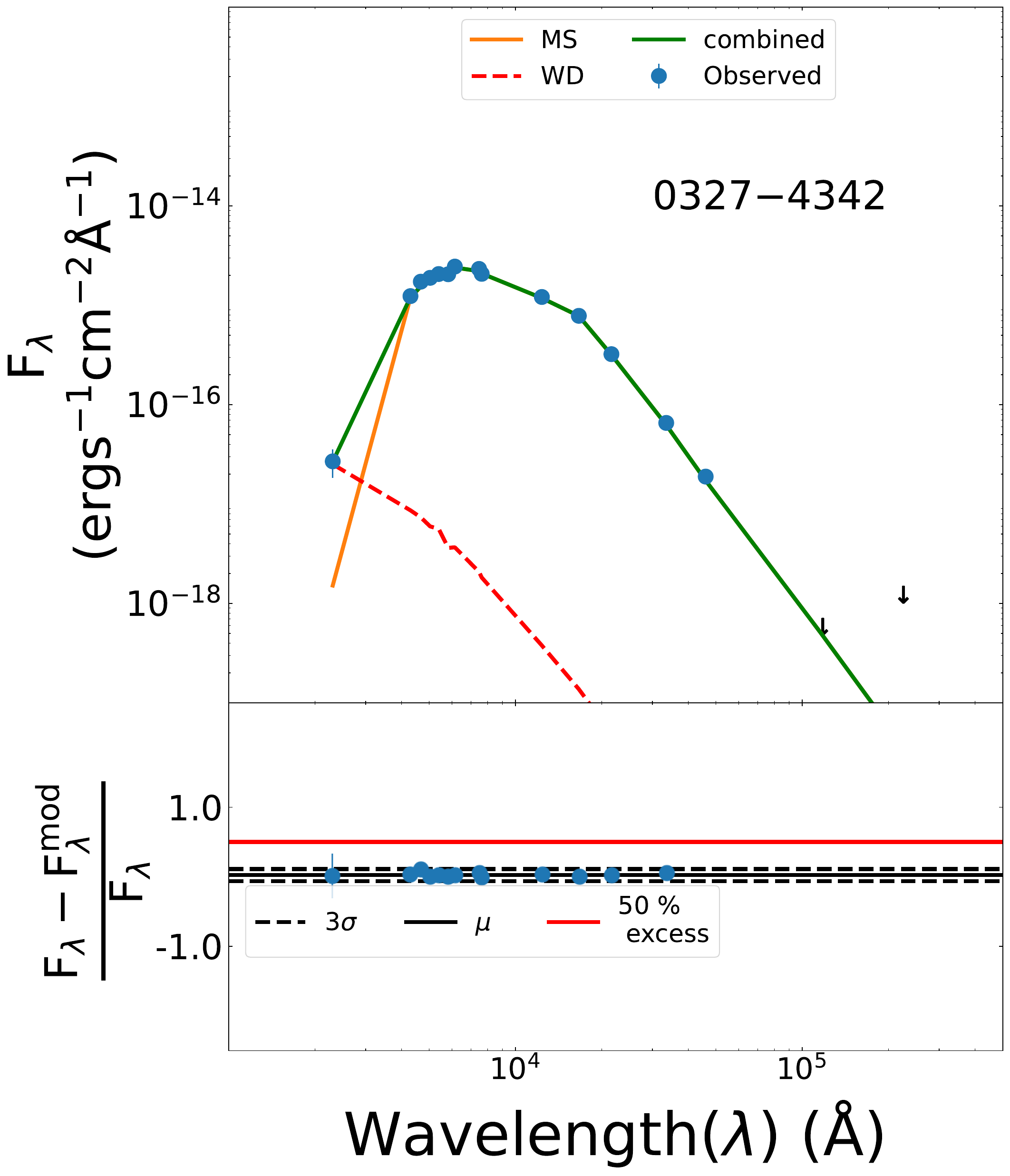}
\figsetgrpnote{Same as \autoref{fig:sed_wd_all},but for source 0327$-$4342}
\label{fig:sed_wd_all_9}

\figsetgrpend

\figsetgrpstart
\figsetgrpnum{4.10}
\figsetgrptitle{Image for figure 4_10}
\figsetplot{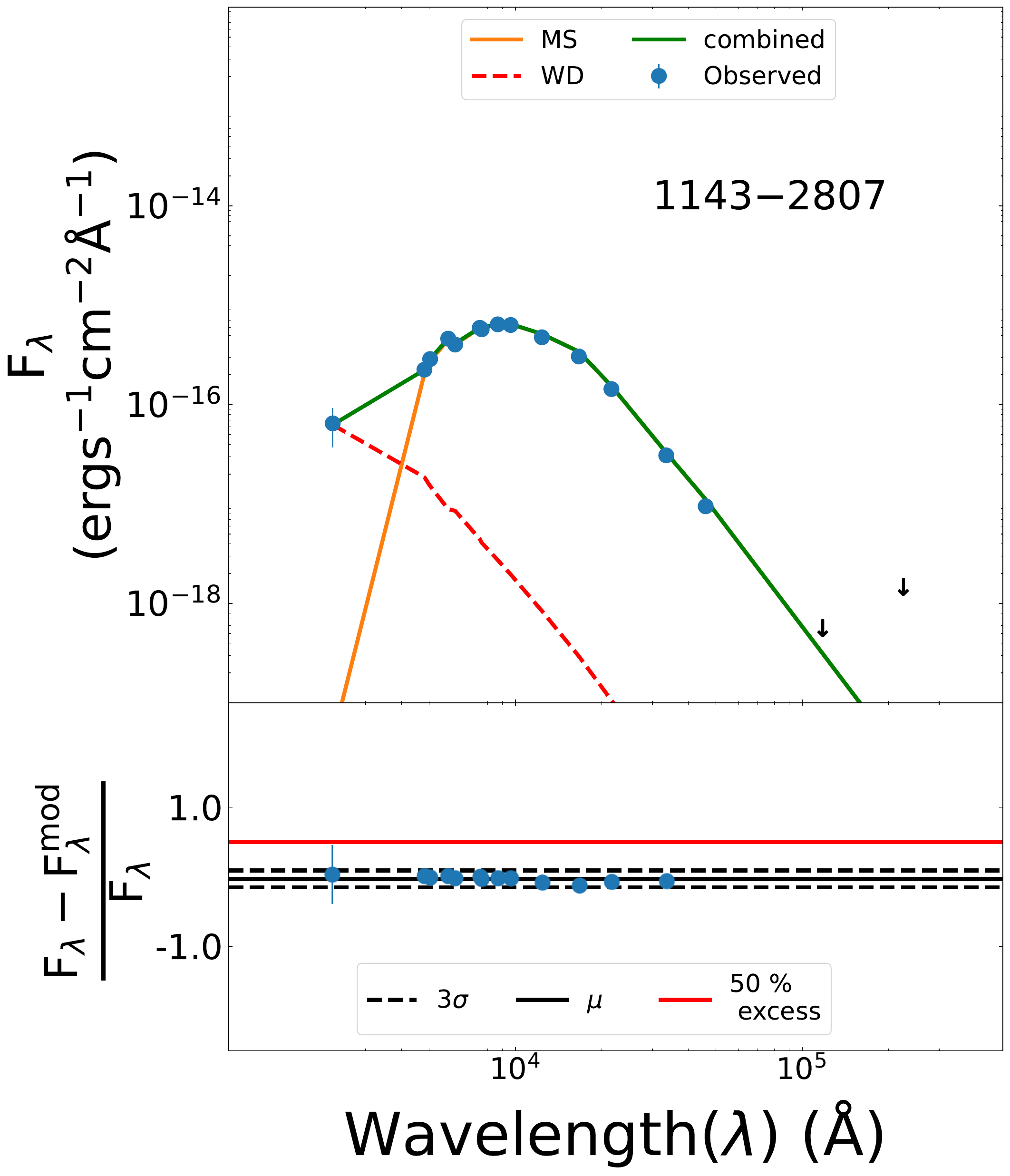}
\figsetgrpnote{Same as \autoref{fig:sed_wd_all},but for source 1143$-$2807}
\label{fig:sed_wd_all_10}

\figsetgrpend

\figsetgrpstart
\figsetgrpnum{4.11}
\figsetgrptitle{Image for figure 4_11}
\figsetplot{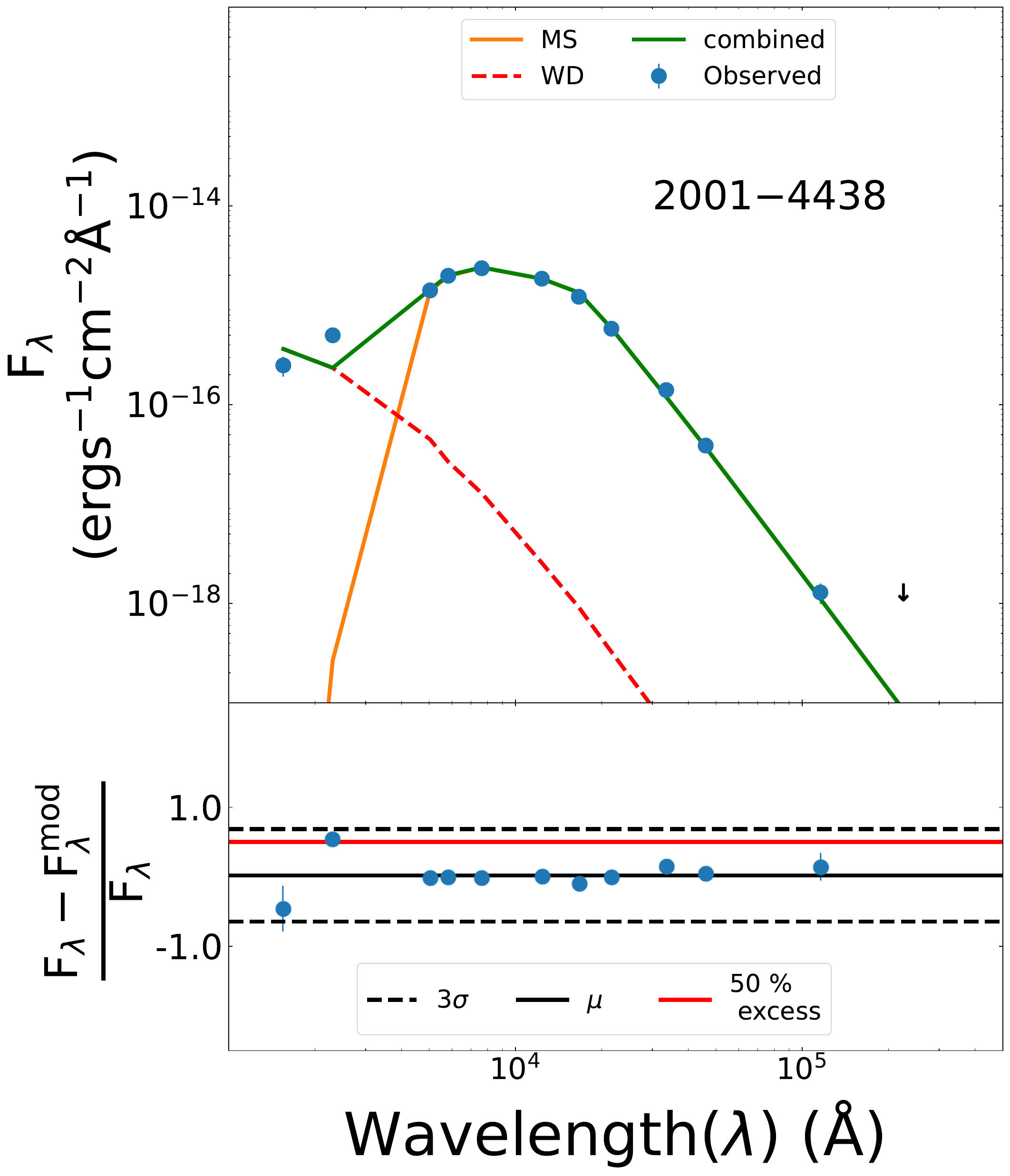}
\figsetgrpnote{Same as \autoref{fig:sed_wd_all},but for source 2001$-$4438}
\label{fig:sed_wd_all_11}

\figsetgrpend

\figsetgrpstart
\figsetgrpnum{4.12}
\figsetgrptitle{Image for figure 4_12}
\figsetplot{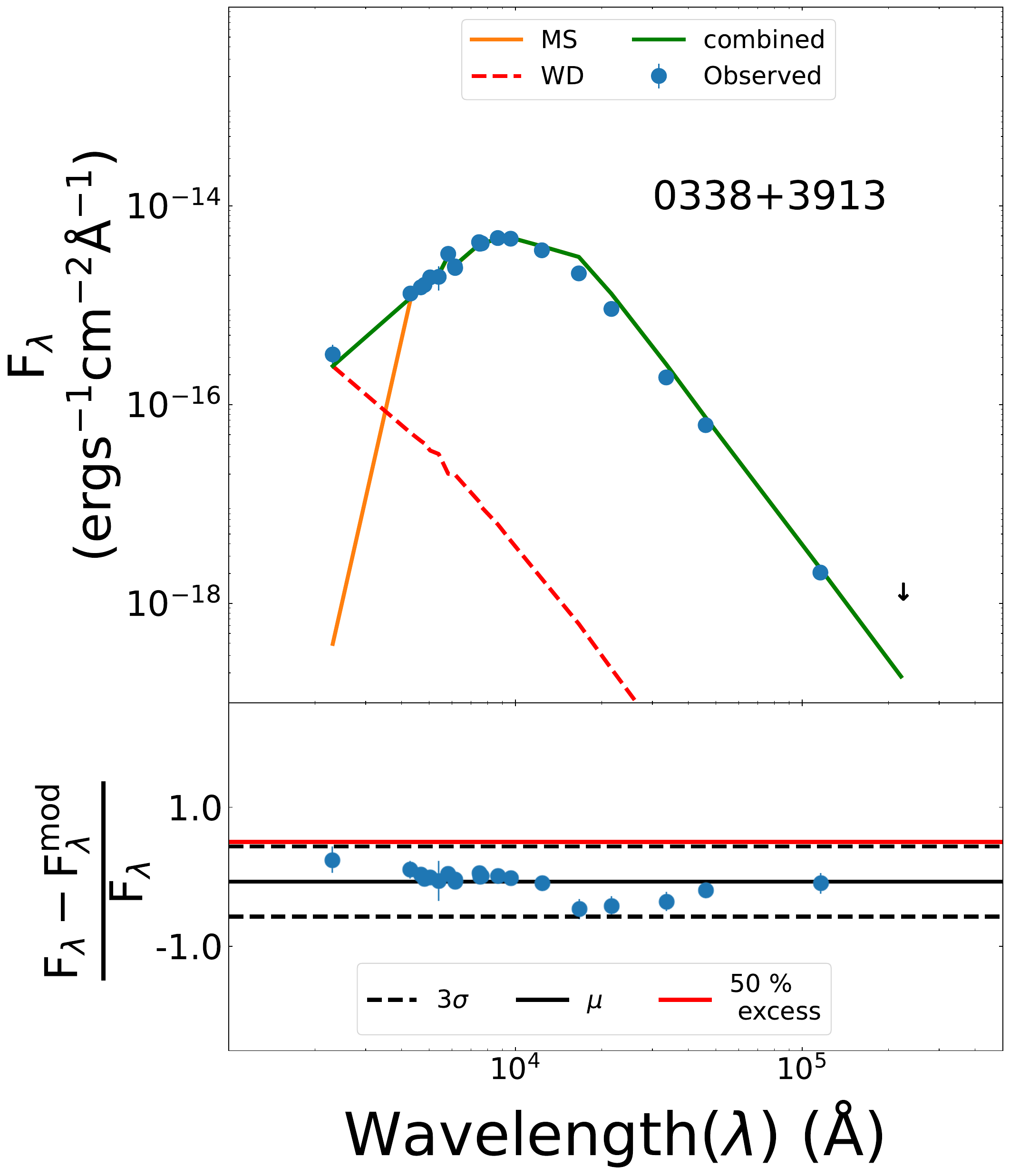}
\figsetgrpnote{Same as \autoref{fig:sed_wd_all},but for source 0338$+$3913}
\label{fig:sed_wd_all_12}

\figsetgrpend

\figsetgrpstart
\figsetgrpnum{4.13}
\figsetgrptitle{Image for figure 4_13}
\figsetplot{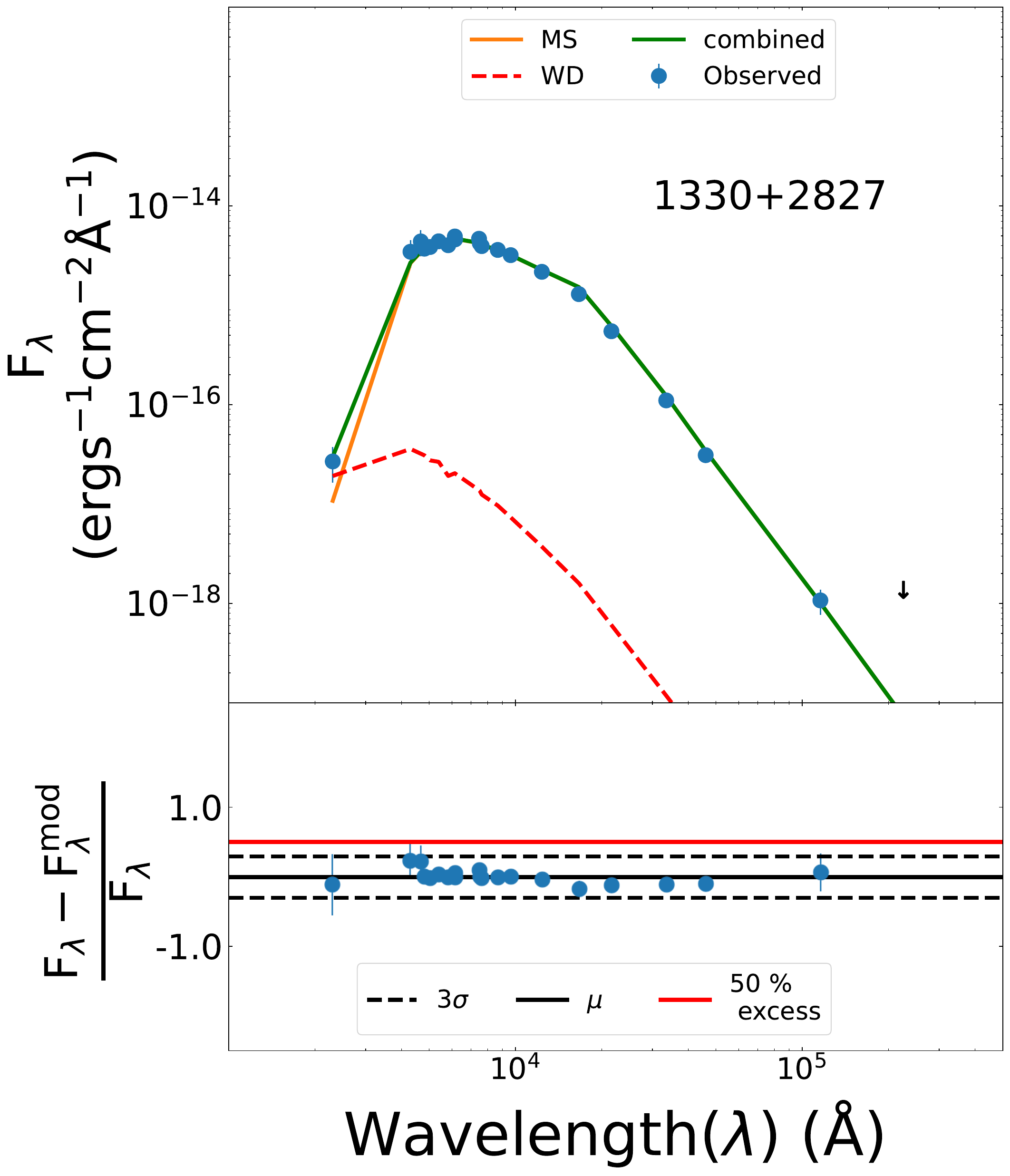}
\figsetgrpnote{Same as \autoref{fig:sed_wd_all},but for source 1330$+$2827}
\label{fig:sed_wd_all_13}

\figsetgrpend

\figsetgrpstart
\figsetgrpnum{4.14}
\figsetgrptitle{Image for figure 4_14}
\figsetplot{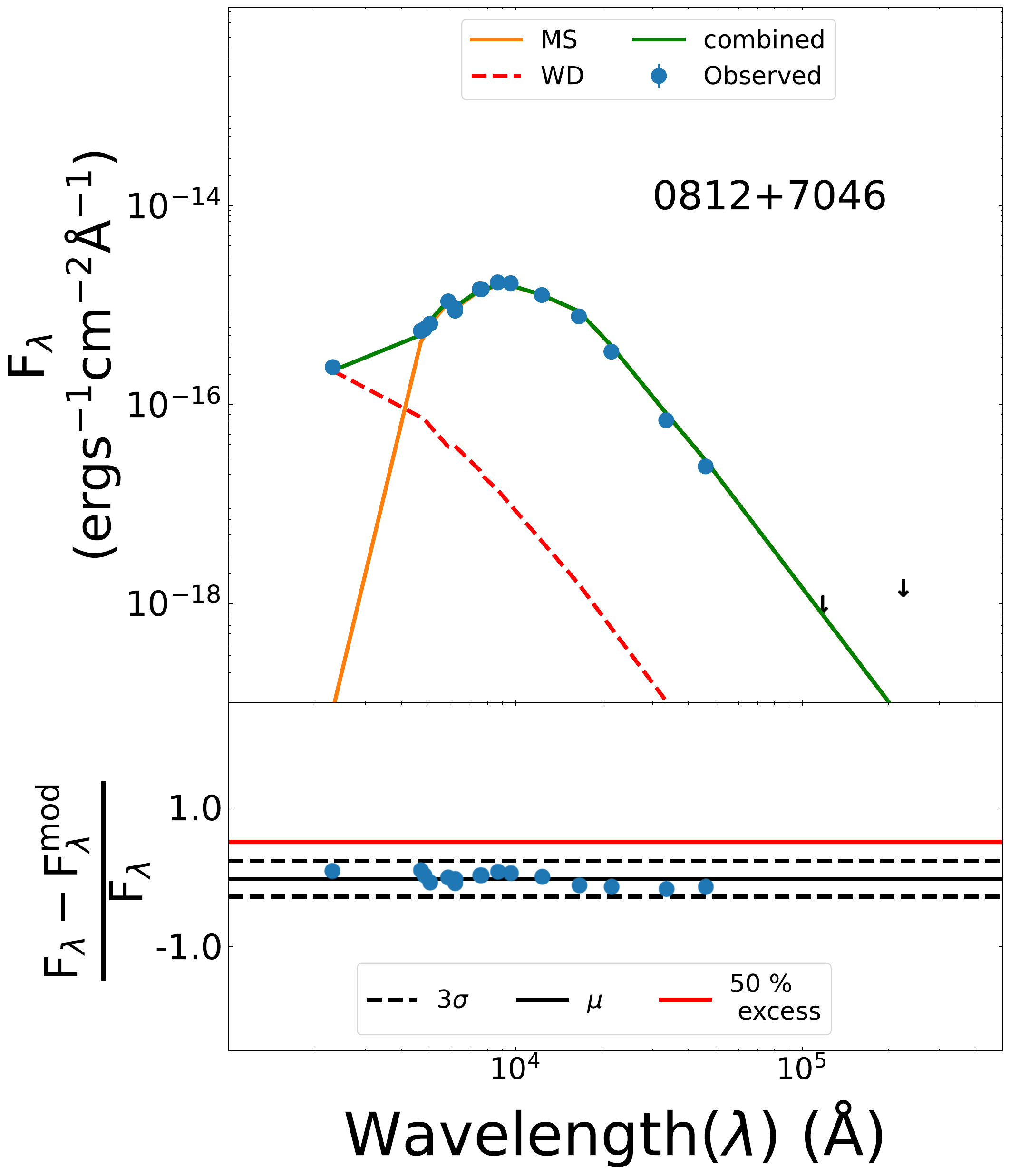}
\figsetgrpnote{Same as \autoref{fig:sed_wd_all},but for source 0812$+$7046}
\label{fig:sed_wd_all_14}

\figsetgrpend

\figsetgrpstart
\figsetgrpnum{4.15}
\figsetgrptitle{Image for figure 4_15}
\figsetplot{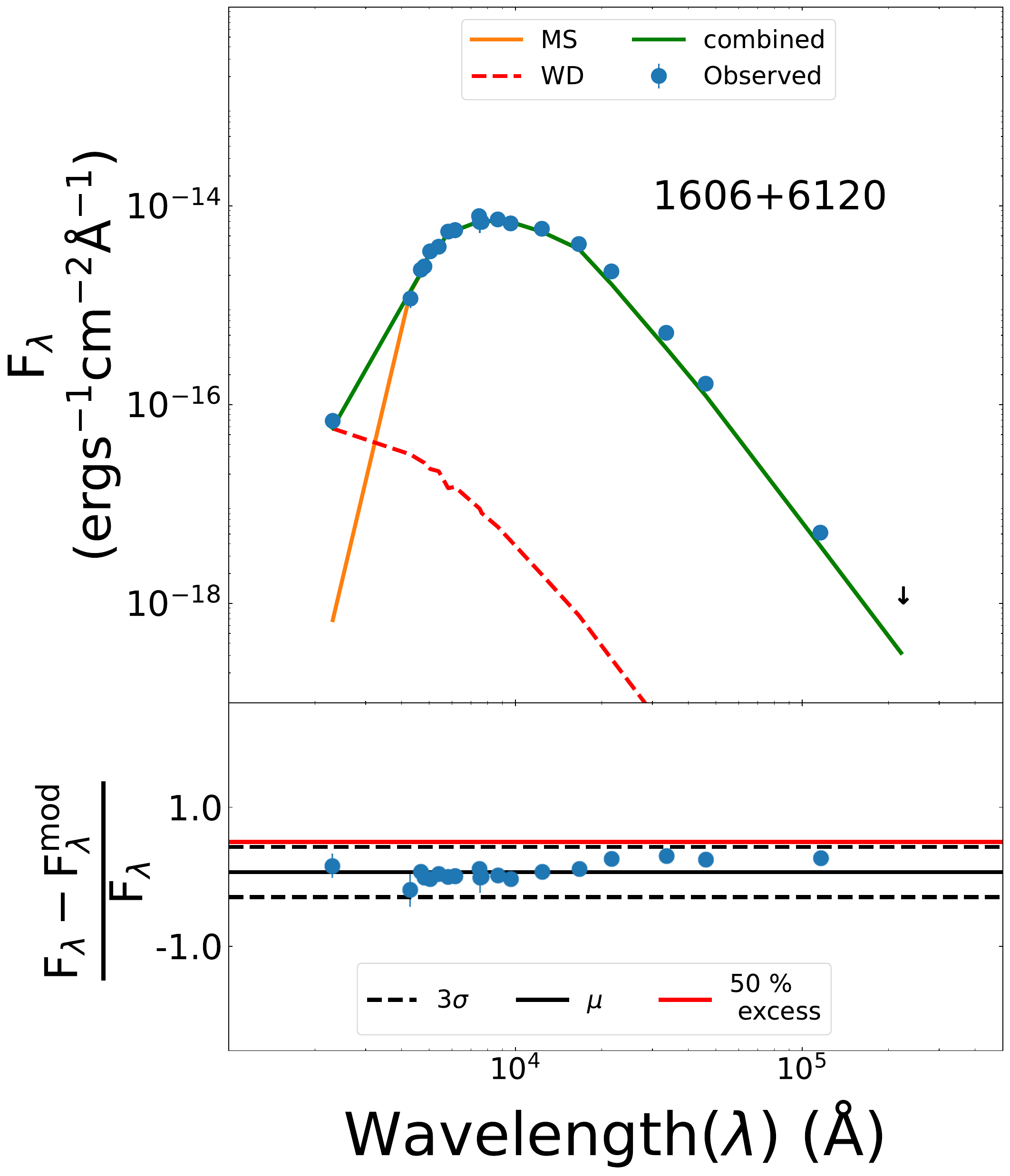}
\figsetgrpnote{Same as \autoref{fig:sed_wd_all},but for source 1606$+$6120}
\label{fig:sed_wd_all_15}

\figsetgrpend

\figsetend
\begin{deluxetable*}{cccccccc}
\tablecaption{Properties of the candidate WDs.
\label{tab:wd_data}
}
\tabletypesize{\scriptsize}
\tablehead{
\colhead{Object} & \colhead{UV} & \colhead{$\teff$} & \colhead{$\lbol$}&
\colhead{$\log\ (g/\acceleration)$}& $\mdr$ & \% Vgf$_b$ & \colhead{SHA23} \\
\colhead{} & \colhead{} & \colhead{$(10^3\,\rm{K})$} & \colhead{$(L_\odot)$} &
\colhead{} & \colhead{$(\msun)$} & Improvement$^\tablenotemark{a}$ & \colhead{Classification} 
}
\startdata
     0812$+$7046& N & 12.00 $\pm$    0.12 & 0.0014 $\pm$ 0.0001 &  9.00 $\pm$    0.12 &     0.17 $\pm$       0.06 &  87 &       WD \\
    1220$+$5841 & N &  7.00 $\pm$    0.12 & 0.0029 $\pm$ 0.0004 &  9.00 $\pm$    0.12 &     1.28 $\pm$       0.06 &  41 &       NS \\
    2106$-$5218 & N & 10.00 $\pm$    0.12 & 0.0023 $\pm$ 0.0003 &  6.00 $\pm$    0.12 &     0.34 $\pm$       0.13 &  43 &       WD \\
   0824$+$2300 & B & 17.50 $\pm$    0.12 & 0.0021 $\pm$ 0.0003 &  9.00 $\pm$    0.12 &     0.47 $\pm$       0.07 &  71 &       WD \\
   0709$+$7052 & B & 16.00 $\pm$    0.12 & 0.0036 $\pm$ 0.0017 &  9.00 $\pm$    0.12 &     0.67 $\pm$       0.11 &  86 &       NS \\
   2338$-$7152 & N & 80.00 $\pm$    0.50 & 0.0014 $\pm$ 0.0001 &  9.00 $\pm$    0.12 &     0.40 $\pm$       0.03 &  99 &       WD \\
   0124$+$0758 & N &  6.75 $\pm$    0.12 & 0.0031 $\pm$ 0.0002 &  8.00 $\pm$    0.12 &     0.18 $\pm$       0.06 &  -50 &       WD \\
   0358$-$8154 & N & 80.00 $\pm$    0.50 & 0.0028 $\pm$ 0.0001 &  9.00 $\pm$    0.12 &     0.35 $\pm$       0.03 &  89 &       WD \\
   0327$-$4342 & N & 13.00 $\pm$    0.12 & 0.0024 $\pm$ 0.0008 &  9.00 $\pm$    0.12 &     0.87 $\pm$       0.03 &  86 &       NS \\
   1143$-$2807 & N & 14.50 $\pm$    0.12 & 0.0005 $\pm$ 0.0001 &  6.00 $\pm$    0.12 &     0.41 $\pm$       0.00 &  76 &       WD \\
   2001$-$4438 & B & 14.50 $\pm$    0.12 & 0.0017 $\pm$ 0.0002 &  9.00 $\pm$    0.12 &     0.41 $\pm$       0.08 &  74 &       WD \\
   0338$+$3913 & N & 17.00 $\pm$    0.12 & 0.0032 $\pm$ 0.0002 &  9.00 $\pm$    0.12 &     0.50 $\pm$       0.06 &  25 &       WD \\
   1330$+$2827 & N &  7.75 $\pm$    0.12 & 0.0105 $\pm$ 0.0030 &  9.00 $\pm$    0.12 &     0.58 $\pm$       0.10 &  24 &       NS \\
   0640$-$2621 & B & 50.00 $\pm$    0.38 & 0.0269 $\pm$ 0.0014 &  6.00 $\pm$    0.12 &     1.18 $\pm$       0.04 &  88 &       NS \\
   1606$+$6120 & N & 10.25 $\pm$    0.12 & 0.0011 $\pm$ 0.0001 &  9.00 $\pm$    0.12 &     0.50 $\pm$       0.04 &  44 &       WD \\
\enddata
\tablenotetext{a}{Improvement of fit by using a composite model, $\rm{(Vgf_{b,1}-Vgf_{b,2})/Vgf_{b,1}}$, where, $\rm{Vgf_{b,x}}$ denotes $\rm{Vgf_b}$ for $x$ components, $x=1,2$. More positive, the higher the improvement. }
\tablecomments{
    \scriptsize
    Properties of the 15 sources with significant UV excess. Errors in $\mdr$ is estimated from $\mlc$ errors. $\Teff$ and $\log\ (g)$ estimates are from SED modeling. Uncertainties denote half of the grid spacing for the SED models. $\lbol$\ measurements and their corresponding errors come from the observed flux and \gaia-estimated distances. N, B denote only NUV, both FUV and NUV flux are available. While the composite fit is visibly better (7th figure of \autoref{fig:sed_wd_all}) than the single component fit for source 0124$+$0758, we do not find a clear improvement in VGf$_b$.     
}
\end{deluxetable*}
We find significant UV excess relative to a single-component model SED for 15 of the 49 sources we have analysed. In 4 of these sources (0824$+$2300, 0709$+$7052, 2001$-$4438, and 0640$-$2621) both FUV and NUV data are available and both fluxes show excess. In case of the others, only the NUV flux is available. Using two-component MS--WD synthetic SEDs we are able to fit all of the 15 sources. \autoref{fig:sed_wd_all} shows the observed SEDs, best-fit models for the combined two-component models, the MS and WD individual contributions, and the residuals for the 15 sources. Clearly, adding a hot component from a WD to the MS star's SED significantly improves the fit and the presence of a WD companion can easily explain the observed excess UV flux. We find the stellar properties including [Fe/H], $\lbol$, $\teff$, and $\log\ (g/\acceleration)$ for both components. The stellar properties of the LCs are summarised in \autoref{tab:all_data} and those for the WDs are shown in \autoref{tab:wd_data}. In all cases the errors in $\Teff$, [Fe/H], and $\log(g/\acceleration)$ denote half of the grid-spacing available in \vosa\ for the SED modelling, whereas, $\lbol$ error denotes propagated errors from uncertainties in the observed flux and \gaia's distance measurement.  

\begin{figure}
    \centering
    \plotone{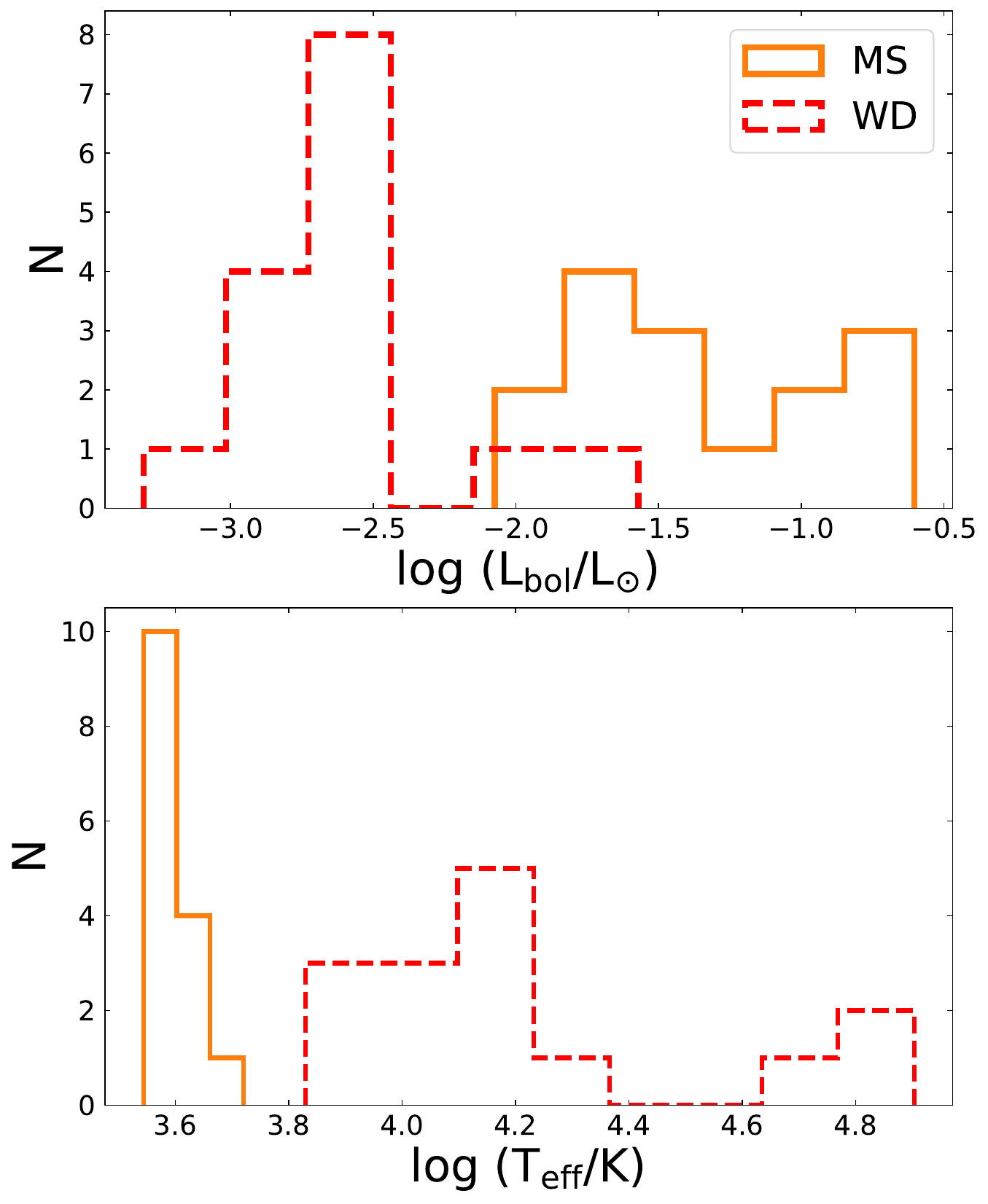}
    \caption{Probability distribution functions for $\lbol$ (top) and $\teff$ (bottom) for the WDs (red dashed) and MS stars (orange solid) for sources showing significant UV excess. 
    }
    \label{fig:lbol_teff_wd}
\end{figure}
\autoref{fig:lbol_teff_wd} shows the distributions of $\lbol$ and $\teff$ for the MS and the WD in these 15 sources. 
As expected, distributions for $\teffwd$ and $\tefflc$ are well separated with medians $\teffwd/\rm{K}=14500\pm7125$ and $\tefflc/\rm{K}=5750\pm 1750$. Whereas, the medians for $\lbolwd/L_\odot=0.0024\pm 0.0016$ and $\lbollc/L_\odot=0.51\pm1.50$.\footnote{Errorbars denote $25$ and $75$ percentiles.} All of these sources are well explained by DA type WDs with pure hydrogen atmosphere. 

\begin{figure}
\plotone{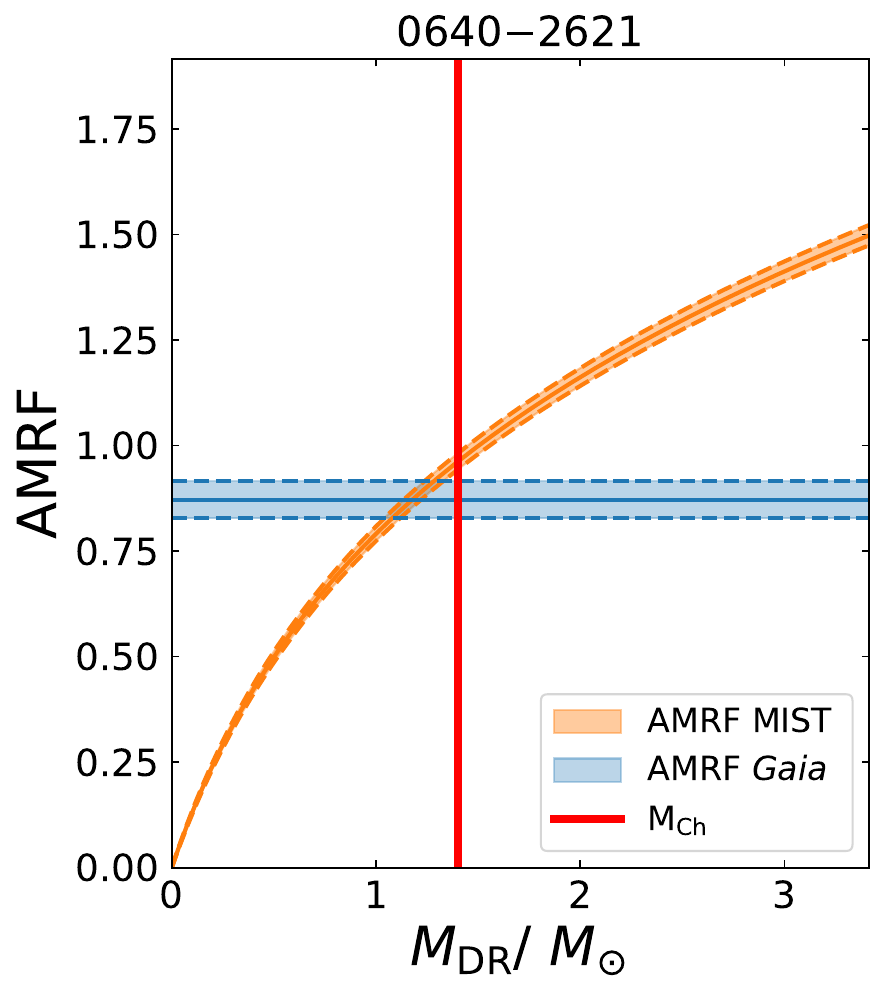}
\caption{AMRF vs $\mdr$ for 0640$-$2621. The orange-shaded region shows the allowed parameter space based on our estimated $\mlc$ (\autoref{sec:LC}). The blue-shaded region shows the estimated AMRF from SHA23. The overlap between the blue and the orange regions is the solution for $\mdr$ that satisfies all available constraints. The red vertical line denotes $\mch$. Our results suggest that source 0640$-$2621 hosts a WD. (Similar figures for the other sources are available online.)}
\label{fig:fm_0640-2621}
\end{figure}

\figsetgrpstart
\figsetgrpnum{6.1}
\figsetgrptitle{Image for figure 6_1}
\figsetplot{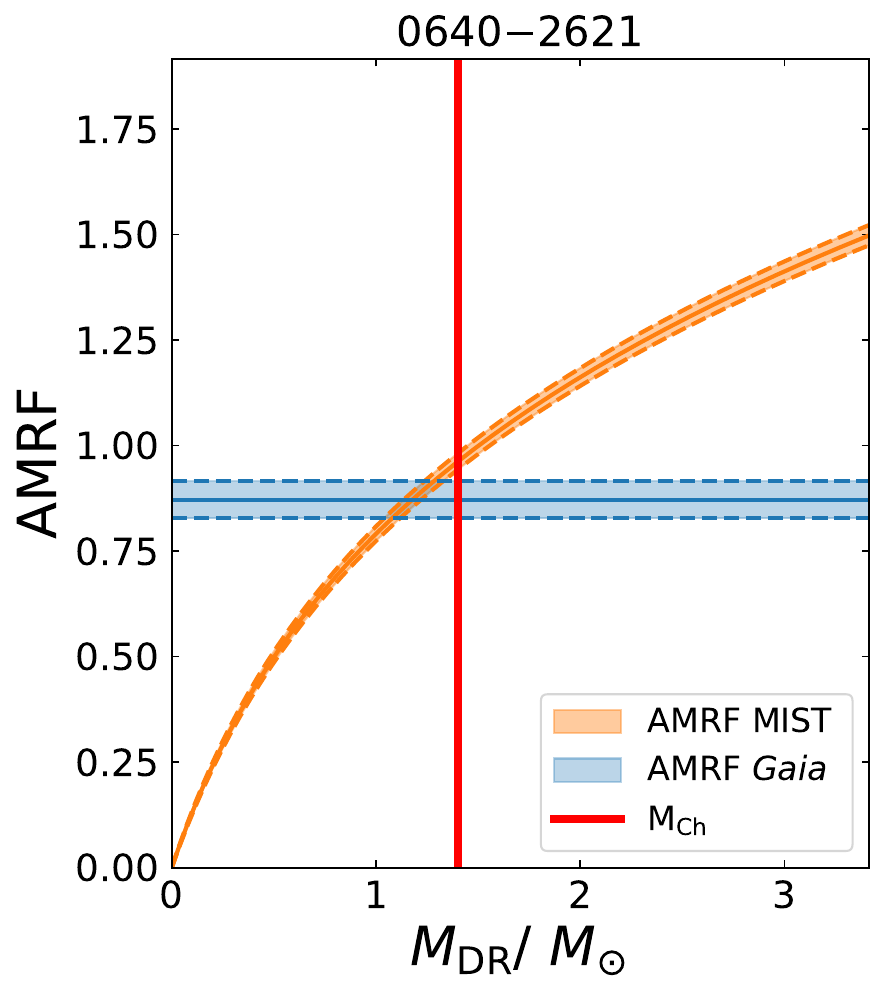}
\figsetgrpnote{Same as \autoref{fig:fm_0640-2621},but for source 0640$-$2621}
\label{fig:fm_1}

\figsetgrpend

\figsetgrpstart
\figsetgrpnum{6.2}
\figsetgrptitle{Image for figure 6_2}
\figsetplot{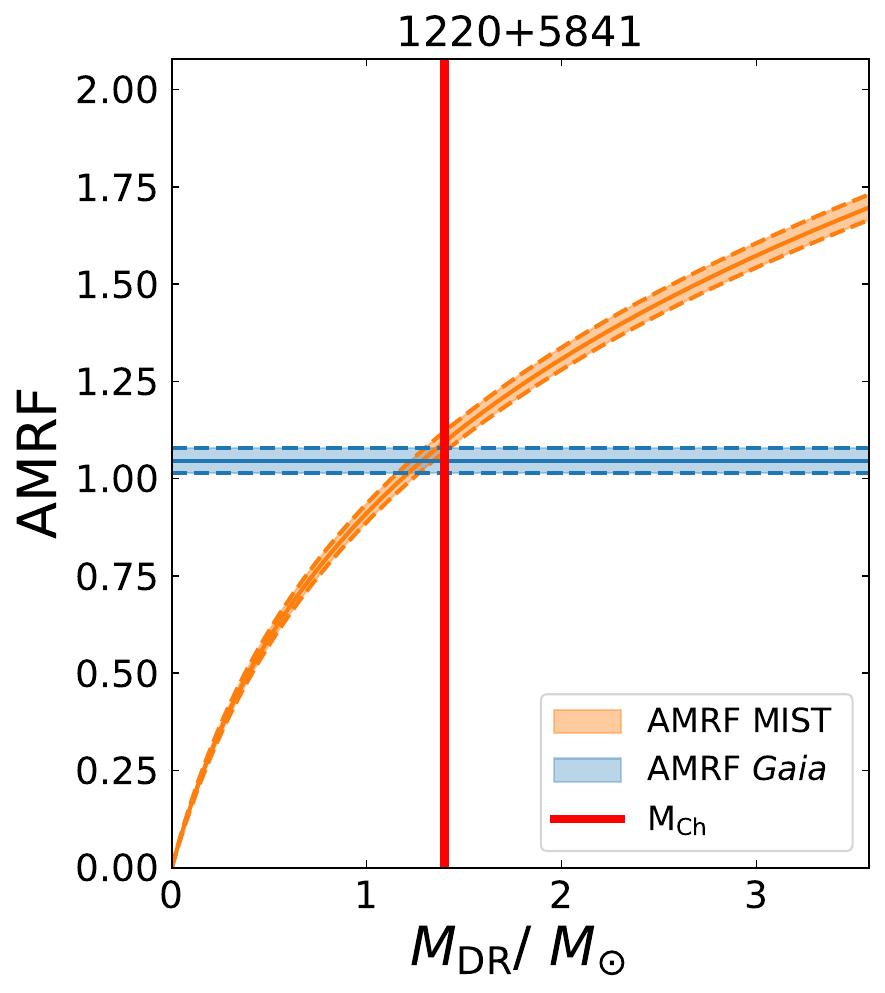}
\figsetgrpnote{Same as \autoref{fig:fm_0640-2621},but for source 1220$+$5841}
\label{fig:fm_2}

\figsetgrpend

\figsetgrpstart
\figsetgrpnum{6.3}
\figsetgrptitle{Image for figure 6_3}
\figsetplot{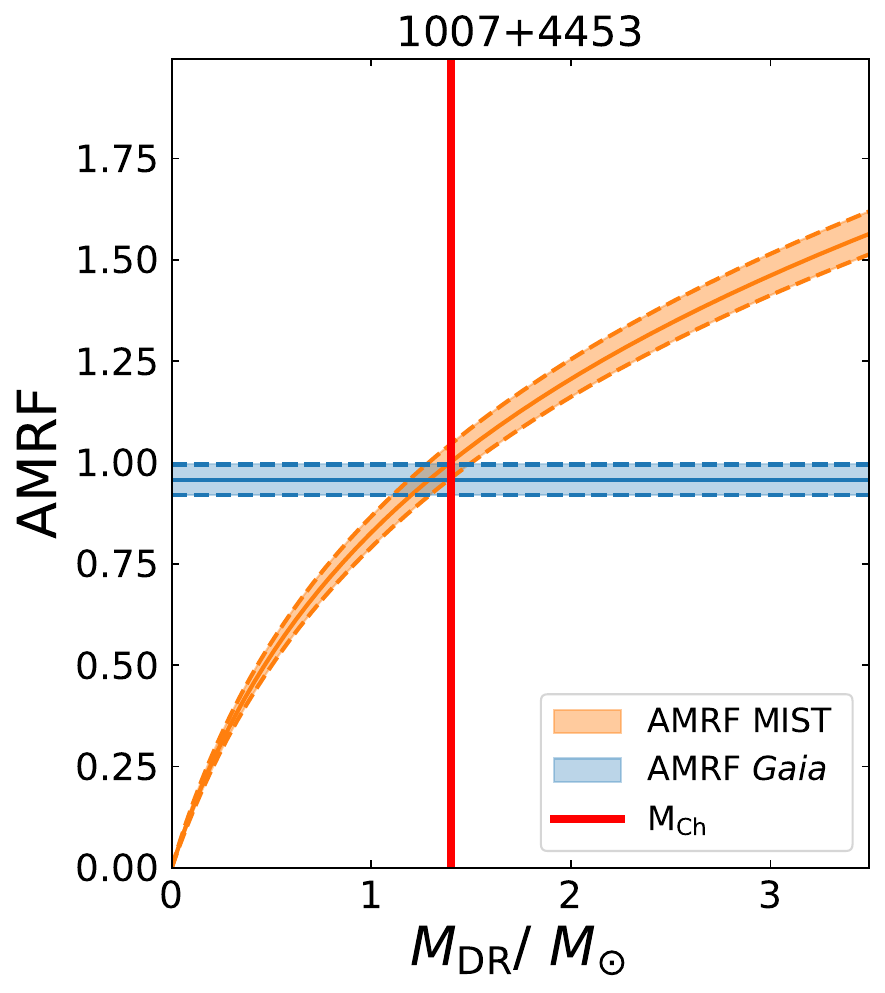}
\figsetgrpnote{Same as \autoref{fig:fm_0640-2621},but for source 1007$+$4453}
\label{fig:fm_3}

\figsetgrpend

\figsetgrpstart
\figsetgrpnum{6.4}
\figsetgrptitle{Image for figure 6_4}
\figsetplot{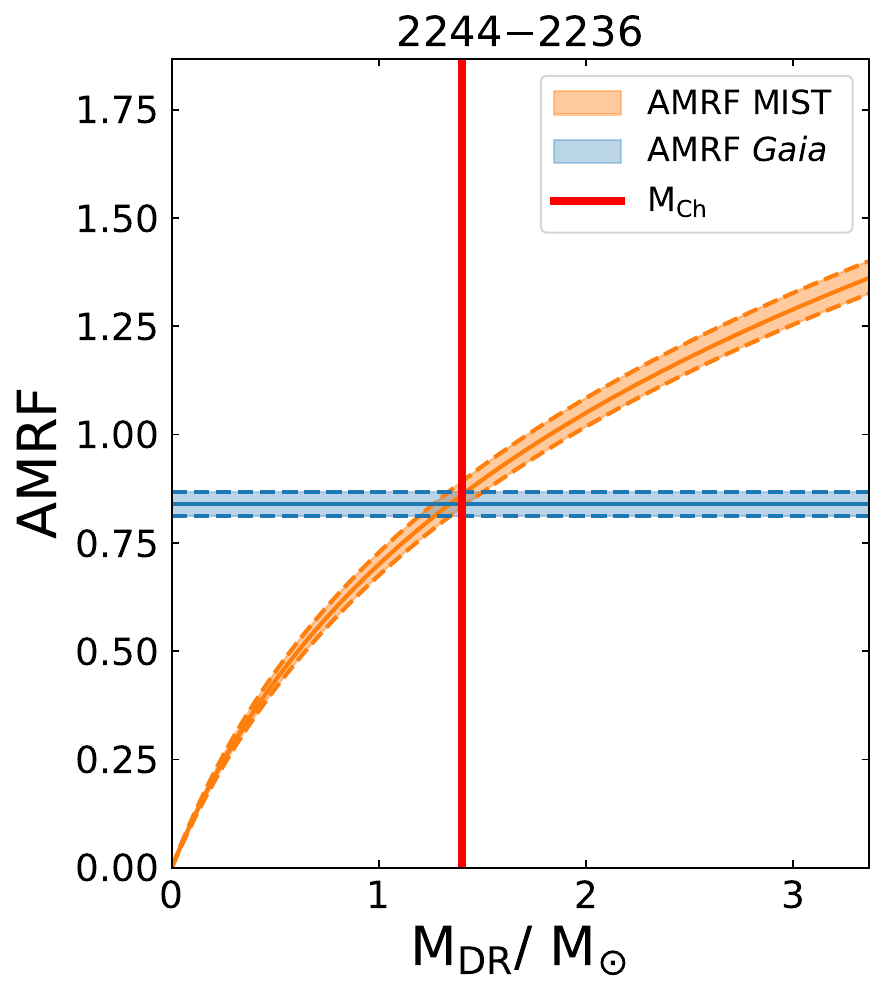}
\figsetgrpnote{Same as \autoref{fig:fm_0640-2621},but for source 2244$-$2236}
\label{fig:fm_4}

\figsetgrpend

\figsetgrpstart
\figsetgrpnum{6.5}
\figsetgrptitle{Image for figure 6_5}
\figsetplot{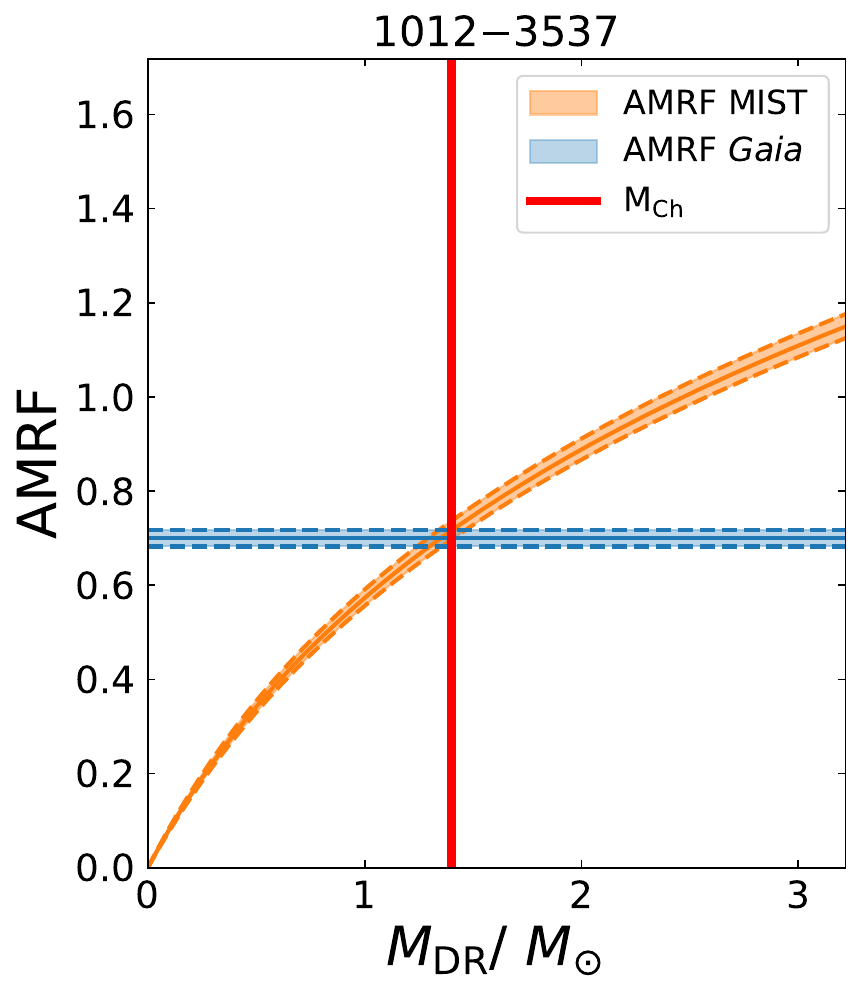}
\figsetgrpnote{Same as \autoref{fig:fm_0640-2621},but for source 1012$-$3537}
\label{fig:fm_5}

\figsetgrpend

\figsetgrpstart
\figsetgrpnum{6.6}
\figsetgrptitle{Image for figure 6_6}
\figsetplot{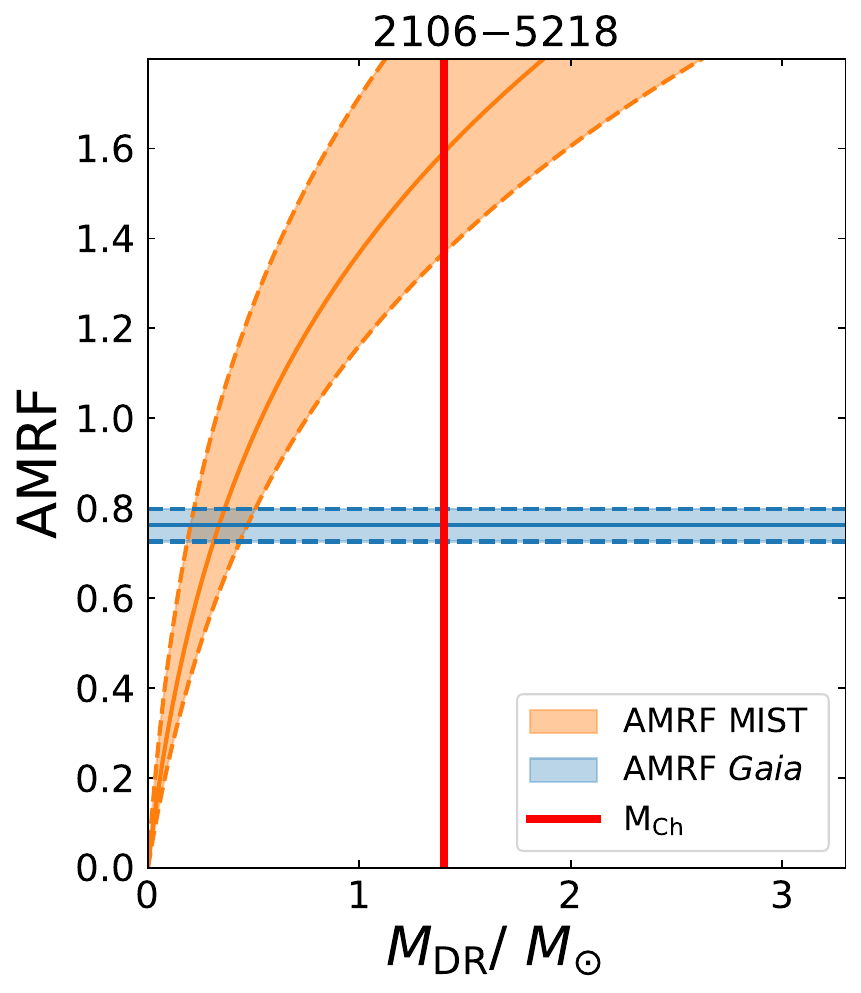}
\figsetgrpnote{Same as \autoref{fig:fm_0640-2621},but for source 2106$-$5218}
\label{fig:fm_6}

\figsetgrpend

\figsetgrpstart
\figsetgrpnum{6.7}
\figsetgrptitle{Image for figure 6_7}
\figsetplot{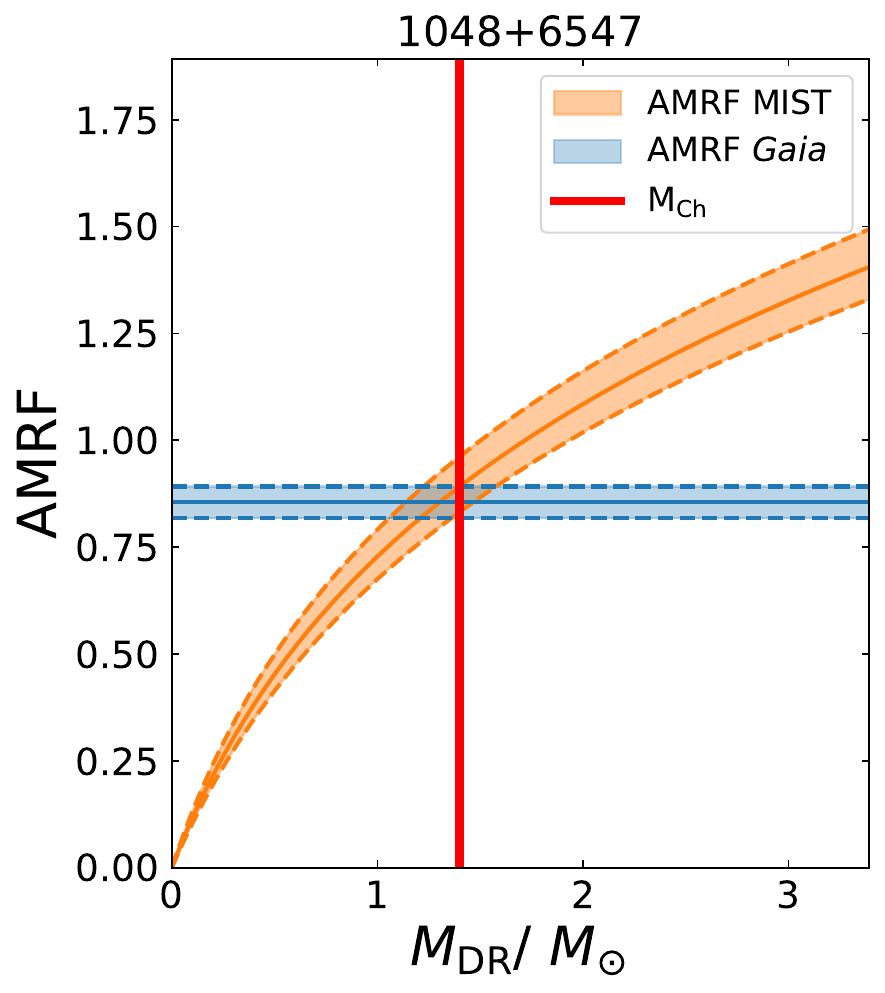}
\figsetgrpnote{Same as \autoref{fig:fm_0640-2621},but for source 1048$+$6547}
\label{fig:fm_7}

\figsetgrpend

\figsetgrpstart
\figsetgrpnum{6.8}
\figsetgrptitle{Image for figure 6_8}
\figsetplot{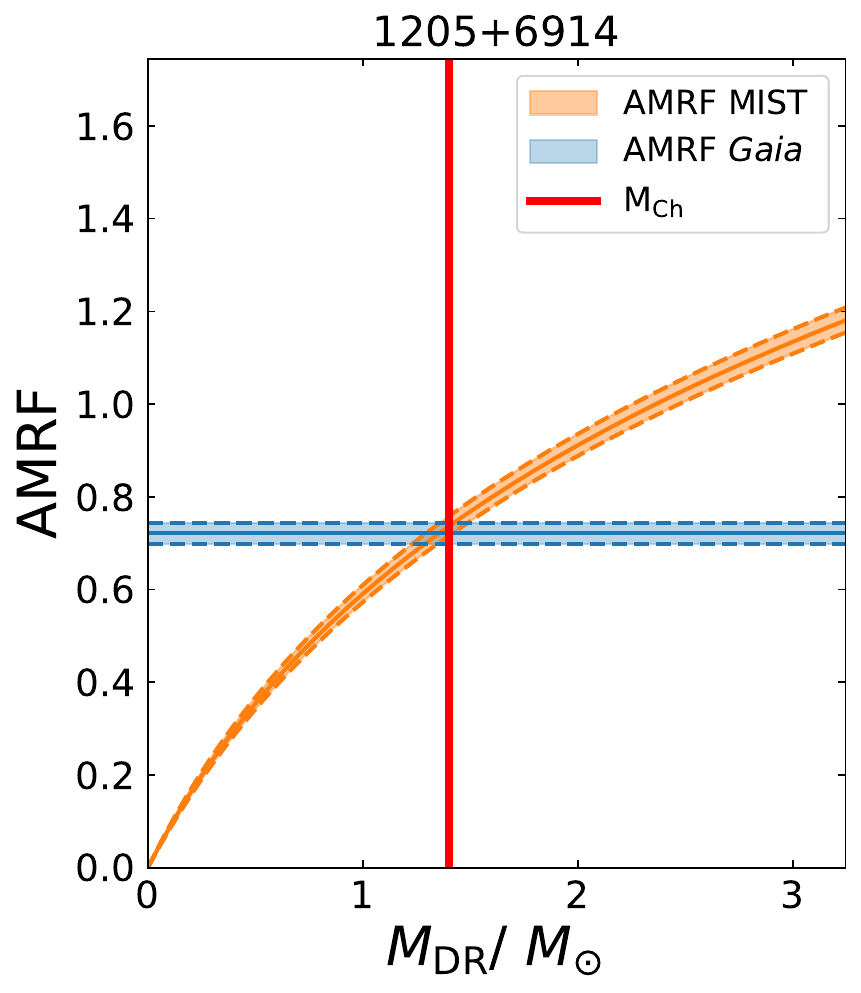}
\figsetgrpnote{Same as \autoref{fig:fm_0640-2621},but for source 1205$+$6914}
\label{fig:fm_8}

\figsetgrpend

\figsetgrpstart
\figsetgrpnum{6.9}
\figsetgrptitle{Image for figure 6_9}
\figsetplot{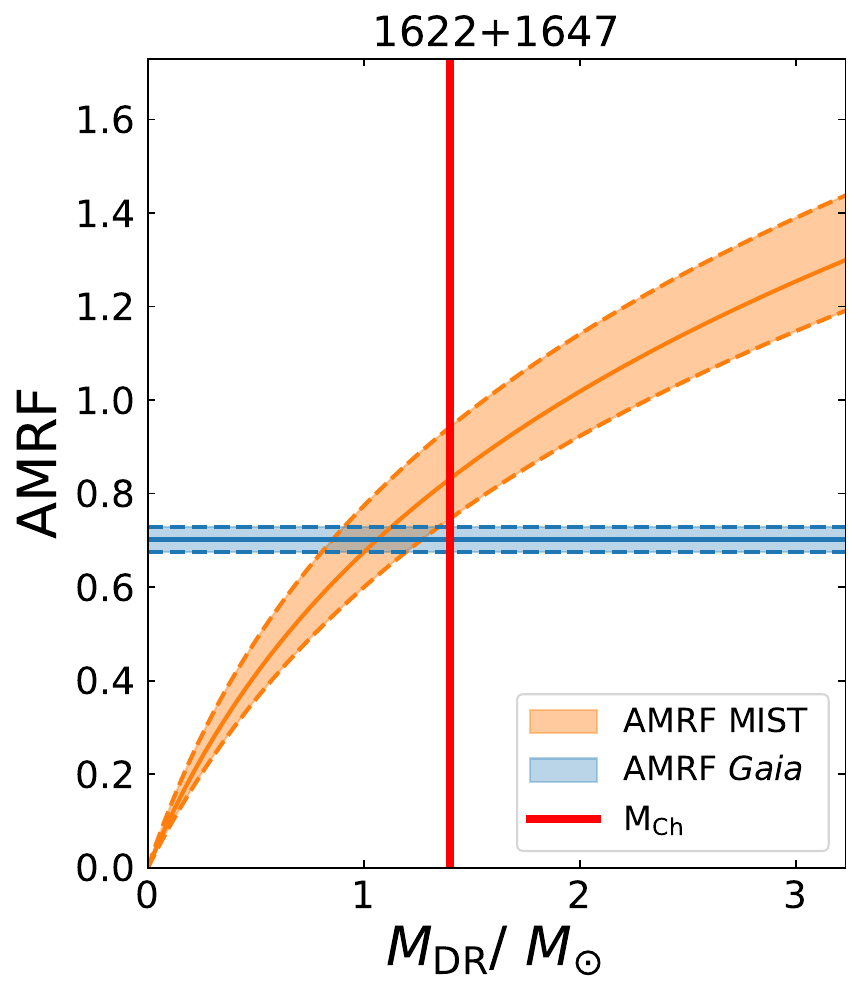}
\figsetgrpnote{Same as \autoref{fig:fm_0640-2621},but for source 1622$+$1647}
\label{fig:fm_9}

\figsetgrpend

\figsetgrpstart
\figsetgrpnum{6.10}
\figsetgrptitle{Image for figure 6_10}
\figsetplot{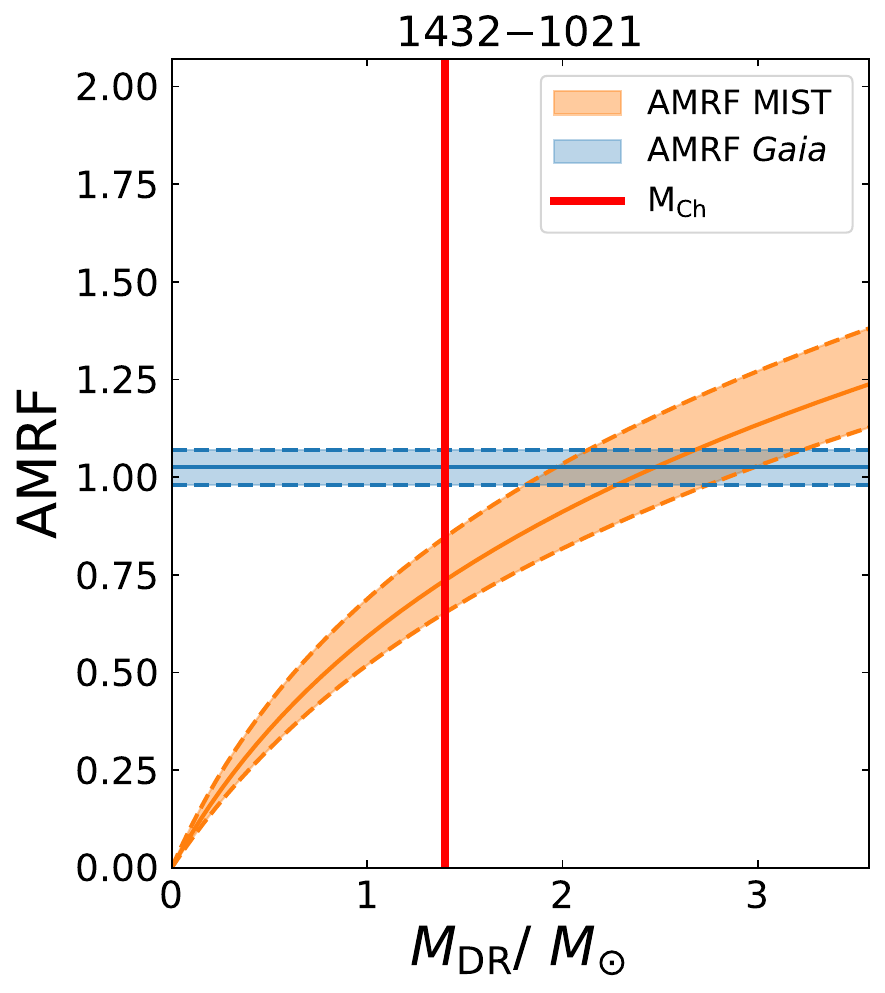}
\figsetgrpnote{Same as \autoref{fig:fm_0640-2621},but for source 1432$-$1021}
\label{fig:fm_10}

\figsetgrpend

\figsetgrpstart
\figsetgrpnum{6.11}
\figsetgrptitle{Image for figure 6_11}
\figsetplot{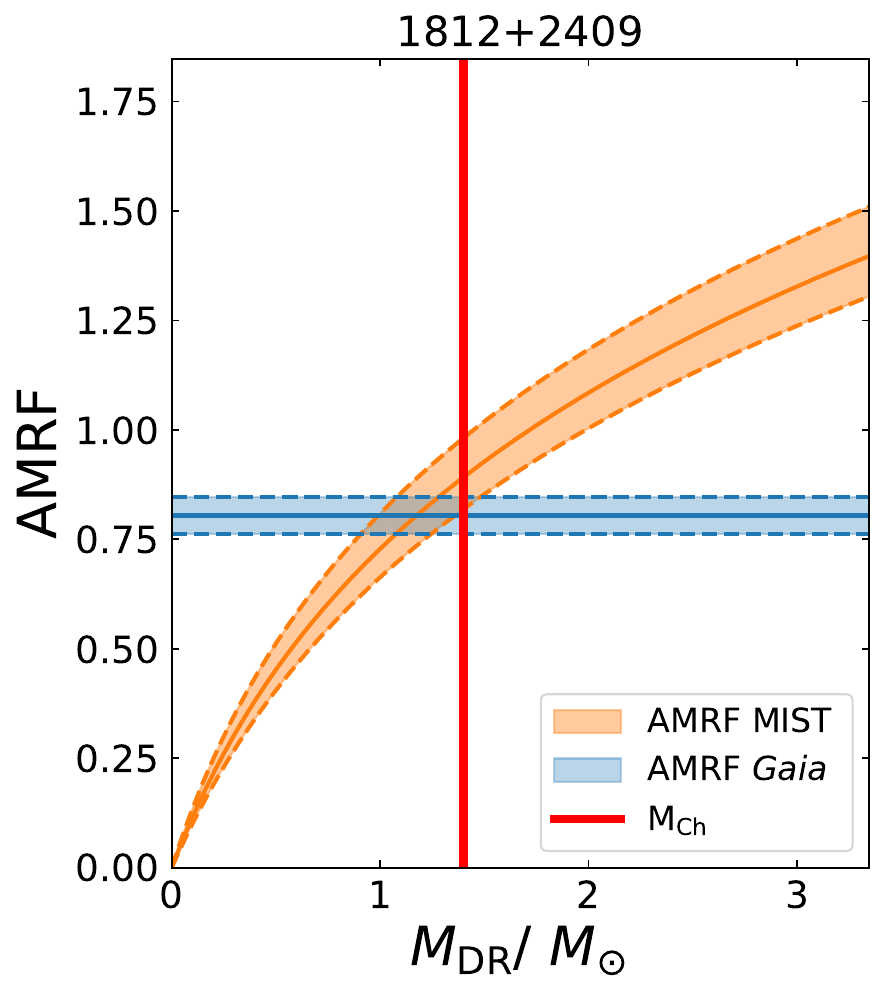}
\figsetgrpnote{Same as \autoref{fig:fm_0640-2621},but for source 1812$+$2409}
\label{fig:fm_11}

\figsetgrpend

\figsetgrpstart
\figsetgrpnum{6.12}
\figsetgrptitle{Image for figure 6_12}
\figsetplot{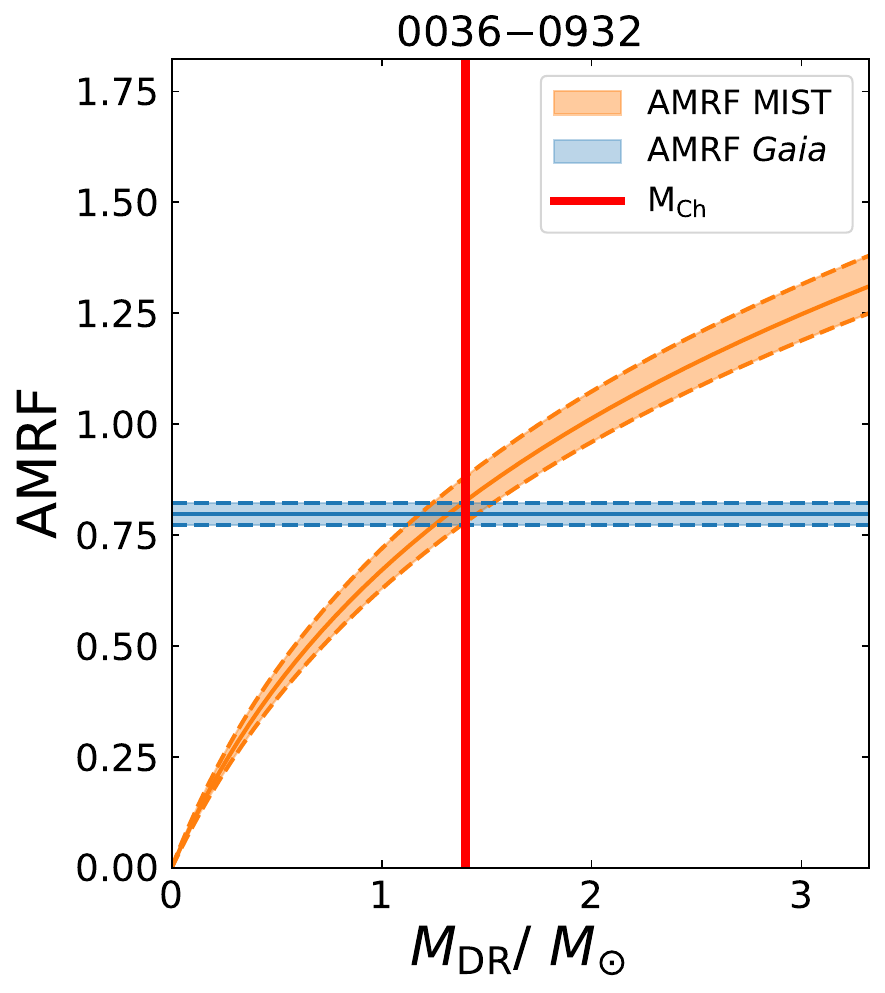}
\figsetgrpnote{Same as \autoref{fig:fm_0640-2621},but for source 0036$-$0932}
\label{fig:fm_12}

\figsetgrpend

\figsetgrpstart
\figsetgrpnum{6.13}
\figsetgrptitle{Image for figure 6_13}
\figsetplot{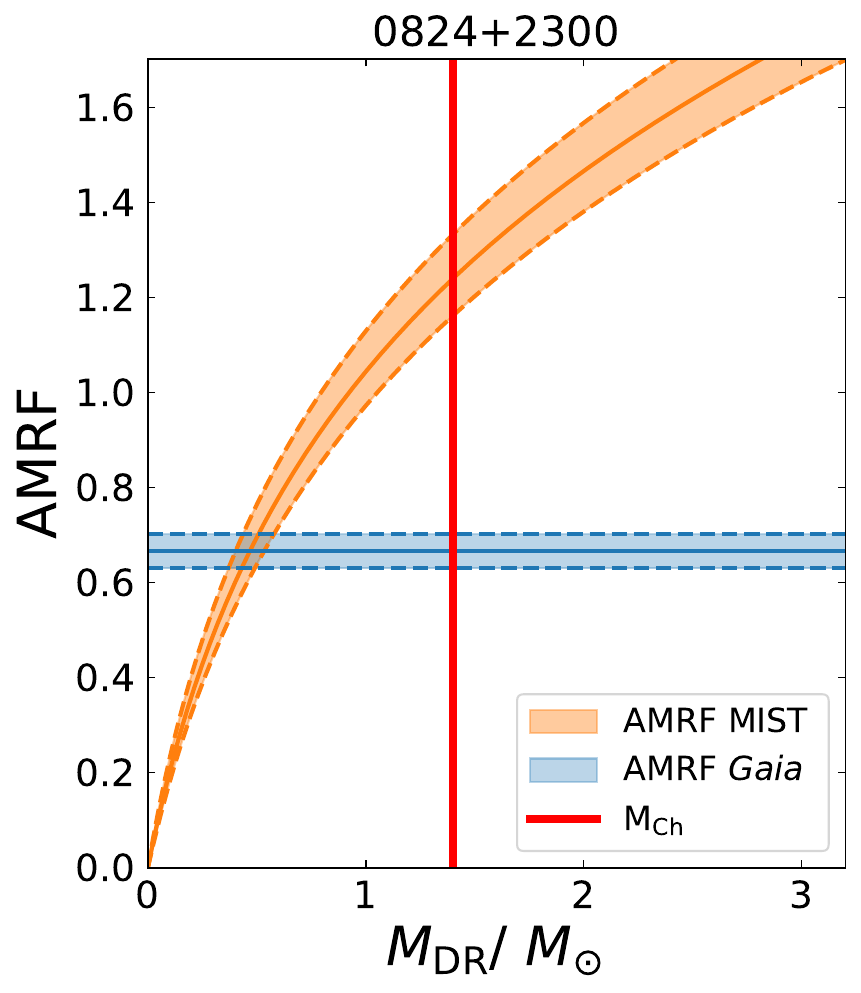}
\figsetgrpnote{Same as \autoref{fig:fm_0640-2621},but for source 0824$+$2300}
\label{fig:fm_13}

\figsetgrpend

\figsetgrpstart
\figsetgrpnum{6.14}
\figsetgrptitle{Image for figure 6_14}
\figsetplot{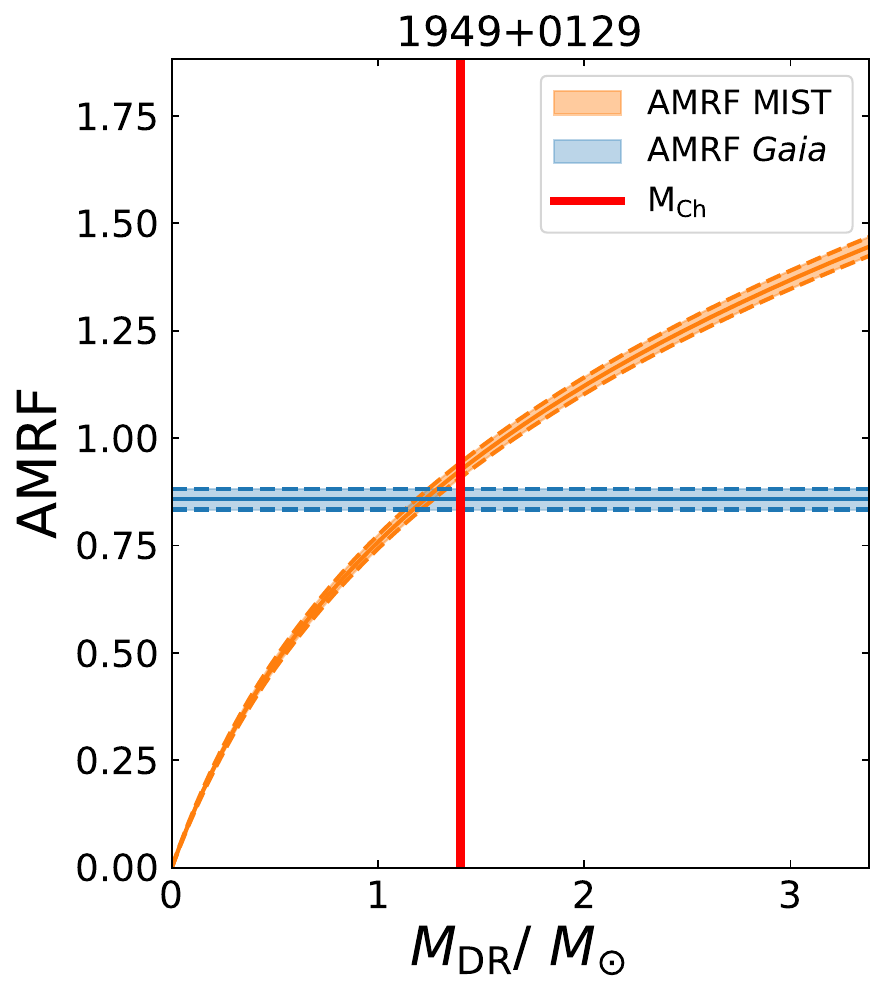}
\figsetgrpnote{Same as \autoref{fig:fm_0640-2621},but for source c1949$+$0129}
\label{fig:fm_14}

\figsetgrpend

\figsetgrpstart
\figsetgrpnum{6.15}
\figsetgrptitle{Image for figure 6_15}
\figsetplot{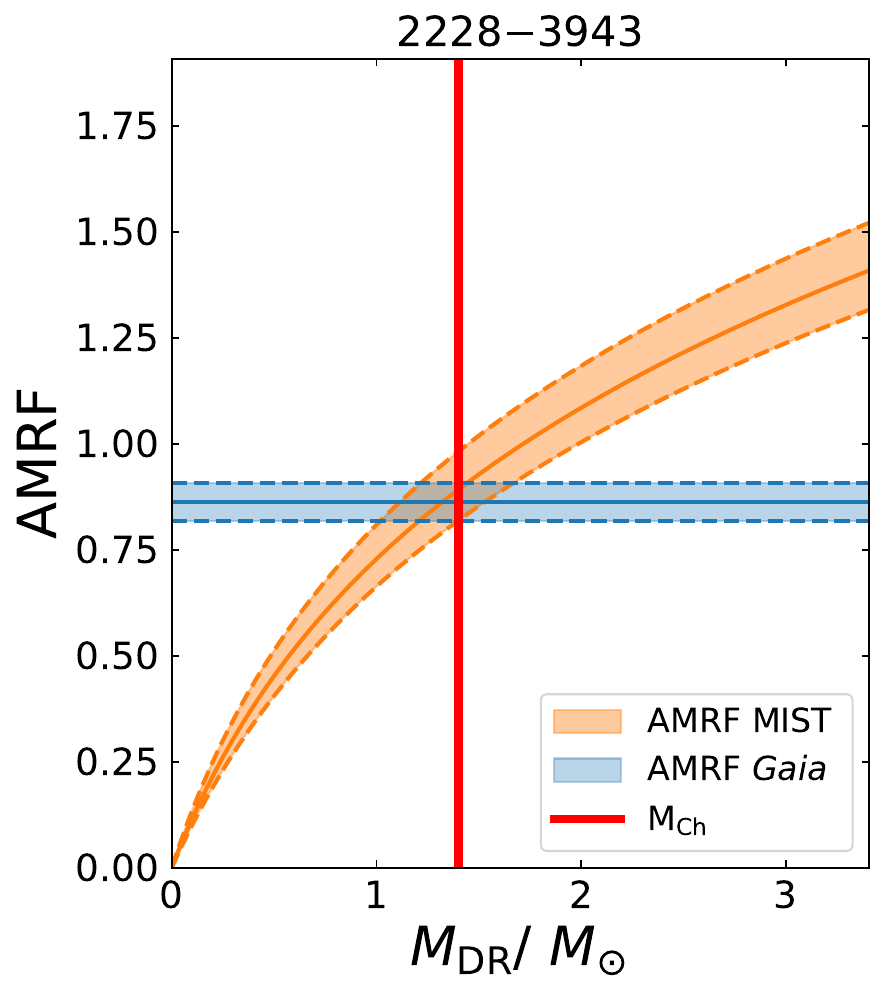}
\figsetgrpnote{Same as \autoref{fig:fm_0640-2621},but for source 2228$-$3943}
\label{fig:fm_15}

\figsetgrpend

\figsetgrpstart
\figsetgrpnum{6.16}
\figsetgrptitle{Image for figure 6_16}
\figsetplot{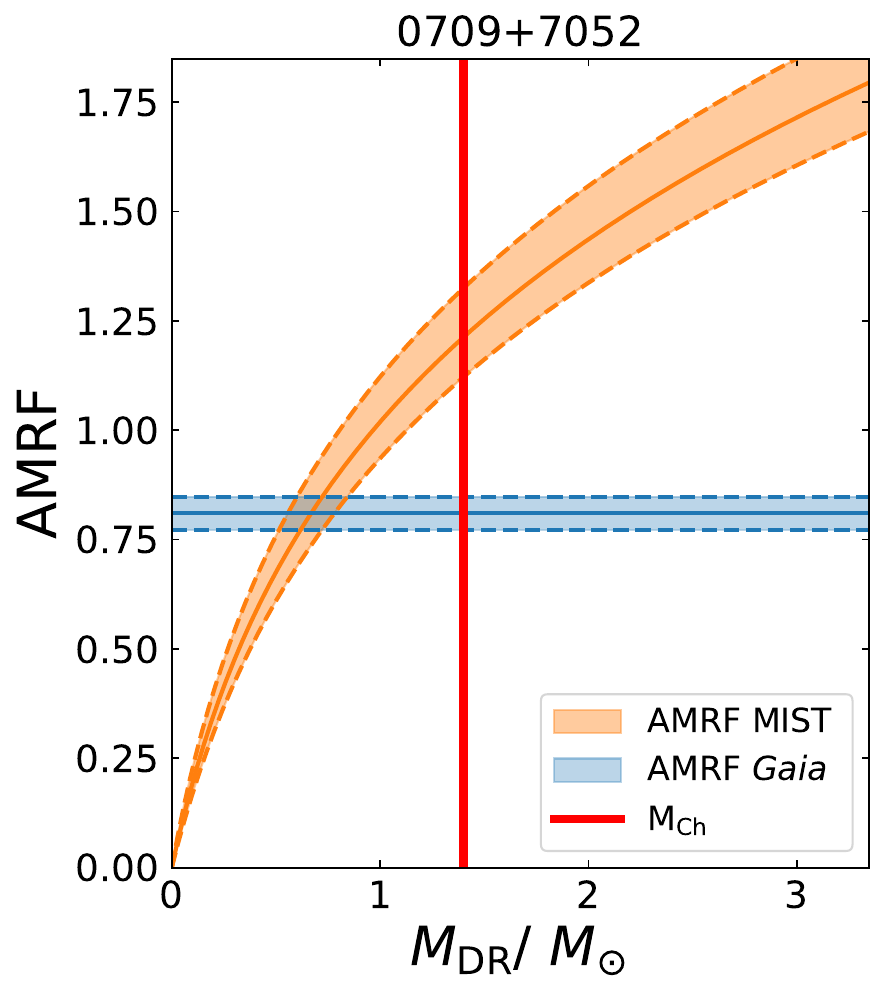}
\figsetgrpnote{Same as \autoref{fig:fm_0640-2621},but for source 0709$+$7052}
\label{fig:fm_16}

\figsetgrpend

\figsetgrpstart
\figsetgrpnum{6.17}
\figsetgrptitle{Image for figure 6_17}
\figsetplot{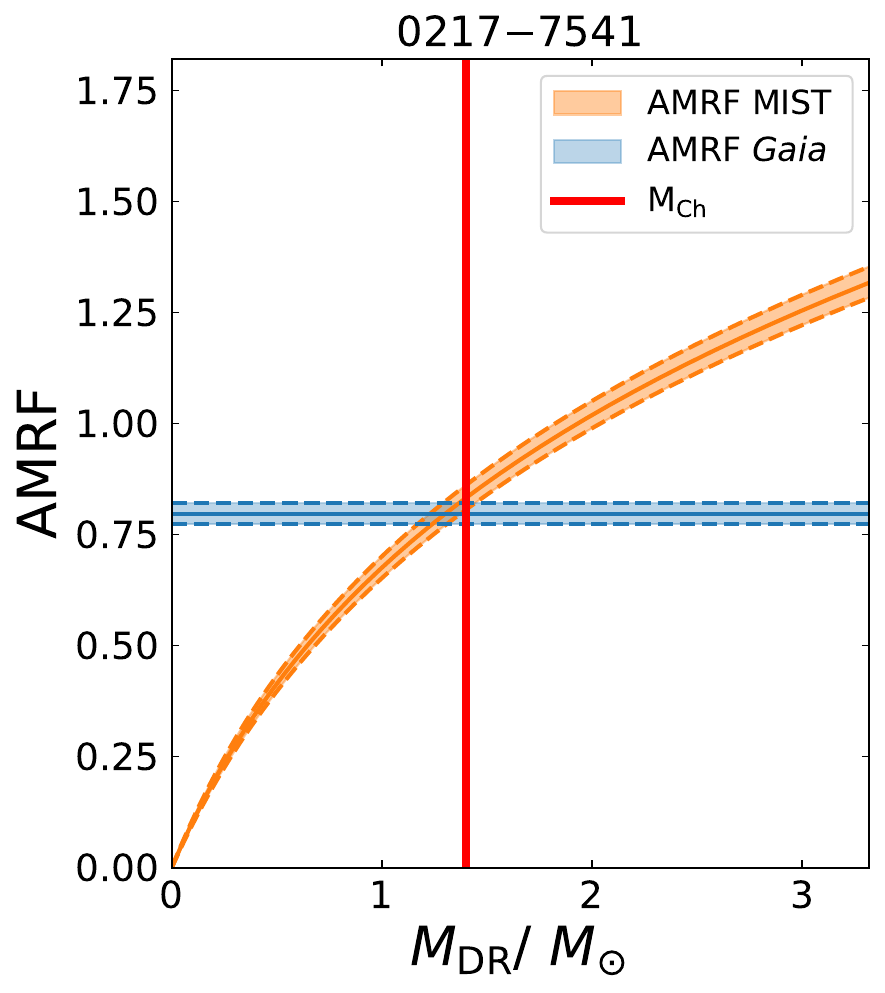}
\figsetgrpnote{Same as \autoref{fig:fm_0640-2621},but for source 0217$-$7541}
\label{fig:fm_17}

\figsetgrpend

\figsetgrpstart
\figsetgrpnum{6.18}
\figsetgrptitle{Image for figure 6_18}
\figsetplot{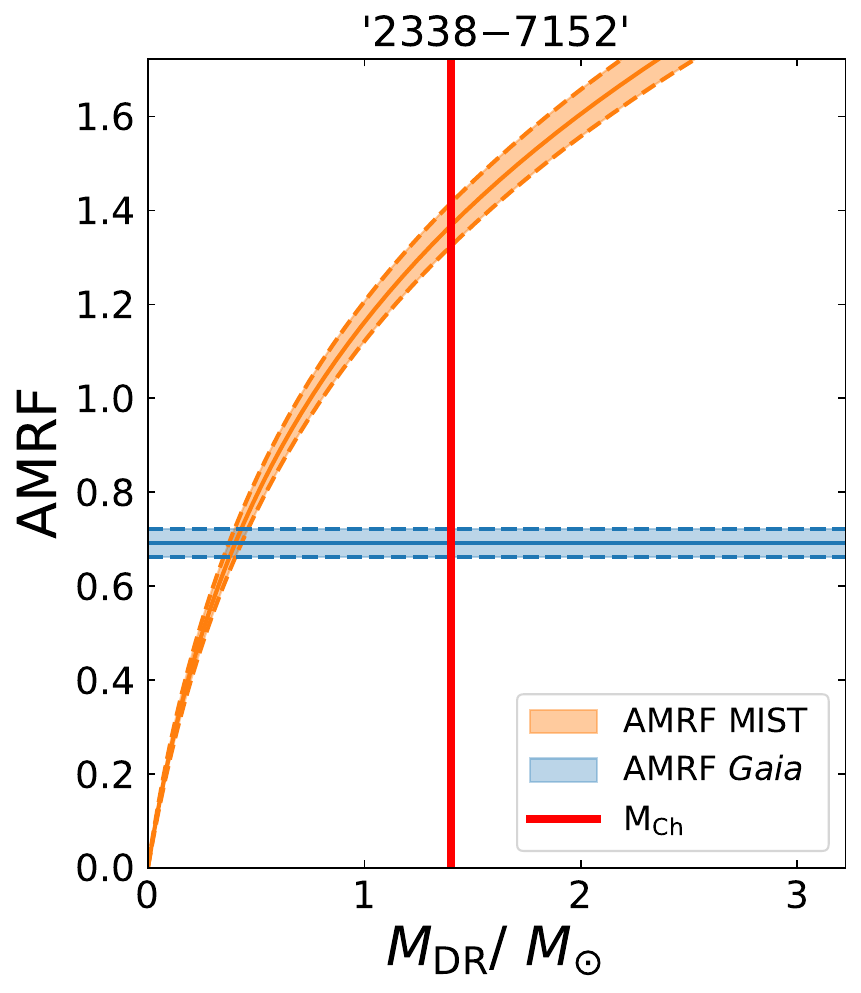}
\figsetgrpnote{Same as \autoref{fig:fm_0640-2621},but for source 2338$-$7152}
\label{fig:fm_18}

\figsetgrpend

\figsetgrpstart
\figsetgrpnum{6.19}
\figsetgrptitle{Image for figure 6_19}
\figsetplot{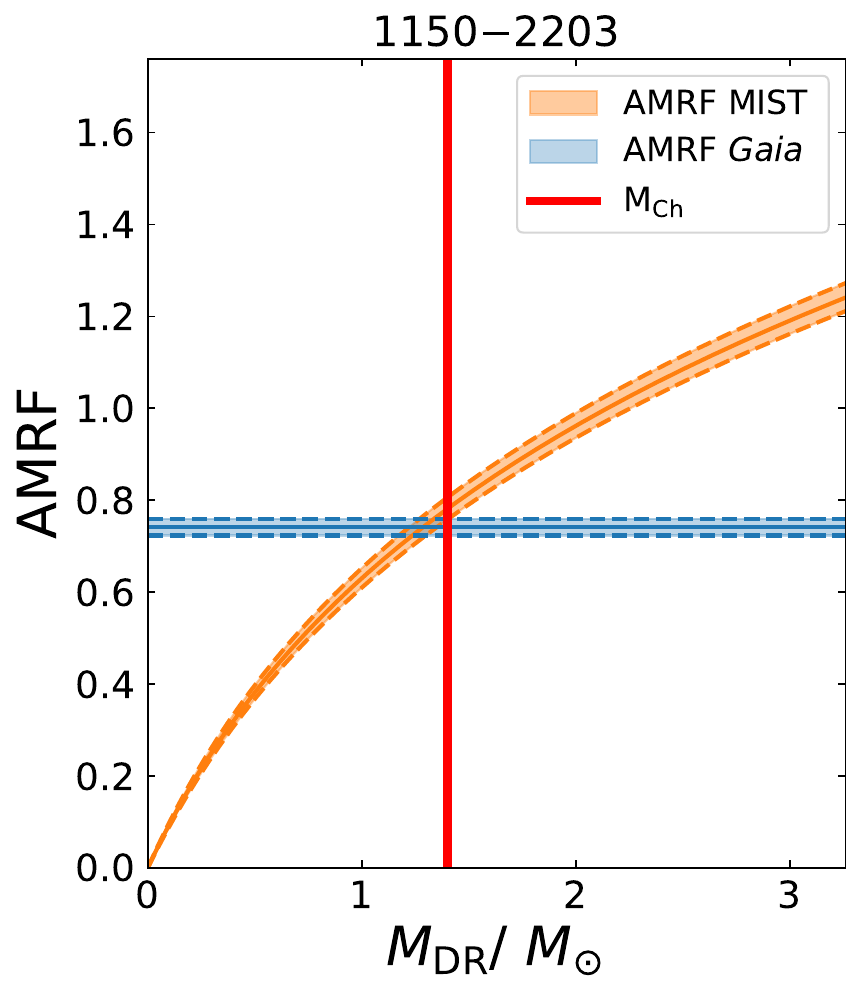}
\figsetgrpnote{Same as \autoref{fig:fm_0640-2621},but for source 1150$-$2203}
\label{fig:fm_19}

\figsetgrpend

\figsetgrpstart
\figsetgrpnum{6.20}
\figsetgrptitle{Image for figure 6_20}
\figsetplot{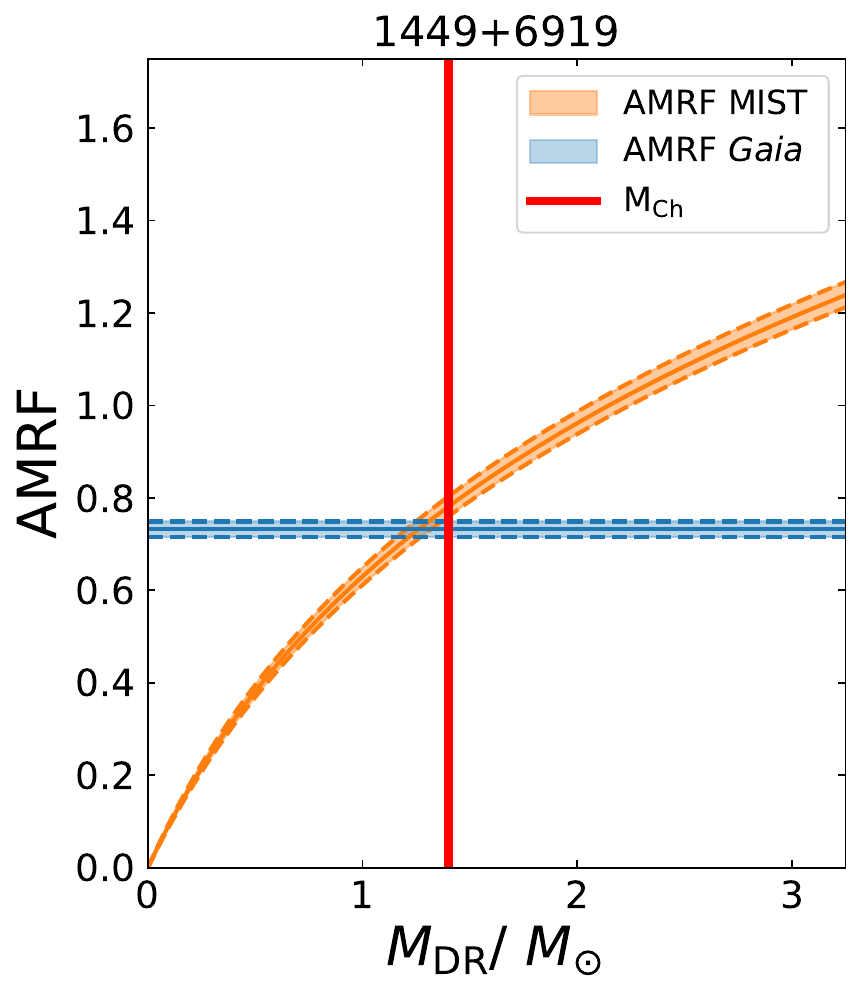}
\figsetgrpnote{Same as \autoref{fig:fm_0640-2621},but for source 1449$+$6919}
\label{fig:fm_20}

\figsetgrpend

\figsetgrpstart
\figsetgrpnum{6.21}
\figsetgrptitle{Image for figure 6_21}
\figsetplot{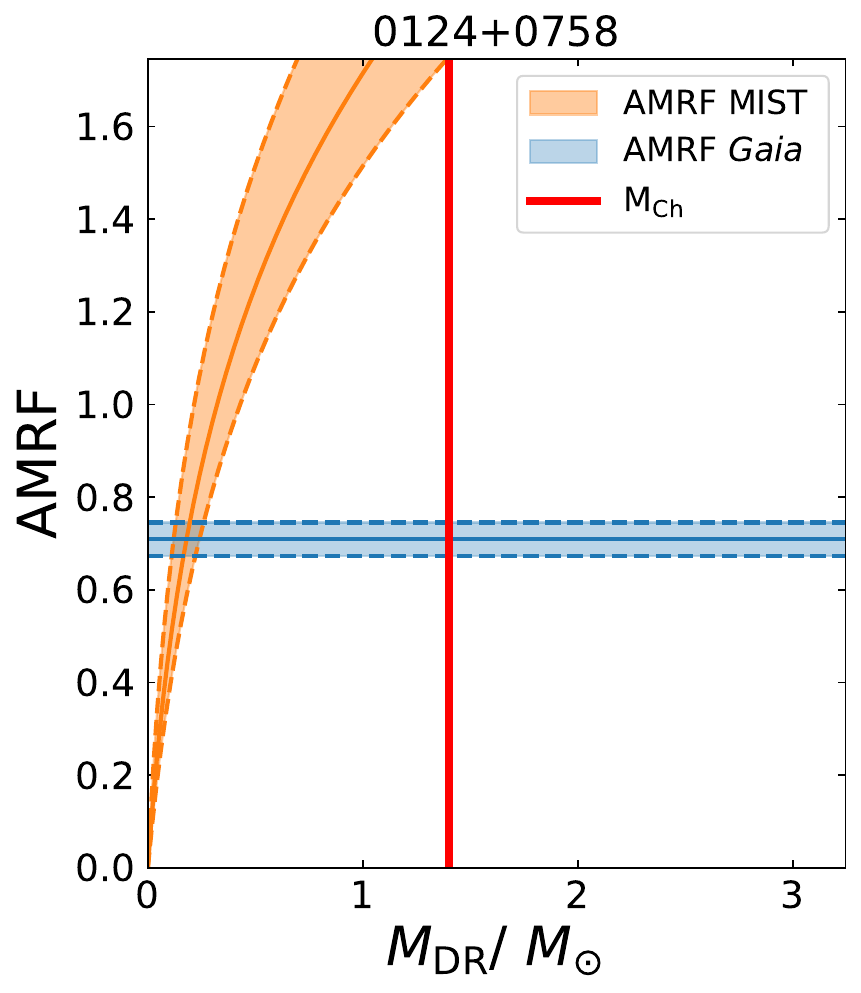}
\figsetgrpnote{Same as \autoref{fig:fm_0640-2621},but for source 0124$+$0758}
\label{fig:fm_21}

\figsetgrpend

\figsetgrpstart
\figsetgrpnum{6.22}
\figsetgrptitle{Image for figure 6_22}
\figsetplot{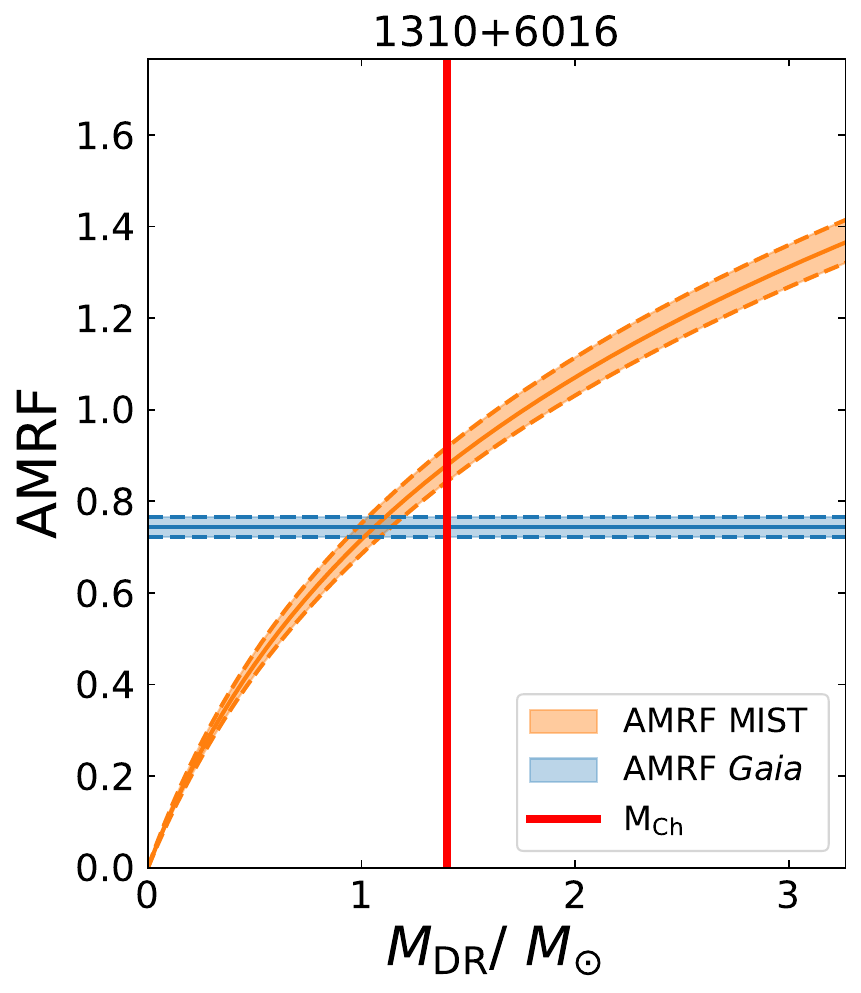}
\figsetgrpnote{Same as \autoref{fig:fm_0640-2621},but for source 1310$+$6016}
\label{fig:fm_22}

\figsetgrpend

\figsetgrpstart
\figsetgrpnum{6.23}
\figsetgrptitle{Image for figure 6_23}
\figsetplot{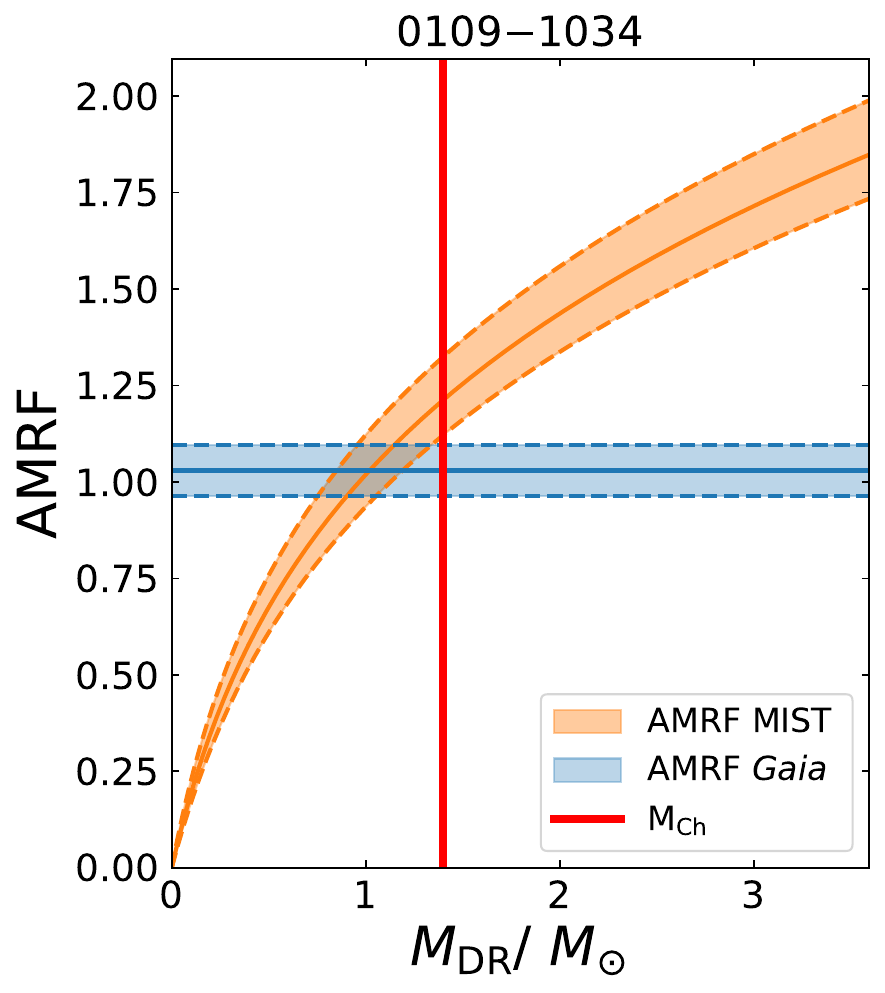}
\figsetgrpnote{Same as \autoref{fig:fm_0640-2621},but for source 0109$-$1034}
\label{fig:fm_23}

\figsetgrpend

\figsetgrpstart
\figsetgrpnum{6.24}
\figsetgrptitle{Image for figure 6_24}
\figsetplot{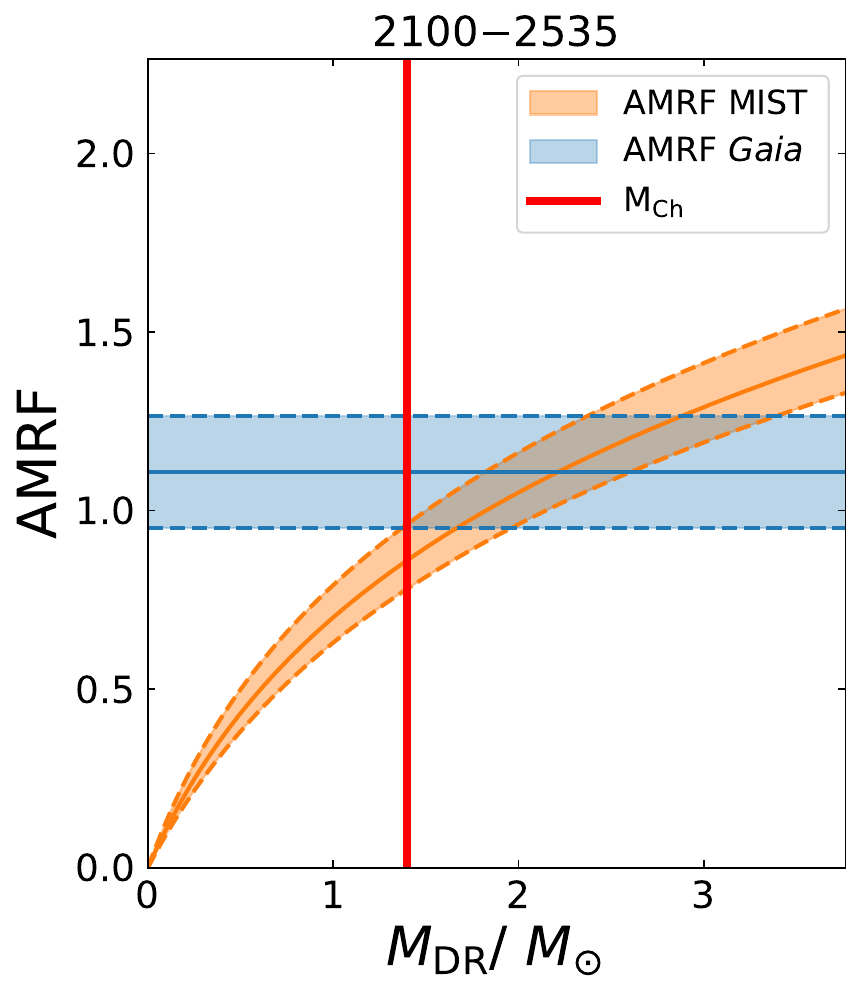}
\figsetgrpnote{Same as \autoref{fig:fm_0640-2621},but for source 2100$-$2535}
\label{fig:fm_24}

\figsetgrpend

\figsetgrpstart
\figsetgrpnum{6.25}
\figsetgrptitle{Image for figure 6_25}
\figsetplot{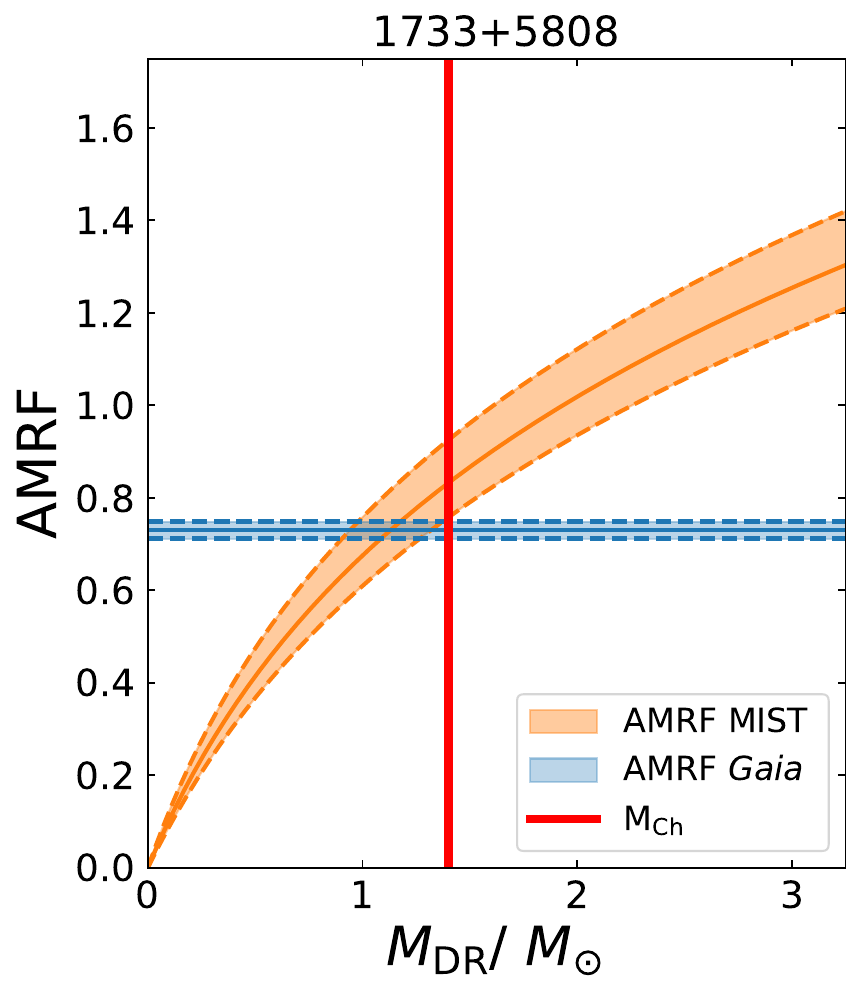}
\figsetgrpnote{Same as \autoref{fig:fm_0640-2621},but for source 1733$+$5808}
\label{fig:fm_25}

\figsetgrpend

\figsetgrpstart
\figsetgrpnum{6.26}
\figsetgrptitle{Image for figure 6_26}
\figsetplot{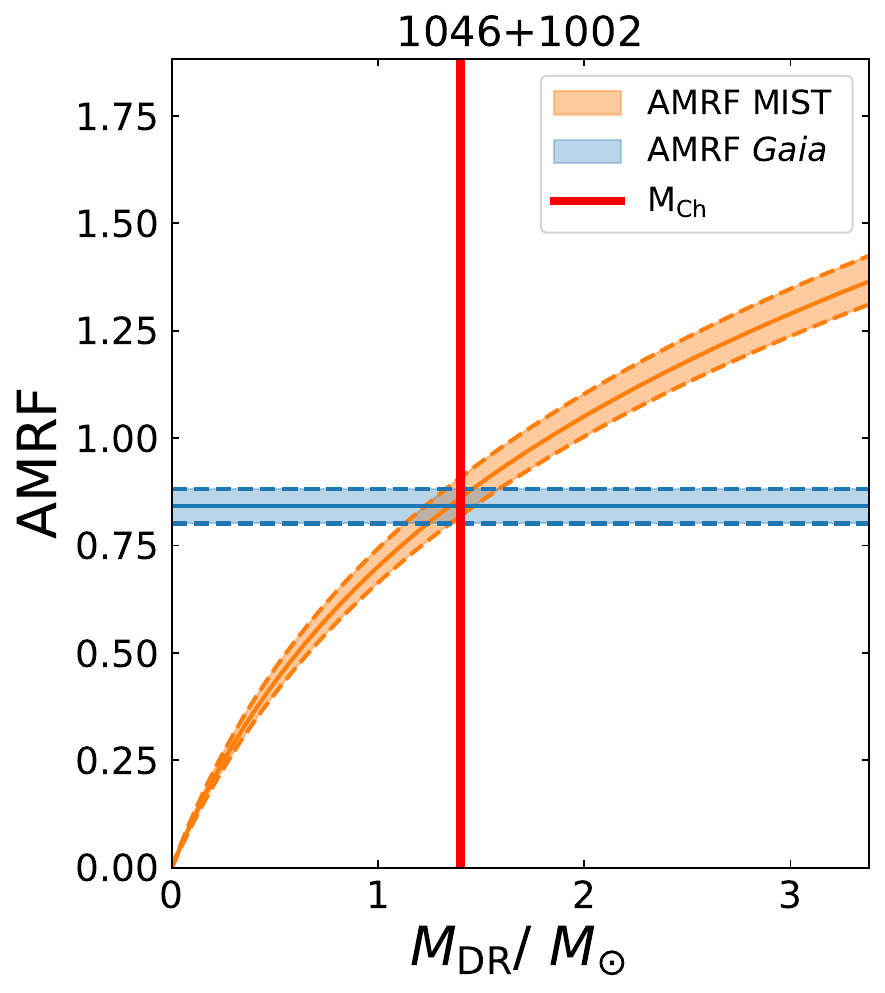}
\figsetgrpnote{Same as \autoref{fig:fm_0640-2621},but for source 1046$+$1002}
\label{fig:fm_26}

\figsetgrpend

\figsetgrpstart
\figsetgrpnum{6.27}
\figsetgrptitle{Image for figure 6_27}
\figsetplot{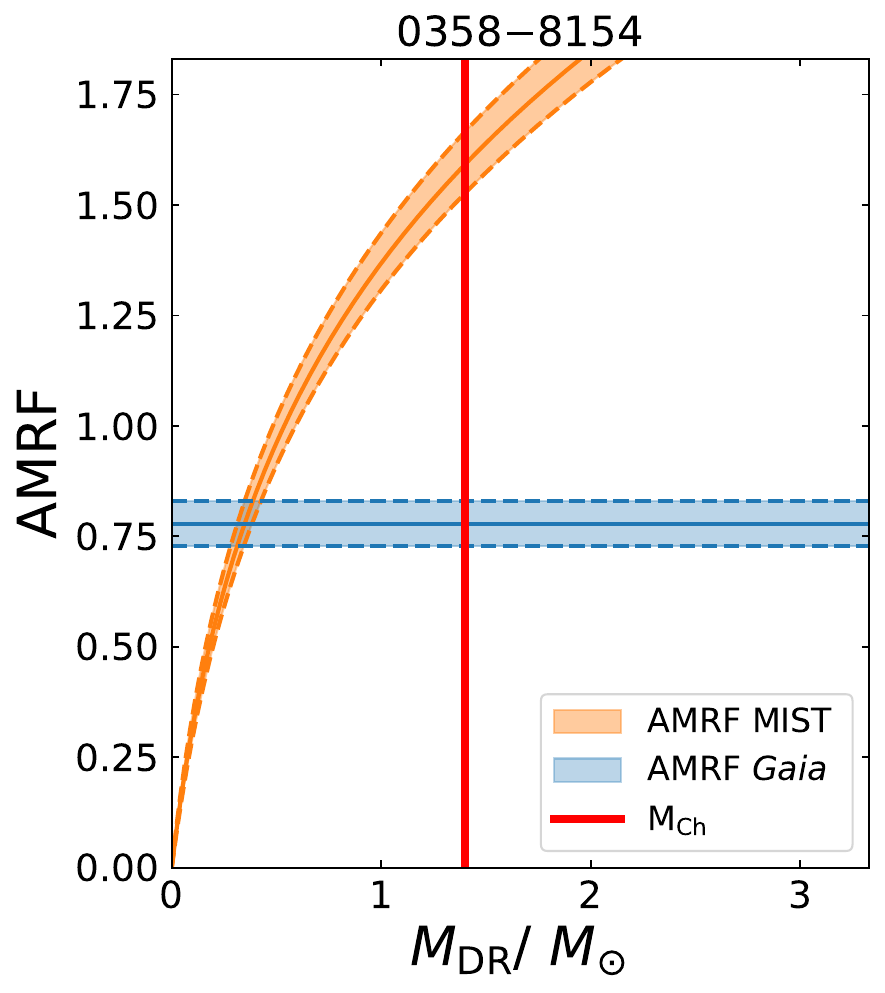}
\figsetgrpnote{Same as \autoref{fig:fm_0640-2621},but for source 0358$-$8154}
\label{fig:fm_27}

\figsetgrpend

\figsetgrpstart
\figsetgrpnum{6.28}
\figsetgrptitle{Image for figure 6_28}
\figsetplot{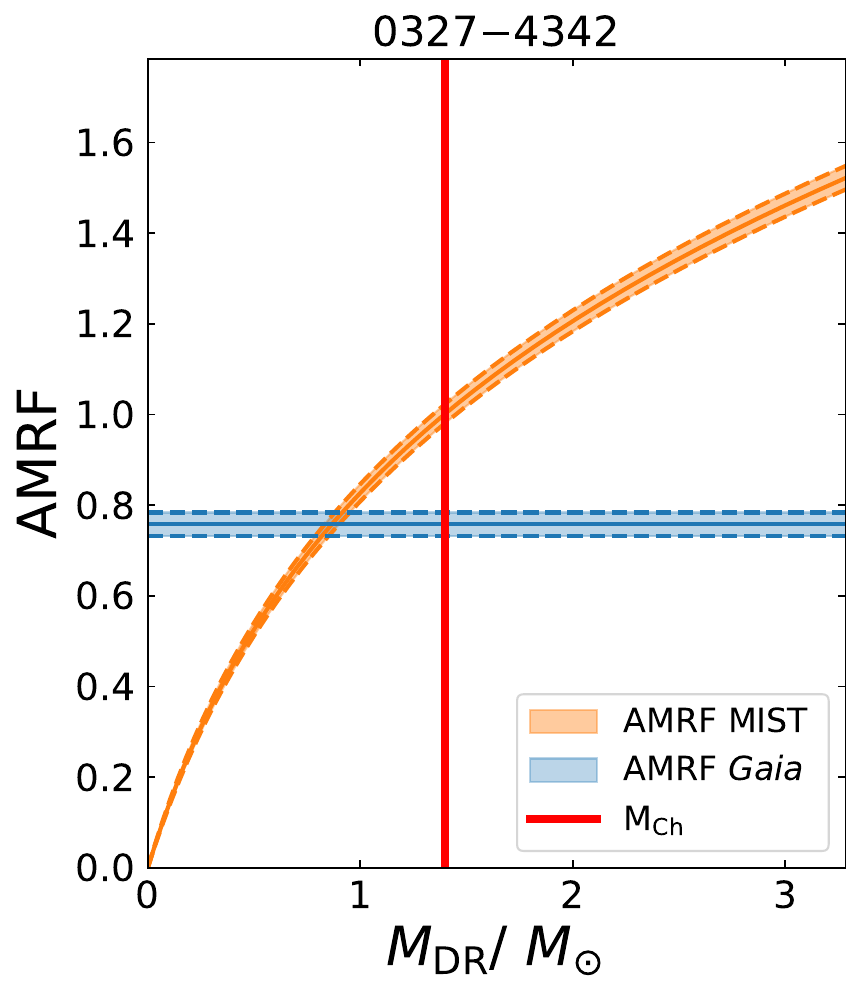}
\figsetgrpnote{Same as \autoref{fig:fm_0640-2621},but for source 0327$-$4342}
\label{fig:fm_28}

\figsetgrpend

\figsetgrpstart
\figsetgrpnum{6.29}
\figsetgrptitle{Image for figure 6_29}
\figsetplot{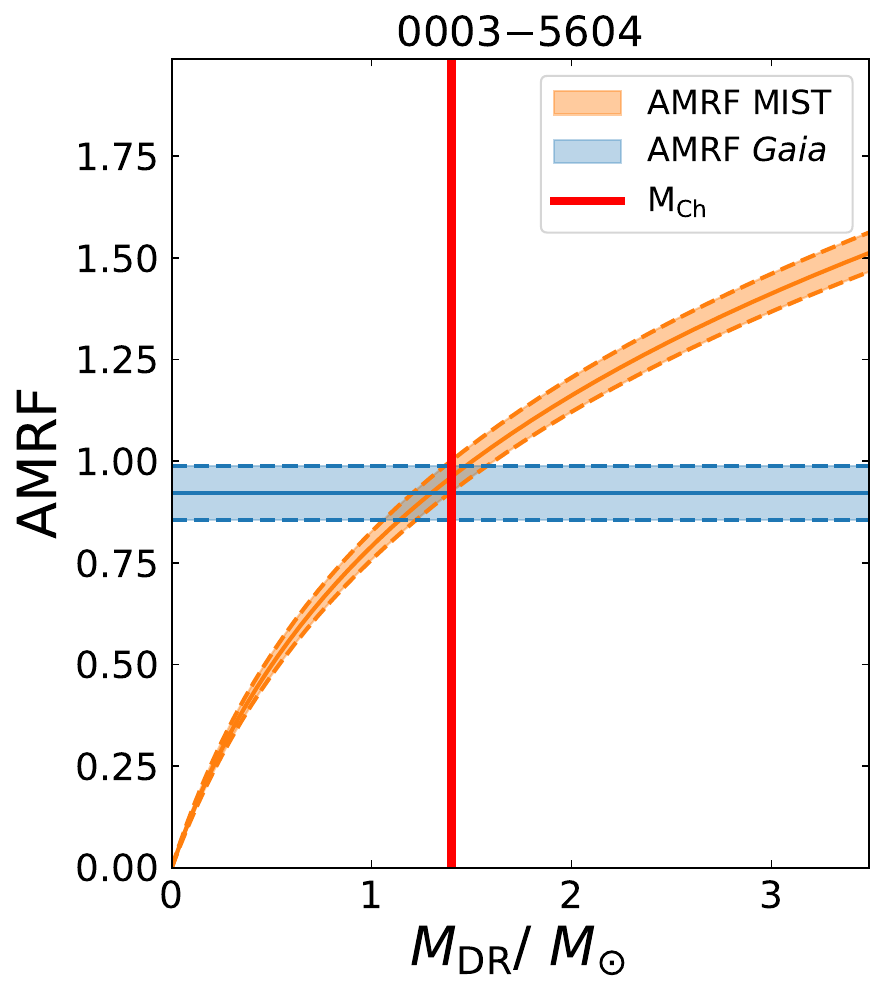}
\figsetgrpnote{Same as \autoref{fig:fm_0640-2621},but for source 0003$-$5604}
\label{fig:fm_29}

\figsetgrpend

\figsetgrpstart
\figsetgrpnum{6.30}
\figsetgrptitle{Image for figure 6_30}
\figsetplot{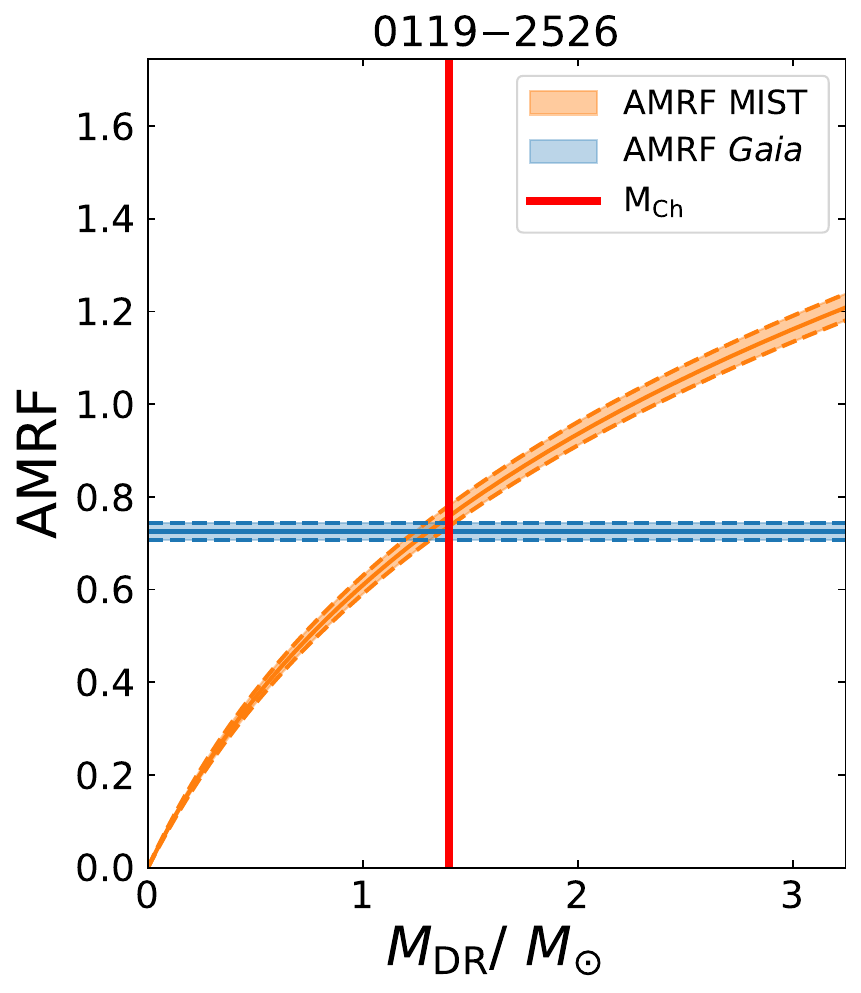}
\figsetgrpnote{Same as \autoref{fig:fm_0640-2621},but for source 0119$-$2526}
\label{fig:fm_30}

\figsetgrpend

\figsetgrpstart
\figsetgrpnum{6.31}
\figsetgrptitle{Image for figure 6_31}
\figsetplot{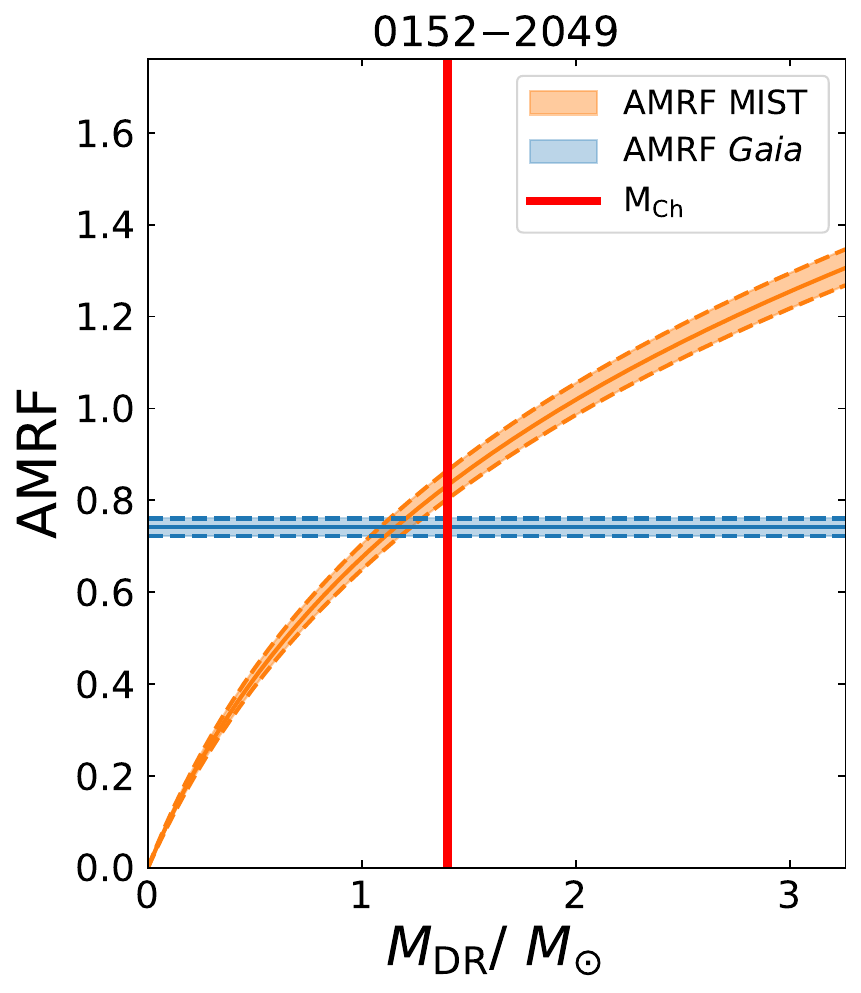}
\figsetgrpnote{Same as \autoref{fig:fm_0640-2621},but for source 0152$-$2049}
\label{fig:fm_31}

\figsetgrpend

\figsetgrpstart
\figsetgrpnum{6.32}
\figsetgrptitle{Image for figure 6_32}
\figsetplot{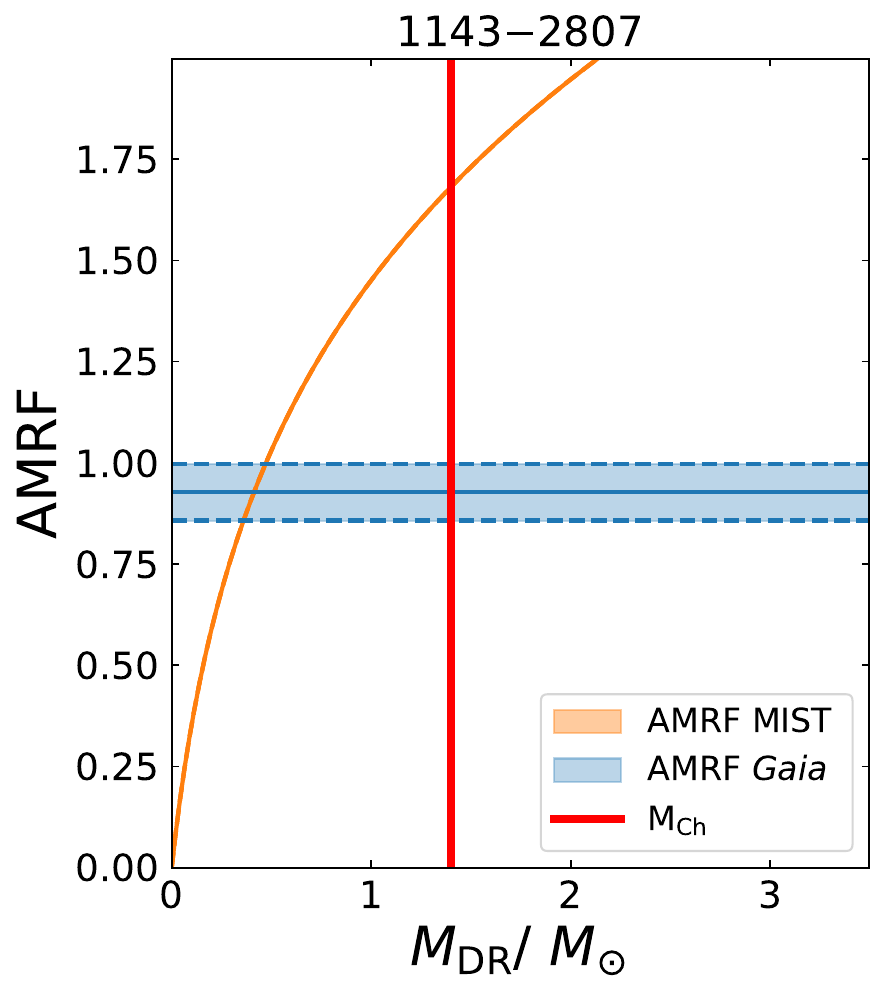}
\figsetgrpnote{Same as \autoref{fig:fm_0640-2621},but for source 1143$-$2807}
\label{fig:fm_32}

\figsetgrpend

\figsetgrpstart
\figsetgrpnum{6.33}
\figsetgrptitle{Image for figure 6_33}
\figsetplot{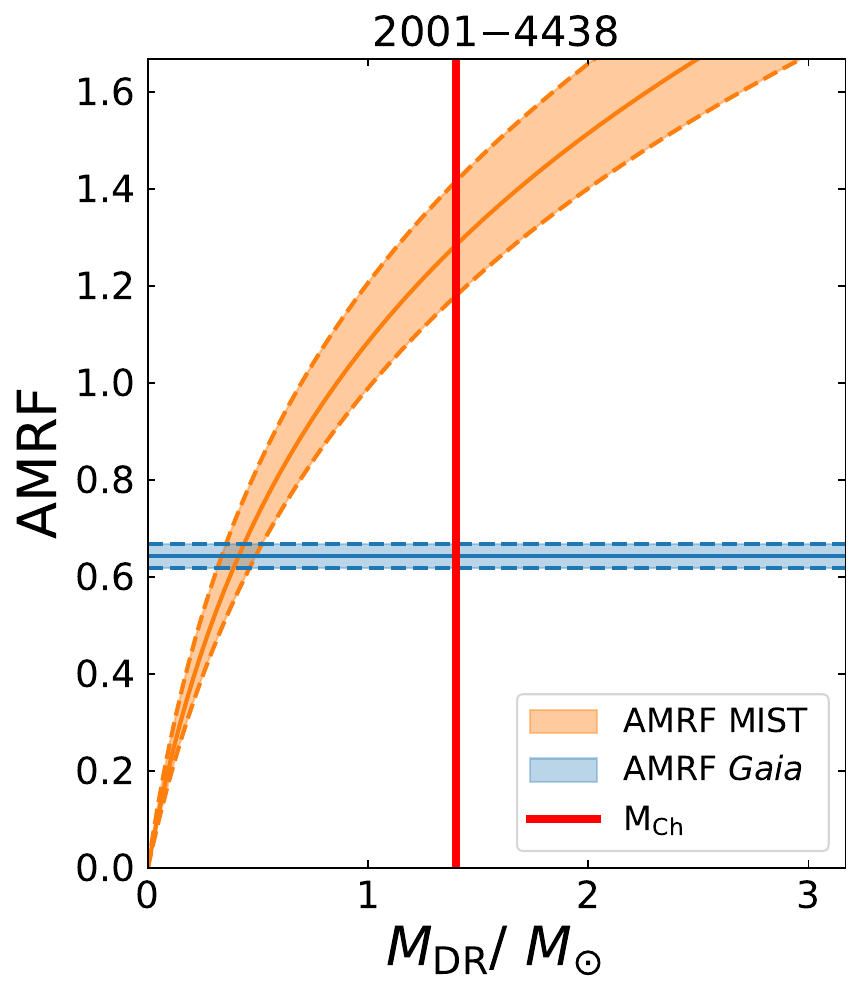}
\figsetgrpnote{Same as \autoref{fig:fm_0640-2621},but for source 2001$-$4438}
\label{fig:fm_33}

\figsetgrpend

\figsetgrpstart
\figsetgrpnum{6.34}
\figsetgrptitle{Image for figure 6_34}
\figsetplot{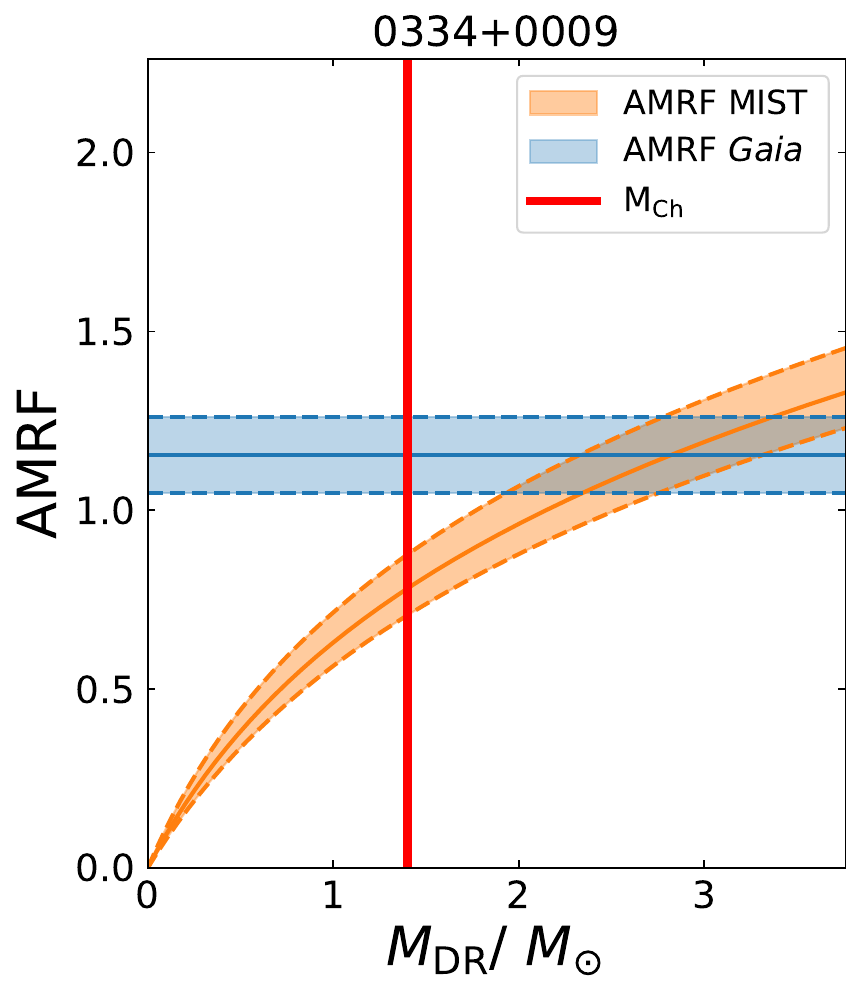}
\figsetgrpnote{Same as \autoref{fig:fm_0640-2621},but for source 0334$+$0009}
\label{fig:fm_34}

\figsetgrpend

\figsetgrpstart
\figsetgrpnum{6.35}
\figsetgrptitle{Image for figure 6_35}
\figsetplot{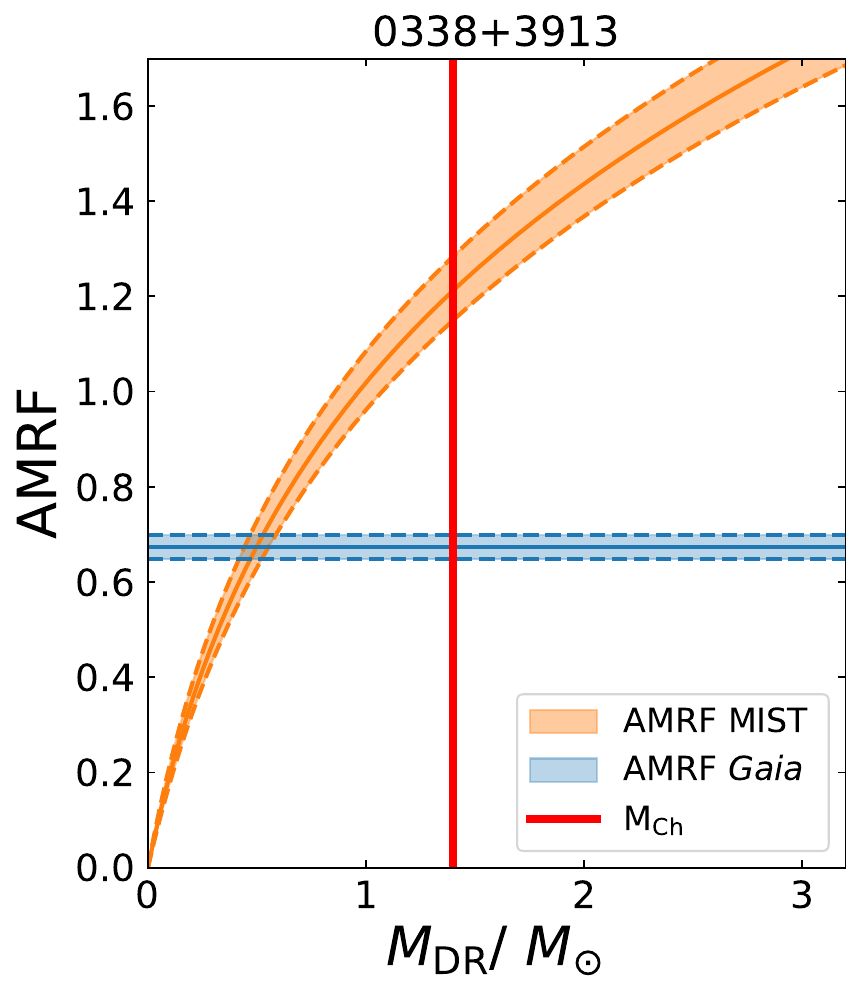}
\figsetgrpnote{Same as \autoref{fig:fm_0640-2621},but for source 0338$+$3913}
\label{fig:fm_35}

\figsetgrpend

\figsetgrpstart
\figsetgrpnum{6.36}
\figsetgrptitle{Image for figure 6_36}
\figsetplot{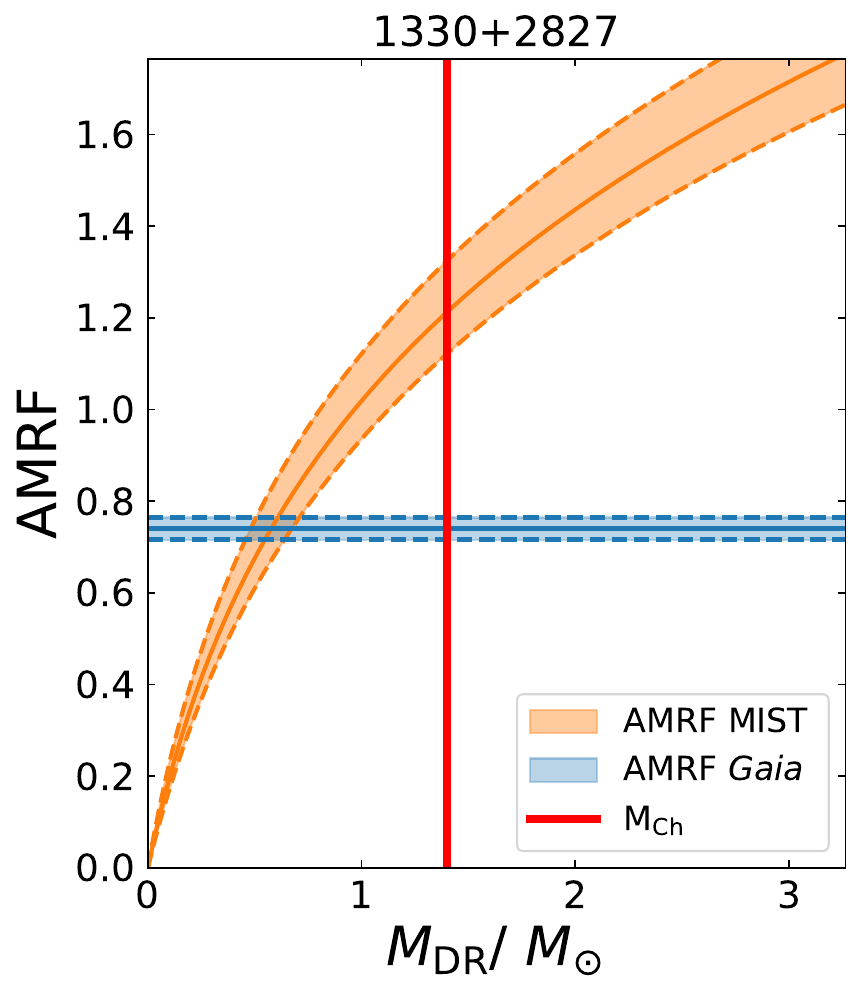}
\figsetgrpnote{Same as \autoref{fig:fm_0640-2621},but for source 1330$+$2827}
\label{fig:fm_36}

\figsetgrpend

\figsetgrpstart
\figsetgrpnum{6.37}
\figsetgrptitle{Image for figure 6_37}
\figsetplot{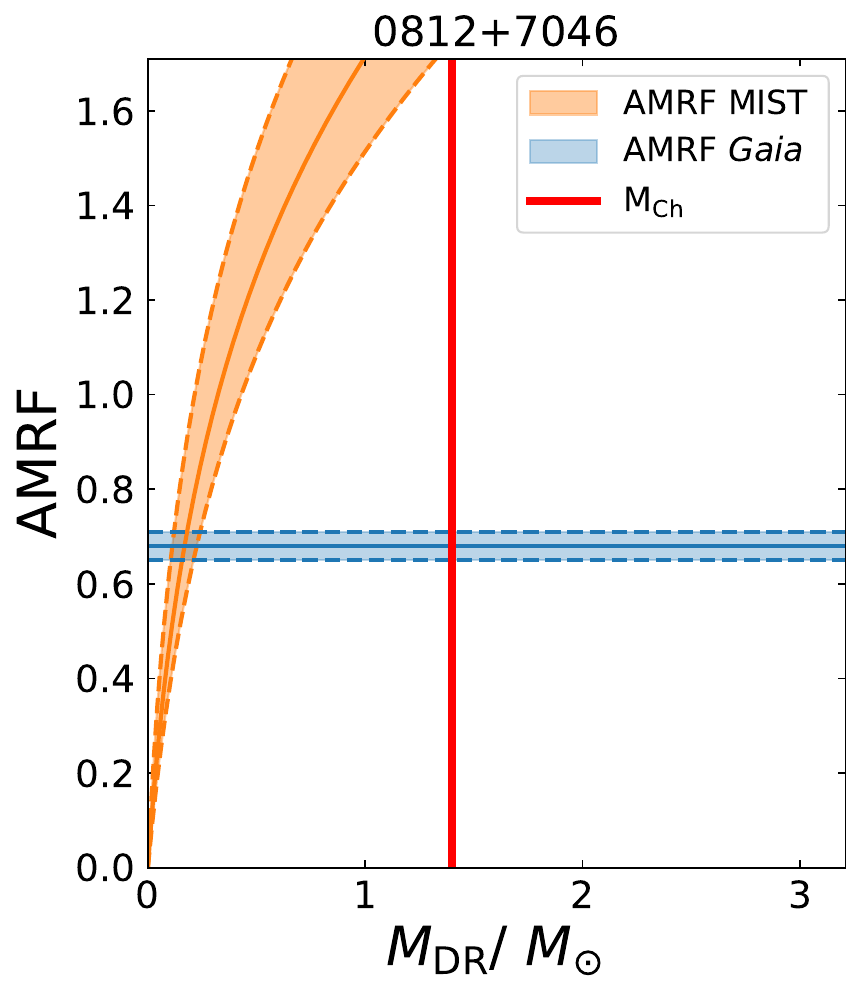}
\figsetgrpnote{Same as \autoref{fig:fm_0640-2621},but for source 0812$+$7046}
\label{fig:fm_37}

\figsetgrpend

\figsetgrpstart
\figsetgrpnum{6.38}
\figsetgrptitle{Image for figure 6_38}
\figsetplot{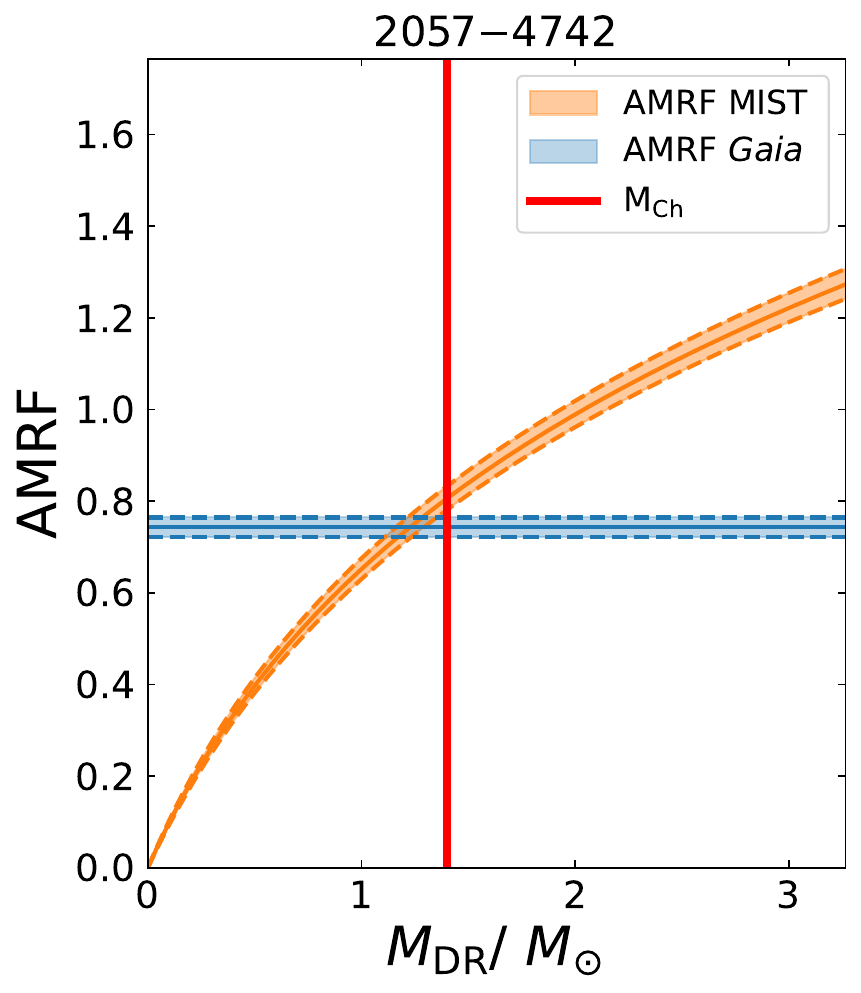}
\figsetgrpnote{Same as \autoref{fig:fm_0640-2621},but for source 2057$-$4742}
\label{fig:fm_38}

\figsetgrpend

\figsetgrpstart
\figsetgrpnum{6.39}
\figsetgrptitle{Image for figure 6_39}
\figsetplot{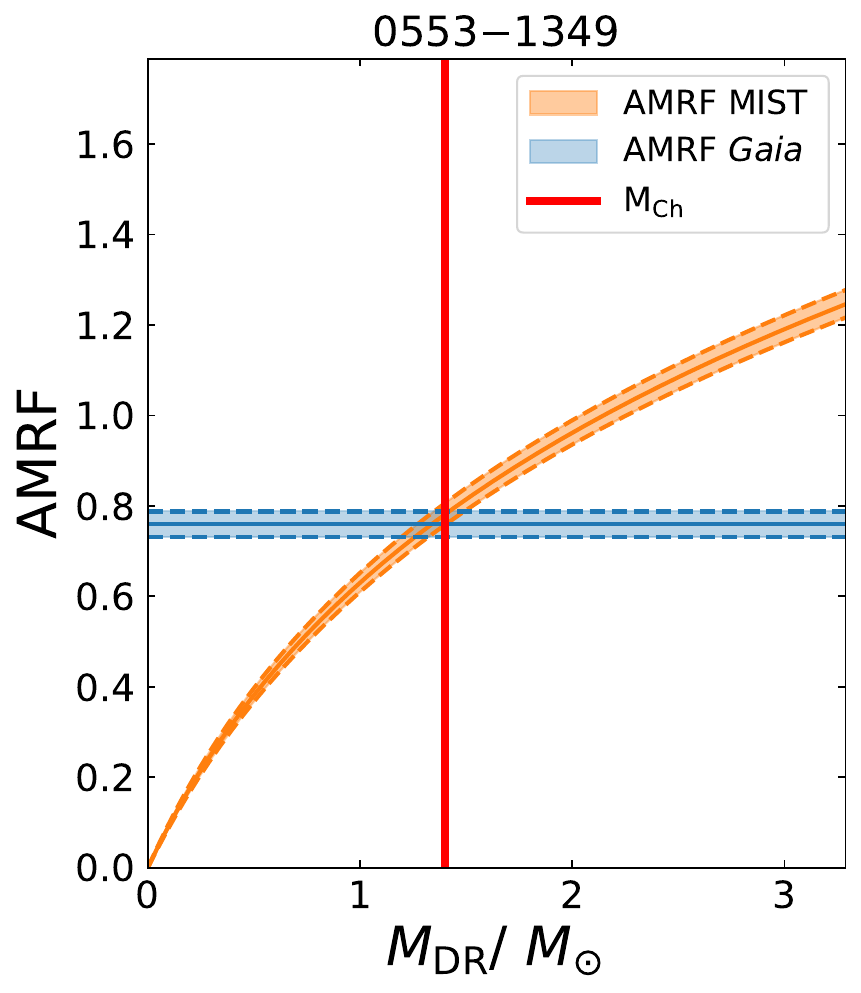}
\figsetgrpnote{Same as \autoref{fig:fm_0640-2621},but for source 0553$-$1349}
\label{fig:fm_39}

\figsetgrpend

\figsetgrpstart
\figsetgrpnum{6.40}
\figsetgrptitle{Image for figure 6_40}
\figsetplot{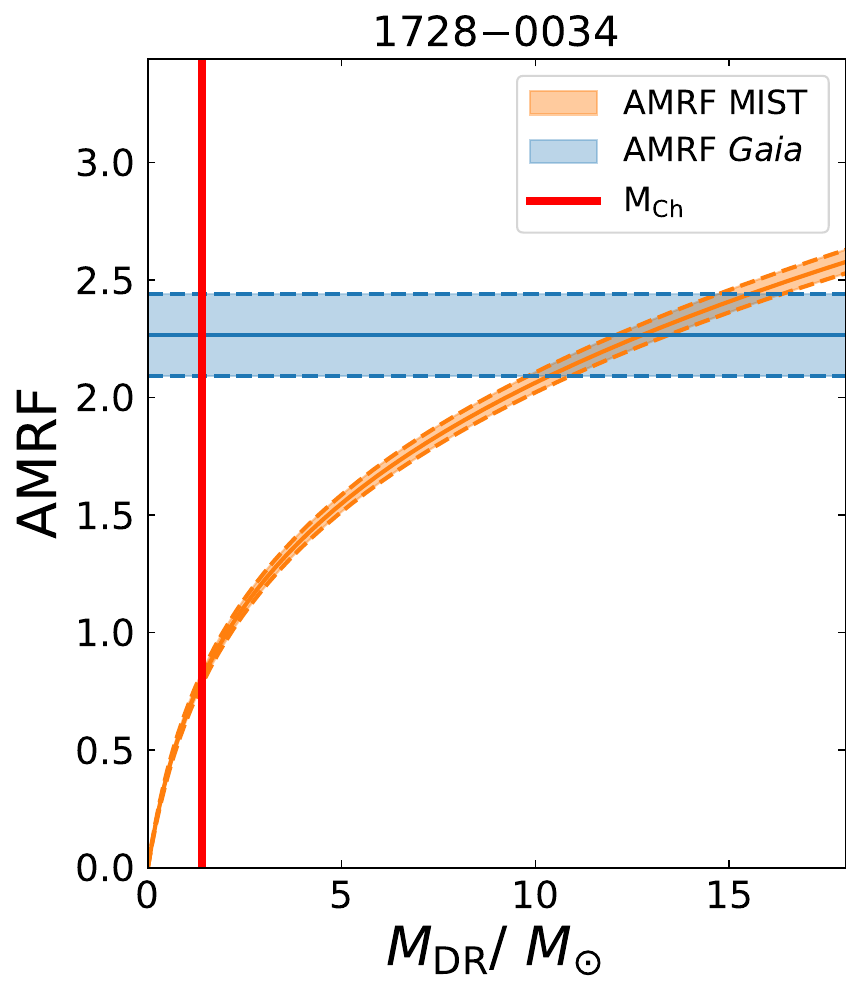}
\figsetgrpnote{Same as \autoref{fig:fm_0640-2621},but for source 1728$-$0034}
\label{fig:fm_40}

\figsetgrpend

\figsetgrpstart
\figsetgrpnum{6.41}
\figsetgrptitle{Image for figure 6_41}
\figsetplot{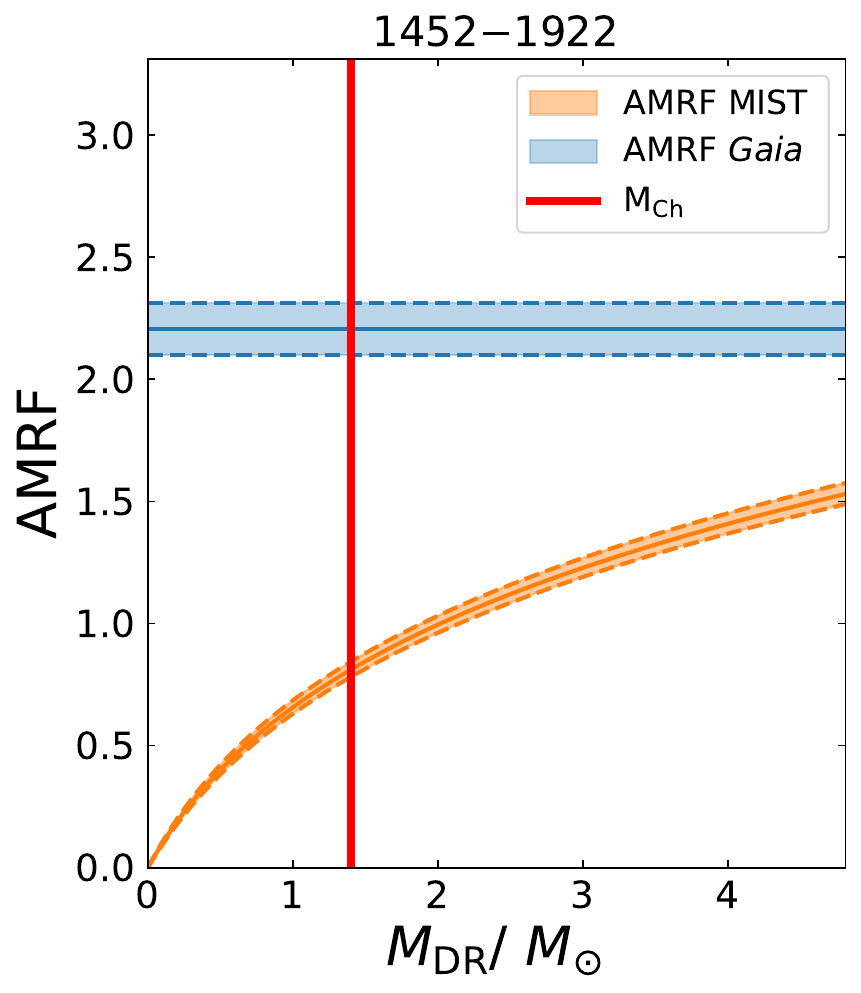}
\figsetgrpnote{Same as \autoref{fig:fm_0640-2621},but for source 1452$-$1922}
\label{fig:fm_41}

\figsetgrpend

\figsetgrpstart
\figsetgrpnum{6.42}
\figsetgrptitle{Image for figure 6_42}
\figsetplot{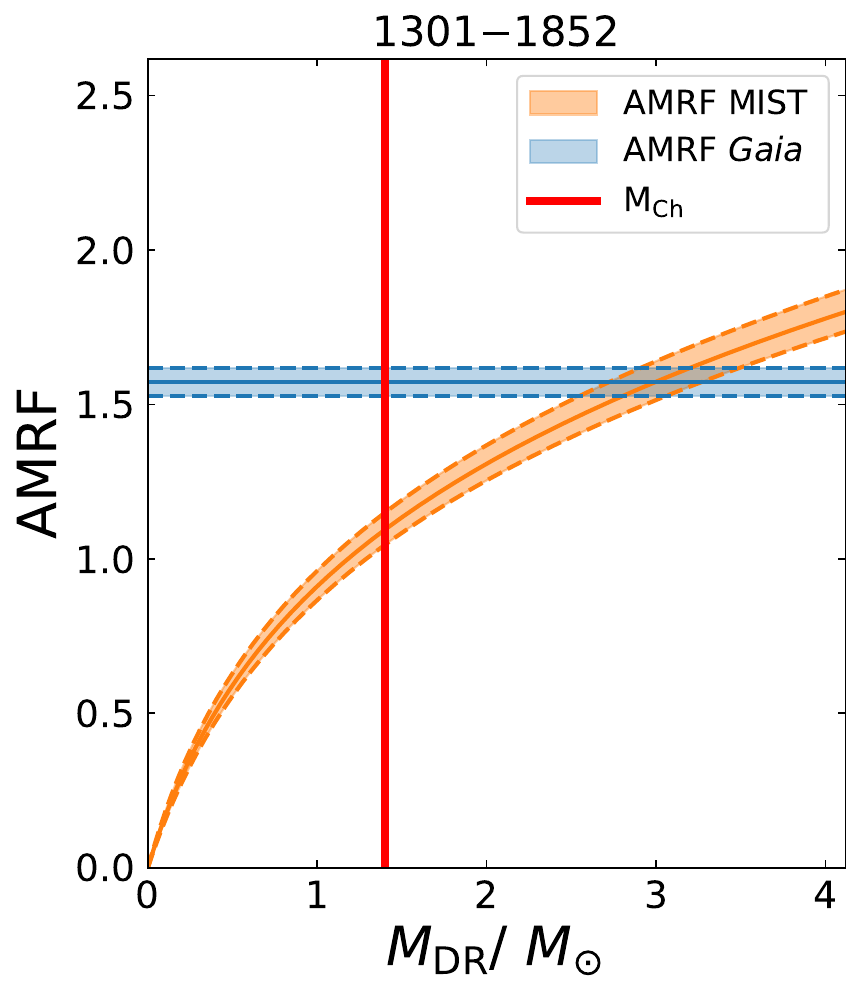}
\figsetgrpnote{Same as \autoref{fig:fm_0640-2621},but for source 1301$-$1852}
\label{fig:fm_42}

\figsetgrpend

\figsetgrpstart
\figsetgrpnum{6.43}
\figsetgrptitle{Image for figure 6_43}
\figsetplot{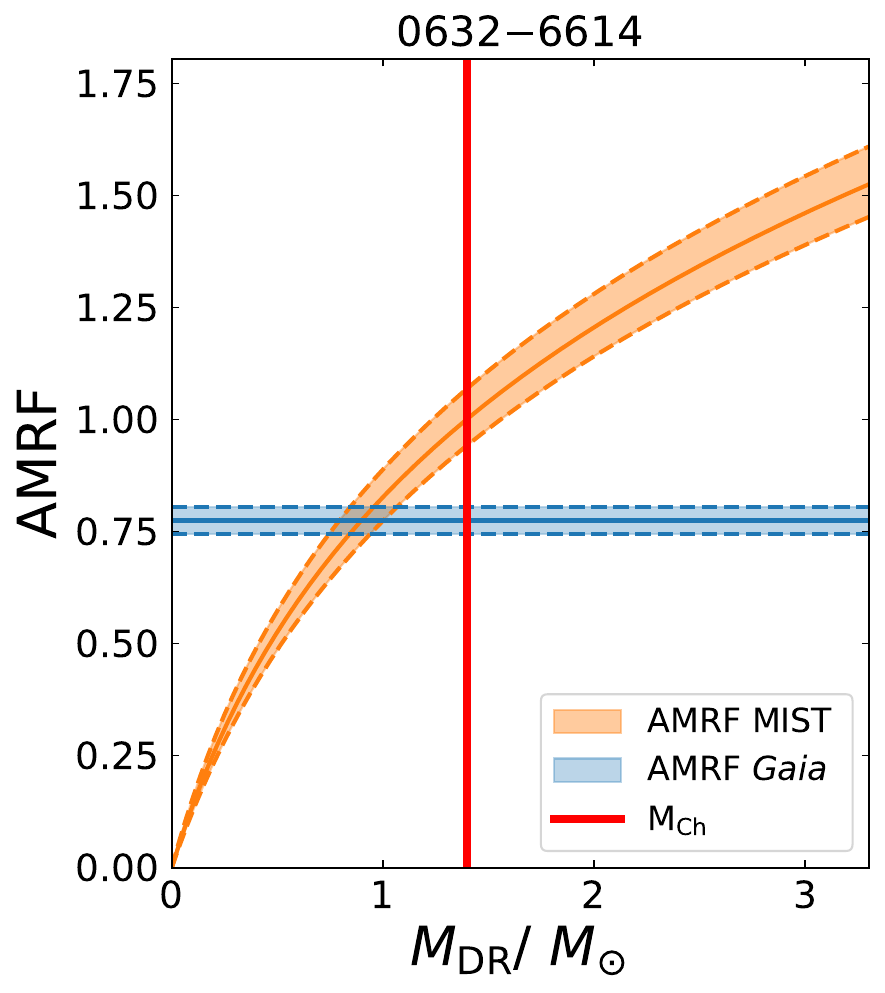}
\figsetgrpnote{Same as \autoref{fig:fm_0640-2621},but for source 0632$-$6614}
\label{fig:fm_43}

\figsetgrpend

\figsetgrpstart
\figsetgrpnum{6.44}
\figsetgrptitle{Image for figure 6_44}
\figsetplot{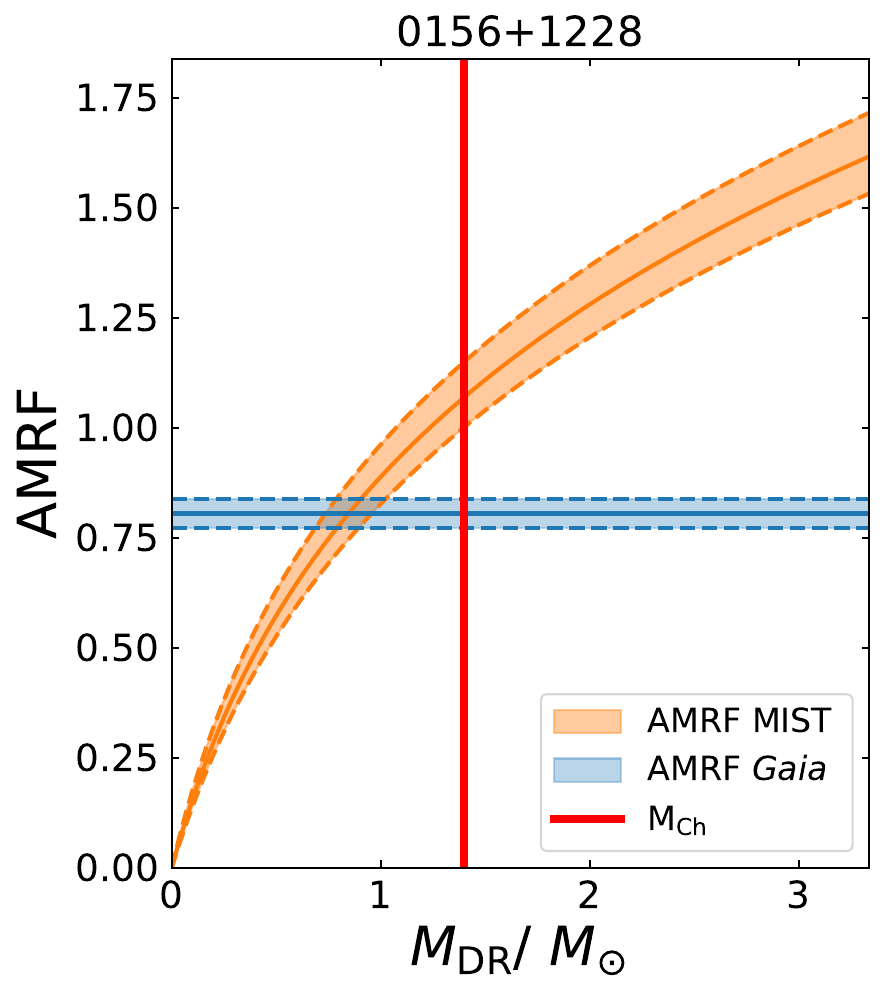}
\figsetgrpnote{Same as \autoref{fig:fm_0640-2621},but for source 0156$+$1228}
\label{fig:fm_44}

\figsetgrpend

\figsetgrpstart
\figsetgrpnum{6.45}
\figsetgrptitle{}
\figsetplot{}
\figsetgrpnote{Same as \autoref{fig:fm_0640-2621},but for source 1606$+$6120}
\label{fig:fm_45}

\figsetgrpend

\figsetgrpend

\figsetgrpstart
\figsetgrpnum{6.46}
\figsetgrptitle{}
\figsetplot{}
\figsetgrpnote{$f_{M}$  vs $\mdr$ for 1007$+$3408. The orange-shaded region shows the allowed parameter space based on our estimated $\mlc$ (\autoref{sec:LC}). The blue-shaded region shows the estimated $f_{M}$ from ATF22. The overlap between the blue and the orange regions is the solution for $\mdr$ that satisfies all available constraints. The red vertical line denotes $\mch$.}
\label{fig:fm_46}

\figsetgrpend

\figsetgrpend

\figsetgrpstart
\figsetgrpnum{6.47}
\figsetgrptitle{}
\figsetplot{}
\figsetgrpnote{Same as \autoref{fig:fm_46},but for source 2033$+$0758}
\label{fig:fm_47}

\figsetgrpend

\figsetend
Incidentally, all of these 15 sources belong to the SHA23 catalog which provides constraints on the AMRF (\autoref{eq:amrf}) from \gaia's astrometric solutions for these candidate binaries. Adopting the constraints of $\lbollc$, $\tefflc$, and [Fe/H] from SED modeling, and using MIST stellar evolution models we constrain $\mlc$ (\autoref{S:methods_mass}). Using $\mlc$ and the constraints on the AMRF we estimate $\mdr$ for these sources (\autoref{S:methods_mass}). \autoref{fig:fm_0640-2621} shows AMRF vs $\mdr$ using source 0640-2621 as an example. The orange shaded region shows the allowed AMRF values as a function of $\mdr$ adopting our $\mlc$ constraints. The blue shaded region shows the AMRF measured by SHA23 from \gaia\ DR3. The overlap between the two shaded regions satisfy all available constraints. Clearly, the estimated $\mdr$ for source 0640$-$2621 is below $\mch$ denoted by the vertical red line. Indeed, we find that the estimated $\mdr$ for all of these 15 sources are significantly below $\mch$. While $\mdr/\msun<1$ for most sources, source 1220$+$5841 exhibits the highest $\mdr/M_\odot=1.28\pm0.06$ with source 0640$-$2621 a close second, $\mdr/\msun=1.18\pm0.04$ (\autoref{tab:wd_data}). This bolsters our belief that the easiest way to explain the UV excess for these sources is that the DRs in them are indeed WDs. 

Interestingly, sources 0812$+$7046, 2106$-$5218, 0124$+$0758, and 0358$-$8154 with $\mdr/\msun<0.4$ are likely so-called extremely low-mass WDs (ELMWD). ELMWDs cannot be created via a single star's evolution simply because the universe is not old enough for the progenitors to evolve off the MS \citep{iben1990,iben1997}. ELMWDs are usually observed in compact binaries, the orbital period set by the requirement of mass transfer via Roche-lobe overflow, typically with a WD or a NS companion \citep{Brown_2020}. If indeed, these sources host ELMWDs, these would be very interesting sources to study in detail since the LCs are MS stars and the orbital periods are $\porb/\days>5.6\times10^2$, significantly larger compared to the boundary predicted by the requirement of mass transfer via RLOF \citep{rappaport_1995,tauris_1999,Lin_2011,istrate_2014,istrate_2016}. These four ELMWD candidates from our analysis show eccentricity values around 0.1 and a period of 500 - 1000 days. These eccentricities and orbital periods may indicate creation via dynamical processes \cite{Khurana_2023}.
\begin{figure}
    \epsscale{1.1}
    \plotone{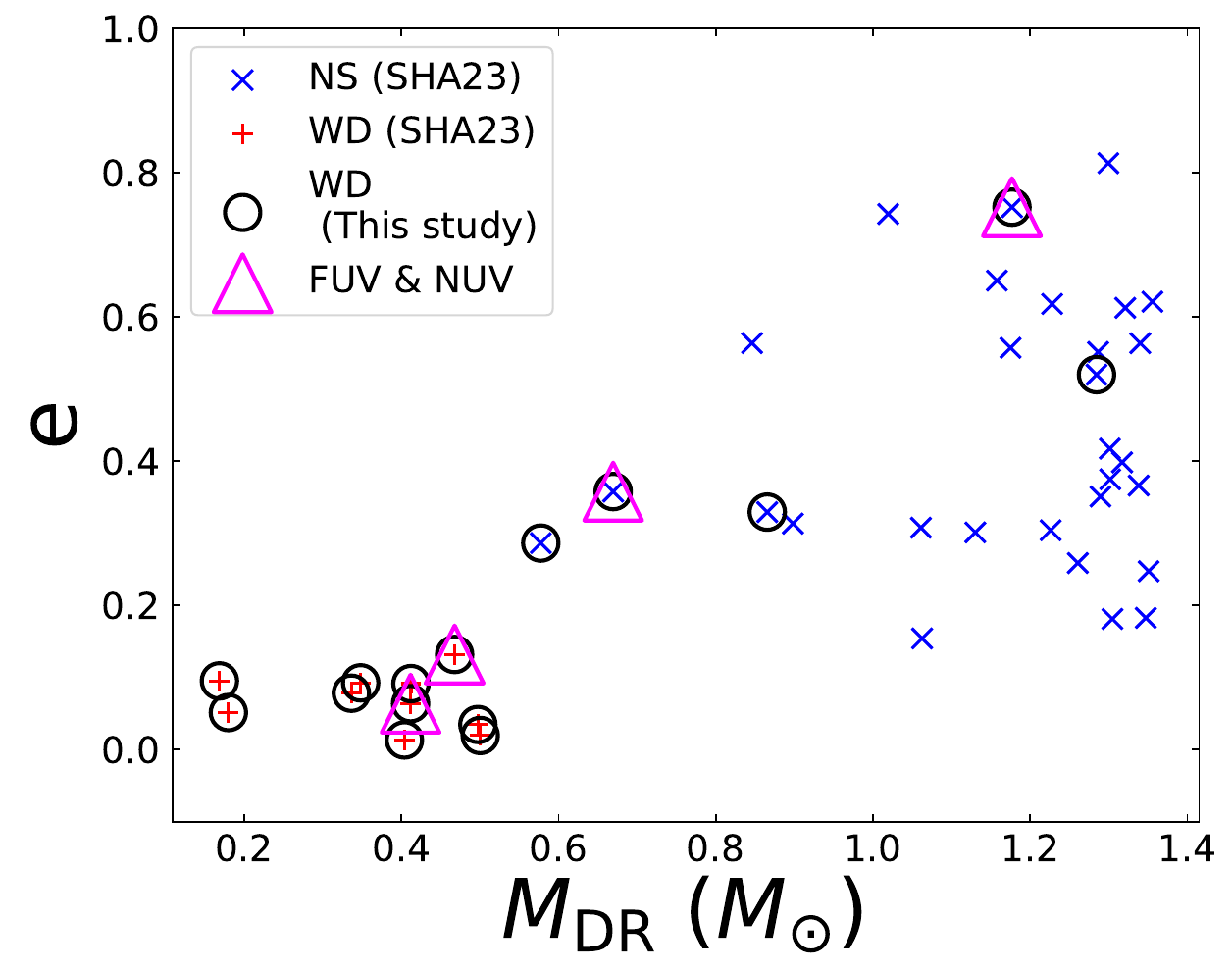}
    \caption{Eccentricity vs $\mdr$ for the 49 sources with identified UV counterparts. Circles denote the 15 sources with significant UV excess. Red + (blue x) denotes sources identified as WDs (NSs) in the SHA23 catalog. In all cases, we use $\mdr$ estimated by this study (\autoref{sec:masses}) and the eccentricities given by SHA23. All the magenta triangles show where we have both FUV and NUV data.  
    }
    \label{fig:e_mdr}
\end{figure}
Interestingly, SHA23, found that the candidate sources hosting DRs with estimated $\mdr/M_\odot<2.1$ likely constitute a mixture of gaussian distributions with peaks that are reasonably well separated in $\mdr$. Based on the gaussian mixture model SHA23 classified these sources into NSs and WDs. Of course, this is a probabilistic classification and can have large uncertainties. Based on Gaussian mixtures, SHA23 classified 0709+7052 and 0640$-$2621 as NSs. Both 0709$+$7052 and 0640$-$2621 show significant excess both in the NUV and FUV. We find $\mdr/\msun=0.7$ ($1.2$) for 0709$+$7052 (0640$-$2621). While mass alone cannot be a clear classifier between WDs and NSs, together with the UV excess both in the FUV and NUV for these sources would suggest that these sources were wrongly classified as NSs by SHA23. We also find significant UV excess in NUV and masses below the NS mass range for 1220$+$5841, 0327$-$4342, and 1330$+$2827  though SHA23 classified them as NSs. The other 10 sources showing significant UV excess and fitted well by a WD as the DR, were also identified as WDs by SHA23. 

\autoref{fig:e_mdr} shows eccentricity vs $\mdr$ for the 15 sources showing significant UV excess and compares them with all SHA23 sources shown in \citep[figure 8][]{Shahaf2023}. Clearly, the misidentified sources are near the parameter space where the NSs and the WDs are well mixed. This is likely the reason why the gaussian mixture model failed to predict the nature of these two sources correctly. This underlines the power of multi-wavelength followup and SED analysis to identify or confirm the nature of DRs in candidate systems identified by \gaia.  

\subsection{Sources without significant UV excess}
\label{sec:LC}
We do not find significant (\autoref{sec:sed_fit}) UV excess in the other 34 sources. In these cases, we are able to fit a single-component MS star model to the observed SEDs and estimate $\Teff$, [Fe/H], $\lbol$, and $\log(g/\acceleration)$ for the LC. We estimate $\mlc$ for all of these sources using the constraints on the stellar properties and MIST models except for sources 1433$-$0114 and 0336$+$1419.  
Source 1433$-$0114 has been previously identified as a hot sub-dwarf \citep{sub_dwarf_2017,sub_dwarf_2017_2}. Being a hot sub-dwarf, source 1433$-$0114 lies in between the MS and the WD regions on a Hertzprung-Russel diagram (HRD). Due to its potentially complex formation history and its unusual position in the HRD \citep{Heber_2016}, we do not attempt to estimate $\mlc$ for 1433$-$0114. In case of 0336$+$1419, we could not find enough MIST isochrones satisfying the constraints on [Fe/H], $\lbollc$, and $\tefflc$.  
All stellar properties as well as $\mlc$ (when available) are summarised in \autoref{tab:all_data}. 

\subsection{Estimated masses}
\label{sec:masses}
\begin{centering}
\begin{figure*}
\plottwo{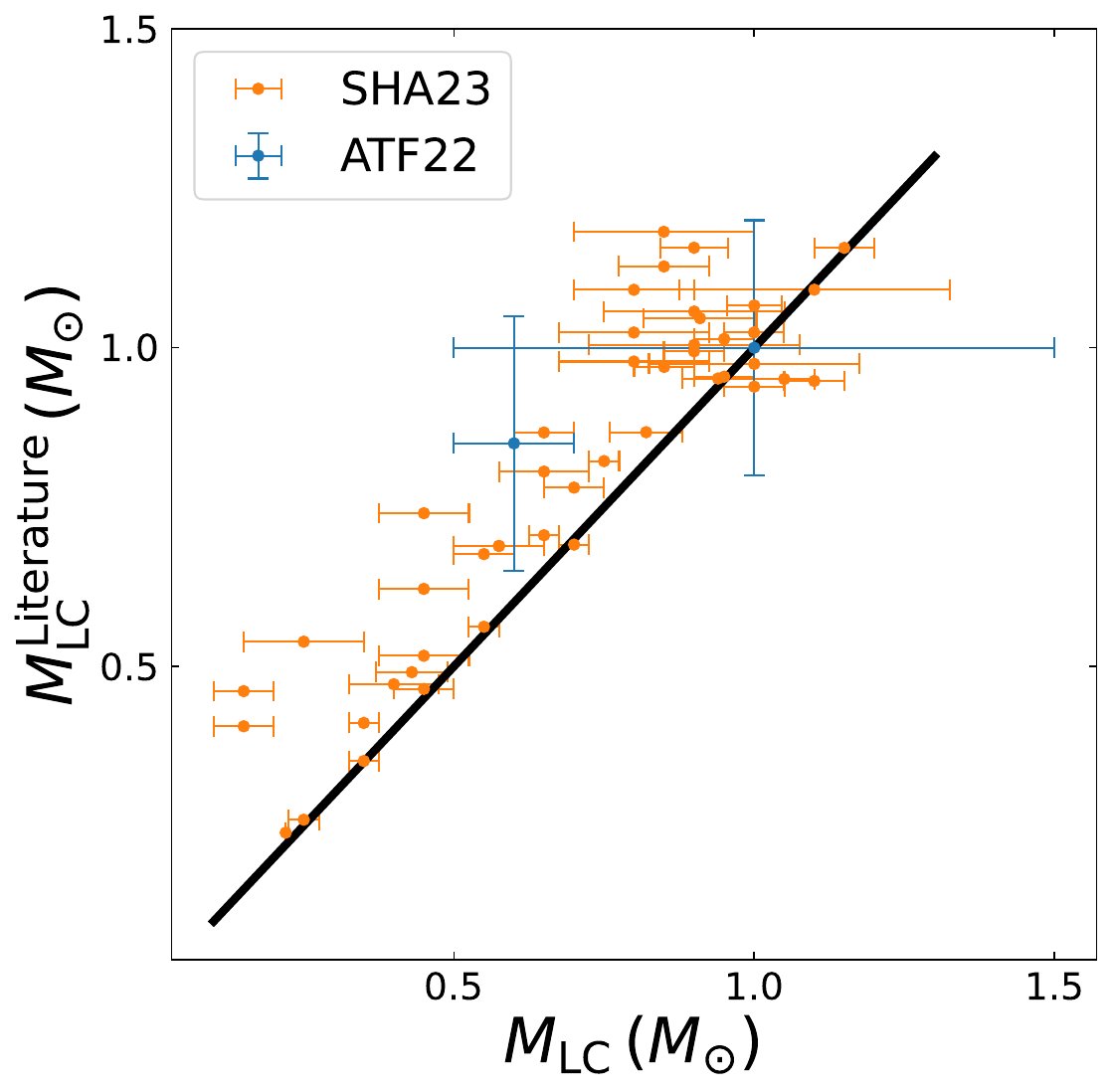}{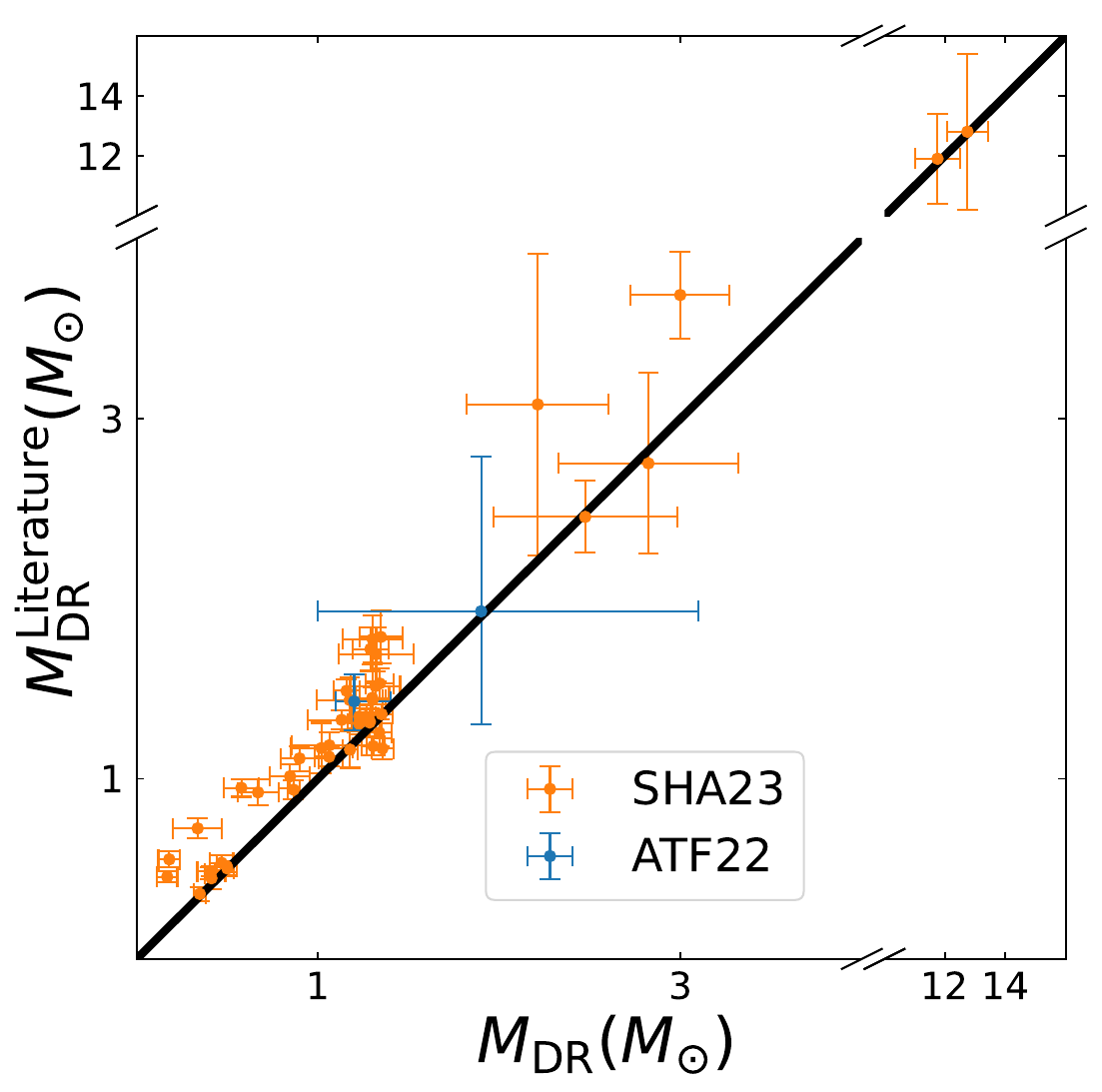}
    \caption{Comparison of mass estimated in this study vs those reported in the SHA23 and ATF22 catalogs for the LCs (left) and DRs (right). Blue (orange) dots show candidates from the ATF22 (SHA23) catalogs. Black solid line shows the $x=y$ line. While more or less consistent, our estimated $\mlc$ and $\mdr$ are typically slightly lower. 
    }
    \label{fig:lc_dr_mass_comparison}
\end{figure*}
\end{centering}
Our estimates provide independent measurements of $\mlc$ employing completely different methods compared to the estimated, or often, adopted masses available in the literature. For example, ATF22 estimated $\mlc$ of 1433$-$0114 using the UCO Lick spectra and Apsis and assumed an ad-hoc uncertainty of $0.1\,\msun$. The reported masses of the other ATF22 sources were simply adopted to be between $0.63$ and $1\,\msun$ based on their nominal locations on the \gaia\ CMD. SHA23 directly used \gaia's mass estimates, which are derived using \texttt{PARSEC} isochrones and \gaia's color and magnitude. In contrast, we estimate $\mlc$ uniformly for all our analysed sources using [Fe/H], $\lbollc$, and $\tefflc$ constrained by SED modelling and \MIST\ stellar evolution models. Instead of depending only on \gaia\ colors and magnitude, SED modeling takes into account flux from UV to NIR to constrain the stellar properties which is then used to constrain the mass. In spite of the different methods, our $\mlc$ measurements are more or less in agreement with those estimated or adopted previously in the ATF22 and SHA23 catalogs (left panel \autoref{fig:lc_dr_mass_comparison}). Nevertheless, we find that our estimated $\mlc$ is usually slightly lower than the previously adopted or estimated values. 
Note that these small differences in the estimated $\mlc$ actually push the corresponding $\mdr$ a little lower than was reported in the SHA23 and ATF22 catalogs (right panel \autoref{fig:lc_dr_mass_comparison}; \autoref{eq:amrf}, \ref{eq:fm}).     

For sources 1728$-$0034 and 1452$-$1922, our estimated $\mdr/\msun\gtrsim10$  
and they show no UV excess. These are good candidates for wide-orbit BH--LC binaries predicted to be present in large numbers in the Milky Way \citep[e.g.,][]{Breivik2017,Chawla2022}. Interestingly, source 1728$-$0034, popularly called the \gaia\ BH1, has already been identified through radial velocity and astrometric measurements to likely host a BH with $\mdr/\msun = 9.8\pm0.2$ \citep{gaia_bh1} to $11.9^{+2.0}_{-1.6}$ \citep{Chakrabarti2022}. The proximity of our estimated mass for 1728$-$0034 and these recently estimated spectroscopic masses provides further confidence in our estimates. For 1452$-$1922, we estimate $\mdr/\msun=11.75\pm0.75$ relative to $11.9\pm1.5$ by SHA23. Both measurements suggest that this could be a candidate BH-LC binary. Although, recently \citet{el_badry_2022} ruled out this candidate claiming that the astrometric solution for this source may be spurious. We strongly encourage further followup on this source via RV or multiple wavelengths such as radio and X-ray. 

While significant uncertainties still exist in the maximum NS mass \citep[e.g.,][]{bh_mass_fun_1,bh_fn3,bh_fn_2}, sources such as 1432$-$1021, 2100$-$2535, 0334$+$0009, 1301$-$1852 exhibit $\mdr$ within the so-called mass gap between NSs and BHs \citep{fryer_2012,Belczynski_2012} if it exists. Sources such as 0632$-$6614 and 0156$+$1228 have $\mdr<\mch$ and 1007$+$4453, 2244$-$2236, and 1012$-$3537 are very close to $\mch$. We do not find signficant UV excess for these sources. Sources that are not clearly BHs or mass gap objects, or identified as WDs because of UV excess in this study, have $\mdr$ ranging from $1.0$ $\msun$ to $1.4$  $\msun$. These might be good candidates for low mass NSs. However, it is hard to characterise the nature of a DR based on mass only. Even if clear mass boundaries do exist in nature, significant uncertainties remain in identifying them \citep{fryer_2012,Belczynski_2012,griffith2021, Fryer_2022, patton2022}. Only NUV data is available for these sources. Although, these sources have been classified as NSs in the SHA23 catalog, followup observation to obtain FUV flux may help ascertain whether the DRs in these sources may actually be hotter and fainter WDs compared to those where excess in NUV is detected. Hence, it remains unclear whether they are very faint WDs or low-mass NSs. 
In any case, in order to clearly characterize all candidate sources into BH, NS, or WD, we encourage followup observations in multiple wavelengths including UV and radio. 

\section{Summary and Discussion}\label{sec:discussion}
We have identified the UV counterparts in the archival \galex\ data for 49 of the 187 candidate sources in the ATF22 and SHA23 catalogs expected to host unseen DR companions to luminous stars observed by \gaia\ (\autoref{fig:combined_andrews}, \autoref{tab:all_data}). All of these sources are expected to host a MS star as the LC and have long $\porb/\days\gtrsim45$. Hence, there is little chance for ongoing mass transfer at present. We further find optical and IR fluxes for these sources by cross-matching with the archival data of APASS, PanSTARRS, 2MASS, and ALLWISE. We construct the SEDs from UV to IR for each of these sources and constrain the stellar parameters of the LCs including $\lbol$, $\teff$, and $\log\ g$ using \vosa\ by taking into account \gaia's distance, metallicity, and extinction constraints (\autoref{sec:LC}, \autoref{fig:sed_example_excess}, \autoref{fig:sed_example_no_excess}). 

Below we summariese our key findings. 
\begin{itemize}
    \item Fifteen of the 49 sources show significant UV excess which can be explained if the DR is a WD (\autoref{fig:sed_wd_all}). 
    \item Five of these 15 WDs were classified as NSs by SHA23. Two of these 5 show excess both in NUV and FUV. This shows that the gaussian mixture model of SHA23 can have large uncertainties. Two of these WDs are squarely within SHA23's modeled gaussian for NSs (\autoref{fig:e_mdr}).  
    \item Our estimated $\mlc$ and $\mdr$ (\autoref{fig:lc_dr_mass_comparison}) are somewhat lower but more or less consistent with those estimated or adopted by SHA23 and ATF22. 
    \item We find four sources with $\mdr/\msun\leq0.4$ with significant UV excess (\autoref{tab:wd_data}). These may be the so called ELMWDs. If so, these are extremely interesting sources to study since their $\porb$ is much larger compared to expectations \citep[e.g.,][]{Lin_2011}. Moreover, these have LCs while most observed ELMWDs have other WDs or NSs as companions \citep[e.g.,][]{Brown_2020}. 
    \item We find two WD candidates with $\mdr/\msun\approx1.2$, close to $\mch$. One of them show significant UV excess both in NUV and FUV (\autoref{tab:wd_data}).
\end{itemize}
We caution that NUV and FUV data is available for only 5 (4 show excess) of the 49 sources we have analysed. 
 Furthermore, based on our adopted stringent criteria we have ignored some sources showing less significant NUV excess. Finding FUV flux constraints for our analysed sources may significantly improve the constraints on the WD properties as well as make the analysis more complete by allowing us to identify hotter and fainter WDs. Although unlikely at the levels found in our candidates based on the LC ages and stellar properties, chromospheric activity of the LC may also create an excess in the UV flux \citep{lorenzo_2016,gomes_2021}. Wider availability of both NUV and FUV fluxes should help in this regard as well. RV follow-up may put stronger constraints on $\mdr$, especially because of the possibility of spurious astrometric solutions \citep[e.g.,][]{El-Badry2022} even when the significance and goodness of fit are acceptable. Deep radio observations may also be very useful for these sources, especially to clearly identify NSs if they are pulsating. 

In summary, the sources presented in the SHA23 and ATF22 catalogs can be very interesting candidates as potential wide DR-LC binaries and it will be really interesting to clearly identify the nature of the DRs they host. Our work shows a relatively simple and inexpensive way to characterise such sources.     
\begin{acknowledgements}
AG acknowledges support from TIFR’s graduate fellowship. PKN acknowledges TIFR's postdoctoral fellowship.  SC acknowledges support from the Department of Atomic Energy, Government of India, under project no. 12-R\&D-TFR-5.02-0200 and RTI 4002.
\end{acknowledgements}

\textit{Software} : \vosa\ \citep{Bayo2008}, Python 3 \citep{python3}, Numpy \citep{numpy_scipy_matp}, scipy \citep{numpy_scipy_matp}, matplotlib \citep{numpy_scipy_matp}, pandas \citep{numpy_scipy_matp} and Astropy \citep{astropy:2013, astropy:2018}

\bibsep=0 pt
{ \footnotesize
\bibliographystyle{apj}
\bibliography{reference}
}
\end{document}